\documentclass[a4paper,11pt]{article}
\pdfoutput=1 
\usepackage{jcappub} 
\usepackage[T1]{fontenc} 
\usepackage{multirow}
\usepackage{cprotect}
\usepackage{subfig}
\newcommand{\ion}[2]{#1\,\textsc{#2}}

\newcommand{\orcid}[1]{\href{https://orcid.org/#1}{\includegraphics[width=10pt]{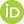}}}

\title{\boldmath Rapid follow-up observations of infant supernovae with the Gran Telescopio Canarias}

\author[a,b,1]{Llu\'is~Galbany\note{Corresponding author. l.g@csic.es}\orcid{0000-0002-1296-6887},}
\author[b,a]{Claudia P. Guti\'errez\orcid{0000-0003-2375-2064},}
\author[a,c]{Lara~Piscarreta\orcid{0009-0006-4637-4085},}
\author[a,b]{Alaa~Alburai\orcid{0009-0007-2731-5562},}
\author[a,b]{Noor~Ali\orcid{0009-0002-6802-8045},}
\author[a,d]{Dane~Cross\orcid{0000-0002-6356-7085},}
\author[a,b]{Maider~Gonz\'alez-Ba\~nuelos\orcid{0009-0006-6238-3598},}
\author[a,b]{Cristina~Jim\'enez-Palau\orcid{0000-0002-4374-0661},}
\author[a,b]{Maria~Kopsacheili\orcid{0000-0002-3563-819X},}
\author[e,f]{Tom\'as~E.~M\"uller-Bravo\orcid{0000-0003-3939-7167},}
\author[b,a]{Kim~Phan\orcid{0000-0001-6383-860X},}
\author[a,b]{Ramon~Sanfeliu\orcid{0009-0007-0178-2875},}
\author[g]{Maximilian~D.~Stritzinger\orcid{0000-0002-5571-1833},}
\author[h]{Chris~Ashall\orcid{0000-0002-5221-7557},}
\author[i]{Eddie~Baron\orcid{0000-0001-5393-1608},}
\author[j]{Gast\'on~Folatelli\orcid{0000-0001-5247-1486},}
\author[j]{Melina Bersten\orcid{0000-0002-6991-0550},}
\author[h]{Willem~Hoogendam\orcid{0000-0003-3953-9532},}
\author[k]{Saurabh~Jha\orcid{0000-0001-8738-6011},}
\author[l]{Thomas~de~Jaeger\orcid{0000-0001-5955-2502},}
\author[m]{Alexei~V.~Filippenko\orcid{0000-0003-3460-0103},}
\author[m]{Thomas~G.~Brink\orcid{0000-0001-5955-2502},}
\author[n,o]{D.~Andrew~Howell\orcid{0000-0003-4253-656X},}
\author[p]{and Daichi~Hiramatsu\orcid{0000-0002-1125-9187}}

\affiliation[a]{Institute of Space Sciences (ICE, CSIC), Campus UAB, Carrer de Can Magrans, s/n, E-08193 Barcelona, Spain}
\affiliation[b]{Institut d'Estudis Espacials de Catalunya (IEEC), 08860 Castelldefels (Barcelona), Spain}
\affiliation[c]{European Southern Observatory, Karl-Schwarzschild-Strasse 2, 85748 Garching bei München, Germany}
\affiliation[d]{Institut de F\'isica d’Altes Energies (IFAE), Campus UAB, 08193 Bellaterra (Barcelona), Spain}
\affiliation[e]{School of Physics, Trinity College Dublin, The University of Dublin, Dublin 2, Ireland}
\affiliation[f]{Instituto de Ciencias Exactas y Naturales (ICEN), Universidad Arturo Prat, Chile}
\affiliation[g]{Department of Physics and Astronomy, Aarhus University, Ny Munkegade 120, DK-8000 Aarhus C, Denmark}
\affiliation[h]{Institute for Astronomy, University of Hawai‘i, 2680 Woodlawn Drive, Honolulu, HI 96822, USA}
\affiliation[i]{Homer L. Dodge Dept. of Physics and Astronomy, U. of Oklahoma, Rm 100 440 W. Brooks, Norman, OK 73019-2061, USA}
\affiliation[j]{Instituto de Astrof\'isica de La Plata (IALP), CCT-CONICET-UNLP. Paseo del Bosque S/N, B1900FWA, La Plata, Argentina}
\affiliation[k]{Dept. of Physics and Astronomy, Rutgers, the State U. New Jersey,136 Frelinghuysen Road, Piscataway, NJ08854-8019, USA}
\affiliation[l]{Sorbonne Universit\'e, CNRS/IN2P3, LPNHE, F-75005, Paris, France}
\affiliation[m]{Department of Astronomy, University of California, Berkeley, CA 94720-3411, USA}
\affiliation[n]{Las Cumbres Observatory, 6740 Cortona Dr, Suite 102, Goleta, CA 93117-5575, USA}
\affiliation[o]{Department of Physics, University of California, Santa Barbara, CA 93106-9530, USA}
\affiliation[p]{Center for Astrophysics, Harvard \& Smithsonian, 60 Garden Street, Cambridge, MA 02138-1516, USA}

\abstract{The first few hours of a supernova (SN) contain significant information about the progenitor system. The most modern wide-field surveys that scan the sky repeatedly every few days can discover all kinds of transients in those early epochs.
At such times, some progenitor footprints may be visible, elucidating critical explosion parameters and helping to distinguish between leading explosion models.
A dedicated spectroscopic classification programme using the optical spectrograph OSIRIS mounted on the Gran Telescopio Canarias was set up to try to obtain observations of supernovae (SNe) at those early epochs.
With the time awarded, we obtained spectra of 10 SN candidates, which we present here. Half of them were thermonuclear SNe, while the other half were core-collapse SNe. Most (70\%) were observed within the first six days of the estimated explosion, with two being captured within the first 48\,hr. We present a characterization of the spectra, together with other public ancillary photometry from the Zwicky Transient Facility (ZTF) and the Asteroid Terrestrial-impact Last Alert System (ATLAS). 
This project shows the need for an accompanying rapid-response spectroscopic programme for existing and future deep photometric wide-field surveys located at the right longitude to be able to trigger observations in a few hours after the discovery of the SN candidate.
}

\begin{document} 
\maketitle
\flushbottom



\section{Introduction}

Supernovae (SNe) mark the termination of massive stars and some white dwarfs (WDs), in some cases releasing as much luminous energy as the Sun produces over its entire lifetime. As cauldrons of nucleosynthesis, SNe provide the interstellar medium with heavy elements, while their enormous kinetic energy ($\sim$10$^{51}$ erg) drives galaxy evolution (e.g., \citep{1960ApJ...132..565H,2000Natur.403..727B}). Moreover, given their high intrinsic peak luminosity, SNe serve as excellent cosmological distance indicators used to map out the expansion history of the universe \citep{1998AJ....116.1009R,1999ApJ...517..565P,2020MNRAS.495.4860D}.

Besides direct progenitor detections in archival images \citep{2009ARA&A..47...63S,2017RSPTA.37560277V}, and nebular spectra \citep{2012MNRAS.420.3451M,2022ApJ...928..151F}, much of what we know about SNe and their progenitors is based on observations of objects obtained a few days or weeks after the explosion. These SNe were discovered by searches that repeatedly observed a number of targeted galaxies (e.g., the Lick Observatory Supernova Search, LOSS, \citep{2001ASPC..246..121F}; the CHilean Automatic Supernova sEarch, CHASE, \citep{2009AIPC.1111..551P}) or by search programs that repeatedly scanned large sections of the sky (e.g., the Sloan Digital Sky Survey II -- SN Survey, SDSS-II/SN, \citep{2008AJ....135..338F}; the Palomar Transient Factory, PTF, \citep{2009PASP..121.1334R}; the Nearby SN Factory, \citep{2002SPIE.4836...61A}) combined with other dedicated follow-up programmes to obtain dense, multiwavelength observations of nearby SNe (e.g., the Carnegie Supernova Project, CSP, \citep{2017AJ....154..211K,2019PASP..131a4001P}; the Center for Astrophysics SN programme, CfA, \citep{2009ApJ...700..331H,2012ApJS..200...12H}).

More recently, to expand our understanding of the various SN types and the physics of their explosions and progenitor systems, wide-field untargeted surveys such as the All-Sky Automated Survey for Supernovae (ASAS-SN; \citep{2014ApJ...788...48S}), the Asteroid Terrestrial Impact Last Alert System (ATLAS; \citep{2018PASP..130f4505T, Smith2020}), the Panoramic Survey Telescope and Rapid Response System (Pan- STARRS; \citep{2016arXiv161205560C}), and the Zwicky Transient Facility (ZTF; \citep{2019PASP..131a8002B, Graham2019}), are regularly obtaining high-cadence (in some cases intraday) observations and discovering SNe and other transients hours to days after the explosion. This dense cadence has led to the discovery of a multitude of new types of transient phenomena (e.g., superluminous SNe, fast-and-faint transients, and relativistic transients), revealed a population of transients at all luminosity scales, and yielded less biased results compared to previous targeted surveys.

It is through early-phase observations that we can estimate critical explosion parameters (previously out of reach with traditional studies), delineate between leading explosion models, and study the local environment of cosmic explosions (e.g., \citep{2021arXiv210807278F,2022ApJ...932L...2A}). In particular for SNe, this corresponds to the epochs when the photosphere is close to the outer layers of the exploding progenitor, uncovering more insights about its nature. In the literature, there has been a lack of SN early-epoch spectra given the difficulty of discovering transients right after explosion, but this situation has recently evolved, thanks to the advent of wide-field sky surveys. Works addressing this have been published in recent years (e.g., \citep{2016ApJ...818....3K,2023ApJ...952..119B,2024ApJ...970..189J,2025A&A...693A.307Y}).

With this in mind, we started a spectroscopic programme at the 10.4\,m Gran Telescopio  Canarias (GTC) at the Roque de los Muchachos Observatory in the Canary Islands to obtain early spectra of SN candidates within 48\,hr of the explosion that complements imaging surveys. 
Thanks to the large aperture of the GTC, we are able to reach fainter magnitudes in the first few days after the SN first light and obtain spectra having higher signal-to-noise ratios with reasonable exposure times.
These observations probe the outermost layers of the ejecta, helping to distinguish between explosion and progenitor scenarios and to answer questions such as the origin of the bimodality in the early phase $B-V$ colors of Type Ia SNe (SNe~Ia, \citep{2018ApJ...864L..35S}), the pre-SN masses of the progenitors of stripped-envelope core-collapse (CC) SNe \citep{2020A&A...634A..21S}, and the outermost structure and environment of the red supergiant progenitors of Type II SNe \citep{2018NatAs...2..808F}. This early-time spectroscopy has the potential to offer new insights into the composition of the outer layers of SN progenitors right before the explosion, as well as yield new clues regarding the underlying physics of different types of SNe.

Additionally, a parallel goal of the programme was to feed potential SN~Ia candidates for  Supernovae in the near-InfraRed Avec Hubble (SIRAH; \citep{2020hst..prop16234J,sirah}), a \textit{Hubble Space Telescope (HST)} Cycle 27+28 and NOAO/Gemini programme to obtain near-infrared (NIR) {\t HST} spectra of a sample of 24 Hubble-flow ($0.02<z<0.07$) target-of-opportunity SNe~Ia, supplemented by the Gemini NIRI and FLAMINGOS2 imaging data to obtain the full NIR light curves. The main goals of SIRAH are to (i) build NIR spectrophotometric templates that will serve as a foundation for the Nancy Grace Roman Space Telescope SN survey \citep{2021arXiv211103081R}, (ii) yield an NIR-only SN~Ia measurement of the Hubble constant (H$_0$), and (iii) constrain dark energy together with data from the RAISINs surveys \citep{2022ApJ...933..172J}, producing a NIR Hubble diagram from redshift $z = 0.02$ to $z = 0.6$ all calibrated on the WFC3/IR photometric system. Once an SN candidate is classified and identified as an early SN~Ia, an {\it HST} trigger is sent with enough anticipation that a spectrum, which typically requires 1--2 weeks for execution, is obtained around the epoch of maximum brightness. Under our GTC programme, we contributed two SNe~Ia to SIRAH (SNe 2020jgl and 2020kyx). Their ZTF and ATLAS light curves and spectra are presented here. 

The paper is organized as follows. In Section \S\ref{sec:strat}, we describe the strategy followed to select our targets. Section \S\ref{sec:data} explains the GTC data reduction and how we collected publicly available multiband optical light curves. We analyze the collected data in Section \S\ref{sec:analysis} by determining the SN spectral typing, estimating the time of first light and the maximum brightness epoch, characterizing their light curves, determining the GTC spectral epoch, and measuring SN photospheric velocities. Finally, in Section \S\ref{sec:conc}, we summarize our work and discuss improvements for future similar early SN programmes.

\renewcommand{\arraystretch}{1.25}
\begin{table*}[!t]
\caption{SNe Properties. }
\begin{center}
\resizebox{\textwidth}{!}{ 
\begin{tabular}{rlllllcrrc}
\hline \hline
Name         & IAU Name & Type    & R.A. (J2000)   & Dec (J2000)   & Host Galaxy & $E(B-V)$ MW (mag) & $t_{\rm obs}-t_{\rm lnd}^{\rm reported}$ & $t_{\rm obs}-t_{\rm disc}$ & TNS Report \\ \hline 
ZTF20aaxvzja & 2020itj  & IIb     & 14:34:48.394 & +08:10:10.42 & CGCG 047-108   & 0.0211$\pm$0.0006 & 16h39.94m   & 15h05.84m & \cite{tns_2020itj} \\ 
ZTF20aaynrrh & 2020jfo  & II      & 12:21:50.480 & +04:28:54.05 & M61            & 0.0194$\pm$0.0001 & 4d15h06.93m & 16h51.43m & \cite{tns_2020jfo} \\ 
ATLAS20lti   & 2020jgl  & Ia      & 09:28:58.426 & $-$14:48:19.88 & MCG -02-24-027 & 0.0585$\pm$0.0016 & 1d14h12.80m & 14h34.39m & \cite{tns_2020jgl} \\ 
ATLAS20lts   & 2020jhf  & Ia      & 13:14:53.440 & +27:00:31.10 & UGC 8325       & 0.0103$\pm$0.0006 & 3d12h07.48m & 1d11h48.75m & \cite{tns_2020jhf} \\ 
ZTF20abaovyz & 2020kjt  & II      & 18:43:16.457 & +71:40:36.60 & WISEA J184316.29+714032.6 & 0.0462$\pm$0.0011 & 1d16h07.03m & 18h58.49m & \cite{tns_2020kjt_2020kku} \\   
ZTF20abapyxl & 2020kku  & Ia      & 17:32:33.723 & +15:44:09.17 & WISEA J173233.56+154410.2 & 0.0909$\pm$0.0017 & 20h02.58m & 18h28.06m & \cite{tns_2020kjt_2020kku} \\ 
ZTF20abbhyxu & 2020kyx  & Ia      & 16:13:45.510 & +22:55:14.38 & CGCG 137-064   & 0.0768$\pm$0.0029 & 1d21h57.10m & 22h02.15m & \cite{tns_2020kyx} \\ 
ZTF20abbplei & 2020lao  & Ic-BL   & 17:06:54.600 & +30:16:17.40 & CGCG 169-041   & 0.0432$\pm$0.0010 & 21h07.81m & 18h45.72m & \cite{tns_2020lao} \\ 
ZTF20abisitz & 2020nny  & II      & 22:58:34.228 & +15:10:15.07 & IC 1461        & 0.0384$\pm$0.0006 & 1d20h26.74m & 17h12.89m & \cite{tns_2020nny} \\ 
ZTF20abtctgr & 2020rlj  & Ia-91bg & 23:01:09.614 & +23:29:14.00 & WISEA J230109.61+232903.7 & 0.1098$\pm$0.0036 & 1d20h59.64m & 21h55.85m & \cite{tns_2020rlj} \\ \hline 
\end{tabular}}
\end{center}
\label{tab:SNe_properties}
\end{table*}

\begin{figure*}[t]
  \centering
  \includegraphics[width=0.3\textwidth]{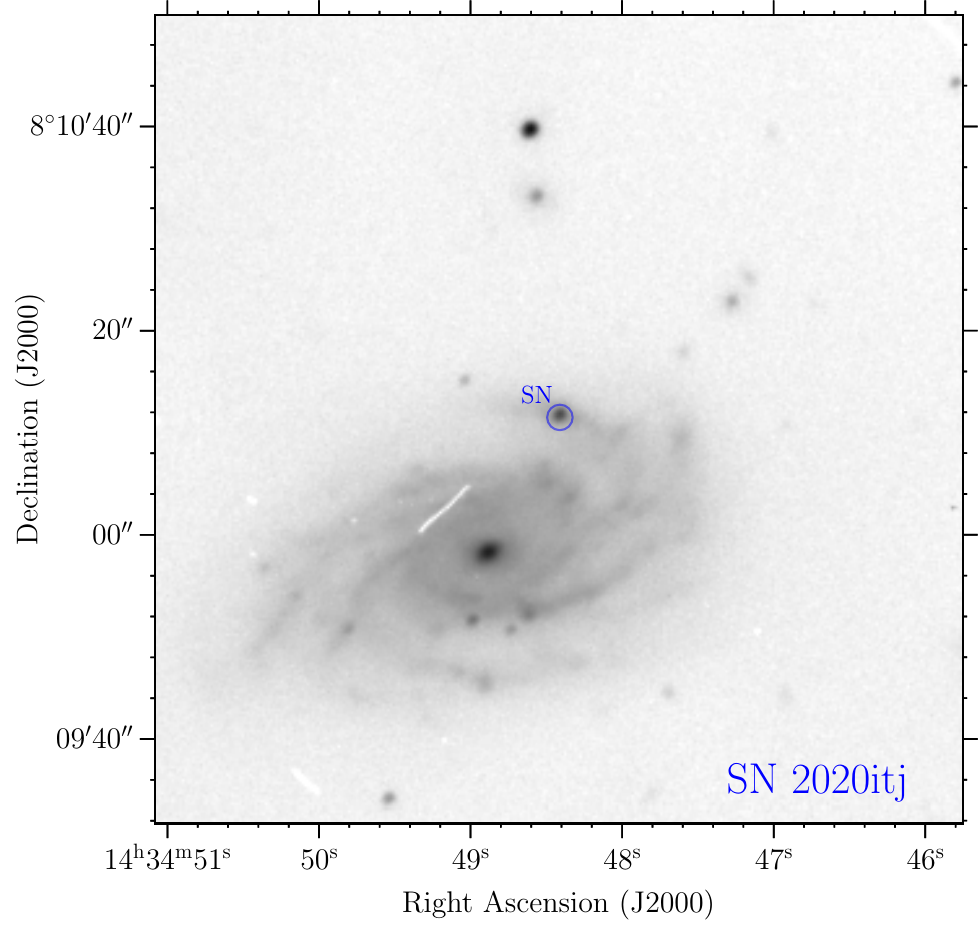}
  \hspace{0.1cm}
  \includegraphics[width=0.3\textwidth]{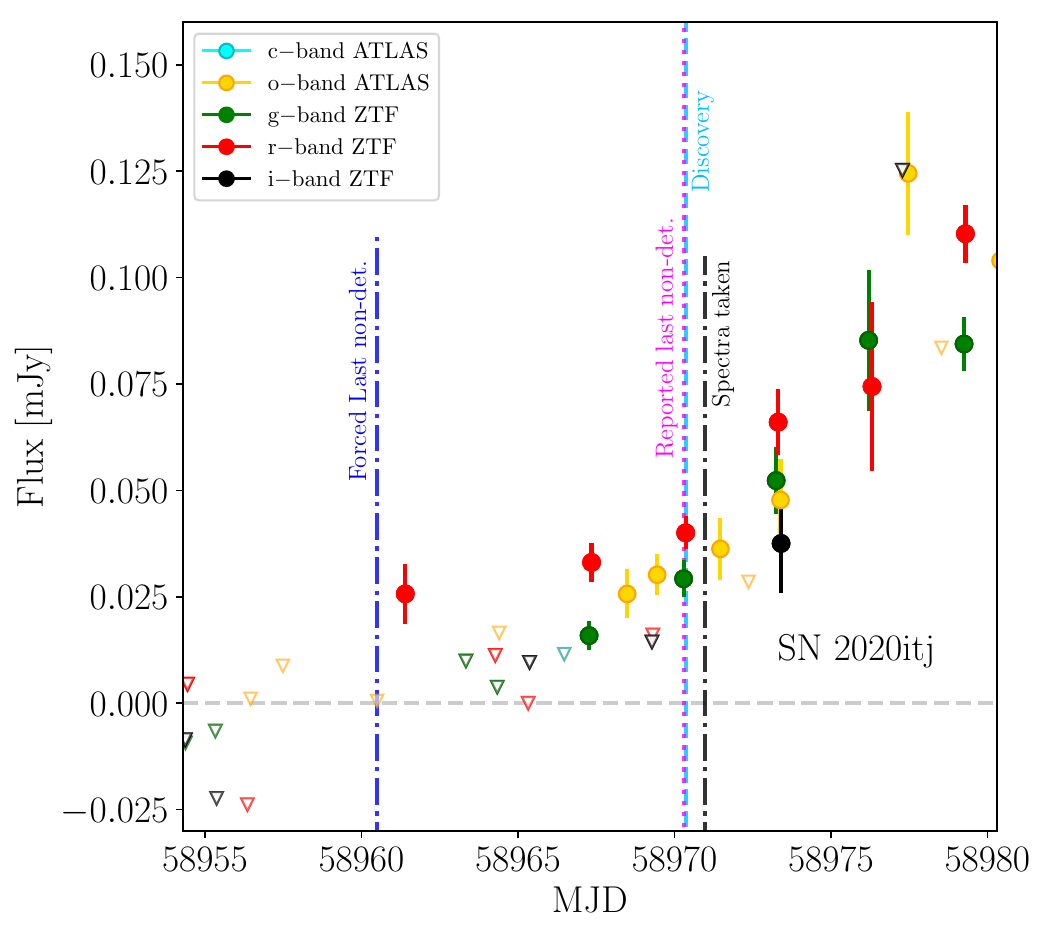}
  \hspace{0.1cm}
  \includegraphics[width=0.3\textwidth]{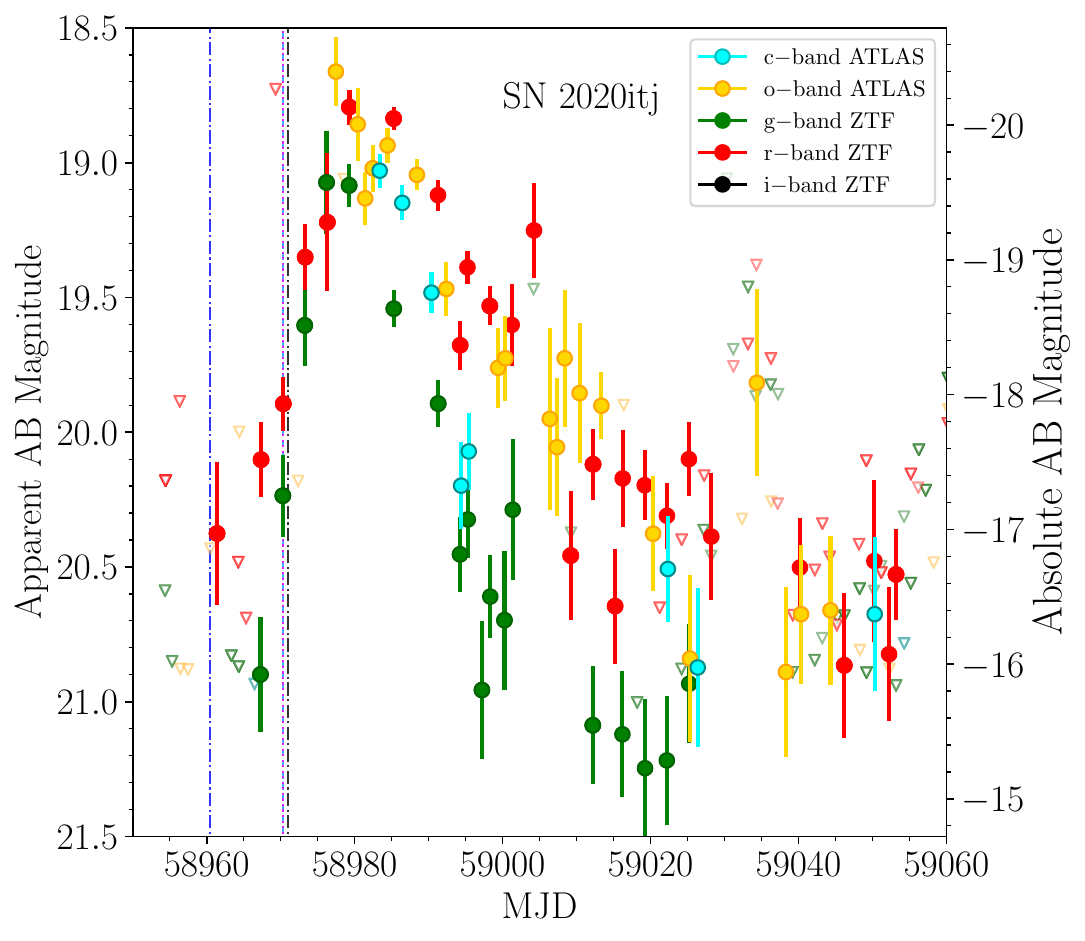}
  \includegraphics[width=0.3\textwidth]{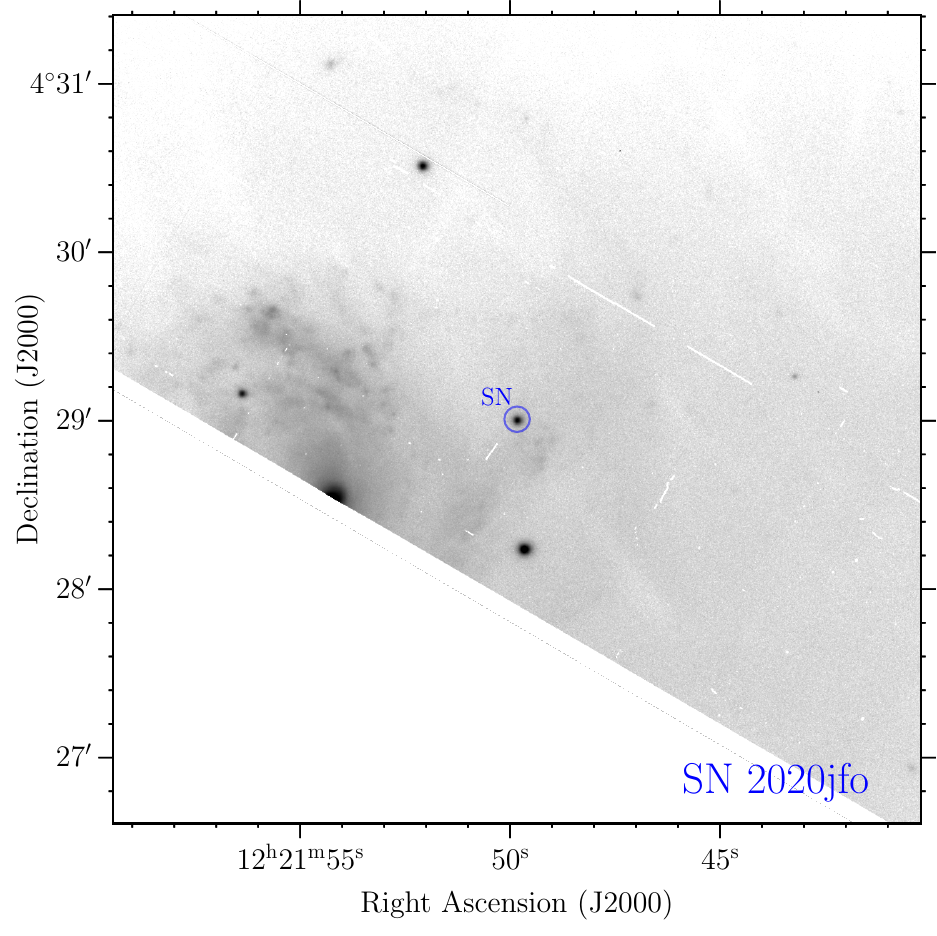}
  \hspace{0.1cm}
  \includegraphics[width=0.3\textwidth]{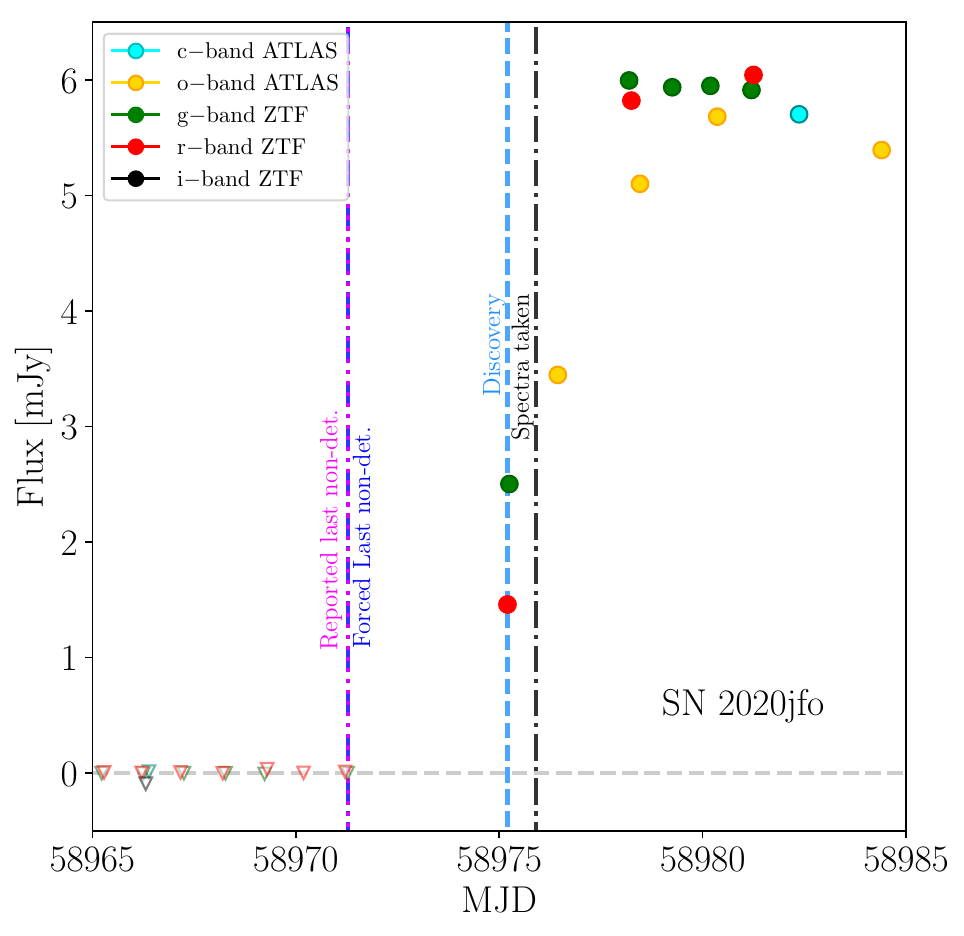}
  \hspace{0.1cm}
  \includegraphics[width=0.3\textwidth]{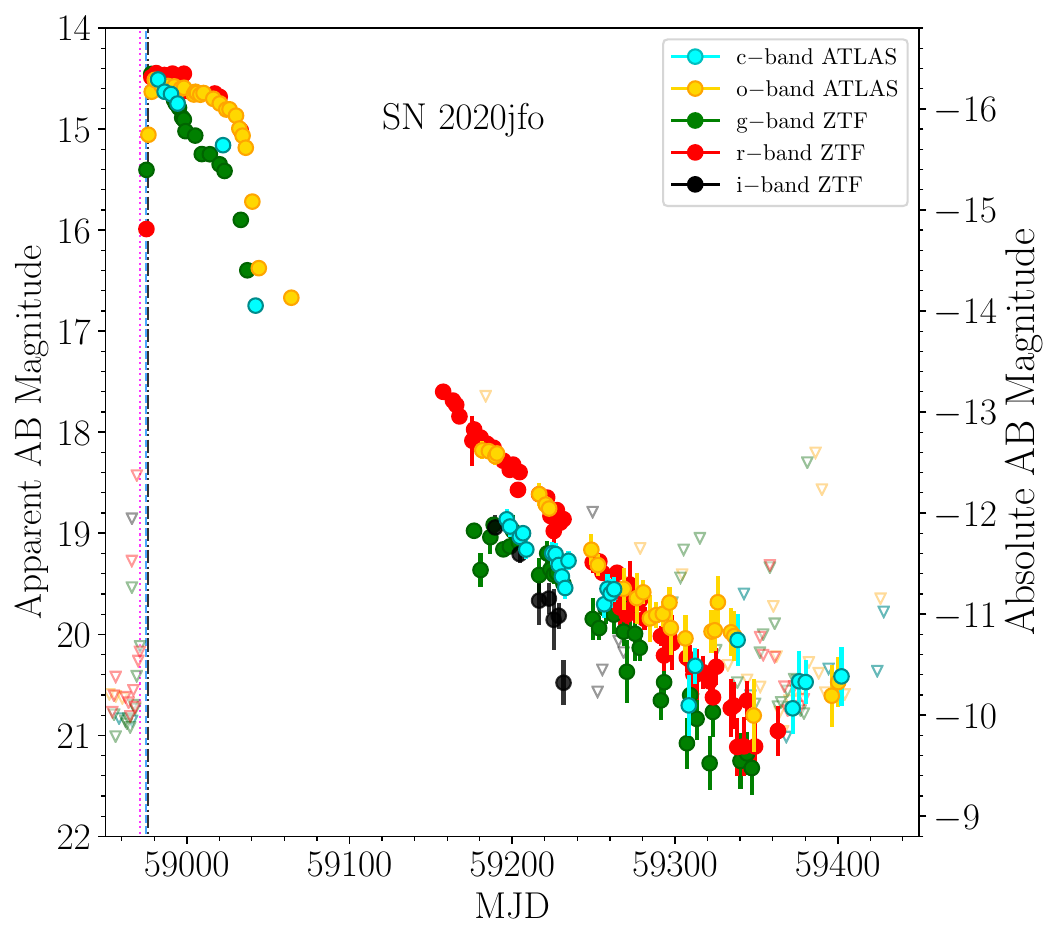}
  \includegraphics[width=0.3\textwidth]{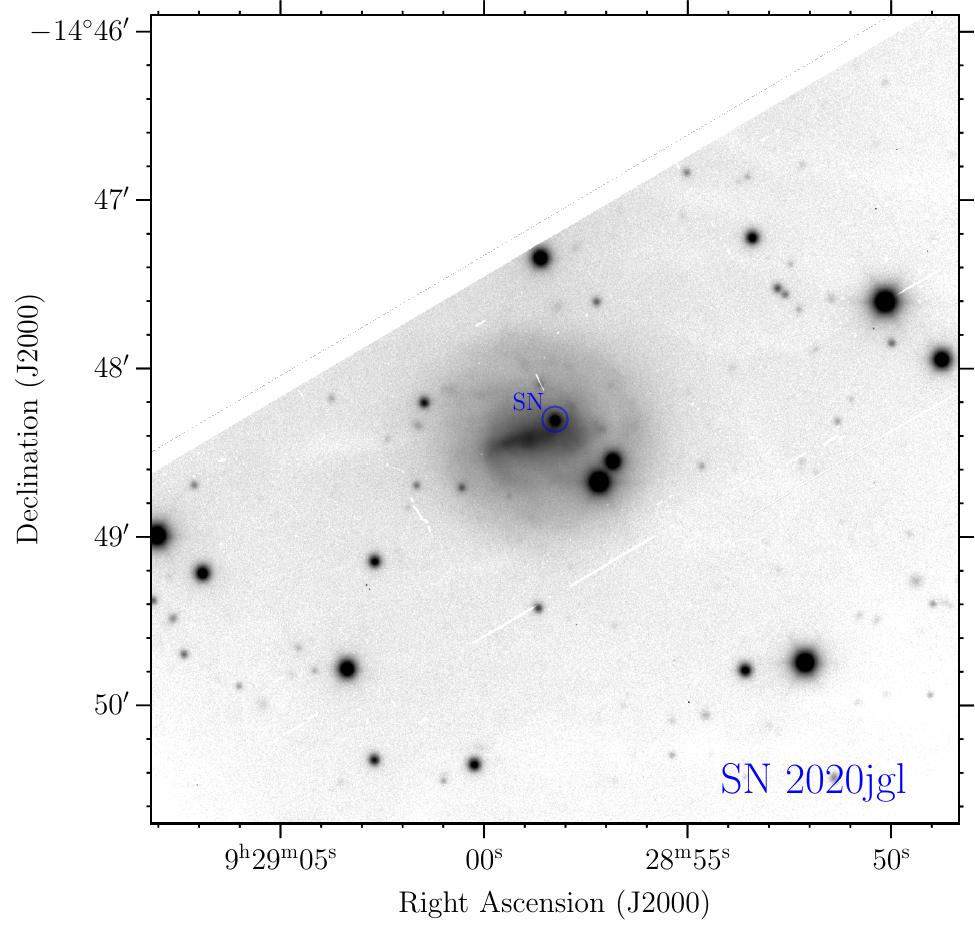}
  \vspace{0.1cm}
  \includegraphics[width=0.3\textwidth]{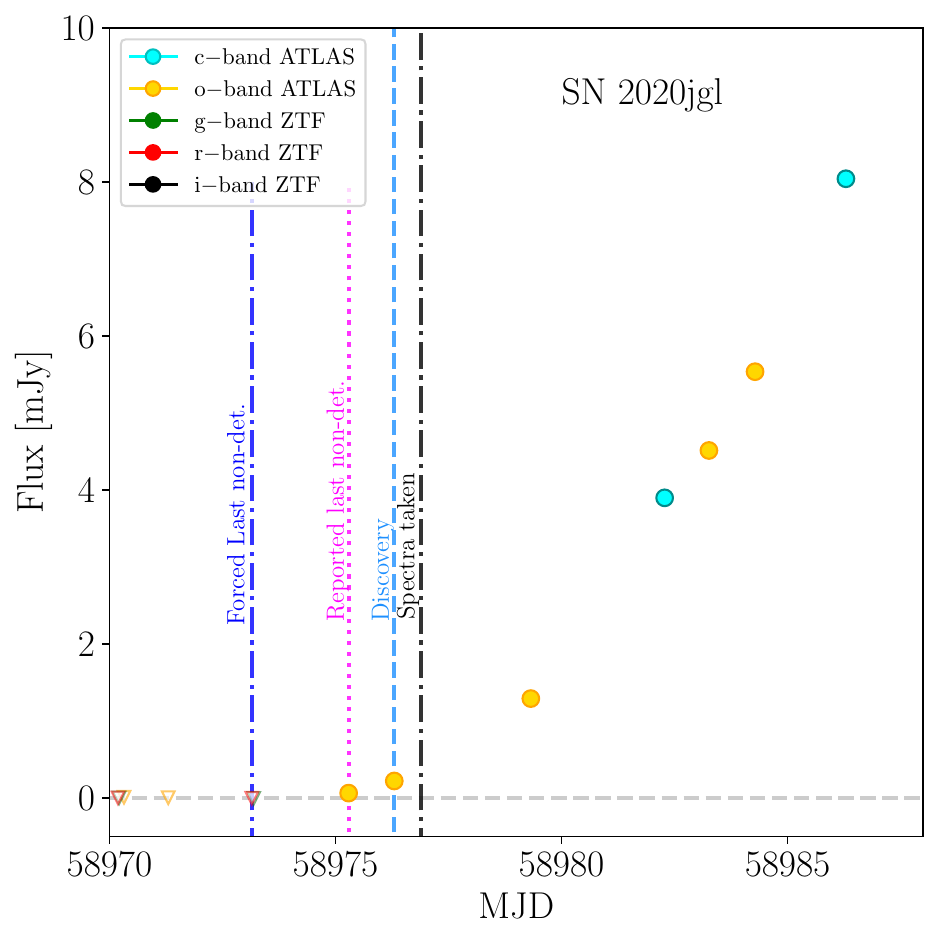}
  \hspace{0.1cm}
  \includegraphics[width=0.3\textwidth]{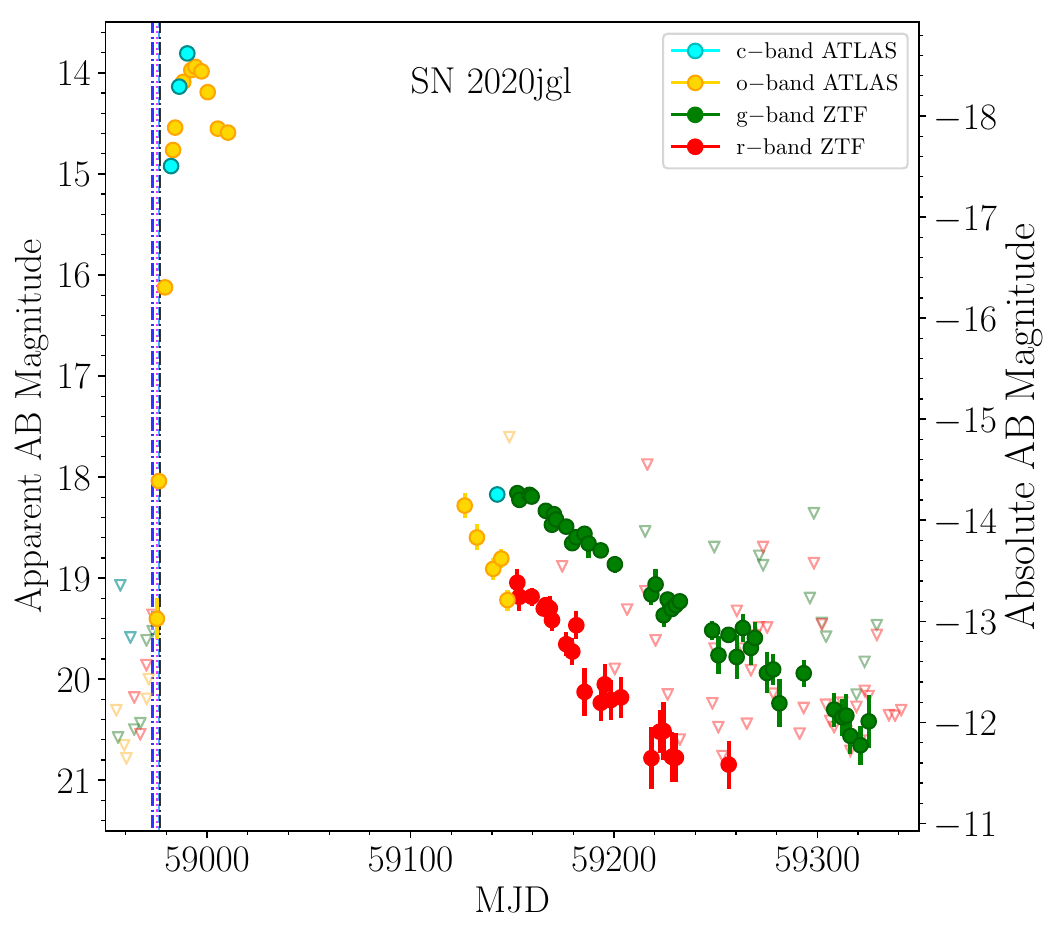}
  \caption{SN finder and combined light curves in the observer's frame. \textit{Left panels:} Host-galaxies images for our SNe sample. The SNe are marked with blue circles. \textit{Middle panels:} Early-time light curves of our SNe in flux units. The circles represent detections, whereas the triangles are considered nondetections. Vertical lines denote the last nondetection from forced photometry (dashed-dotted blue), the reported last nondetections at the time of discovery (dotted magenta), the time of discovery (dashed clear blue), and the time when each spectrum was taken with GTC/OSIRIS (dashed-dotted black). \textit{Right panels:} SN light curves in magnitude space (apparent magnitude marked in the left ordinate axis and the absolute magnitude in the right ordinate axis; see Section \ref{subsec:lcs} for information on distances). Downward-facing triangles denote 3$\sigma$ upper limits. }
  \label{fig:finders_lcs}
\end{figure*} 

\begin{figure*}\ContinuedFloat
  \centering  
  \includegraphics[width=0.3\textwidth]{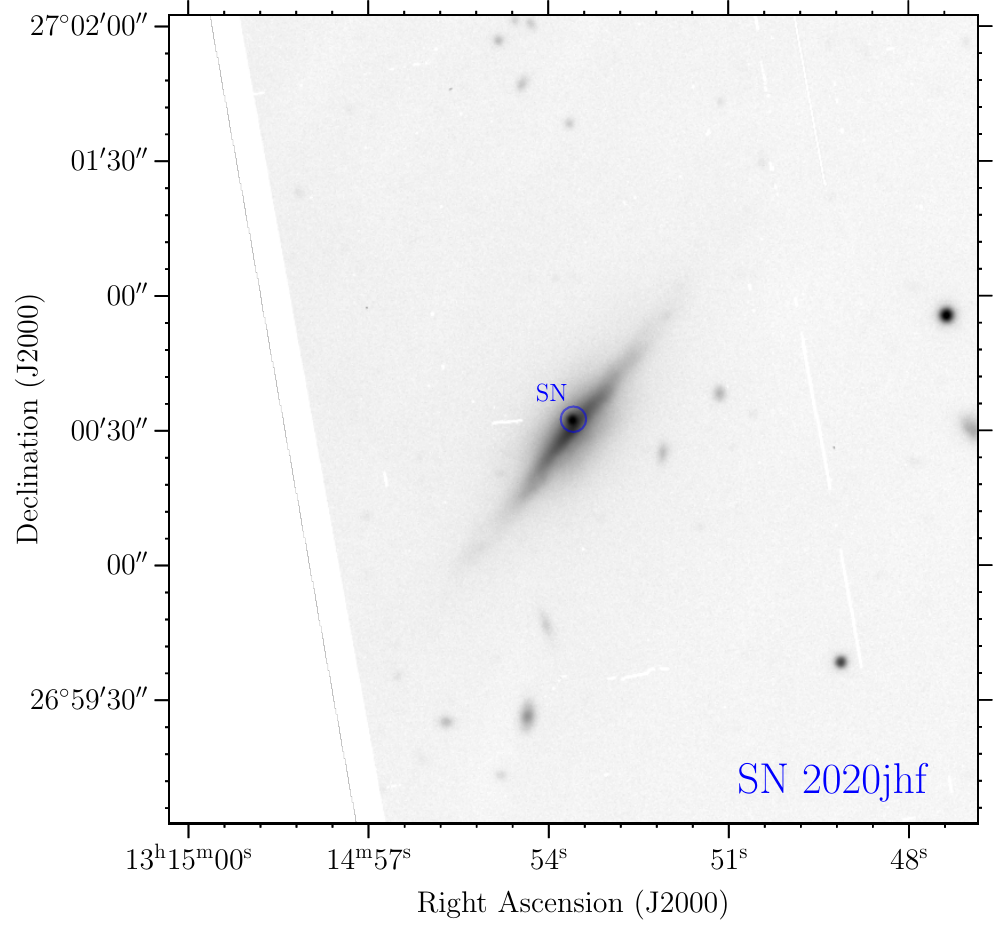}
  \hspace{0.1cm}
  \includegraphics[width=0.3\textwidth]{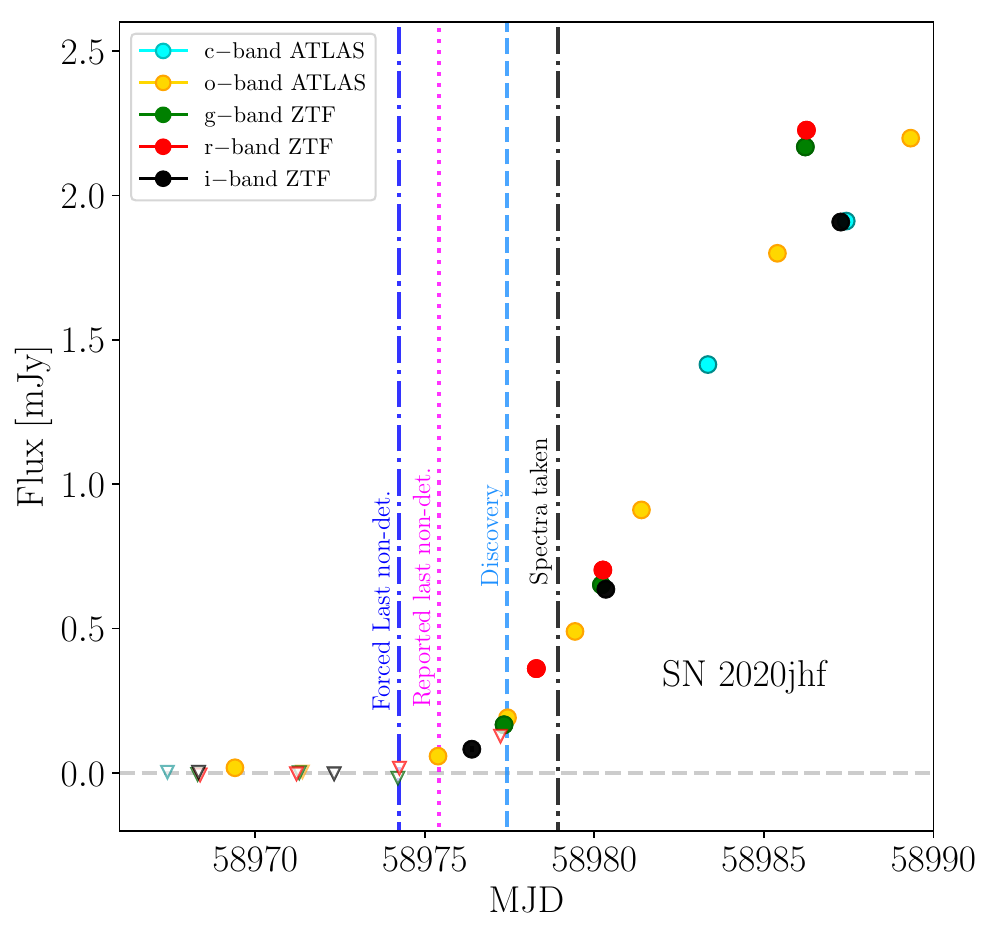}
  \hspace{0.1cm}
  \includegraphics[width=0.3\textwidth]{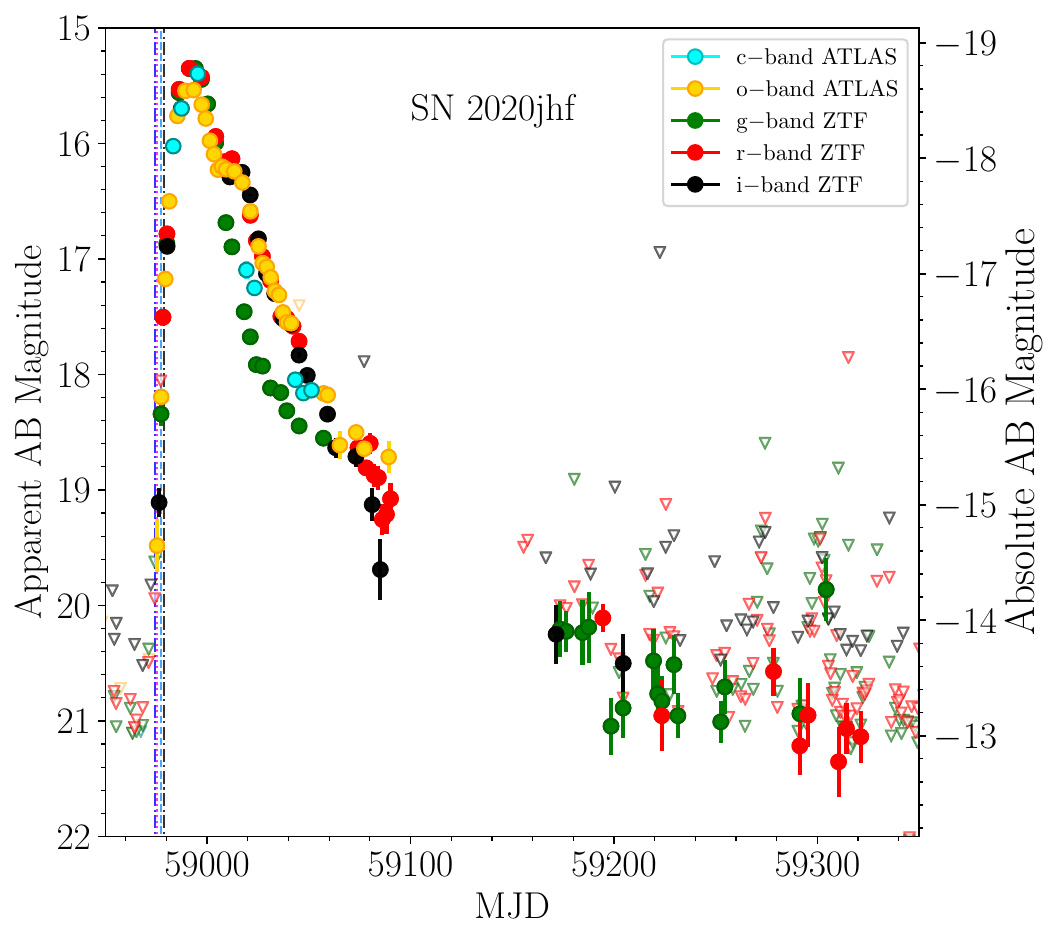}
  \includegraphics[width=0.3\textwidth]{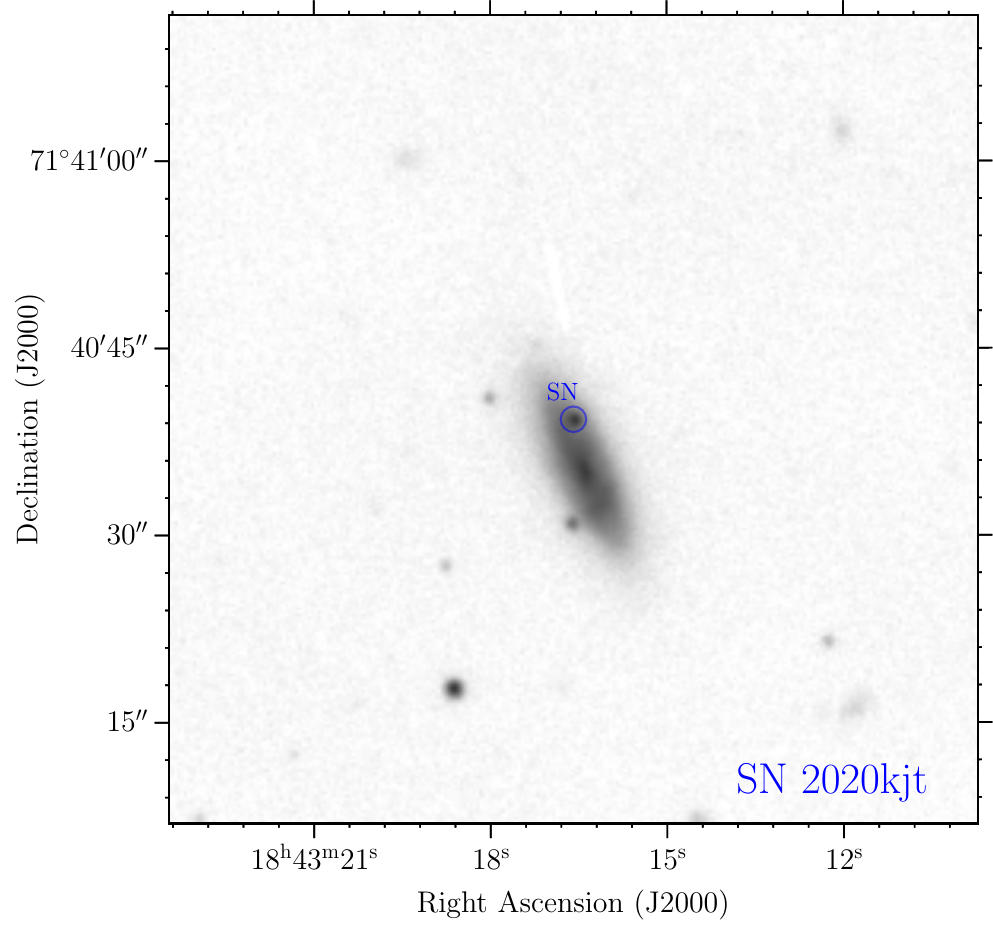}
  \hspace{0.1cm}
  \includegraphics[width=0.3\textwidth]{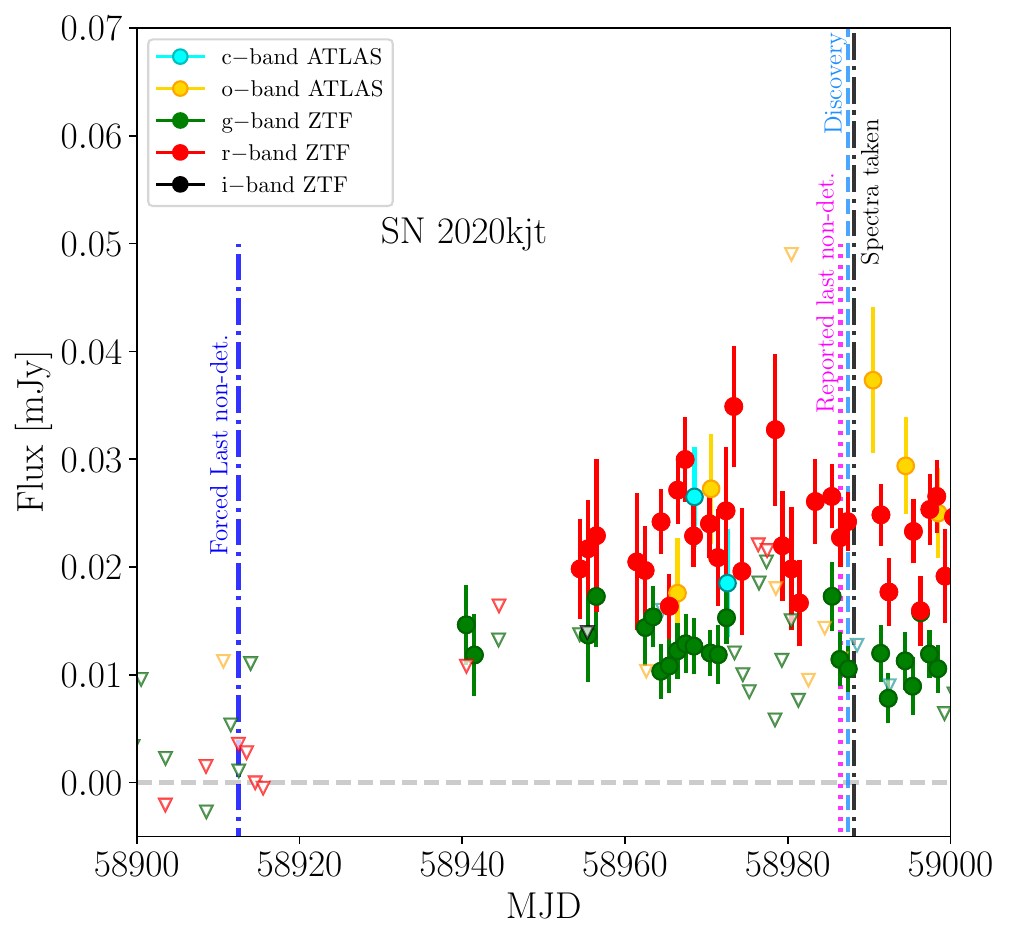}
  \hspace{0.1cm}
  \includegraphics[width=0.3\textwidth]{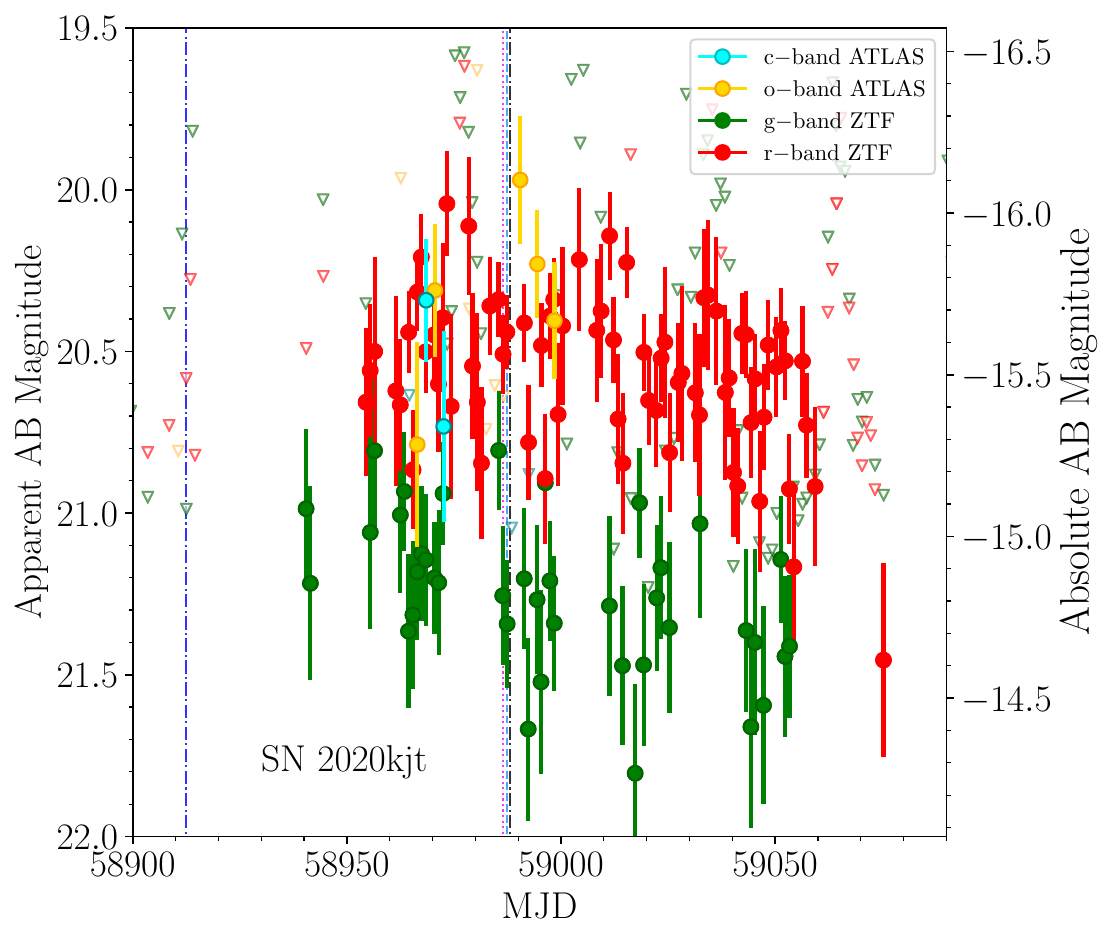}
  \includegraphics[width=0.3\textwidth]{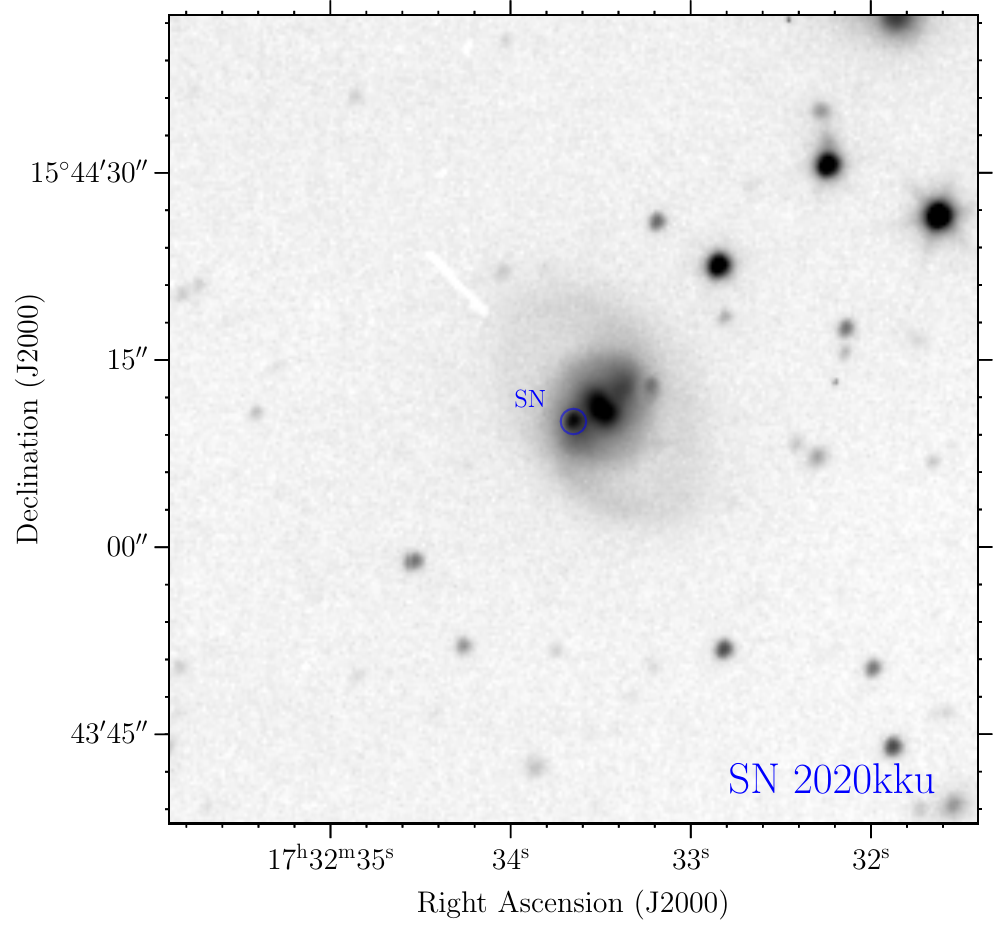}
  \hspace{0.1cm}
  \includegraphics[width=0.3\textwidth]{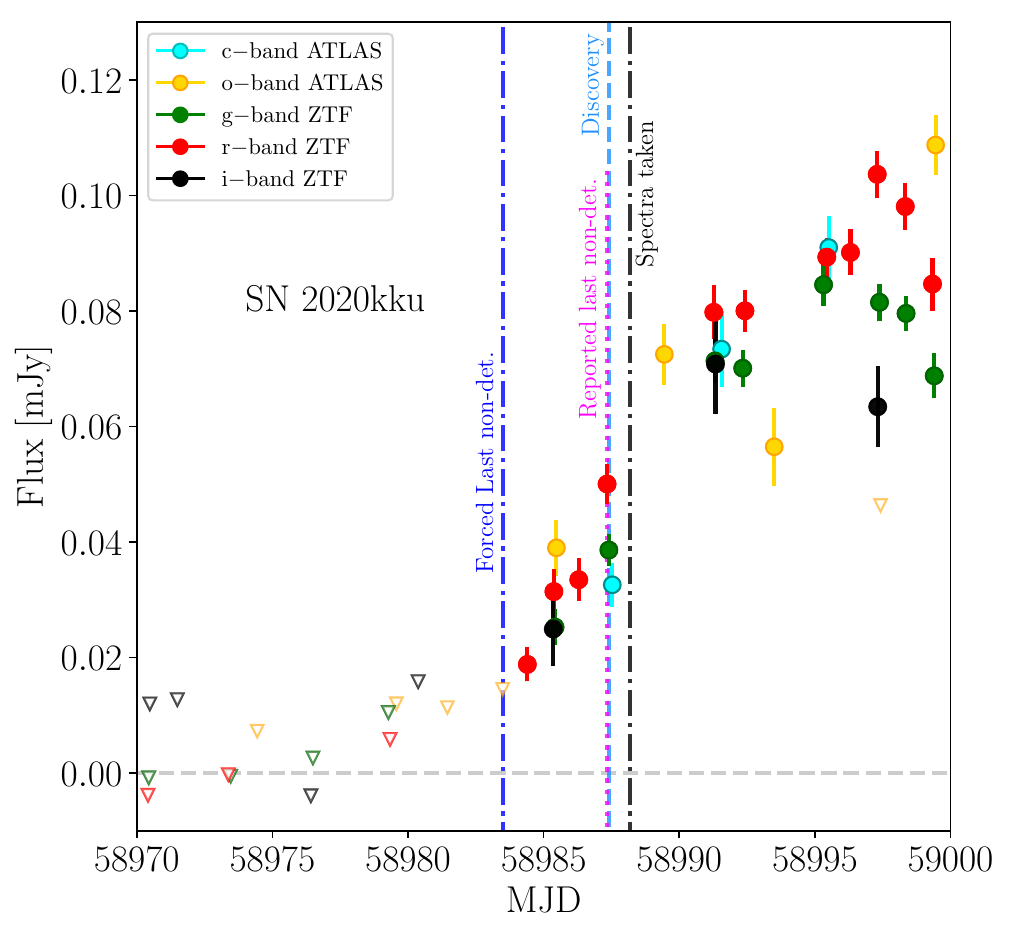}
  \hspace{0.1cm}
  \includegraphics[width=0.3\textwidth]{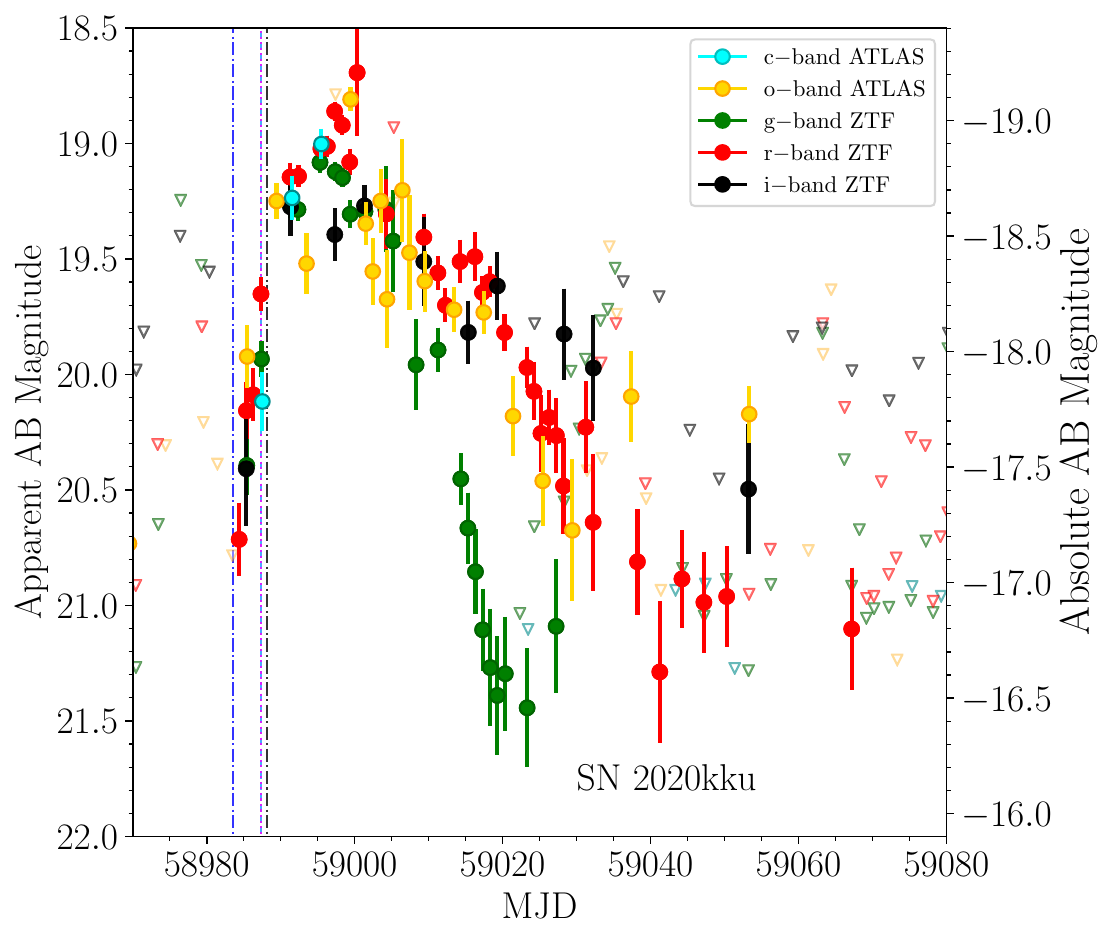}
  \caption{\textit{(cont.)}}
\end{figure*} 

\begin{figure*}\ContinuedFloat
  \centering  
  \includegraphics[width=0.3\textwidth]{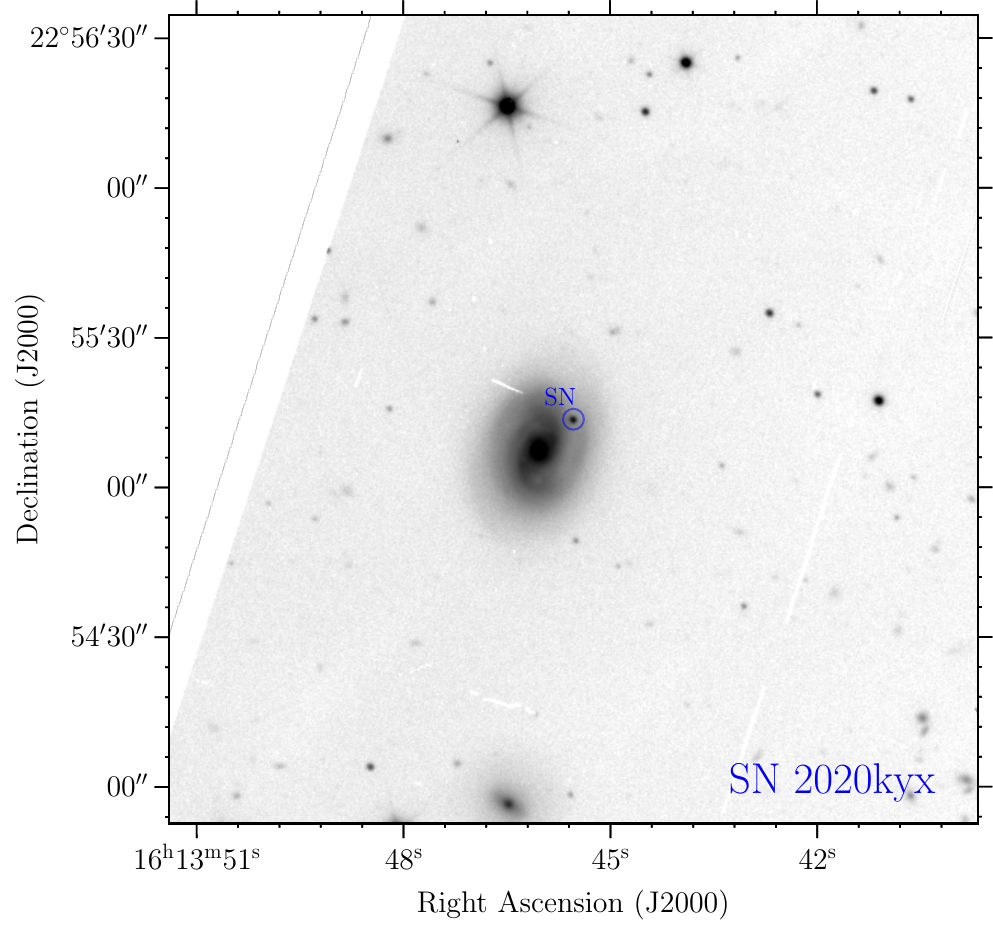}
  \hspace{0.1cm}
  \includegraphics[width=0.3\textwidth]{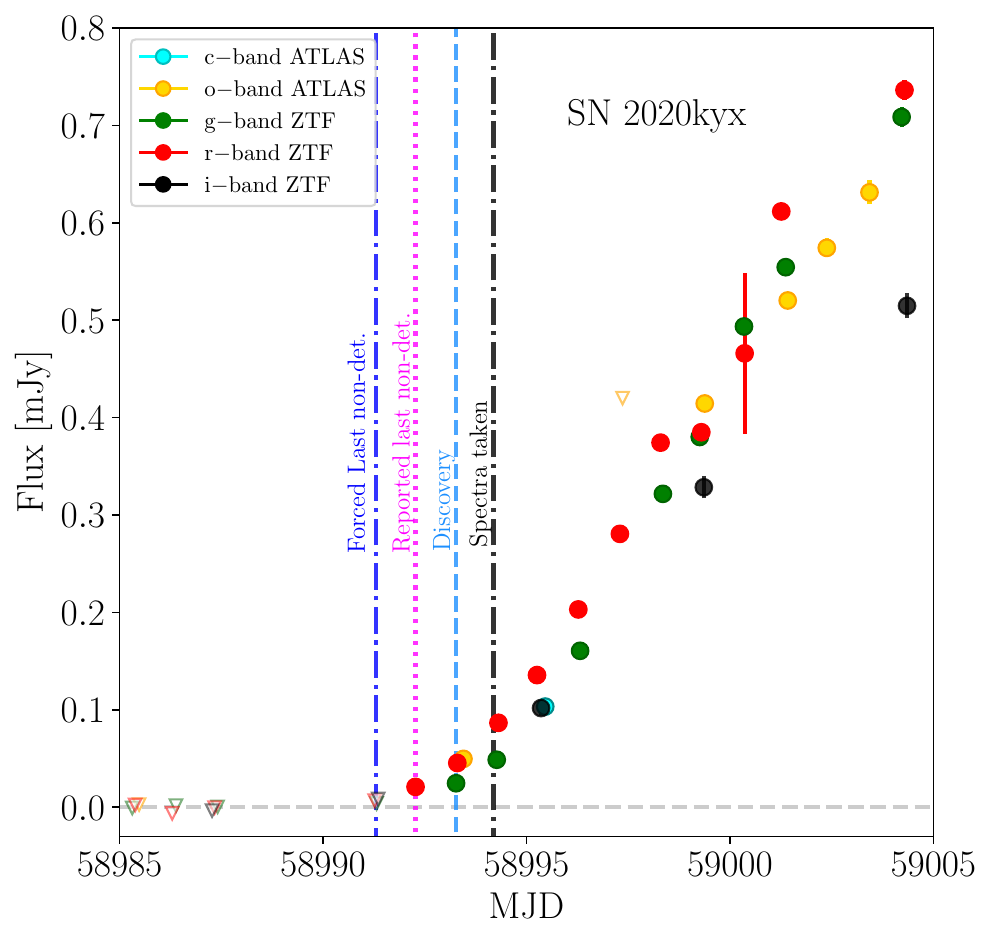}
  \hspace{0.1cm}
  \includegraphics[width=0.3\textwidth]{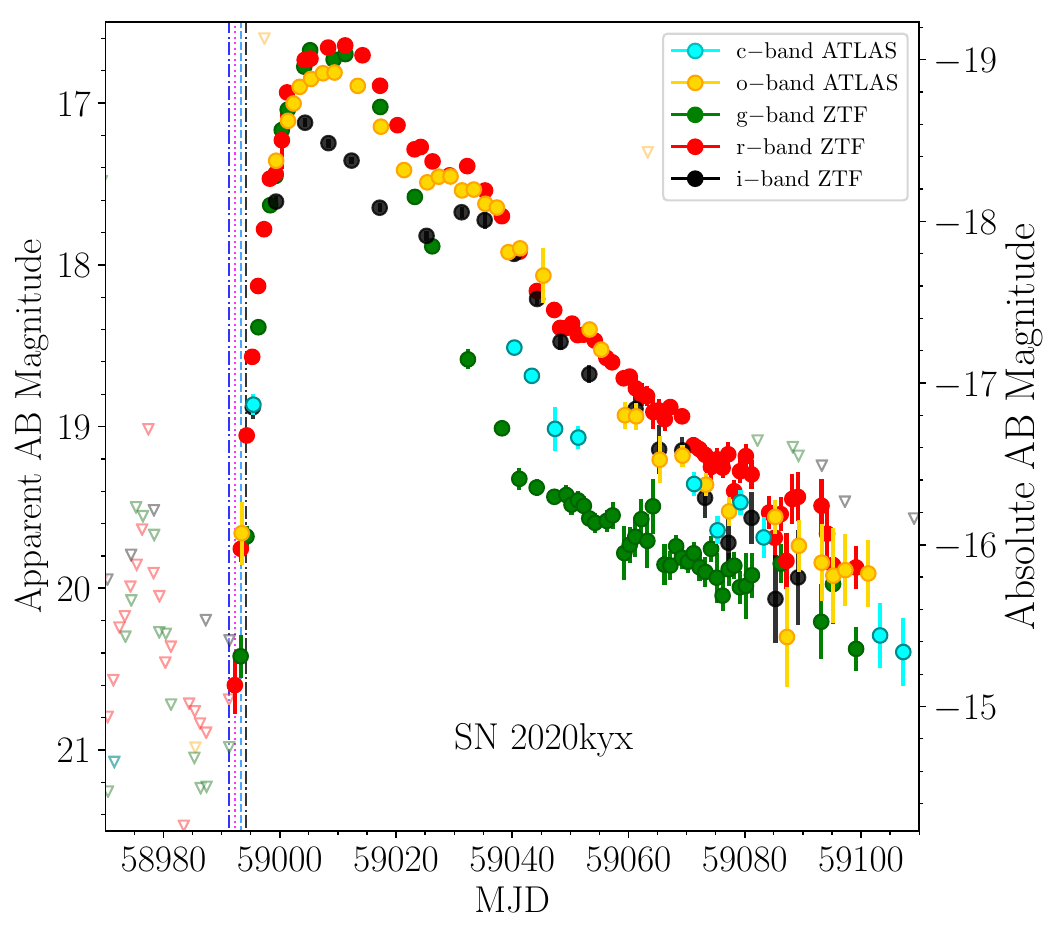} 
  \includegraphics[width=0.3\textwidth]{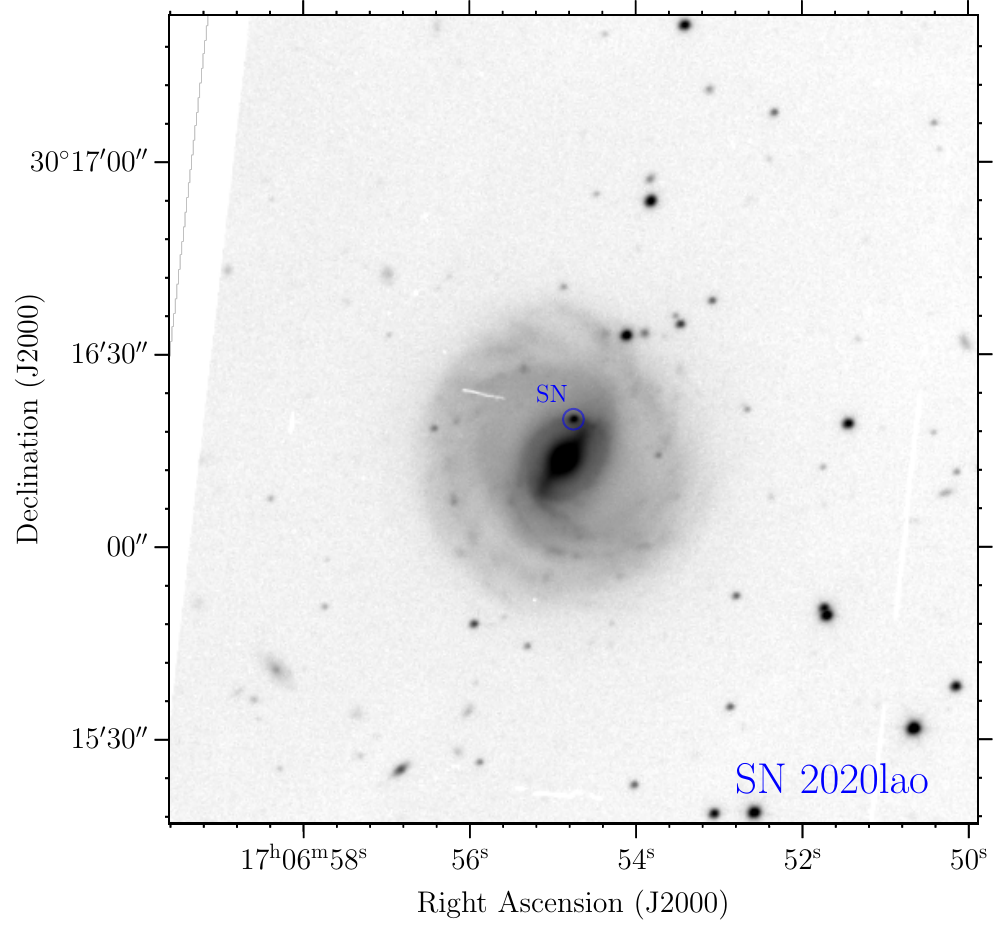}
  \hspace{0.1cm}
  \includegraphics[width=0.3\textwidth]{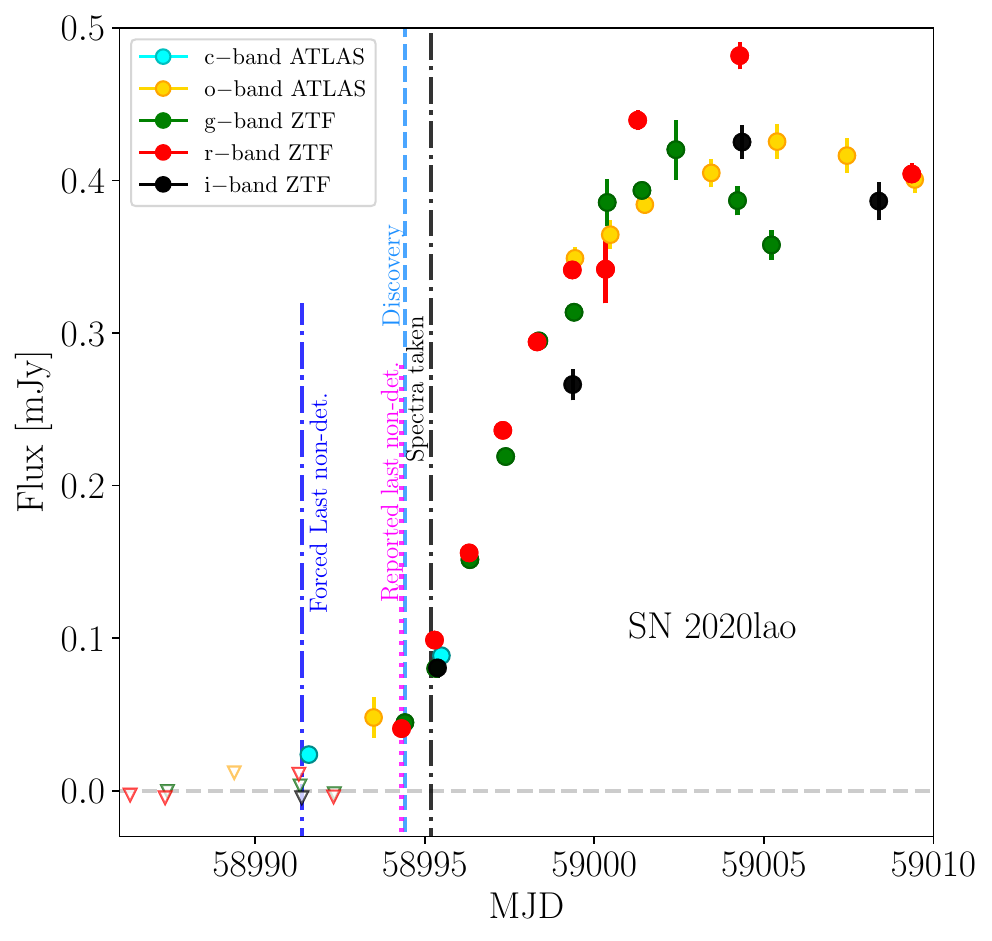}
  \hspace{0.1cm}
  \includegraphics[width=0.3\textwidth]{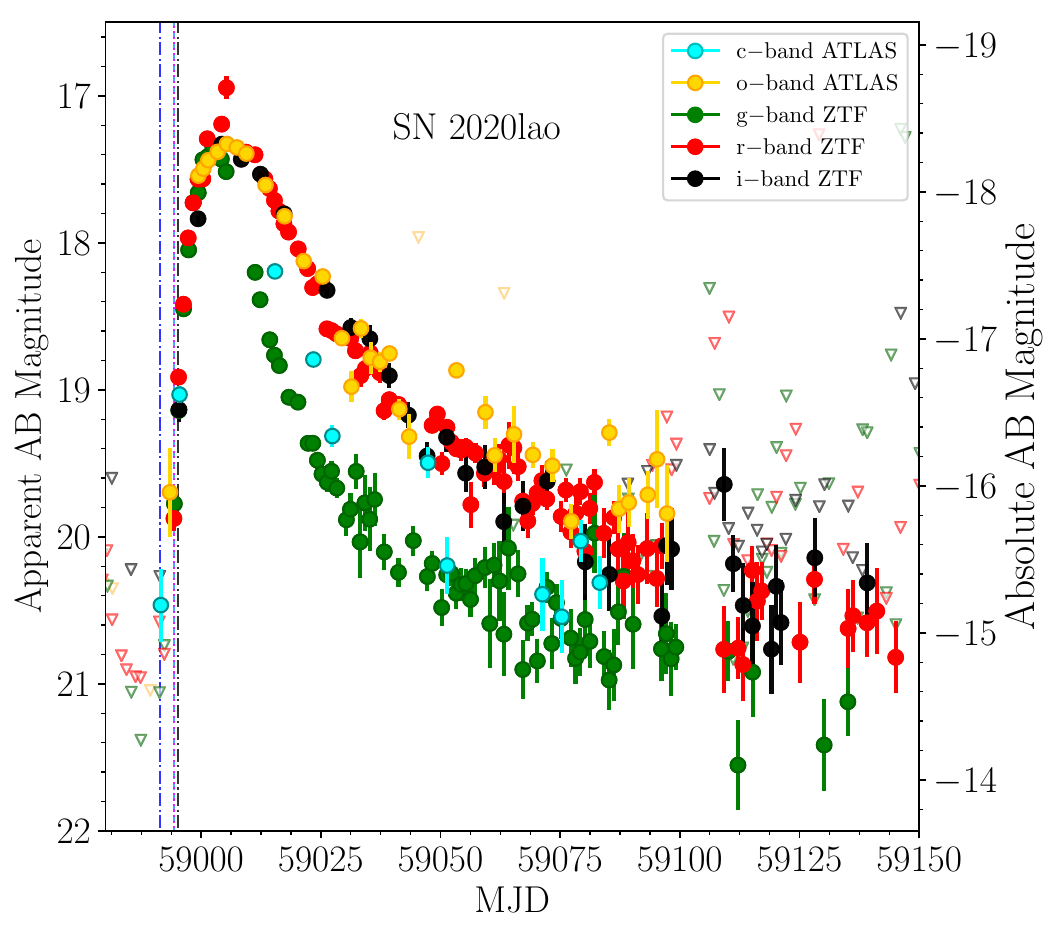}
  \includegraphics[width=0.3\textwidth]{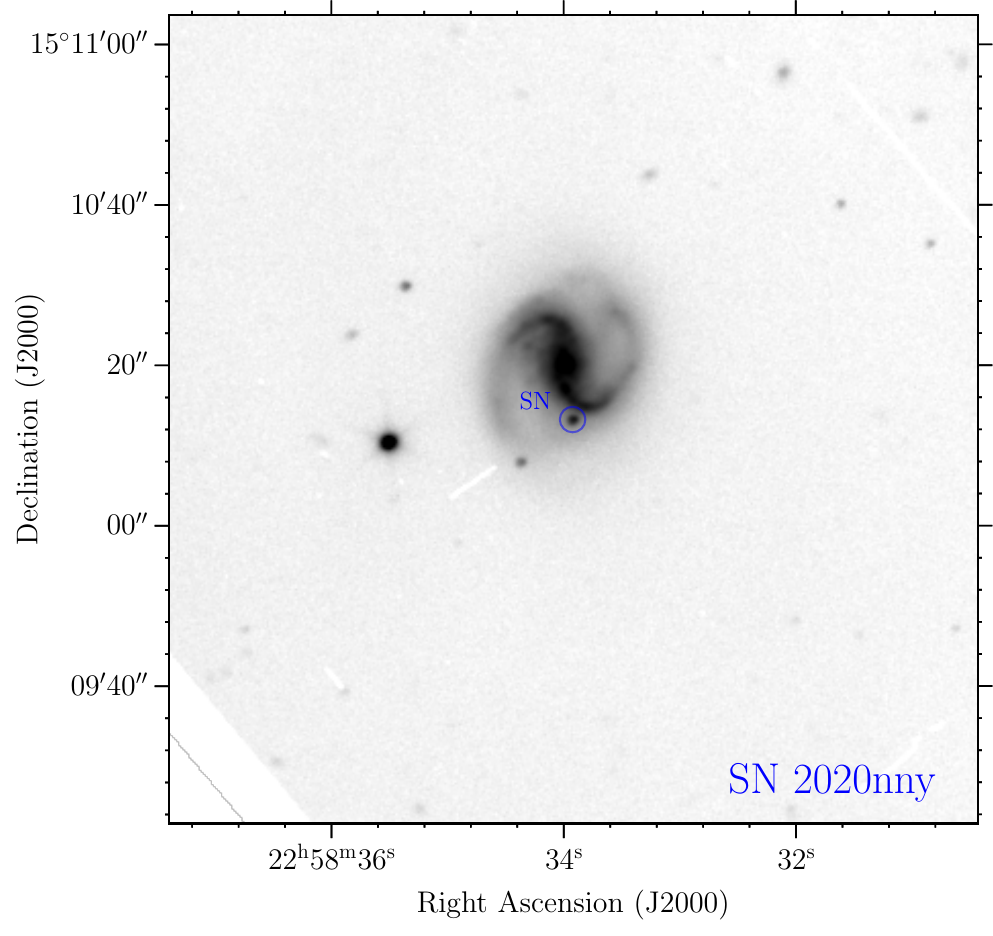}
  \hspace{0.1cm}
  \includegraphics[width=0.3\textwidth]{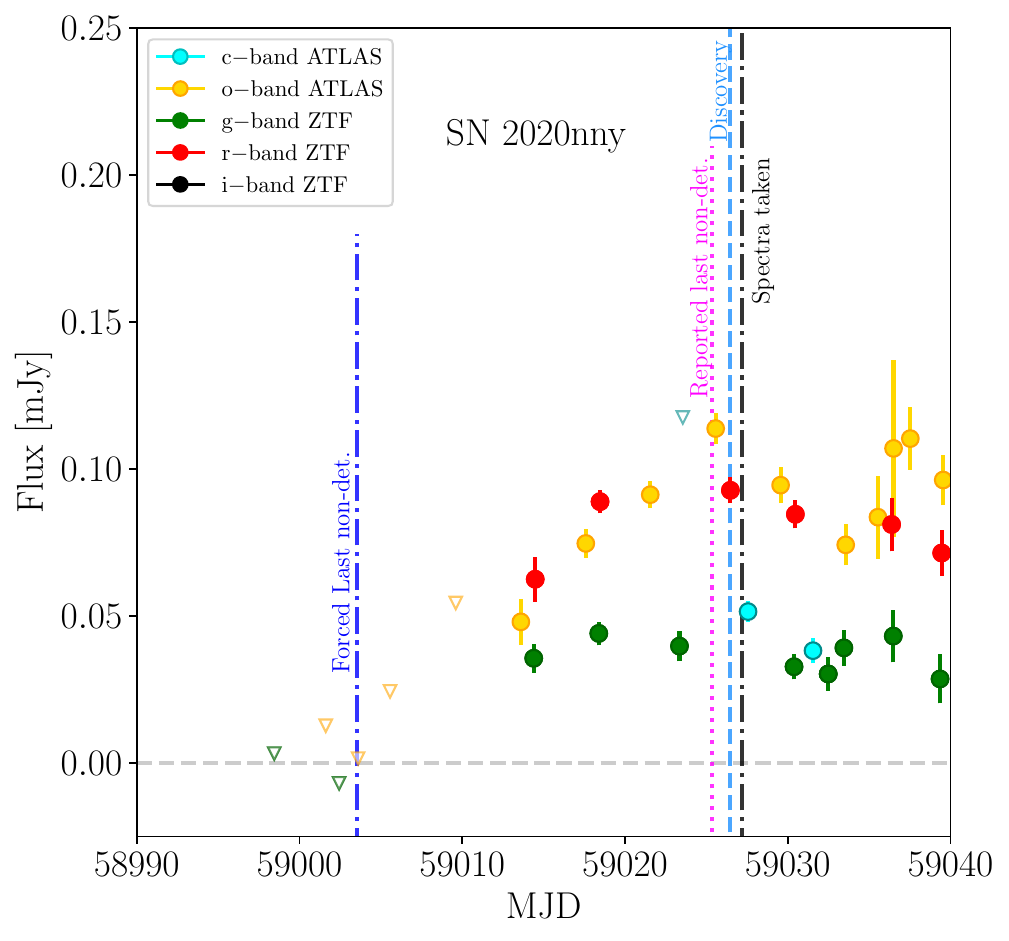}
  \hspace{0.1cm}
  \includegraphics[width=0.3\textwidth]{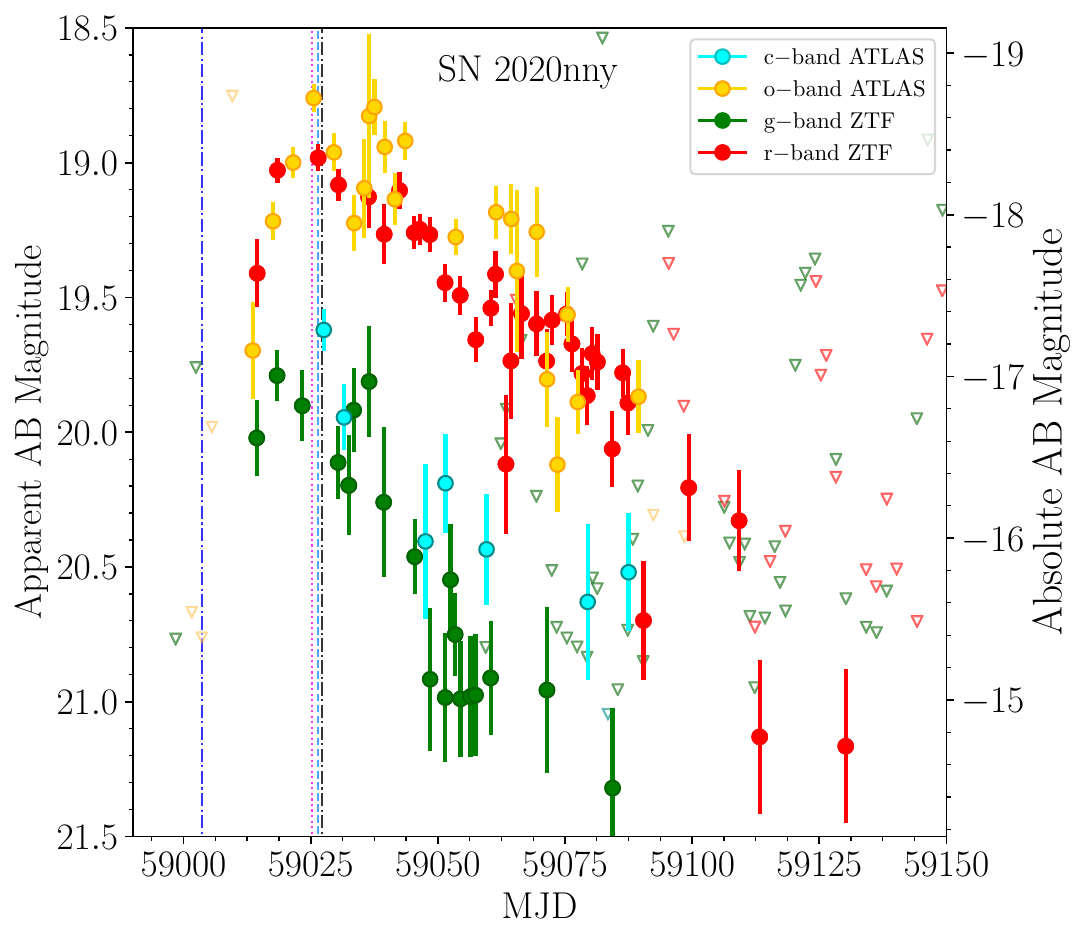} 
  \includegraphics[width=0.3\textwidth]{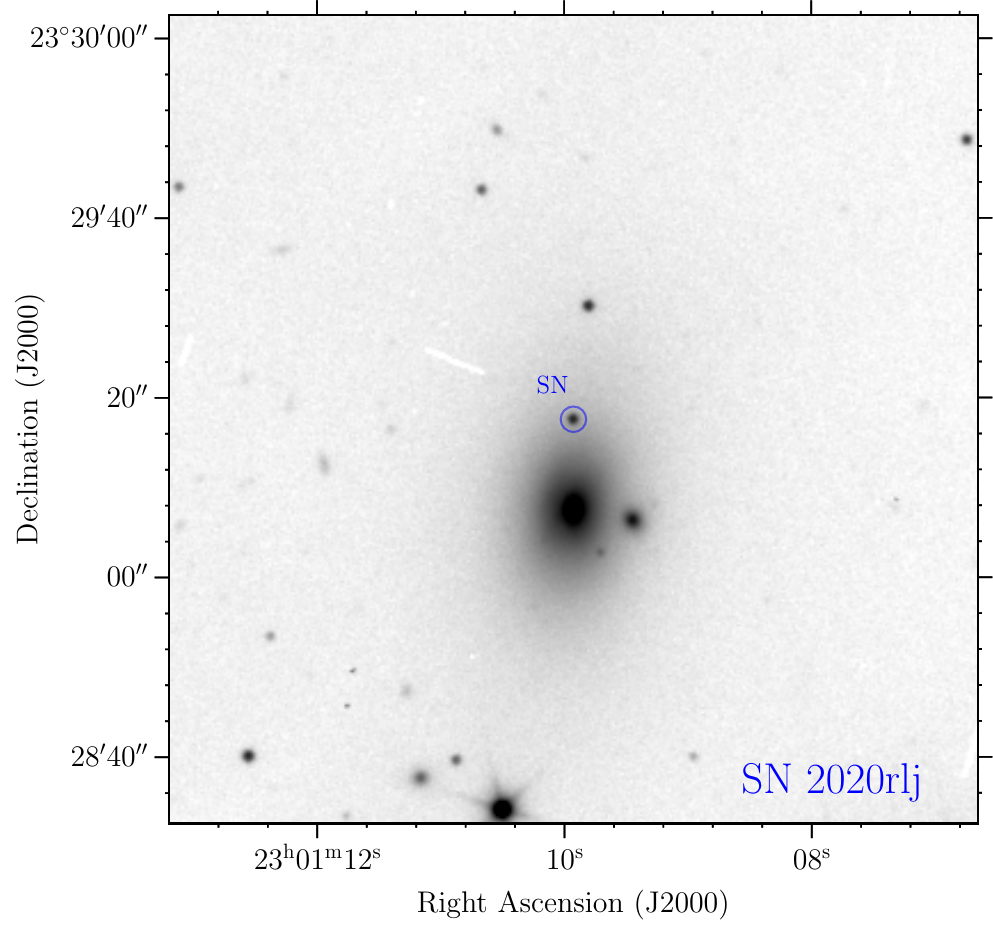}
  \hspace{0.1cm}
  \includegraphics[width=0.3\textwidth]{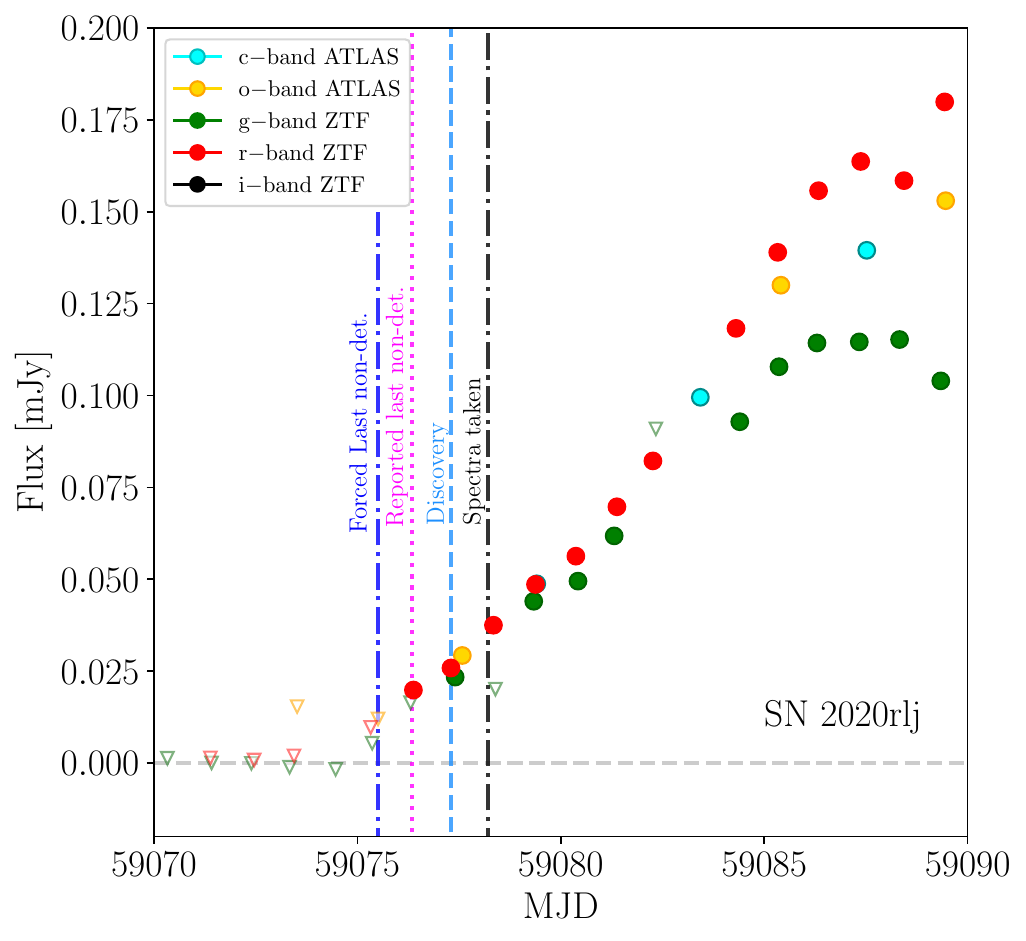}
  \hspace{0.1cm}
  \includegraphics[width=0.3\textwidth]{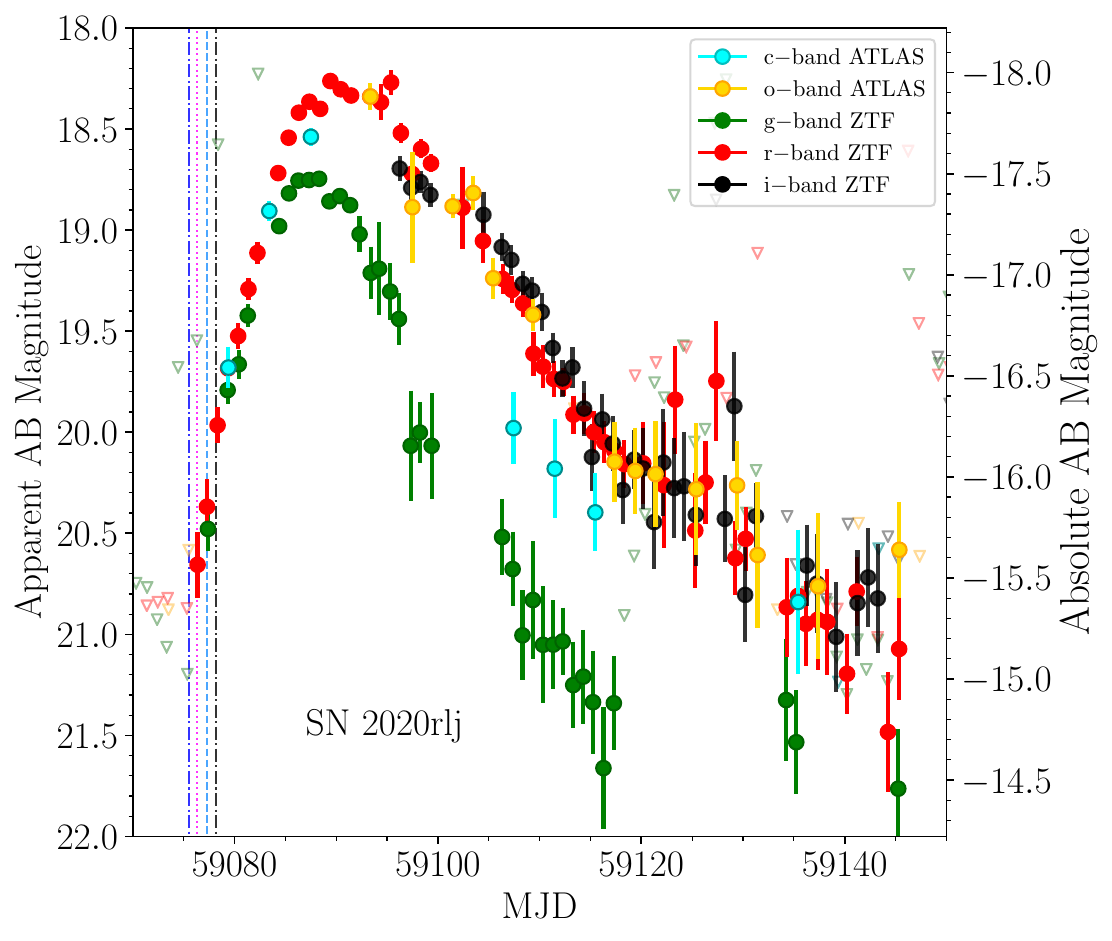}
  \caption{\textit{(cont.)}}
\end{figure*} 


\section{Strategy}\label{sec:strat}

Our programme ran intermittently from May to August 2020. The baseline approach we followed to select our candidates was using The Automatic Learning for the Rapid Classification of Events (ALeRCE) Alert Broker \citep{2021AJ....161..242F} through its Supernova Hunter tool\footnote{\href{https://snhunter.alerce.online/}{https://snhunter.alerce.online/}}. Every day, right after observations by ZTF at the Palomar Observatory were finished (about 11\,hr UTC in May 2020; 13\,hr local CET), we checked the ALeRCE SN Hunter for new SN candidates. Usually, about a hundred objects pass the criteria to be considered reliable SN candidates. They were visually checked to select those that fulfill two criteria: (i) there was a ZTF difference-imaging nondetection of the candidate from the previous night, so the object had a high probability of having exploded within the past 24\,hr, and (ii) it was clearly associated with a spatially resolved host galaxy, so the host  redshift could be used to determine the absolute magnitude of the SN and check whether it was consistent with a typical young SN brightness. Only in a few cases (SN~2020jgl  and SN~2020jhf), when ZTF did not provide candidates and the ALeRCE SN Hunter list was empty, we directly checked for newly reported SN candidate discoveries at the Transient Name Server (TNS\footnote{\href{https://www.wis-tns.org/}{https://www.wis-tns.org/}}) page. ATLAS reported these new objects. Also, in three cases (SNe~2020jfo, 2020jhf, and 2020kjt), we did not follow criterion (i), and we triggered observations anyway, given the faint reported discovery magnitude. As we will show in Section~\ref{sec:analysis}, two of them ended up being early-time spectra.

Most days, there were no objects that fulfilled both criteria. But when a good candidate appeared, we immediately triggered optical spectroscopy with the Optical System for Imaging and low-Intermediate-Resolution Integrated Spectroscopy (OSIRIS) at the 10.4\,m GTC using our target-of-opportunity (ToO) programme, {\it Follow-up of infant supernovae within 48h from explosion} (GTC52-20A). For that, the Phase 2 form\footnote{\href{https://gtc-phase2.gtc.iac.es/science/F2/}{https://gtc-phase2.gtc.iac.es/science/F2/}} was filled in with the ALeRCE finding chart, the SN candidate coordinates, and exposure times that varied from 500\,s to 900\,s in each of the two grisms used (R1000B and R1000R).

Observations were typically executed the same night, ensuring that fewer than 24\,hr had passed since discovery. There is one exception, SN~2020jhf, that could not be observed on the same night of the trigger owing to weather conditions, and it was observed the following night, $\sim 35$\,hr after discovery. In total, we triggered 10 times. Table \ref{tab:SNe_properties} summarizes the information of the 10 SNe observed under the programme and presented here: ZTF/ATLAS and IAU names, SN type (see Section \ref{subsec:typing}), coordinates, host-galaxy name, Milky Way line-of-sight reddening $E(B-V)$ obtained from IRSA\footnote{ \href{https://irsa.ipac.caltech.edu/applications/DUST/}{NASA/IPAC Infrared Science Archive}} Galactic Dust Reddening and Extinction map \citep{Schlafly_Finkbeiner_2011}, the time between the spectroscopic observation and the reported last nondetection ($t_{\rm obs}-t_{\rm lnd}$), the time between the spectroscopic observation and the discovery epoch ($t_{\rm obs}-t_{\rm disc}$), and the respective astronote reference since all classifications were publicly reported in TNS. Table \ref{tab:obs_info} summarizes additional observational information.

Three SNe (SNe~2020itj, 2020kku, and 2020lao) had intranight initially reported nondetections. As described in Section \ref{subsec:lcs}, these reported nondetections resulted in detections {\it a posteriori} using forced photometry. The left column of Figure \ref{fig:finders_lcs} shows the SN location in its host galaxy from OSIRIS acquisition images. 


\section{Data}\label{sec:data}

The long-slit spectroscopic observations of our sample were taken with the OSIRIS instrument mounted in the GTC 10.4\,m telescope. 
Data were conducted using the R1000B and R1000R grisms with contiguous wavelength ranges 3630--7500 \r{A} and 5100--10,000 \r{A} and a dispersion of 2.12--2.62\,\AA\,pixel$^{-1}$, respectively, and a spectral resolution of about 1100.

In this section, we summarize the reduction process for the spectroscopic data and introduce the public photometric measurements retrieved from the ZTF and ATLAS surveys that will be used in the following analysis.


\subsection{Data reduction}\label{sec:reduction}

We conducted the reduction process with a Python-based pipeline developed by our team, which utilizes version 1.11.0 of \texttt{PypeIt} \citep{pypeit_2,pypeit_1}. \texttt{PypeIt}\footnote{\href{https://github.com/pypeit/PypeIt}{Python Spectroscopic Data Reduction Pipeline (\texttt{PypeIt})}} is a Python package designed to reduce spectroscopic data from several different spectrographs. The pipeline performed the following reduction steps: bias subtraction, flat-field correction, wavelength calibration (using HgAr, Ne, and Xe comparison lamps as reference), sky subtraction, and flux calibration. For the telluric correction, we assessed the H$_{2}$O and O$_{2}$ telluric absorption features from the standard star's spectrum obtained on the same night of each SN observation. A spectrum equal to unity was created except in the wavelength ranges of the above telluric absorption. By dividing the SN spectrum by this transmission spectrum, we obtained each SN's final one-dimensional telluric-corrected flux-calibrated spectrum. 

\begin{table*}
\caption{Redshifts obtained from the spectral lines coming from the galaxy hosts.}
\begin{center}
\resizebox{\textwidth}{!}{ 
\begin{tabular}{r|ccccccc|cc|c}
\hline \hline
\multicolumn{1}{l|}{\multirow{2}{*}{IAU Name}} & \multicolumn{7}{c|}{$\lambda_\text{observed}$ (\r{A})} & \multicolumn{2}{c|}{$z_\text{host}$} & \multirow{2}{*}{$\mu$ (mag)} \\ \cline{2-10}
\multicolumn{1}{l|}{} & \multicolumn{1}{c}{H$\beta$} & \multicolumn{1}{c}{[O III]} & \multicolumn{1}{c}{[O III]} & \multicolumn{1}{c}{Na I D} & \multicolumn{1}{c}{H$\alpha$} & \multicolumn{1}{c}{[S II]} & \multicolumn{1}{c|}{[S II]} & \multicolumn{1}{c}{SN spec} & \multicolumn{1}{c|}{Host spec} &  \\ \hline
2020itj & 5022.66 & 5123.75 & 5172.99 & $-$     & 6780.16 & 6938.99 & 6954.41 & 0.0332$\pm$0.0001 & 0.03295$^{a}$ & 35.802 \\
2020jfo & $-$     & $-$     & $-$     & 5923.83 & 6598.36 & $-$     & $-$     & 0.0053$\pm$0.0001 & 0.00522$^{b}$ & 30.810$\pm$0.200 \\
2020jgl & $-$     & $-$     & $-$     & $-$     & $-$     & $-$     & $-$     & $-$               & 0.00676$^{b}$ & 32.428$\pm$0.089 \\
2020jhf & $-$     & $-$     & $-$     & 5985.92 & 6664.92 & $-$     & $-$     & 0.0157$\pm$0.0001 & 0.01544$^{a}$ & 34.127 \\
2020kjt & 5040.27 & $-$     & 5191.70 & $-$     & 6808.13 & 6967.42 & 6982.07 & 0.0372$\pm$0.0002 & $-$           & 36.072 \\
2020kku & $-$     & $-$     & $-$     & 6385.03 & $-$     & $-$     & $-$     & 0.0835$\pm$0.0004 & 0.08356$^{c}$ & 37.901 \\
2020kyx & $-$     & $-$     & $-$     & $-$     & $-$     & $-$     & $-$     & $-$               & 0.03192$^{a}$ & 35.731 \\
2020lao & $-$     & $-$     & $-$     & $-$     & 6764.89 & 6924.02 & 6939.43 & 0.0309$\pm$0.0001 & 0.03081$^{a}$ & 35.653 \\
2020nny & 5010.28 & $-$     & $-$     & 6074.94 & 6765.18 & 6922.74 & 6937.57 & 0.0308$\pm$0.0001 & 0.03084$^{a}$ & 35.655 \\
2020rlj & $-$     & $-$     & $-$     & $-$     & $-$     & $-$     & $-$     & $-$               & 0.03976$^{b}$ & 36.221 \\ \hline
\end{tabular}}
\end{center}
Spectroscopic redshift from $^{a}$SDSS, $^{b}$NED, $^{c}$DESI.
\label{tab:redshift}
\end{table*}

\begin{figure}[t]
\centering
    \includegraphics[width=0.8\textwidth]{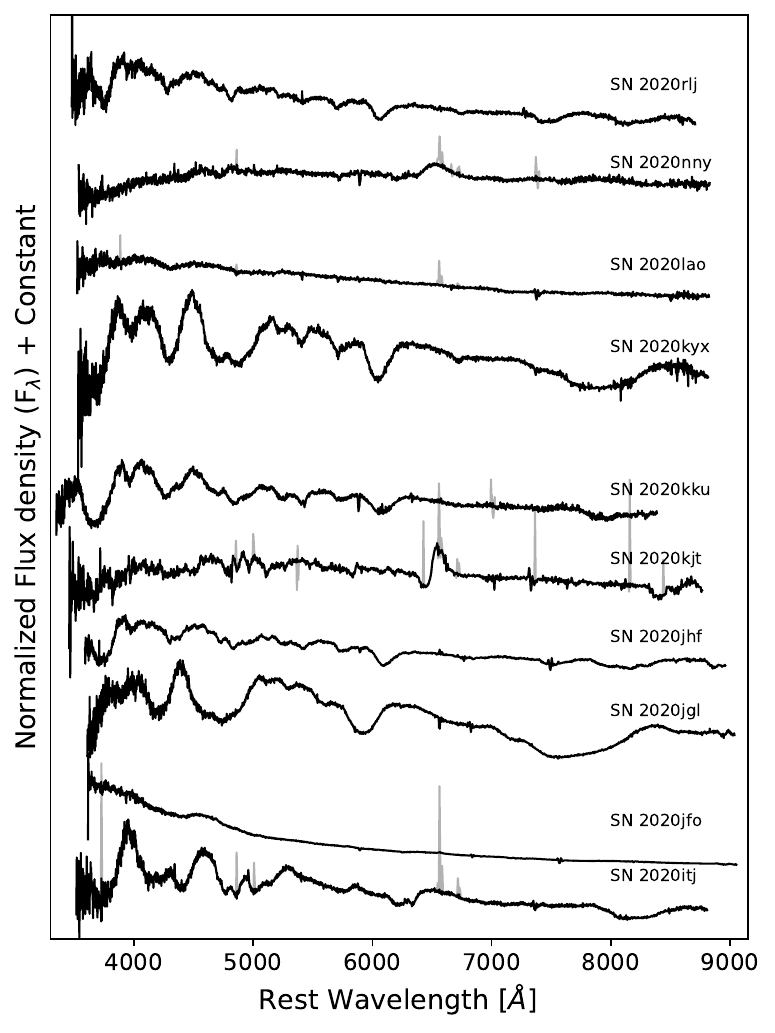}
    \caption{GTC/OSIRIS spectra of the 10 SNe in our dataset. Each spectrum has been corrected for Milky Way reddening and redshift, and thus is in the rest-wavelength frame. The spectra were cleaned for narrow emission lines from the host galaxy and some spikes due to cosmic rays. Each spectrum was normalized by the flux average in the wavelength range 4500--5000\,\r{A} and added an arbitrary constant for clarity.}
\label{fig:all_spectra}
\end{figure}


\subsection{Redshift and distance}

We estimated the redshift of each SN by measuring the host galaxy's narrow spectral lines present in the SN spectra. The following lines were considered: Balmer H$\alpha$ and H$\beta$ at 6562.81 and 4861.35 \r{A}, respectively, [O III] doublet at 4958.91\,\r{A} and 5006.84\,\r{A}, Na~I~D doublet at 5889.95 and 5895.92\,\r{A}, and [S II] doublet at 6716.44 and 6730.82\,\r{A}. All of these spectral features are in emission except for Na~I~D, which is in absorption. We used IRAF\footnote{IRAF is distributed by the National Optical Astronomy Observatories, which are operated by the Association of Universities for Research in Astronomy, Inc., under a cooperative agreement with the U.S. National Science Foundation.} (Image Reduction and Analysis Facility, \citep{IRAF}) to fit a Gaussian profile to the lines and estimate their wavelength center. Then, we computed the redshift values considering the shift of each line from its rest-frame wavelength. The final redshift value and respective uncertainty were taken as the mean of the different values and the standard deviation, respectively. For the case of SN 2020kku, we only measured the Na~I~D absorption line; thus, we took the wavelength sampling interval as a conservative estimate of the uncertainty associated with the final redshift value, which we then obtained through error propagation. 

Table \ref{tab:redshift} shows the central wavelengths of the lines detected in each SN along with the final computed redshift. SNe 2020jgl, 2020kyx, and 2020rlj lacked spectral lines from their respective galaxy hosts. Along with the redshift values estimated in this work, Table \ref{tab:redshift} shows public spectroscopic redshift measurements of the respective hosts. The redshifts estimated from the SN spectra include the contribution of the galaxy's rotational velocity, so the final redshift values that we adopt throughout this work are the ones coming from the galaxy's spectra. The only exception is SN 2020kjt, where no public spectroscopic redshift was found. 

We determined the distance modulus of the SN host galaxies assuming a flat-$\Lambda$CDM model with H$_{0} = 70$ km s$^{-1}$ Mpc$^{-1}$ and $\Omega_{m,0} = 0.3$ and using the redshifts measured above for all SNe, except for SN 2020jfo and SN 2020jgl. For these two SNe, we obtained redshift-independent distances from the NASA/IPAC Extragalactic Database (NED\footnote{\href{https://ned.ipac.caltech.edu/}{https://ned.ipac.caltech.edu/}}) since they are very nearby, and the contribution of peculiar velocities may be significant when using cosmological redshifts. SN 2020jfo exploded in the spiral galaxy M61, which is at $z = 0.005224$ according to NED. There are several distance measurements for this nearby host. In this work, we assume the distance of $14.51 \pm 1.38$\,Mpc ($\mu = 30.81 \pm 0.20$\,mag) obtained by \citet{Bose_Kumar_2014} considering redshift-independent methods based on optical data of SN 2008in. \citet{Sollerman_2021}, \citet{Ailawadhi_2023}, and \citet{Kilpatrick_2023} also assume the same distance value whereas \citet{Teja_2022} find a distance of $16.45 \pm 2.69$\,Mpc ($\mu = 31.08 \pm 0.36$\,mag), which is in agreement with the value found here within the uncertainties. The host of SN 2020jgl is the face-on spiral galaxy MCG-02-24-027 located at $z=0.006758$ (NED). No distance measurements were available in the literature; thus, we obtained the distance modulus, $\mu= 32.428 \pm 0.089$\,mag, from fitting SNooPy \citep{Burns_2011_SNooPy} to the SN light curve (see Section \ref{sec:lc}).

For clarity, we masked the host-galaxy contamination lines from the spectra for the following analysis. The final reduced and corrected spectra of all 10 SNe are presented in Figure \ref{fig:all_spectra}.


\subsection{Light curves}\label{subsec:lcs}

We obtained forced photometry of the 10 SNe in our sample from both the ATLAS \citep{2018PASP..130f4505T,ATLAS_2} survey in the orange (\textit{o}, 5600--8200\,\r{A}) and cyan (\textit{c}, 4200--6500\,\r{A}) bands\footnote{\url{https://fallingstar-data.com/forcedphot/}} and from the ZTF \citep{ZTF} survey in the \textit{g} (3676--5614\,\r{A}), \textit{r} (5498--7394\,\r{A}), and \textit{i} (6871--8965\,\r{A}) bands\footnote{\url{https://ztfweb.ipac.caltech.edu/cgi-bin/requestForcedPhotometry.cgi}}.

ATLAS consists of four telescopes, which automatically scan the sky several times every night, resulting in several photometric measurements. ZTF can also have several data points in each band per night. Aiming to obtain a clearer perception of the light curves' behavior, we performed a binning of the photometric data. Thus, the forced photometry flux measurements and uncertainties retrieved on the same night, using the same band, were combined as the weighted average and the mean deviation, respectively. Then, we produced the light curves in magnitude space following the steps described in the ATLAS and ZTF documentation. We converted flux counts into flux-density units (AB magnitude system; \cite{1983ApJ...266..713O}) as follows:
\begin{eqnarray}
f_{AB} &=& f_{\rm counts} \times 10^{-0.4(ZP-ZP_{AB})}, \\
\sigma_{f_{AB}} &=& | 10^{-0.4(ZP-ZP_{AB})} \times \sigma_{f_{\rm counts}} |,
\end{eqnarray}
where $f_{counts}$ and $\sigma_{f_{counts}}$ are the flux and its respective uncertainty in count units, $ZP$ is the survey zero-point, and $ZP_{AB} = 8.90$ to ensure the flux is in Janskys (Jy) units. We then compute the AB magnitude and its respective uncertainties,
\begin{eqnarray}
m_{det} &=& - 2.5\, {\rm log}_{10}f + ZP_{AB}, \\
\sigma_{m} &=& 1.0857\, \sigma_{f}/f,
\end{eqnarray}
where $f$ is the flux in Jy and $\sigma_{f}$ the flux uncertainty. We used the distance modulus presented in Table~\ref{tab:redshift} to convert observed apparent magnitudes into absolute magnitudes. 

Observations with a signal-to-noise ratio (S/N) greater or equal to 3 were considered as detections; otherwise, they were considered nondetections and reported as 3$\sigma$ upper-limit magnitudes as
\begin{align}
m_{\rm ndet} = - 2.5 \, {\rm log}_{10}\left( 3\sigma_{f} \right) + ZP_{AB}.
\end{align}
Therefore, we define the last nondetection as the data point immediately preceding the first detection with a deeper magnitude. Table~\ref{tab:tspec_texp} (second and third columns, respectively) presents the forced photometry last nondetection and first detection estimated for our sample.

\begin{table*}
\caption{Summary of last nondetections, first detections, times of first light, and spectra epochs.}
\centering
\resizebox{\textwidth}{!}{ 
\begin{tabular}{rcccccccc}
\hline \hline
\multirow{2}{*}{IAU Name} & Reported & Forced & Forced & \multicolumn{2}{c}{Time of first light} & \multirow{2}{*}{Spectrum Date} & \multicolumn{2}{c}{Spectrum Epoch} \\
 & last nondetection & last nondetection &first detection& Midpoint & Early LC Fitting &  & Midpoint & Early LC Fitting \\ \hline
2020itj & 58970.29 & 58960.50 & 58961.90 &{\bf 58961.20 $\pm$ 0.70}& $-$                              & 58971.00 $\pm$ 0.01 & {\bf 9.49 $\pm$ 0.68} & $-$                   \\
2020jfo & 58971.28 & 58971.28 & 58975.20 &     58973.24 $\pm$ 1.96 &{\bf 58974.72 $\pm$ 0.19}         & 58975.92 $\pm$ 0.01 &      2.66 $\pm$ 1.95  & {\bf 1.18 $\pm$ 0.19} \\
2020jgl & 58975.30 & 58973.15 & 58975.30 &     58974.23 $\pm$ 1.08 &{\bf 58974.31 $\pm$ 1.14}         & 58976.91 $\pm$ 0.01 &      2.66 $\pm$ 1.07  & {\bf 2.57 $\pm$ 1.13} \\
2020jhf & 58975.41 & 58974.25 & 58975.41 &     58974.83 $\pm$ 0.58 &{\bf \textit{58974.76 $\pm$ 0.20}}& 58978.93 $\pm$ 0.01 &      4.03 $\pm$ 0.57  & {\bf 4.10 $\pm$ 0.20} \\
2020kjt & 58986.47 & 58912.50 & 58940.50 &{\bf 58926.50 $\pm$14.00}& $-$                              & 58988.15 $\pm$ 0.01 &{\bf 59.44 $\pm$ 13.50}& $-$                   \\
2020kku & 58987.34 & 58983.50 & 58984.40 &     58983.95 $\pm$ 0.45 &{\bf \textit{58981.81 $\pm$ 0.86}}& 58988.18 $\pm$ 0.01 &      3.90 $\pm$ 0.42  & {\bf 5.87 $\pm$ 0.79} \\
2020kyx & 58992.27 & 58991.30 & 58992.27 &     58991.79 $\pm$ 0.49 &{\bf \textit{58990.71 $\pm$ 0.33}}& 58994.20 $\pm$ 0.01 &      2.33 $\pm$ 0.47  & {\bf 3.37 $\pm$ 0.32} \\
2020lao & 58994.31 & 58991.37 & 58991.58 &     58991.48 $\pm$ 0.11 &{\bf 58993.14 $\pm$ 0.18}         & 58995.21 $\pm$ 0.01 &      3.62 $\pm$ 0.11  & {\bf 2.00 $\pm$ 0.17} \\
2020nny & 59025.34 & 59003.57 & 59013.60 &     59008.59 $\pm$ 5.02 &{\bf 59009.94 $\pm$ 0.95}         & 59027.20 $\pm$ 0.01 &     18.05 $\pm$ 4.87  &{\bf 16.73 $\pm$ 0.92} \\
2020rlj & 59076.34 & 59075.50 & 59076.34 &     59075.92 $\pm$ 0.42 &{\bf \textit{59074.31 $\pm$ 0.41}}& 59078.23 $\pm$ 0.01 &      2.22 $\pm$ 0.40  & {\bf 3.76 $\pm$ 0.39} \\
\hline
\end{tabular}}
{\small Numbers in boldface indicate the final time of first light.
Numbers in italics indicate cases where the early light-curve method provides an estimate earlier than the forced last nondetection. This occurs because the image was not deep enough to properly detect the SN, but the early fit is considered reliable.}
\label{tab:tspec_texp}
\end{table*}

The resulting light curves in flux (middle panels) and magnitude (right panels) space are presented in Figure~\ref{fig:finders_lcs}, along with their respective finding charts (left panels). As seen in the middle panels, the reported last nondetection resulted in most cases being an actual detection in forced photometry. The only exception is SN~2020jfo, in which both estimations coincide. Generally, the last nondetection was found a few days earlier. The median of this difference is 3.5 days; however, there are some cases where the difference is larger. The most extreme case is SN~2020kjt, which has a difference of $\sim74$ days. This is followed by SN~2020nny ($\sim22$ days) and SN~2020jfo ($\sim11$ days). 

Given SN~2020kjt's apparent faintness, its analysis is complicated. Thus, the last nondetection is hard to define because none of the upper limits are deeper than the actual first detection. However, our estimated nondetection seems to be accurate, considering the spectrum phase obtained from the spectral matching (Section~\ref{subsec:tspec}). In this object, the first detection from the forced photometry also appears many days before the reported discovery date ($\sim50$ days earlier). Therefore, the spectrum obtained from our programme turns out to be very old ($\sim50$--70 days from the explosion). 


\section{Analysis}\label{sec:analysis}

\subsection{Typing}\label{subsec:typing}

\begin{figure*}
  \centering  
  \includegraphics[width=0.48\textwidth]{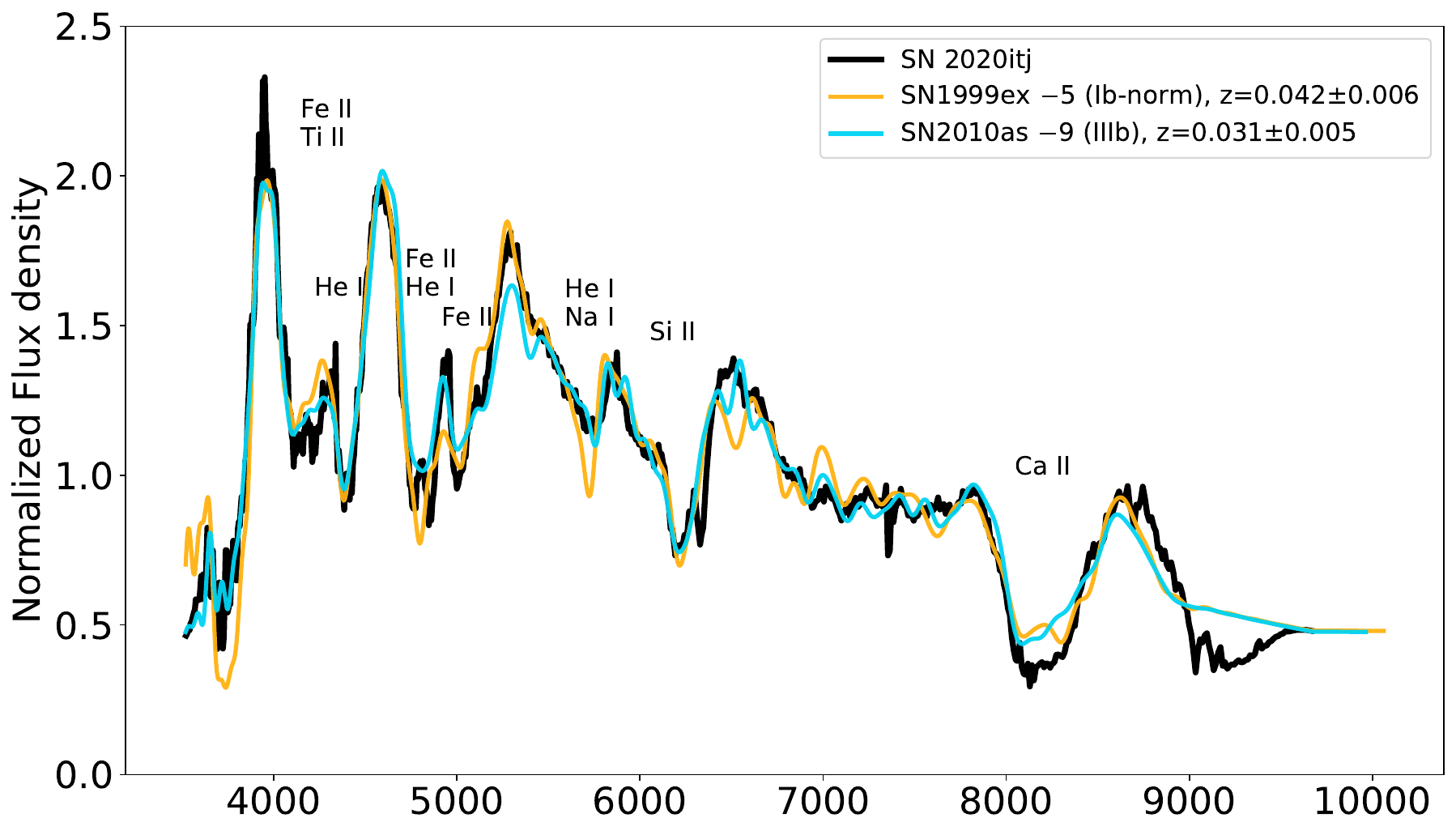}
  \includegraphics[width=0.48\textwidth]{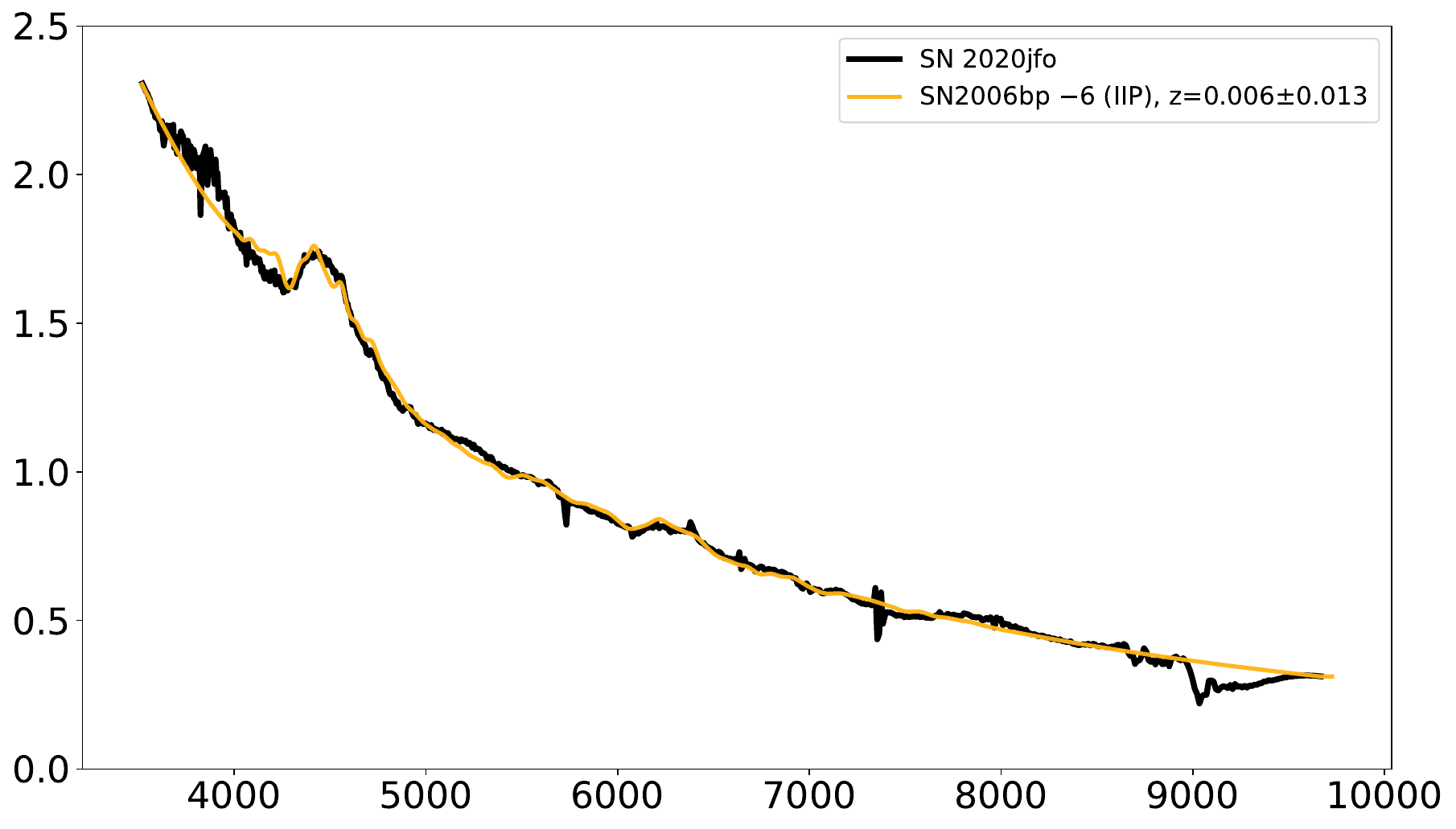}
  \includegraphics[width=0.48\textwidth]{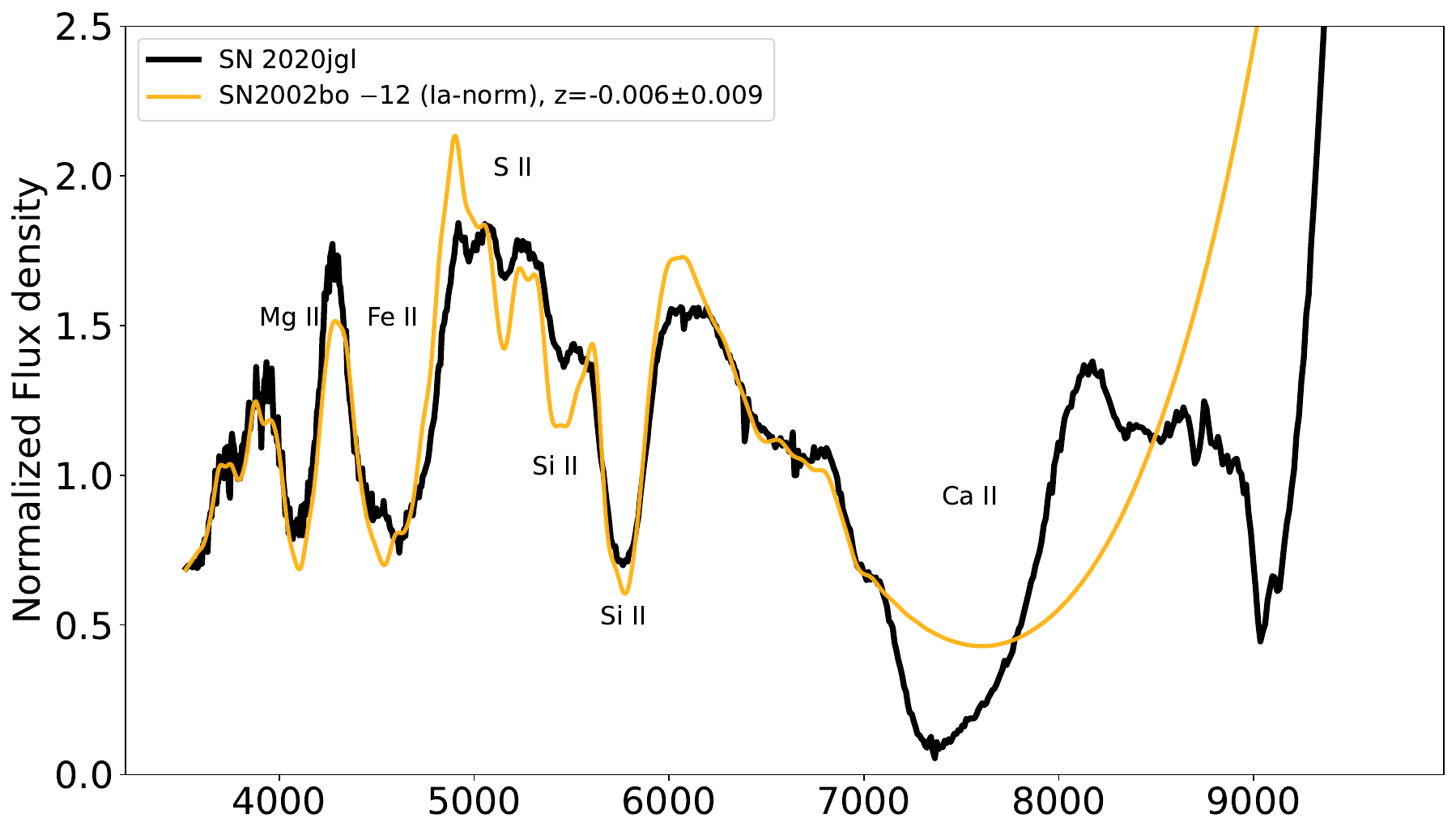}
  \includegraphics[width=0.48\textwidth]{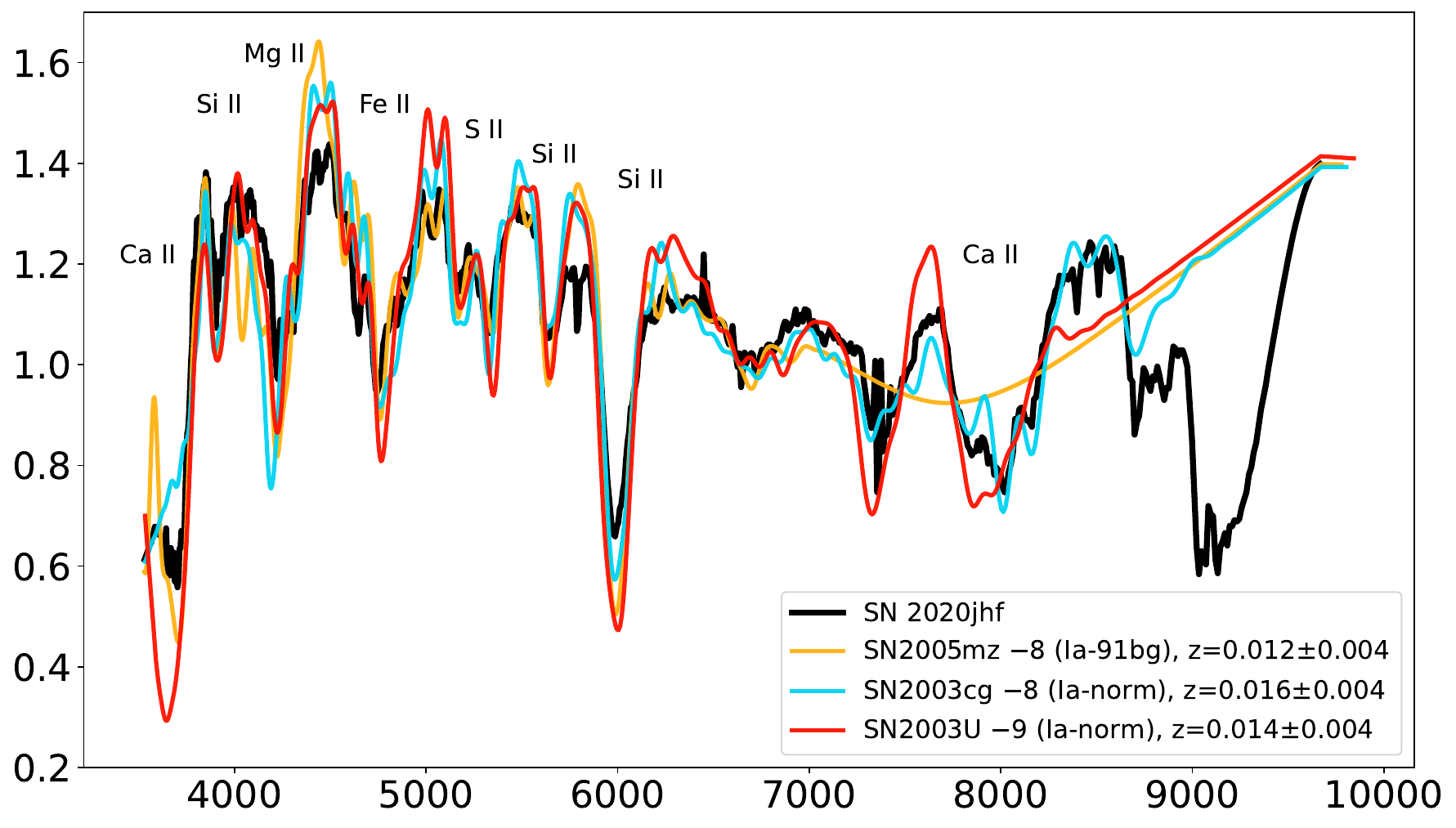}
  \includegraphics[width=0.48\textwidth]{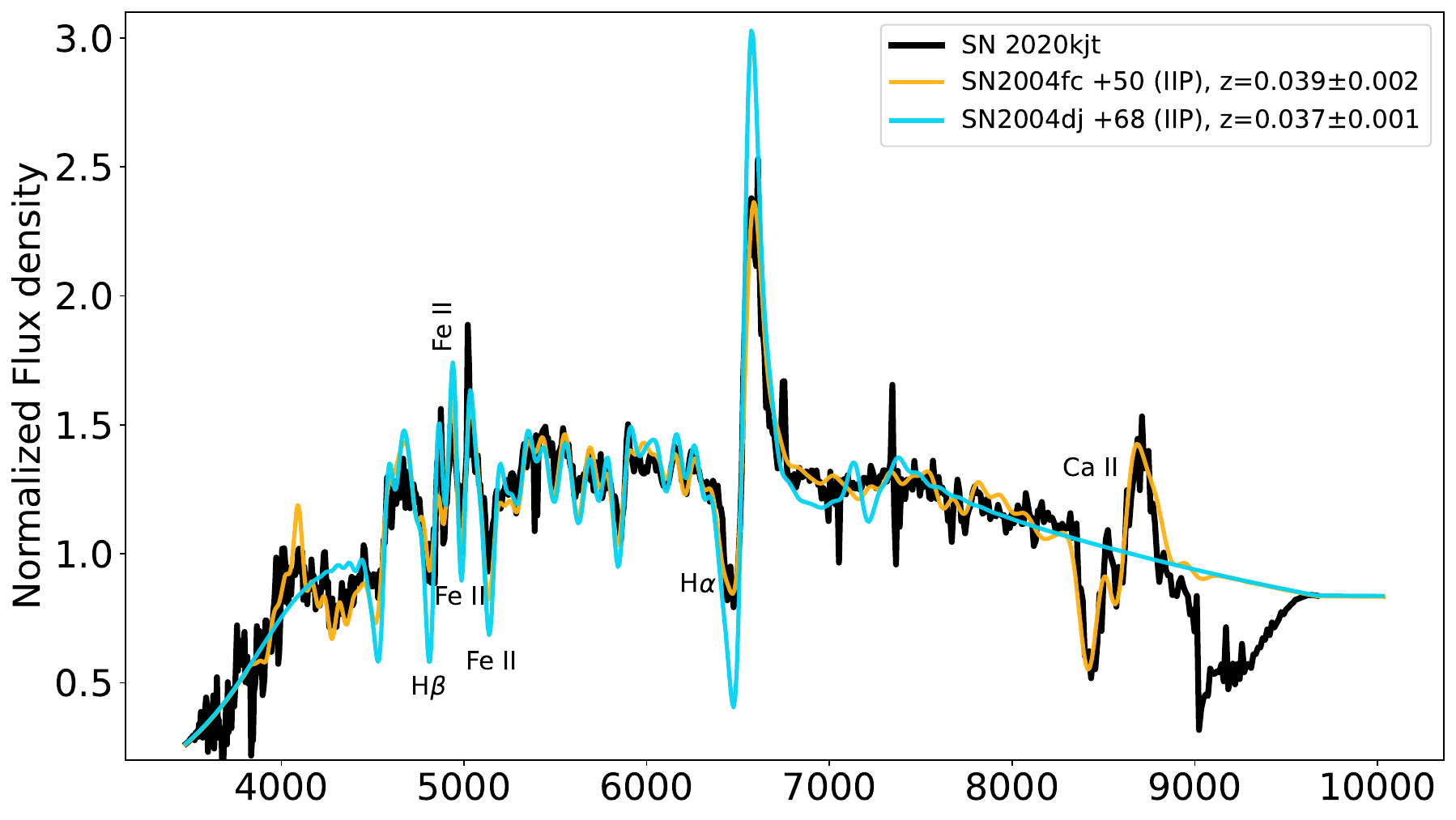}
  \includegraphics[width=0.48\textwidth]{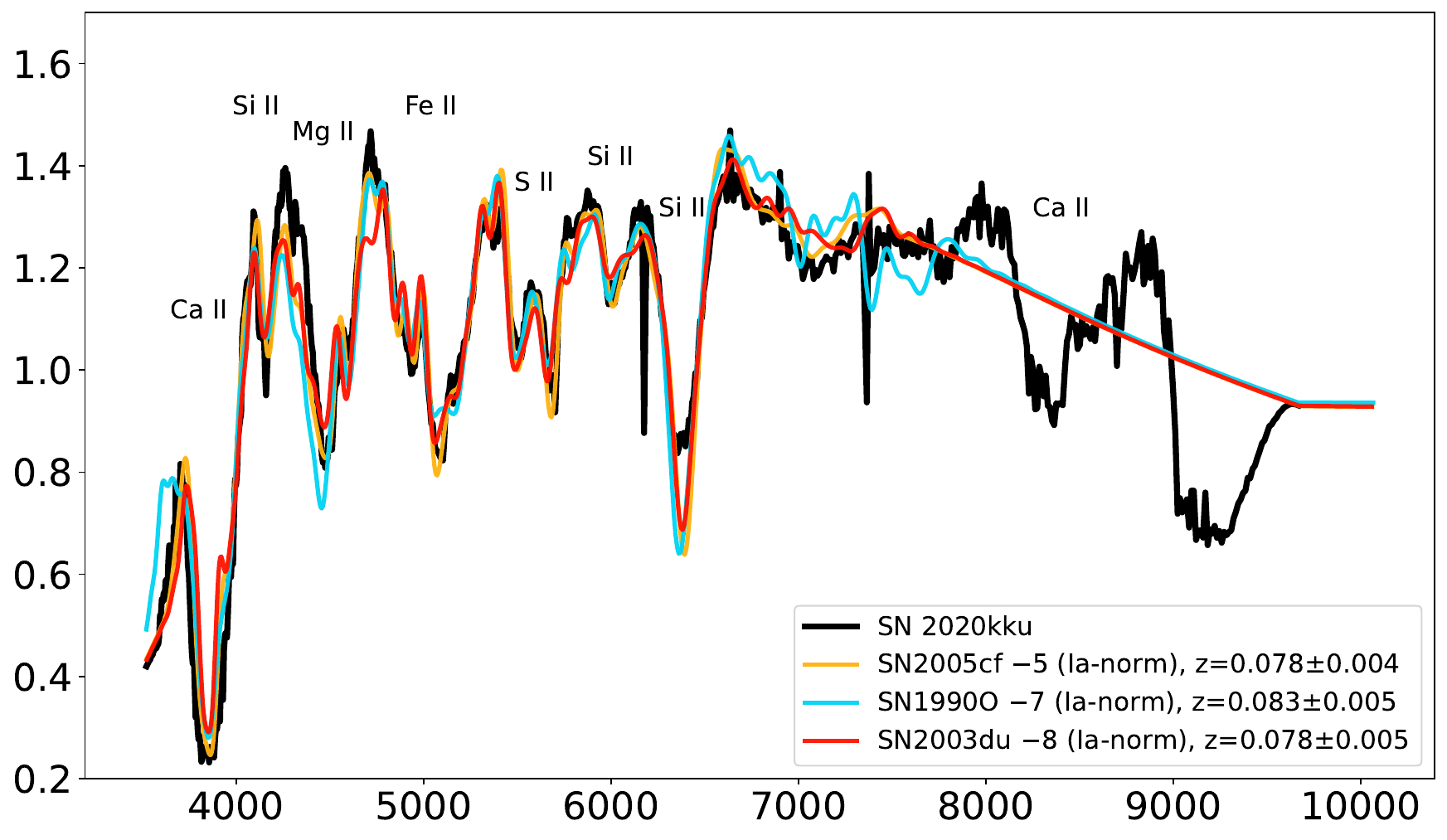}
  \includegraphics[width=0.48\textwidth]{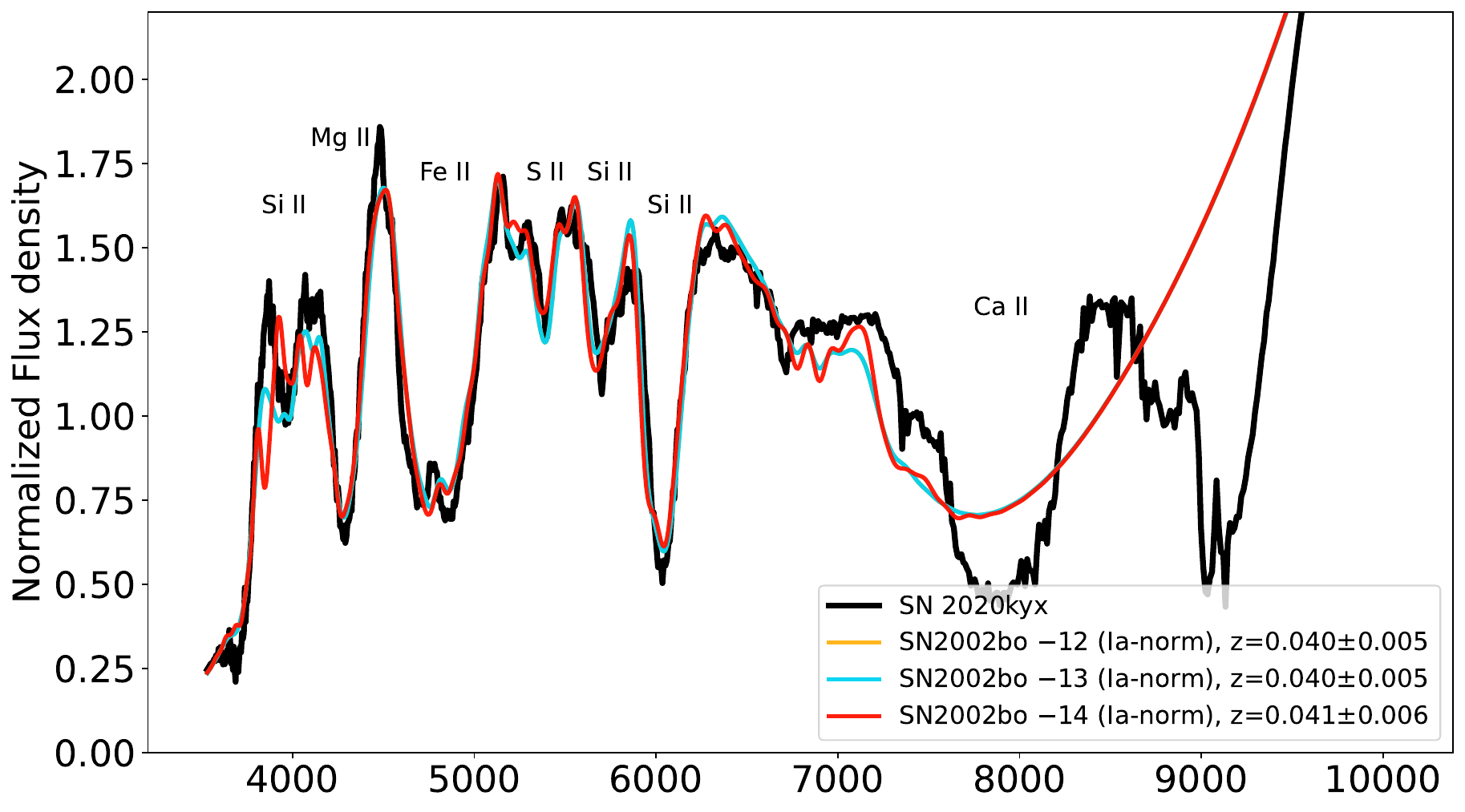}
  \includegraphics[width=0.48\textwidth]{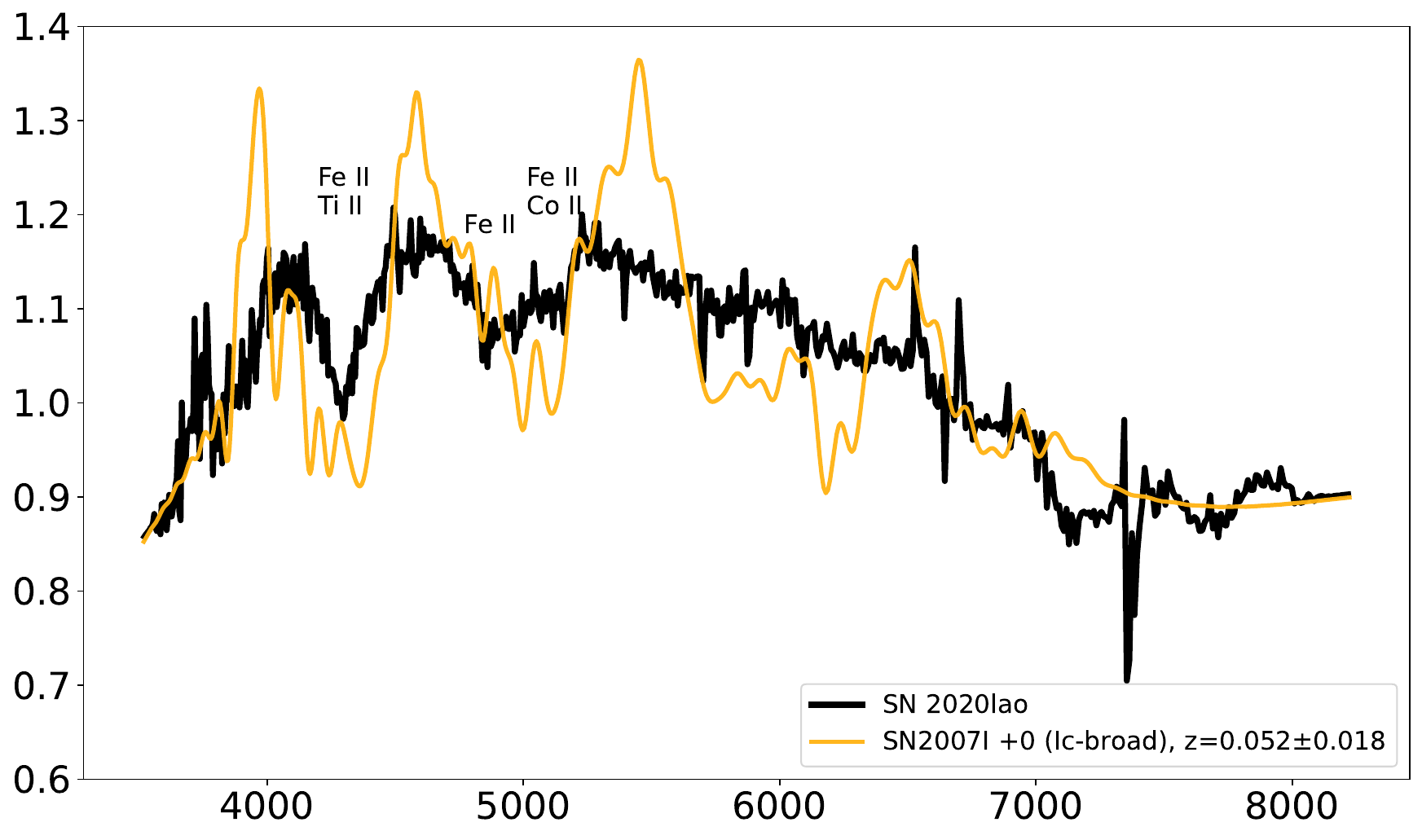}
  \includegraphics[width=0.48\textwidth]{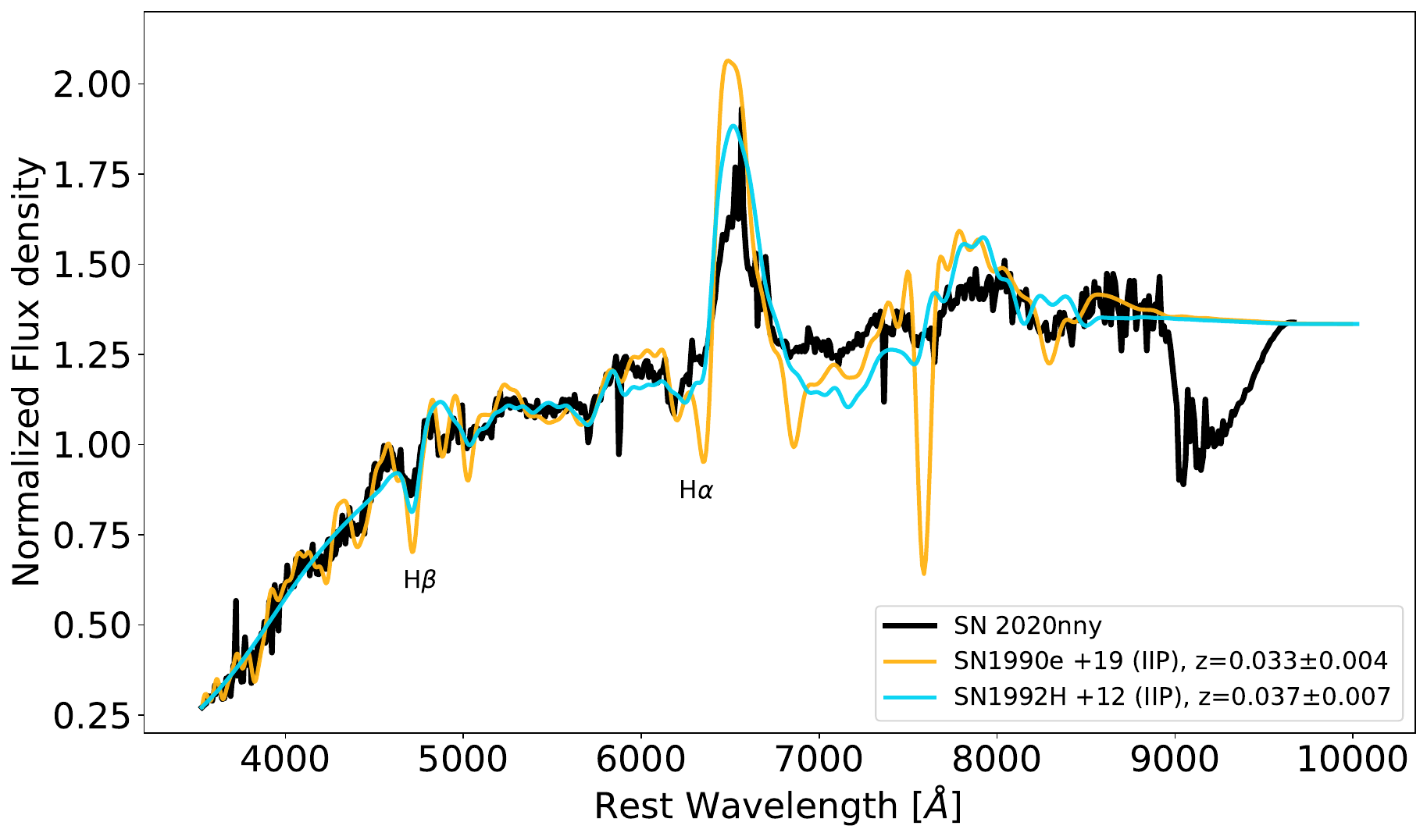}
  \includegraphics[width=0.48\textwidth]{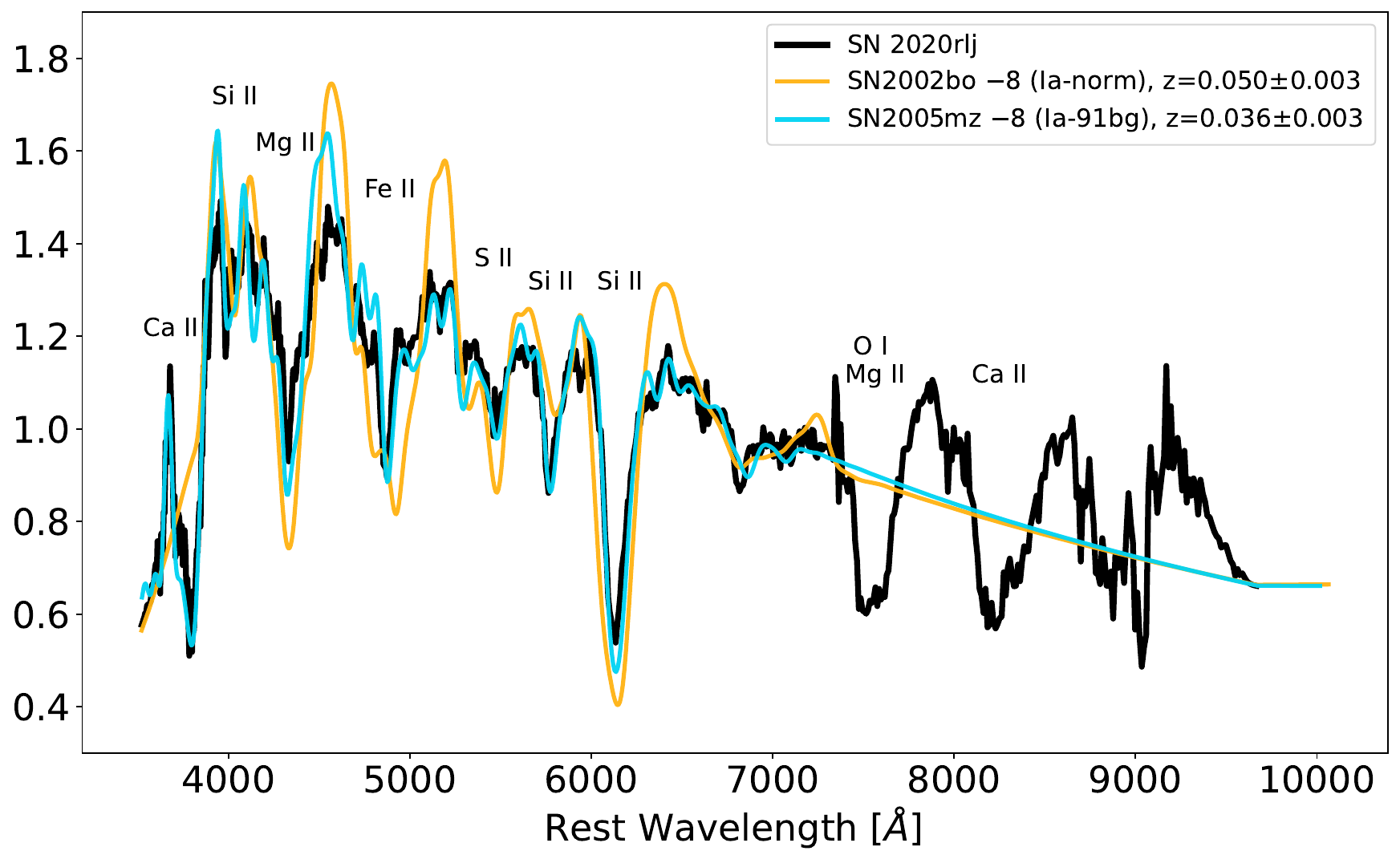}
  \caption{Best spectral template matches from SNID for each of the 10 SNe in our sample. Each fit label includes the SN name compared, the epoch from the time of maximum light, the type, and the redshift. Common spectral features of each SN type are represented in the different plots.}
  \label{fig:snid_fits}
\end{figure*}

To determine the SN type and its epoch, we used the Supernova Identification code (SNID, \citep{SNID}) as it performs a cross-correlation to spectra in a database including thousands of observed SN spectra of different types and at different epochs. Figure \ref{fig:snid_fits} summarizes a few of the best template fits obtained for each SN in our sample. Below, we comment on the different spectral matchings considering the light curves (Figure \ref{fig:finders_lcs}) and the redshift of each SN (Table \ref{tab:redshift}).

\begin{itemize}
    \item SN 2020itj resembles Type Ib SN~1999ex and Type IIb SN~2010as at 5 and 9 days before reaching maximum brightness, respectively. With a measured redshift of 0.033, the redshift fit of SN 2010as (0.031) aligns more closely than that of SN 1999ex (0.042). Upon visual examination of the early-time light curve of SN 2020itj, both best fits offer a reliable estimate of the epoch, as it seems that the SN~2020itj spectrum was obtained about 10 days before the peak.
    \item SN 2020jfo exhibits the closest match with Type II SN 2006bp, approximately 6 days before reaching maximum brightness. The redshift of the comparison aligns well, within the margin of error, with the redshift estimated for SN 2020jfo (0.006 and 0.00522, respectively), and the epoch also appears to align well with the information obtained from the light curve.
    \item SN~2020jgl has a good match with the normal Type Ia SN~2002bo, 12 days before maximum at redshift $\sim$0.006. However, there are not many spectra earlier than $-$12 days in the SNID available database; thus, as we will show later (Section \ref{ref:jgl}), this spectrum could be even earlier. The estimated redshift is also consistent with that found by SNID.
    \item SN~2020jhf has as best matches to Type Ia SN 1991bg-like SN~2005mz, Type Ia SN~2003cg, and Type Ia SN~2003U at about 8--9 days before maximum light. There is no clearly favorable fit, although we show later (Section \ref{sec:lc}) that this SN may be a transitional from normal to subluminous (e.g., visible secondary peak, not-so-low light-curve stretch).
    \item SN2020kjt is best matched to Type II SNe 2004fc and 2004dj at 50 and 68 from explosion, respectively. 
    Both fits are similarly good, with SN~2004fc probably being a bit better, and consistent with an SN~II in the middle of the plateau phase.
    \item SN~2020kku is best matched with Type~Ia SNe 2005cf, 1990O, and  2003du at 5, 7, and 8 days before maximum light, all similarly providing a reasonable fit to a normal SN~Ia about a week before maximum.
    \item SN2020kyx is best matched with SN 2002bo at different epochs: 12, 13, and 14 days before maximum brightness. The three spectra of SN 2002bo do not change considerably and are typical of a normal SN~Ia about two weeks before maximum.
    \item SN~2020lao does not have a good match with any template present in the SNID database, but we overplotted the spectrum of the Type Ic broad-lined SN~2007I at maximum light. While the fit does not perfectly match, it has similar features at similar wavelengths as SN~2020lao. We will present in a separate paper a detailed analysis of this SN, confirming it is effectively a broad-lined Type Ic SN (Stritzinger et al. in prep.).
    \item The best SNID fits for SN~2020nny are SNe II 1990E and 1992H at 19 and 12 days after maximum light, respectively. 
    The main feature of the spectrum, broad H$\alpha$ emission, aligns well with that of SN~2020nny and corresponds to a similar epoch and redshift.
    \item SN~2020rlj has as best matches normal Type Ia SN~2002bo and SN 1991bg-like Type Ia SN~2005mz, both 8 days before their respective time of maximum. The SN 1991bg-like spectrum (e.g., \cite{1992AJ....104.1543F}) provides a better match, and both the redshift and the epoch match that of SN~2020rlj.
\end{itemize}

In summary, we conclude that the dataset collected under our programme and presented in this work is composed of 5 SNe~Ia (SN 2020jgl, SN 2020jhf, SN 2020kku, SN 2020kyx, and SN 2020rlj), 1 SN IIb (SN 2020itj), 1 SN Ic-BL (SN 2020lao), and 3 SNe~II (SN 2020jfo, SN 2020kjt, and SN 2020nny). 

\subsection{Epoch of first light}\label{subsec:t0}

\begin{figure}[!t]
\centering
    \includegraphics[width=0.6\textwidth]{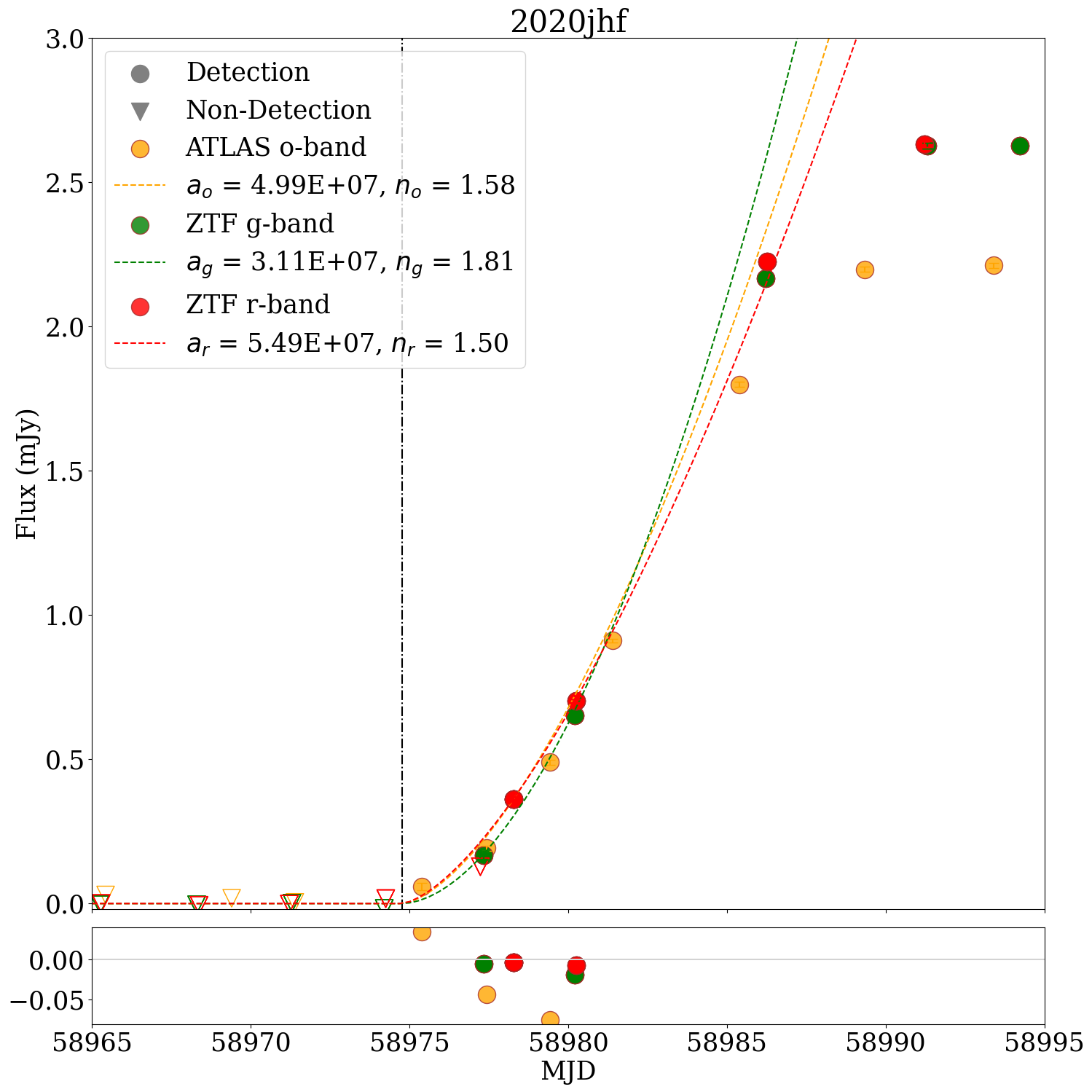}
    \caption{Early observer-frame light curve of SN 2020jhf fitted to different power laws considering each band. The black line denotes the estimated time of first light, $t_0 = 58974.76$. The estimated coefficients $a$ and $n$ from Equation \ref{eq:fireball} are shown in the legend. The ATLAS $c$ band did not have sufficient data points to fit a curve. The bottom panel exhibits the residuals in the fitting range, where the dotted grey line is the fit for each band.}
\label{fig:2020jhf_fireball}
\end{figure}
\begin{figure}[!t]
\centering
    \includegraphics[width=0.49\textwidth]{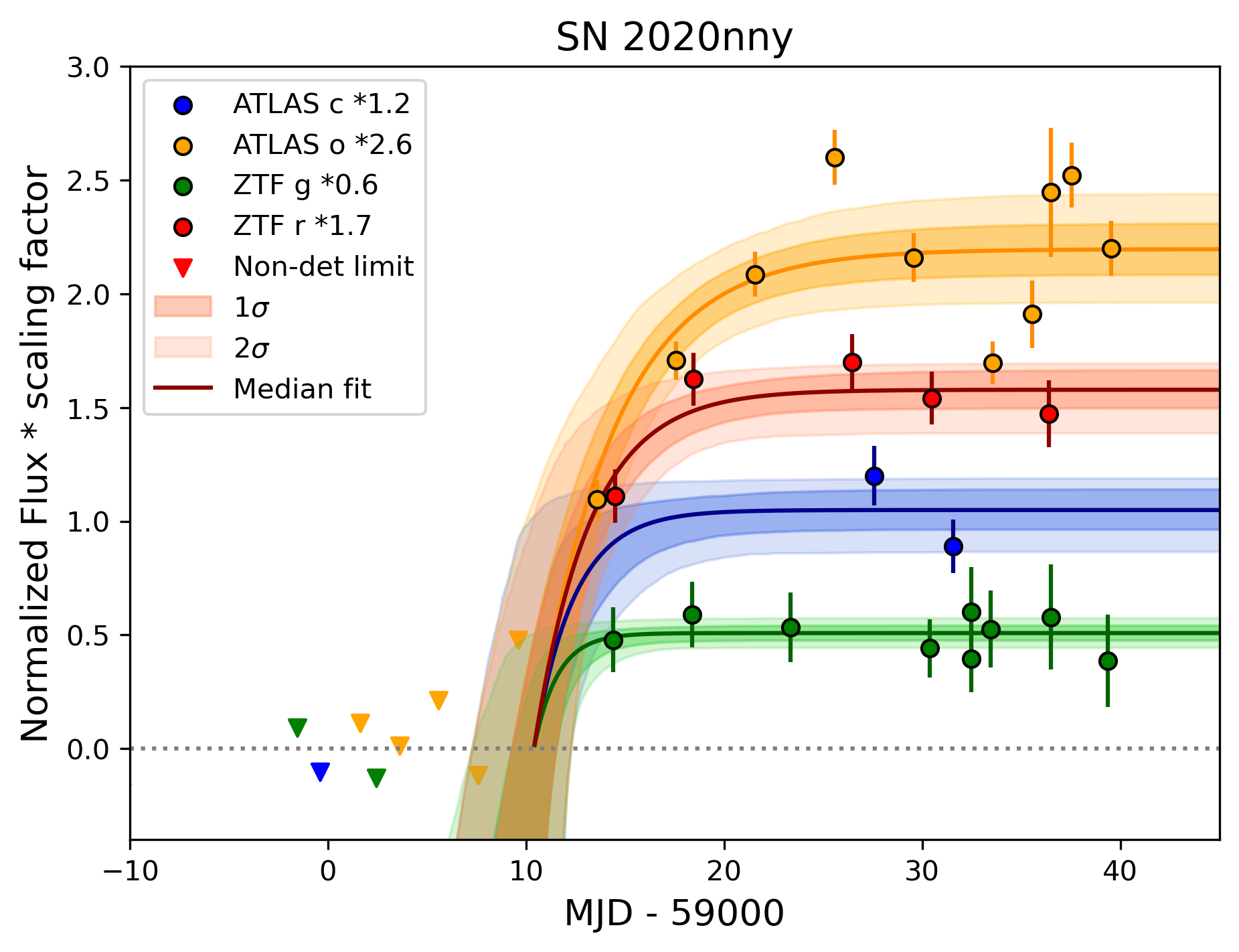}
    \includegraphics[width=0.49\textwidth]{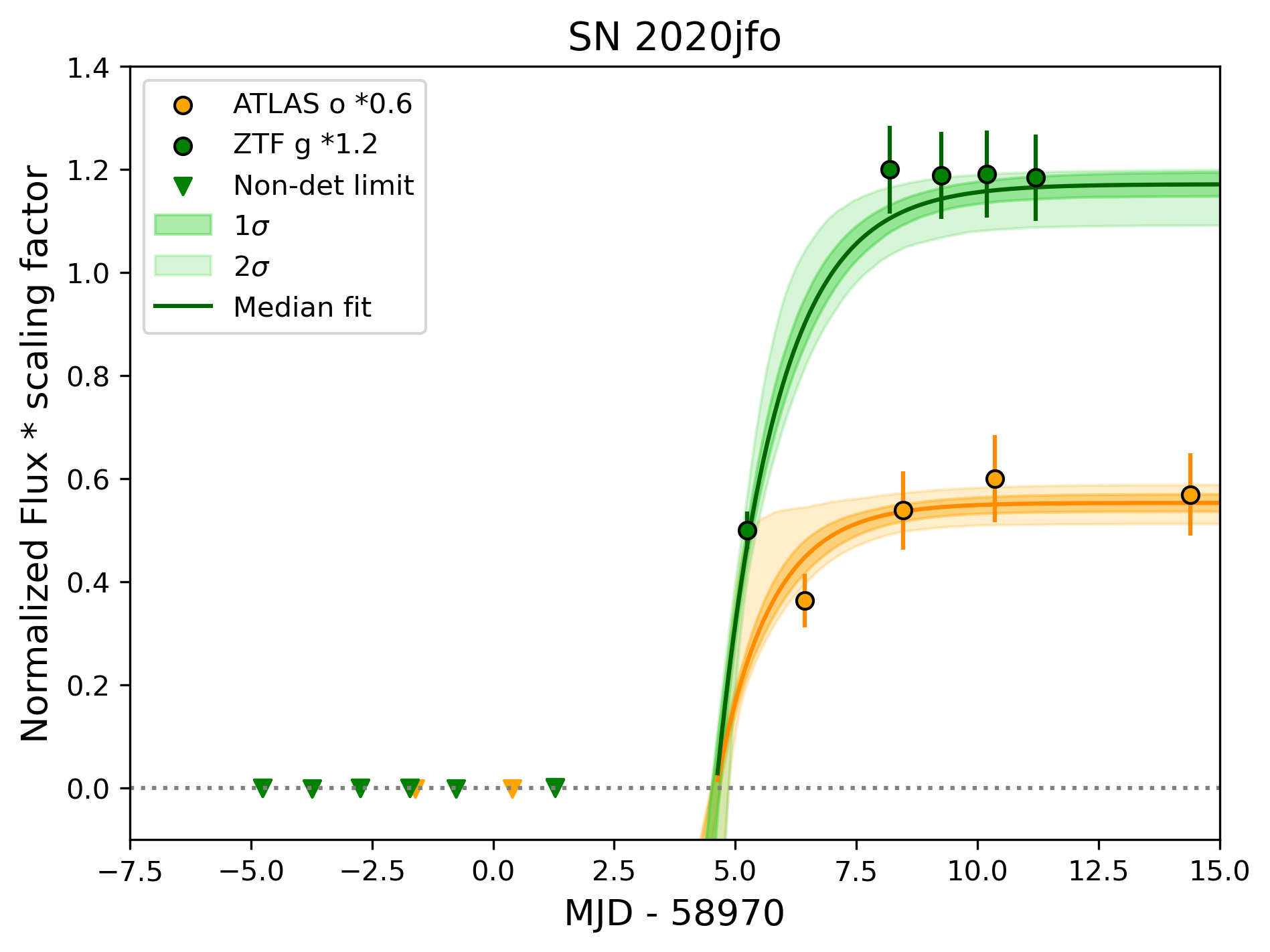}
\caption{Inverse exponential fit on the early observer-frame light curves of SN~2020nny  (left) and SN~2020jfo (right). The dark solid lines show the mean fit, and the shaded regions denote the 1$\sigma$ and 2$\sigma$ error regions. The flux values in the different bands were normalized and then rescaled for better clarity.}
\label{fig:2020nny_exponential}
\end{figure}

We implemented two different methods to estimate the epoch of first light for the SNe in our dataset. The first is to compute the midpoint between the epoch of the last nondetection and the epoch of the first detection (e.g., \citep{Gutierrez_2017}). The uncertainty associated with this value is equal to (MJD$_{\rm 1st~det}-$MJD$_{\rm non-det}$)/2. However, the accuracy of this method strongly depends on the cadence of the observations. 

To try to constrain the epoch of first light better, we implement a second method, which is to fit the early rise of the light curves. For SNe~Ia and SN 2020lao (Type Ic), we fit a power law to all bands simultaneously, considering the measurements from the last nondetection up to half of the maximum flux and their respective flux uncertainties \citep{GonzalezGaitan_2015},
\begin{equation}\label{eq:fireball}
    f(t) = \begin{cases}
        \alpha\left( t - t_{0} \right)^{n}, & t > t_{0} \\
        0, & t < t_{0},
    \end{cases}
\end{equation}
where $\alpha$ is a normalizing coefficient, $t_{0}$ is the time of first light, and $n$ the index of the power law. We use all available bands for these fits simultaneously, allowing the fit to choose a single $t_0$ value for all bands. We determined whether a band was viable if there were at least two observations between the estimated $t_0$ value and the half-maximum of the light curve. As an example, Figure \ref{fig:2020jhf_fireball} shows the power-law fitting result for SN 2020jhf. In this case, the ALTAS $c$ band did not have sufficient data to perform the fit, so it was not considered. The fits for the remaining 4 SNe Ia and the SN Ic-BL are shown in Figure~\ref{fig:risefits} of Appendix \ref{app}.

In the case of SNe~II, the times of first light are estimated by implementing the method presented by \citet{Geza_2023}. The flux in each photometric band was fitted simultaneously considering the inverse-exponential law
\begin{equation}
    f_{W}(t) = f_{m,W} \left[ 1 - {\rm exp} \left( -\frac{t - t_{0}}{t_{e,W}} \right) \right],
\end{equation}
where $t_{0}$ is the time of first light, $t$  the time, $f_{m,W}$ the peak flux, and $t_{e,W}$ the characteristic rise time in a specific band  $W$. The fitting procedure considers the depth of the nondetections and the fact that the rise time of an SN~II increases for longer wavelengths \citep{GonzalezGaitan_2015}. The fits are then connected through $t_{0}$, enabling us to obtain a single estimate considering simultaneously the different photometric bands. Figure \ref{fig:2020nny_exponential} demonstrates the exponential fitting of the light curves of SNe 2020nny and 2020jfo. Unfortunately, for SN~2020kjt it was not possible to estimate using the method of \cite{Geza_2023} owing to the large gap of ATLAS and ZTF photometry around the explosion time (see Figure \ref{fig:finders_lcs}). 

Finally, the SN~IIb 2020itj was not fit by any of these methods. Besides the main light-curve peak powered by the radioactive decay of $^{56}$Ni, some SNe~IIb show an additional peak at early epochs due to the cooling of the outer layers of the progenitor star that were superheaded and ejected right after the explosion. SN~2020itj early-time photometry does not completely show an initial bump, but some signatures: the first few points do not increase in brightness exponentially from zero flux. The first detection of SN~2020itj resembles the early cooling initial bump seen in some SNe~IIb (e.g., \citep{Richmond94}), which makes our method useless for determining the time of first light. 

The estimated times of first light considering the different implemented methods are summarized in Table~\ref{tab:tspec_texp}. Regarding the two SNe without rising light-curve fits, the midpoint estimate value is adopted. For the rest, we  used the early light-curve method. The final time of first light is highlighted in bold text. We note that in four cases (SNe~2020jhf, 2020kku, 2020kyx, 2020rlj), the early light-curve method gives an estimate that is earlier than the forced last nondetection. We associate this with the fact that the image was not deep enough to properly get a detection, but trust the early fit. These are indicated in italics in Table \ref{tab:tspec_texp}. As a comparison with already reported values in the literature, for SN 2020jfo we obtained $t_{0} = 58974.72\pm0.19$ MJD, which is in agreement within uncertainties with the estimated value from \citet{Ailawadhi_2023}, $t_{0} = 58973.1\pm1.6$, and also with the value used by \citet{Teja_2022}, $t_{0} = 58973.8 \pm 0.5$, estimated considering the mid epoch between the last nondetection and the first detection of ZTF. 


\subsection{Light-curve parameters}\label{sec:lc}

\begin{figure*}[!t]
  \centering  
  \includegraphics[width=\textwidth]{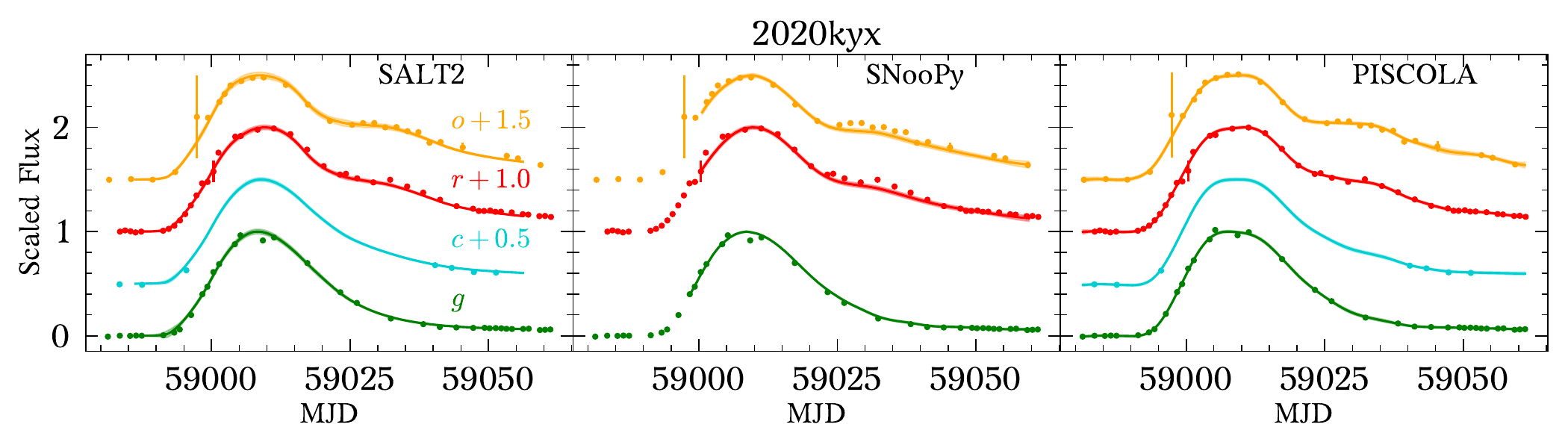}
  \cprotect\caption{Multicolor light-curve fits of SN Ia 2020kyx using SALT2 (left panel), SNooPy (\verb|max_model|; middle panel) and PISCOLA (right panel). The ATLAS $c$ light curve was not included in the SNooPy fit as it produces worse results.}
  \label{fig:snia_fits}
\end{figure*} 

\begin{table*}
\caption{SNe Ia light-curve parameters.}
\centering
\setlength{\tabcolsep}{2pt}

\resizebox{\textwidth}{!}{ 
\begin{tabular}{llcccc|ccc|cccc}
\hline\hline
&           & \multicolumn{4}{c}{SALT2}         & \multicolumn{3}{c}{SNooPy} & \multicolumn{4}{c}{PISCOLA}                \\ \hline
IAU Name & Type      & $t_{\rm max}$ & $B_{\rm max}$ & $x_1$     & $c$       & $t_{\rm max}$ & $B_{\rm max}$ & $s_{BV}$  & $t_{\rm max}$ & $B_{\rm max}$ & $\Delta m_{15}(B)$ & $(B-V)$   \\ \hline

2020jgl  & Ia & 
58993.18 $\pm$ 0.01 & 13.594 $\pm$ 0.002 & -0.095 $\pm$ 0.012 & 0.044 $\pm$ 0.002 &
58993.024 $\pm$ 0.343  & 13.605 $\pm$ 0.018 & 0.967 $\pm$ 0.031 &
58992.116 $\pm$ 0.245 & 13.565 $\pm$ 0.011 & 1.111 $\pm$ 0.018 & 0.054 $\pm$ 0.013  \\

2020jhf  & Ia & 
58991.77 $\pm$ 0.01 & 15.375 $\pm$ 0.002 & -0.922 $\pm$ 0.009 & 0.079 $\pm$ 0.001 &
58991.610  $\pm$ 0.359 & 15.270 $\pm$ 0.016 & 0.883 $\pm$ 0.032 & 
58991.283 $\pm$ 1.265 & 15.773 $\pm$ 0.139 & 1.137 $\pm$ 0.393 & 0.250 $\pm$ 0.130  \\

2020kku  & Ia        & 
58996.45 $\pm$ 0.10 & 18.944 $\pm$ 0.014 & -1.409 $\pm$ 0.109 & 0.104 $\pm$ 0.012 &
58996.294 $\pm$ 0.375 & 19.013 $\pm$ 0.021 & 0.836 $\pm$  0.033 & 
58995.836 $\pm$ 0.348 & 18.996 $\pm$ 0.022 & 1.027 $\pm$ 0.072 & 0.245 $\pm$ 0.046  \\

2020kyx  & Ia        & 
59008.48 $\pm$ 0.05 & 16.445 $\pm$ 0.013 & -0.547 $\pm$ 0.054 & 0.022 $\pm$ 0.009 &
59008.424 $\pm$ 0.354 & 16.533 $\pm$ 0.017 & 0.979 $\pm$ 0.032 & 
59007.233 $\pm$ 0.581 & 16.642 $\pm$ 0.021 & 0.914 $\pm$ 0.045 & 0.272 $\pm$ 0.031  \\

2020rlj  & Ia-91bg$^*$ & 
59088.72 $\pm$ 0.07 & 18.530 $\pm$ 0.014 & -2.814 $\pm$ 0.083 & 0.247 $\pm$ 0.011 &
59088.078 $\pm$ 0.354 & 18.682 $\pm$ 0.017 & 0.476 $\pm$ 0.032 & 
59087.487 $\pm$ 0.257 & 18.664 $\pm$ 0.025 & 2.285 $\pm$ 0.396 & 0.521 $\pm$ 0.036  \\
\hline
\end{tabular}}
$^*$SALT2 is not designed to fit 1991bg-like SNe Ia, thus the reported values are not meaningful but are shown for completeness. SNooPy fits were performed with the \textit{max\_model} and a 1991bg-like template was used for the respective objects. PISCOLA, being data-driven, has no issues fitting 1991bg-like SNe Ia.
\label{tab:SNeIa_parameters}

\end{table*}
\begin{table*}
\caption{SNe light-curve parameters obtained with PISCOLA.}
\centering

\resizebox{\textwidth}{!}{ 
\begin{tabular}{llcc|cc|cc|cc}
\hline\hline
&  & \multicolumn{2}{c}{ZTF-$g$} & \multicolumn{2}{c}{ATLAS-$c$} & \multicolumn{2}{c}{ZTF-$r$} & \multicolumn{2}{c}{ATLAS-$o$}   \\
\hline
IAU Name & Type & $t_{\rm max}$ & $m_{\rm max}$ & $t_{\rm max}$ & $m_{\rm max}$ & $t_{\rm max}$ & $m_{\rm max}$ & $t_{\rm max}$ & $m_{\rm max}$ \\
\hline
2020itj & IIb & 58980.204 $\pm$ 0.412 & 19.173 $\pm$ 0.039 & 58980.780 $\pm$ 0.383 & 18.950 $\pm$ 0.034 & 58981.753 $\pm$ 0.467 & 18.805 $\pm$ 0.030 & 58982.302 $\pm$ 0.649 & 18.873 $\pm$ 0.042 \\
2020jfo & II & 58980.193 $\pm$ 0.408 & 14.456 $\pm$ 0.017 & 58980.120 $\pm$ 0.411 & 14.423 $\pm$ 0.020 & 58980.147 $\pm$ 1.054 & 14.405 $\pm$ 0.029 & 58980.603 $\pm$ 2.244 & 14.479 $\pm$ 0.040 \\
2020jgl & Ia & 58992.846 $\pm$ 0.361 & 13.734 $\pm$ 0.010 & 58994.941 $\pm$ 0.479 & 13.700 $\pm$ 0.014 & 58994.659 $\pm$ 0.397 & 13.654 $\pm$ 0.014 & 58994.863 $\pm$ 0.324 & 13.938 $\pm$ 0.003 \\
2020jhf & Ia & 58992.839 $\pm$ 0.352 & 15.342 $\pm$ 0.004 & 58996.254 $\pm$ 0.941 & 15.392 $\pm$ 0.012 & 58992.631 $\pm$ 0.927 & 15.342 $\pm$ 0.011 & 58991.685 $\pm$ 0.850 & 15.525 $\pm$ 0.013 \\
2020kjt & II & --  -- & --  -- & --  -- & --  -- & --  -- & --  -- & --  -- & --  --  \\
2020kku & Ia & 58995.930 $\pm$ 0.345 & 19.121 $\pm$ 0.019 & 58996.682 $\pm$ 0.563 & 18.987 $\pm$ 0.045 & 58996.923 $\pm$ 0.496 & 18.957 $\pm$ 0.019 & 58996.792 $\pm$ 0.814 & 19.048 $\pm$ 0.054 \\
2020kyx & Ia & 59007.794 $\pm$ 0.815 & 16.689 $\pm$ 0.011 & 59009.449 $\pm$ 1.177 & 16.600 $\pm$ 0.016 & 59010.178 $\pm$ 1.115 & 16.643 $\pm$ 0.012 & 59009.297 $\pm$ 1.569 & 16.820 $\pm$ 0.017 \\
2020lao & Ic-BL & 59002.586 $\pm$ 0.240 & 17.368 $\pm$ 0.016 & 59003.107 $\pm$ 0.364 & 17.166 $\pm$ 0.033 & 59004.807 $\pm$ 0.839 & 17.179 $\pm$ 0.018 & 59005.921 $\pm$ 0.751 & 17.313 $\pm$ 0.027 \\
2020nny & II & 59023.426 $\pm$ 2.037 & 20.294 $\pm$ 0.073 & --  -- & --  -- & 59024.434 $\pm$ 2.356 & 20.027 $\pm$ 0.061 & 59028.879 $\pm$ 1.506 & 18.863 $\pm$ 0.035 \\
2020rlj & Ia-91bg & 59087.800 $\pm$ 0.258 & 18.759 $\pm$ 0.014 & 59088.663 $\pm$ 0.314 & 18.502 $\pm$ 0.019 & 59090.534 $\pm$ 0.647 & 18.309 $\pm$ 0.013 & 59091.718 $\pm$ 0.938 & 18.352 $\pm$ 0.028 \\
\hline
\end{tabular}}
The reported values are in the observer frame, and no corrections are applied. The fits of the SNe II are not completely reliable. For SN 2020jgl, the reported $g$-band and $r$-band values are from SDSS-$g$ and SDSS-$r$, respectively. The values for SN 2020kjt are not reported, as the light-curve rise is not covered.
\label{tab:SNe_parameters}
\end{table*}

To estimate the light-curve parameters of the SNe presented in this work, we use different SN light-curve fitters. SNe Ia are fitted with SALT2 (\citealp{Guy_2005, Guy_2007}, using the implementation in \texttt{sncosmo} \citealp{Barbary_2016_sncosmo}), SNooPy \citep{Burns_2011_SNooPy}, and the updated version \texttt{2.0.0} of PISCOLA \citep{2022MNRAS.512.3266M}. SALT2 uses a template-driven approach, modelling the spectral energy distribution (SED) of an SN Ia as a function of three light-curve parameters: a global flux scaling ($x_0$), time-stretch ($x_1$), and color at maximum light ($c$). SNooPy follows a similar approach, modelling the SED of an SN~Ia as a function of color-stretch \citep[$s_{BV}$;][]{Burns_2014}, plus other parameters depending on the exact model used. PISCOLA uses a data-driven approach for fitting transient light curves, thus not depending on light-curve parameters, based on Gaussian Processes (GPs; \citep{Rasmussen2006}). We report the results of the SNe~Ia fits in Table~\ref{tab:SNeIa_parameters}.

The SALT2 template is only trained with normal SNe Ia, so its fits for the SN 1991bg-like SN~2020rlj are not completely reliable. SNooPy fits were performed using the \verb|max_model|, applying the $K$-correction versions of the $H3$ template \citep{Hsiao_2007} for normal SNe Ia and the 91bg template for SN 1991bg-like SNe Ia. Only epochs between $-30$ and $+50$ days with respect to the optical peak were selected. For SN 2020kyx, we removed the ATLAS $c$ band to produce better results, while for SN 2020jgl we removed both ZTF bands for the same reason. PISCOLA fits were performed in flux space using the GP with a Mat\'ern-5/2 kernel (the default procedure), except for SN~2020itj and SN~2020kku, for which an Exponential-Squared kernel was used, and SN~2020nny, which was fitted in logarithmic space, producing better results.

As an example, in Figure~\ref{fig:snia_fits} we show the results of the SN Ia 2020kyx fit using the three light-curve fitters. For the directly comparable parameters, such as the time and magnitude of rest-frame $B$-band peak ($t_{\rm max}$ and $B_{\rm max}$, respectively), we see that there are some small differences between the fitters. SALT2 and SNooPy provide similar values of $t_{\rm max}$, while PISCOLA estimates the maximum to be a day earlier. In the case of $B_{\rm max}$, all the fitters produce different results, but within $\sim0.1$\,mag. The different results produced by PISCOLA can be attributed to the relatively flat peaks (see the right panel of Figure~\ref{fig:snia_fits}), although this artefact is a consequence of the photometry.
Similar fits for the remaining four SNe~Ia are shown in Figure \ref{fig:snialc} of Appendix \ref{app}.

In addition to the fits of the SNe Ia, given the versatile nature of PISCOLA, we fit the light curves of the five CC~SNe. In Table~\ref{tab:SNe_parameters}, we report the time of maximum light and the peak magnitude for each of the ZTF and ATLAS bands. Note that the values for the SNe~II are not completely reliable owing to the relatively low quality of the photometry. Furthermore, we do not report values for SN 2020kjt, as the light-curve rise was not observed. All fits are shown in Figure \ref{fig:ccsnpiscola} of Appendix \ref{app}.


\begin{figure*}[!t]
\minipage{\textwidth}
\includegraphics[width=0.5\linewidth]{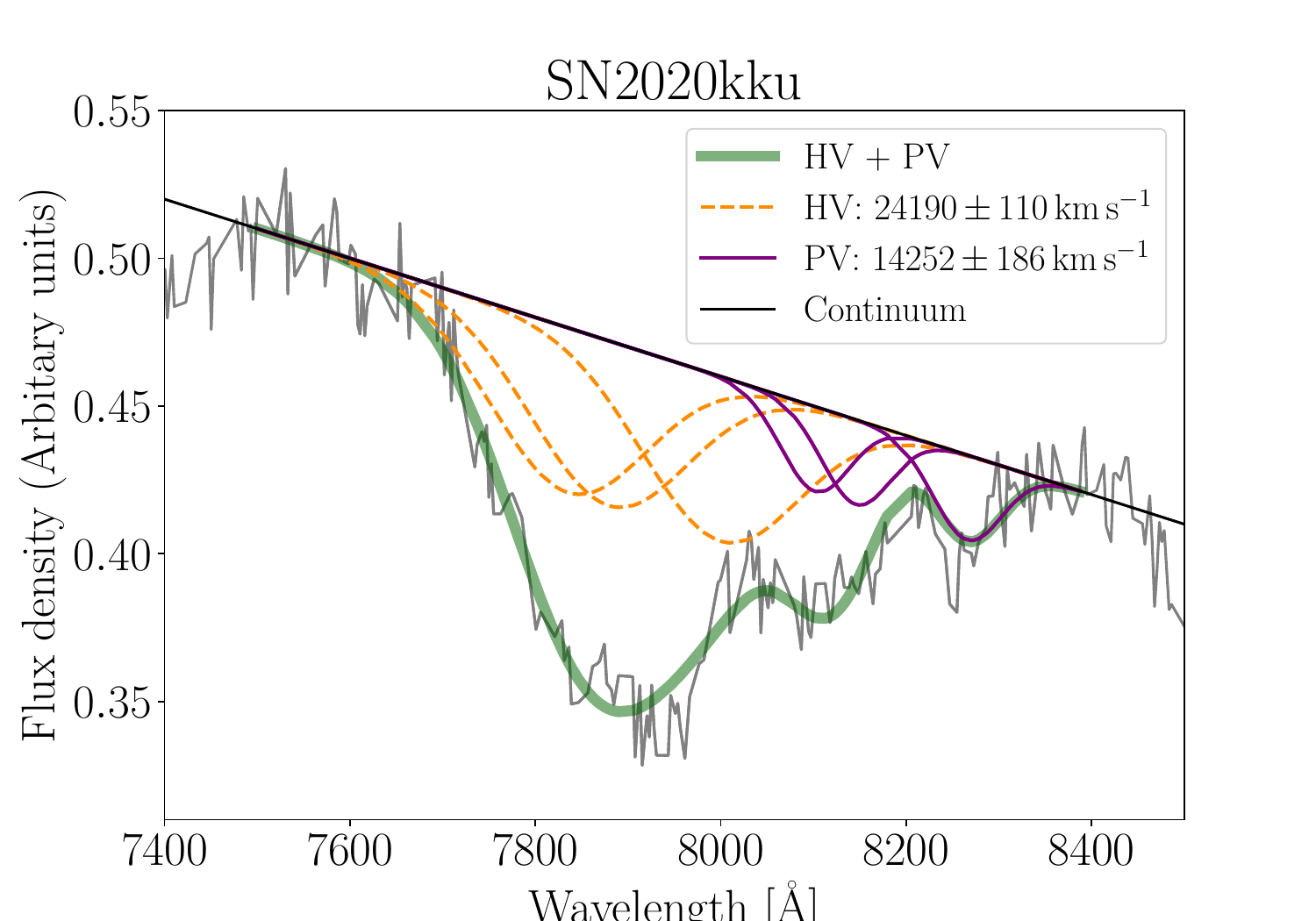}
  \hfil
\includegraphics[width=0.5\linewidth]{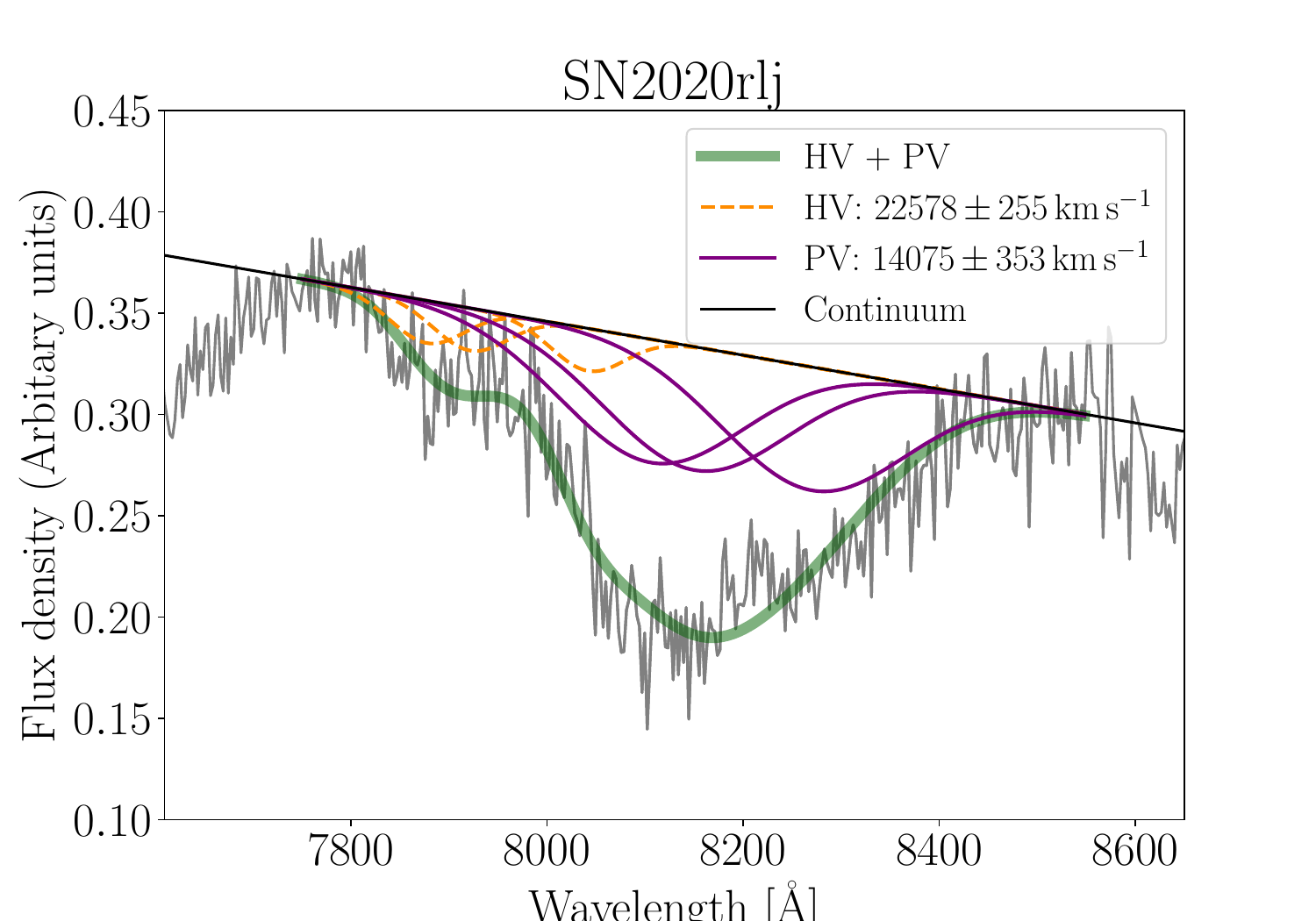}
\caption{Spectral fitting to the \ion{Ca}{II} IR line of SN 2020kku (left) and SN 2020rlj (right). The orange dashed-line Gaussians account for the \ion{Ca}{II} IR triplet of the high-velocity component, while the purple continuous-line Gaussians for the \ion{Ca}{II} IR triplet of the photospheric velocity. The thick green line is the sum of the high and photospheric velocity components, and the black continuous line indicates the continuum.}
\label{fig:CaIR_fits}
\endminipage
\end{figure*}

\subsection{Epochs of the GTC spectra}\label{subsec:tspec}

We determined the epochs of the spectra with respect to the estimated time of first light by using the two methods: midpoint and early fit (see Section \ref{subsec:t0}). Uncertainties in the epochs are estimated accounting for the uncertainty in the estimated time of first light and also the uncertainty of the time at which the spectrum was taken, where we take the time of the middle of the exposure (ranging from 500 to 900\,s) with an uncertainty of half of the exposure time. Epochs are then transformed to and reported in the rest frame in Table \ref{tab:tspec_texp}. The final spectrum dates are highlighted in bold text in the rightmost columns. 

From the 10 objects in our sample, we ended up with only two within 48\,hr from the estimated first light: SN~2020jfo at 1.8 days and SN~2020lao at 2.0 days. Even when SN~2020jfo was one of the two objects that we decided to trigger that had a nondetection earlier than 48\,hr (criterion (i) in Section \ref{sec:strat}), it ended up being our earliest spectrum. Being a Type II SN, the rise time is typically quite short (about a week; see \citep{2015MNRAS.451.2212G}), and given the redshift of its host galaxy, the discovery absolute magnitude was $-14.8$\,mag, consistent with a very young SN. SN~2020lao was a broad-lined SN~Ic at $-15.9$\,mag at discovery. Our spectrum 2 days after the estimated first light is possibly one of the earliest spectra ever taken of an SN Ic-BL with no associated gamma-ray burst (GRB); from the compilation of \cite{2024arXiv241111503F} of SNe~Ic-BL with and without GRB association, the earliest spectrum is of PTF10vgv \citep{2019A&A...621A..71T} obtained 2.33 days from explosion.

\begin{table}[t]
\caption{Rest-frame epochs of the spectra with respect to maximum light (in the $B$ band) and rise time from first light to peak.}
\centering
\setlength{\tabcolsep}{3pt}
\begin{tabular}{lccc}
\hline \hline
SN name & Type & Epoch from maximum & Rise time \\
\hline
2020jgl & Ia & $-$16.16$\pm$0.01 & 18.74$\pm$1.14 \\
2020jhf & Ia &$-$12.65$\pm$0.01 & 16.75$\pm$0.20 \\
2020kku & Ia & $-$7.64$\pm$0.10 & 13.51$\pm$0.87 \\
2020kyx & Ia &$-$13.85$\pm$0.05 & 17.22$\pm$0.33 \\
2020rlj & Ia-91bg &$-$10.10$\pm$0.07 & 13.86$\pm$0.42 \\
\hline
2020lao & Ic-BL & $-$7.16$\pm$0.23 & 9.16$\pm$0.29 \\
\hline
2020itj & IIb & $-$8.91$\pm$0.40 & 18.4$\pm$0.79 \\
\hline
\label{table:epsp}
\end{tabular}
\end{table}

Following these two earliest spectra, we have the five SNe Ia, three with epochs from two to four days (SN~2020jgl at 2.6\,d, SN~2020kyx at 3.4\,d, and SN~2020rlj at 3.8\,d) and two more with spectra from four to six days after the estimated time of first light (SN~2020jhf at 4.1\,d and SN~2020kku at 5.9\,d). SN~2020jhf was the other candidate for which we did not follow the first criterion. In Figure \ref{fig:finders_lcs}, we show a forced-photometry detection in the ZTF $i$ band that, if available when we had to trigger, would have made us discard this object. On the other hand, it would have been a good target one day earlier, and assuming the trigger executed at the same time of the night, we would have obtained a spectrum 3.1\,d after first light, the fourth in our sample.

The SN~2020itj spectrum was obtained nine days after first light. The early light curve of this SN~IIb does not perfectly follow an exponential decline toward the time of first light, but instead, there is a resemblance to the post-shock-breakout (post-SBO) cooling tail typical of some SNe~IIb (see Figure \ref{fig:finders_lcs}). We discuss this further in Section \ref{sec:itj}.

The remaining two objects, SN~2020nny and 2020kjt, have epochs of sixteen and sixty days, respectively, after the estimated first light. We consider these failures of our method. In both cases, the SN was of Type II. SN~2020kjt was discovered during the plateau phase at 20.45\,mag, close to the median magnitude limit of the ZTF survey. Fluctuations in the difference photometry made \cite{2020TNSTR1395....1F} report the detection with a close $r$-band nondetection of 20.65\,mag 21\,hr earlier, but they both turned out to be forced-photometry detections. Similarly, SN~2020nny was discovered at what turned out to be the peak of the light curve at 19.1\,mag in $r$, with a $3\sigma$ nondetection of 20.3\,mag in the $g$~band reported 27\,hr earlier, which also resulted in a detection with forced photometry.

In addition to the epochs with respect to the estimated time of first light, we also report the epoch with respect to maximum light for all SNe Ia and stripped-envelope SNe. For SNe Ia, we used the epoch of the peak as measured with SALT2 (see Section \ref{sec:lc}), while for the SN Ic-BL and the SN IIb, we used the value from PISCOLA measured in the $g$ band. Table \ref{table:epsp} reports the resulting rest-frame epochs. SN~2020jgl is our SN~Ia with the earliest spectrum from maximum light, at an epoch of $-$16.2 days. The following two earliest spectra, from SNe 2020jhf and 2020ky, are at $-$13 and $-$12 days, which are also significantly early, but not as early as we had hoped to achieve with our programme. Spectra for SNe 2020rlj and 2020kku are at $-$10 and $-$7 days, respectively, which are routinely obtained by spectroscopic SN follow-up programmes such as ePESSTO+ \citep{2015A&A...579A..40S}. In Table \ref{table:epsp}, we also report the corresponding rise times for each SN. For SNe Ia, these range from 13 to 19 days, which when compared to tyhe typical distribution from \cite{2020ApJ...902...47M} suggests that our estimated times of first light may not accurately correspond to the actual explosion epochs. 

From all the above, we conclude that the configuration of the discovery survey telescope used in this work (P48, 1.22\,m diameter), even with a great wide-field camera (ZTF, 47 sq. deg), is not deep enough to cover the early epochs at the redshift range of our sample ($z>0.01$). At these redshifts, obtaining spectra of young SNe (within 48\,hr of explosion) is impossible if their brightness during the earliest epochs is significantly below the survey's limiting magnitude.
This is further discussed in Section \ref{sec:conc}.

\subsection{Expansion ejecta velocities}\label{sec:vel}

\begin{table}[t]
\caption{Expansion Velocities in km\,s$^{-1}$.}
\resizebox{\columnwidth}{!}{
\begin{tabular}{lccccc}
\hline \hline
\multicolumn{6}{c}{Type Ia} \\ \hline
 & \multicolumn{1}{c}{\ion{Ca}{II} IR triplet} & \multicolumn{1}{c}{\ion{Ca}{II} IR triplet - HV} & \multicolumn{1}{c}{\ion{Ca}{II} H\&K} & \multicolumn{1}{c}{\ion{Si}{II} $\lambda$5792} & \multicolumn{1}{c}{\ion{Si}{II} $\lambda$6533} \\ \cline{2-6}
2020jgl & 35,800 $\pm$ 420  & $-$              &  5,700 $\pm$ 1,000 & 18,800 $\pm$ 370 & 21,200 $\pm$ 230 \\
2020jhf & 12,100 $\pm$ 170  & 20,100 $\pm$ 120 & 18,000 $\pm$ 690 & 11,600 $\pm$ 400 & 12,600 $\pm$ 140 \\
2020kku & 14,300 $\pm$ 190  & 24,200 $\pm$ 110 & 24,200 $\pm$ 770 & 13,800 $\pm$ 370 & 15,500 $\pm$ 770 \\
2020kyx & 22,600 $\pm$ 1,200 & $-$              & 28,500 $\pm$ 840 & 13,200 $\pm$ 200 & 15,000 $\pm$ 190 \\
2020rlj & 14,100 $\pm$ 360  & 22,600 $\pm$ 260 & 15,600 $\pm$ 230 & 13,600 $\pm$ 170 & 14,200 $\pm$ 250 \\ \hline
\multicolumn{6}{c}{Type II} \\ \hline
 & \multicolumn{1}{c}{H$\alpha$} & \multicolumn{1}{c}{H$\beta$} & \multicolumn{1}{c}{\ion{Na}{I} D $\lambda$5893 } & \multicolumn{1}{c}{\ion{Fe}{II} $\lambda$5018} & \multicolumn{1}{c}{\ion{Fe}{II} $\lambda$5169} \\ \cline{2-6} 
2020kjt &  5,900 $\pm$ 150  &  4,500 $\pm$ 40  &  3,200 $\pm$ 90  & 3,100 $\pm$ 40 & 3,100 $\pm$ 50 \\
2020nny &  9,300 $\pm$ 1000 &  9,000 $\pm$ 60  &  9,000 $\pm$ 10   & 6,800 $\pm$ 90 & 7,200 $\pm$ 170 \\ 
\hline
\multicolumn{6}{c}{Type IIb} \\ \hline
 & \multicolumn{1}{c}{H$\alpha$} & \multicolumn{1}{c}{\ion{He}{I} $\lambda$5876} & \multicolumn{1}{c}{\ion{He}{I} $\lambda$7065}  \\ \cline{2-6}
2020itj & 10,600 $\pm$ 320  &  6,700 $\pm$ 100 & 4,900 $\pm$ 20  \\
\hline
\label{table:velocities}
\end{tabular}}
\end{table}

We have estimated the expansion velocities in the GTC spectra of our ten SNe. We present here the results for eight of them, while those for SN~2020jfo were given by \cite{Ailawadhi_2023} and those for SN~2020lao will be presented by Stritzinger et al. (in prep.).

In the case of the SNe~Ia, a modfied version of the {\sc Spextractor} \citep{2019PhDT.......134P} code was used to measure velocities and pseudoequivalent widths (pEWs). This modified version presented by \citet{2020ApJ...901..154B} has enabled the downsampling of spectral information while ensuring the conservation of the number of photons, in order to reduce computational cost. Additionally, adjustments were made to produce a more representative GP regression model and smoothing for a given spectrum so that it returns reliable measurements of velocities and pEWs. Thus, the code calculates those velocities by converting the wavelength shifts of absorption lines into velocities using the relativistic Doppler formula provided in Equation 6 of \citet{2006AJ....131.1648B}. The ejection velocity is determined by measuring the flux minima of the absorption lines of the spectral features with respect to its rest wavelengths, which are provided and outlined in Table 5 of \citet{2013ApJ...773...53F}. In our analysis, we focused on specific spectral features that are easily identifiable and isolated in the spectra of SNe Ia. These features, given in Table \ref{table:velocities}, include the absorption lines of the \ion{Ca}{II} NIR triplet, \ion{Ca}{II} H\&K, \ion{Si}{II} $\lambda$6533, and \ion{Si}{II} $\lambda$5792, as well as their calculated expansion velocities.

There are three cases (SNe~2020jhf, 2020kku, and 2020rlj) where the \ion{Ca}{II} NIR triplet region (8498\,\AA, 8542\,\AA, 8662\,\AA) presents high-velocity features on top of the photospheric features. To calculate the contribution of both regimes, we fit six Gaussians at the same time: three to estimate the photospheric velocity (PV) and three for the high-velocity (HV) component. Figure~\ref{fig:CaIR_fits} shows an example for SNe 2020kku and 2020rlj. We force the $\sigma$ value and the amplitude of the three Gaussians of the PV components to be equal, and we do the same for the HV components. In addition, a minimum difference of 2000\,km\,s$^{-1}$ between PV and HV was set. All measured velocities are presented in the first rows of Table \ref{table:velocities}.

In the case of CC SNe, Type II and IIb, we computed the relativistic velocities using an alternative method independent of {\sc Spextractor}. First, we choose two neighboring regions for each spectral line, one on each side of the line.  Within these regions, the upper and lower wavelength limits were determined by analyzing the derivative of the flux with respect to the wavelength. Specifically, the numerical derivative was calculated across the region of interest, and the wavelength corresponding to the maximum change in the slope (the maximum or minimum of the derivative) was identified. This point represents the location where the flux transitions most sharply, helping to define the boundaries of the feature. Once the wavelength range for the feature is established, a pseudocontinuum is defined between the two limits and the spectrum is normalized. Finally, a Gaussian profile is fitted to the normalized flux to determine the central wavelength of the feature. Using the Doppler formula with relativistic corrections, we computed the velocities relative to the known rest wavelength of each spectral line. The same process was performed 100 times to determine the uncertainties, applying a shift to the feature limits. The standard error is calculated as the dispersion in the 100 additional velocities. We report the H$\alpha$ and H$\beta$ features for both H-rich (Type II) and H-poor (Type IIb) CC~SNe. Additionally, we report \ion{Fe}{II} lines together with \ion{Na}{I} for SNe II. For the SN IIb 2020itj we report the \ion{He}{I} lines (5876 and 7076\,\AA). All velocities are presented in Table~\ref{table:velocities}.


\section{Earliest spectra}\label{sec:earliest}

From the ten objects in our sample, we succeeded in two cases to obtain a spectrum within 48\,hr from the estimated time of first light, both CC~SNe. Further study of these two objects is left for dedicated publications; the GTC-OSIRIS spectrum of SN~2020jfo has already been presented by \cite{Ailawadhi_2023}, and the spectrum of SN~2020lao, together with a full characterization of the SN, will be presented elsewhere (Stritzinger et al., in prep.).  

The third object is SN~2020jgl, the earliest SN~Ia obtained within our sample at +2.5 days from first light. In order to assess how early is our earliest SN~2020jgl spectrum, we compared it to publicly available spectra of  other SNe~Ia discovered early. We obtained spectra from the Weizmann Interactive Supernova Data Repository (WISeREP\footnote{\href{https://www.wiserep.org/}{https://www.wiserep.org/}}) of several SNe~Ia for which early spectra ($<-14$ days) have been reported:
SN~2009ig \citep{2012ApJ...744...38F},
SN~2011fe \citep{2016ApJ...820...67Z},
SN~2012cg \citep{2012ApJ...756L...7S},
SN~2013dy \citep{2013ApJ...778L..15Z},
SN~2015F \citep{2017MNRAS.464.4476C},
SN~2017cbv \citep{2017ApJ...845L..11H,2022arXiv220707681B},
SN~2019np \citep{2022arXiv220707681B},
SN~2019ein \citep{2020ApJ...897..159P}, 
SN~2019yvq \citep{2020ApJ...898...56M,2021ApJ...919..142B}, 
SN~2020nlb \citep{2024A&A...685A.135W},
SN~2021hpr \citep{2022PASP..134g4201Z},
SN~2021aefx \citep{2022ApJ...933L..45H},
SN~2023bee \citep{2024ApJ...962...17W}, 
and SN~2024epr \citep{Hoogendam25}.

\begin{table}[t]
\caption{Rest-frame epochs of the spectra with respect to maximum light (in the $B$ band) and rise time from explosion to peak.}
\centering
\resizebox{0.6\textwidth}{!}{
\setlength{\tabcolsep}{3pt}
\begin{tabular}{lccccc}
\hline \hline
SN name & Galaxy & redshift & Epoch & method & Ph. velocity\\
\hline
2009ig   & NGC 1015 & 0.008769 & $-$14.4 & Poly   & 21,980$\pm$950 \\
2011fe   & NGC 5457 & 0.000804 & $-$15.9 & Poly   & 15,910$\pm$30 \\
2012cg   & NGC 4424 & 0.001458 & $-$14.8 & Poly   & 22,440$\pm$50 \\
2013dy   & NGC 7250 & 0.003889 & $-$16.0 & Poly   & 17,810$\pm$280 \\
2015F    & NGC 2442 & 0.004890 & $-$14.4 & Poly   & 12,940$\pm$380 \\
2017cbv  & NGC 5643 & 0.003999 & $-$17.5 & SNooPy & 22,450$\pm$6,660 \\ 
2019np   & NGC 3254 & 0.004520 & $-$16.0 & SNooPy & 15,140$\pm$600   \\
2019ein  & NGC 5353 & 0.007755 & $-$14.0 & SALT2  & 23,700$\pm$30\\
2019yvq  & NGC 4441 & 0.009080 & $-$14.4 & Poly   & 21,070$\pm$480 \\
2020nlb  & NGC 4382 & 0.002432 & $-$16.1 & SNooPy & 14,990$\pm$440 \\
2021hpr  & NGC 3147 & 0.009346 & $-$14.2 & Poly   & 21,160$\pm$630  \\
2021aefx & NGC 1566 & 0.005017 & $-$16.6 & Poly   & 27,840$\pm$260 \\
2023bee  & NGC 2708 & 0.006698 & $-$15.7 & SALT3  & 24,380$\pm$320 \\
2024epr  & NGC 1198 & 0.005310 & $-$16.7 & SNooPy & 20,570$\pm$500 \\
\hline
\label{table:earlysp}
\end{tabular}}
\end{table}

\begin{figure}[t]
\centering
    \includegraphics[width=0.6\textwidth]{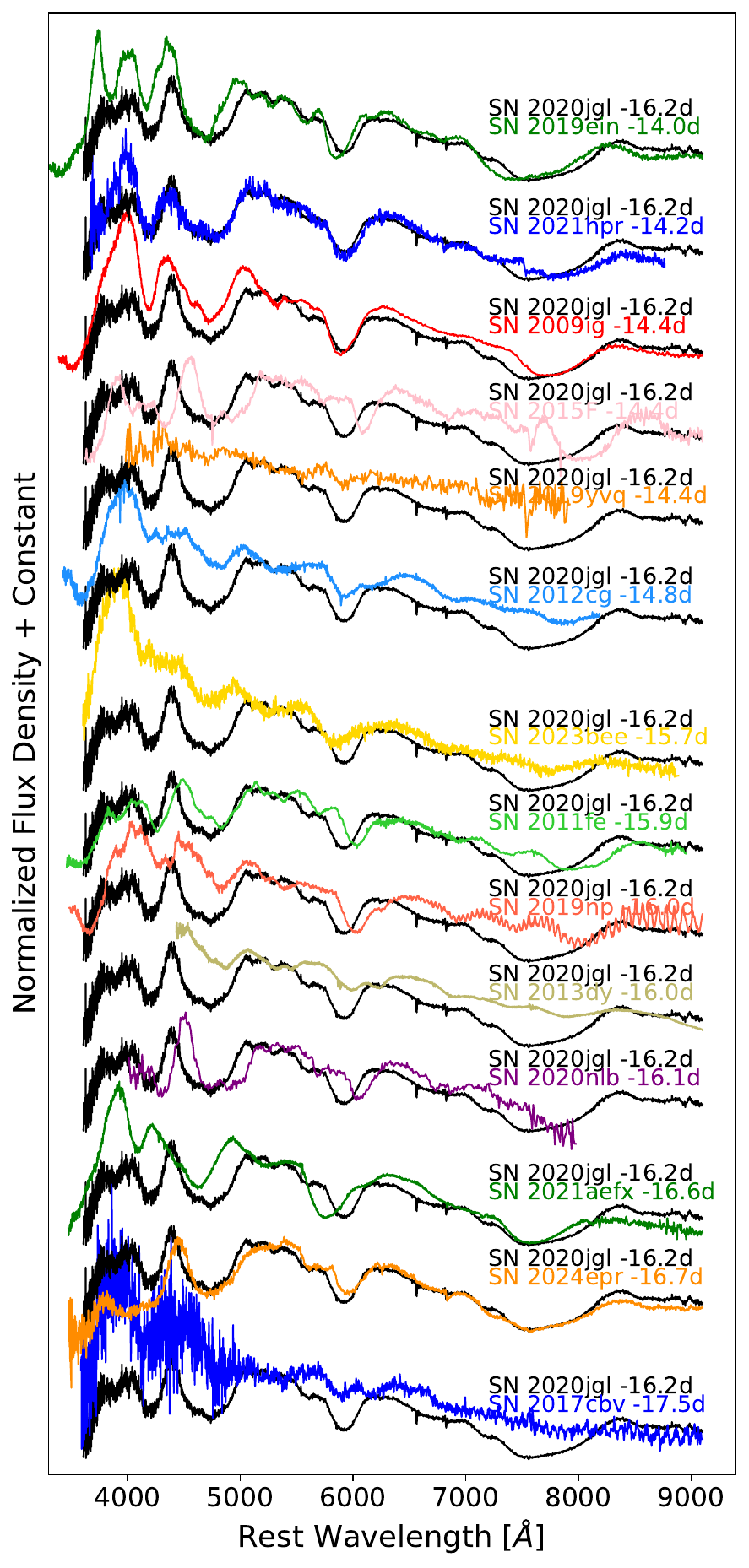}
    \caption{SN~2020jgl spectrum, the earliest presented in this work, compared to that of other SNe Ia with reported very early ($<-14$ days) spectra.} 
\label{fig:comparisonjgl}
\end{figure}

Table \ref{table:earlysp} lists all SNe, their host galaxies and redshift from NED, and epochs obtained from the references above, using different methods, or estimated from maximum-light and spectral epochs. We also run {\sc Spextractor} to measure the Si II $\lambda$6355 velocity, which is reported in the Table. In Figure \ref{fig:comparisonjgl}, we show all these early spectra, sorted by epoch, and compare them to our earliest spectrum of SN~2020jgl, which is at an estimated epoch of $-16.2$ days. Looking at the Figure, it is very clear that SNe Ia are quite heterogeneous in terms of Si II velocity, which is a proxy for the photospheric velocity. SN~2021aefx is the SN with the highest velocity, more than 27,000\,km\,s$^{-1}$ at $-16.6$ days, and SN~2015F is the one with the lowest velocity, about 12,940\,km\,s$^{-1}$ at $-14.4$ days. While SN~2020jgl had a photospheric velocity of 21,200$\pm$230\,km\,s$^{-1}$ (see Section \ref{sec:vel}), there are 8 SNe with higher and 6 with lower velocities. SN~2024epr strongly resembles SN~2020jgl in the optical range redder than 4500\,\AA\ at those epochs. Also remarkable is the large Ca NIR pEW of SN~2020jgl; only SN~2019ein and SN~2024epr exhibit larger values.

A more quantitative analysis of the early epochs of SNe Ia will be a matter for a dedicated study. Here, we just conclude that the SN~2020jgl GTC spectrum presented in this work contributes to increasing the sample of early SN~Ia spectra at epochs $<-14$ days from the epoch of maximum light.


\section{Summary and conclusions}\label{sec:conc}

We have presented the results of a campaign using the Optical System for Imaging and Low-Intermediate-Resolution Integrated Spectroscopy (OSIRIS) mounted on the 10.4\,m Gran Telescopio  Canarias (GTC) with the goal of obtaining spectroscopy of infant SNe within 48\,hr of first light. We successfully acquired data for ten SNe, spanning from 1.2 to 59.4 days after first light. Although only two of the ten spectra were obtained within the first 48\,hr, half were captured within the first four days. Possible reasons for this success ratio are discussed below.

Supplementing the GTC spectra with public photometry from the ZTF and ATLAS surveys, we performed a comprehensive characterization of these ten SNe, providing estimates for the time of first light, time of maximum light, light-curve parameters, and host-galaxy extinction. Using these parameters, we determined the epochs of our spectra relative to both the time of first light and maximum light. Supernova typing was conducted using SNID, identifying five Type Ia, three Type II, one Type IIb, and one Type Ic-BL. The photospheric velocities of the main spectral features were measured and reported for all ten spectra in our GTC sample. A short, more in-depth analysis of two SNe (SN IIb 2020itj and SN Ia 2020jgl), including photometry and spectroscopy from dedicated follow-up campaigns, is presented in Appendix \ref{app:sne}.

Although follow-up observations were triggered within 24\,hr of discovery, several SNe in our sample were ultimately found to be significantly older than initially expected. As shown in Figure \ref{fig:finders_lcs}, forced photometry revealed that some epochs previously classified as nondetections actually contained faint, overlooked flux consistent with early-time emission. These revised detections shifted the inferred explosion dates to earlier times, effectively increasing the estimated age of the SNe at the time of spectroscopy.

In our sample, we identified three Type II SNe. Of these, only one (SN~2020jfo) was observed within 30\,hr of explosion. While this spectrum was taken at an early phase, it does not show the flash-ionized emission lines seen in some Type II SNe shortly after explosion (e.g., \cite{2022ApJ...924...15J,2023ApJ...952..119B}). Instead, it exhibits a distinct bump or ``ledge-like" feature in the 4400–-4800\,\AA\ range, which has been attributed to blueshifted, Doppler-broadened \ion{He}{II} $\lambda$4686 emission. This feature is not associated with prior or ongoing interaction with circumstellar material \citep{2023A&A...677A.105D}. The other two Type II SNe were already several weeks old at the time of observation and, as such, did not meet the early-time goals of this program. Similarly, the Type IIb SN~2020ijt had a spectrum taken more than nine days after explosion, making it difficult to place meaningful constraints on the progenitor. In contrast, the Type Ic-BL SN 2020lao was observed just two days after explosion, providing one of the earliest spectra ever obtained for this class of events. For the five Type Ia SNe in our sample, spectra were obtained between two and six days after explosion. These observations revealed broad, high-velocity features characteristic of Type Ia SNe at these early epochs.

Because many of the spectra were obtained at later phases than initially anticipated, placing meaningful constraints on progenitor mass loss, color, or other physical properties becomes significantly more difficult. Early-time observations are crucial for detecting transient features (such as flash-ionized lines or rapid color evolution) that directly trace the immediate preexplosion environment and the outer layers of the progenitor star. Without such data, much of this valuable diagnostic information is lost, limiting our ability to probe the progenitor systems and their recent mass-loss histories.

This pilot study highlights several key insights. Notably, CC SNe appear to be more affected by forced-photometry corrections than Type Ia SNe. This may stem in part from the lower intrinsic luminosity of CC SNe, which makes them more susceptible to background noise. Additionally, host-galaxy contamination plays a significant role: in most cases, the SNe in our sample are located near the bright central regions of large galaxies. These environments can obscure early-time flux and complicate detection and photometric measurements, underscoring the need for precise image subtraction and optimized follow-up strategies when targeting CC~SNe.

\begin{table}[t]
\centering
\caption{Earliest epochs able to be detected as a function of SN redshift, $s_{BV}$, and telescope/instrument magnitude limit. \label{tab:gold}}
\resizebox{0.6\textwidth}{!}{
\begin{tabular}{ccccc}
\hline \hline
redshift & $M_{\rm lim}$ & $\Delta (t_{\rm 1st - max})$ &  $\Delta (t_{\rm 1st - max})$ & $\Delta (t_{\rm 1st - max})$ \\
 $z$ & & $s_{BV}=0.8$ &  $s_{BV}=1.0$ & $s_{BV}=1.2$ \\
\hline
\multicolumn{5}{c}{mag lim 20.5 (ZTF)}\\
\hline
$0.001$ &  $-$7.66 & $-$14.97 & $-$17.43 & $-$20.00 \\
$0.002$ &  $-$9.17 & $-$14.83 & $-$17.28 & $-$19.86 \\
$0.005$ & $-$11.16 & $-$14.60 & $-$17.01 & $-$19.55 \\
$0.01$  & $-$12.68 & $-$14.30 & $-$16.71 & $-$19.24 \\
$0.02$  & $-$14.20 & $-$13.70 & $-$16.05 & $-$18.53 \\
$0.03$  & $-$15.09 & $-$13.30 & $-$15.61 & $-$18.05 \\
\hline
\multicolumn{5}{c}{mag lim 21.5 (KMTNet)}\\
\hline
$0.002$ &  $-$8.17 & $-$14.93 & $-$17.39 & $-$19.96 \\
$0.005$ & $-$10.16 & $-$14.72 & $-$17.14 & $-$19.69 \\
$0.01$  & $-$11.68 & $-$14.52 & $-$16.93 & $-$19.47 \\
$0.02$  & $-$13.20 & $-$14.12 & $-$16.53 & $-$19.06 \\
$0.03$  & $-$14.09 & $-$13.74 & $-$16.09 & $-$18.59 \\
\hline
\multicolumn{5}{c}{mag lim 22.5 (ATLAS)}\\
\hline
$0.005$ &  $-$9.16 & $-$14.83 & $-$17.28 & $-$19.86 \\
$0.01$  & $-$10.68 & $-$14.66 & $-$17.08 & $-$19.62 \\
$0.02$  & $-$12.20 & $-$14.42 & $-$16.83 & $-$19.37 \\
$0.03$  & $-$13.09 & $-$14.15 & $-$16.57 & $-$19.10 \\
\hline
\multicolumn{5}{c}{mag lim 23.5 (DES)}\\
\hline
$0.005$ &  $-$8.16 & $-$14.93 & $-$17.39 & $-$19.96 \\
$0.01$  &  $-$9.68 & $-$14.77 & $-$17.21 & $-$19.78 \\
$0.02$  & $-$11.20 & $-$14.60 & $-$17.01 & $-$19.54 \\
$0.03$  & $-$12.09 & $-$14.45 & $-$16.86 & $-$19.39 \\
\hline
\multicolumn{5}{c}{mag lim 24.5 (LSST)}\\
\hline
$0.01$  &  $-$8.68 & $-$14.89 & $-$17.34 & $-$19.91 \\
$0.02$  & $-$10.20 & $-$14.71 & $-$17.14 & $-$19.69 \\
$0.03$  & $-$11.09 & $-$14.61 & $-$17.02 & $-$19.56 \\
\hline
\end{tabular}}
\end{table}

\begin{figure*}[t]
\centering
\includegraphics[width=\textwidth]{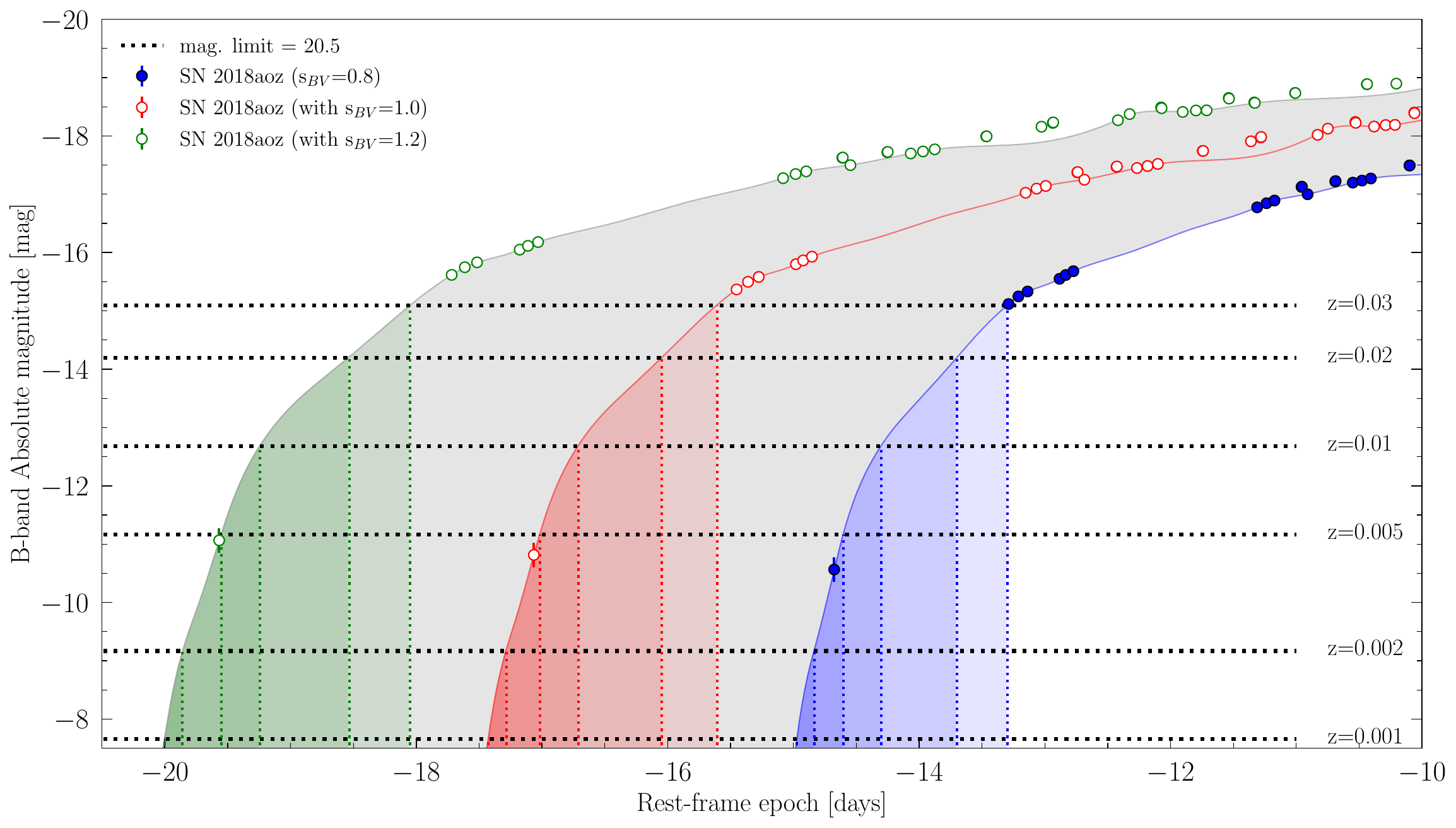}
\caption{Light-curve simulation based on the observed light curve of SN~2018aoz, an SN with a color-stretch parameter $s_{BV} = 0.8$. The original light curve is shown (in blue) alongside simulated versions stretched to represent $s_{BV}$ values of 1.0 (in red) and 1.2 (in green). Horizontal lines indicate the minimum observable magnitude at several redshifts of a telescope-instrument system with a limit of 20.5\,mag.}
\label{fig:gold}
\end{figure*}

Given our low success rate (20\%) relative to the initial goals of the project, we discuss in more detail two possible factors that have not allowed us to obtain earlier SN spectra:
(i) the time of first light is not necessarily the time of explosion owing to the depth of the discovery survey and the distance to the SN; and 
(ii) the difference in longitude between the discovery telescope and the classification spectroscopy telescope introduces a lower limit on the epoch of the first spectrum.

The time of first light is defined here as the moment when an SN becomes bright enough to be detected by a specific telescope and instrument configuration. For instance, an SN may be visible to a large-aperture telescope with a high-efficiency instrument, while a smaller-aperture telescope may fail to detect it. Similarly, owing to higher brightness, earlier epochs of an SN in a nearby galaxy (lower redshift) can be detected compared to one in a more distant galaxy using the same telescope configuration. To visually illustrate these two effects, we present an exercise based on SN~2018aoz \citep{2022NatAs...6..568N}, an SN discovered just an hour after first light, with a color-stretch parameter $s_{BV} = 0.8$. Figure \ref{fig:gold} shows the original light curve of SN~2018aoz alongside two additional light curves representing the same SN, but stretched to simulate SNe with $s_{BV}$ values of 1.0 and 1.2. It is evident that the epochs at which the SN light reaches specific absolute magnitudes vary by a few days across the three light curves. For example, an absolute magnitude of $-10$ is reached at $-15$, $-17$, and $-19$ days, respectively. Assuming a magnitude limit for the discovery survey, horizontal lines in the figure represent the detection thresholds for a telescope-instrument system at various redshifts. The figure shows an example for a limit of 20.5\,mag, comparable to ZTF. Table \ref{tab:gold} provides the corresponding limits for surveys with limits as deep as 24.5\,mag, representative of other current and future surveys (e.g., KMNet, \citep{2016JKAS...49...37K}; ATLAS, \citep{2018PASP..130f4505T}; DES, \citep{2016MNRAS.460.1270D}; LSST, \citep{2019ApJ...873..111I}). The intersections of the light curves and horizontal lines for different redshifts define the earliest epochs at which an SN with a specific $s_{BV}$ can be observed. For example, an SN with $s_{BV} = 1.0$ at a redshift of 0.01 would not be visible earlier than $-17$ days using a system with a  limit of 20.5\,mag. Accounting for the time required to process the discovery image, report it to the community, and set up spectroscopic observations, this illustrates the challenges of obtaining spectra of SNe Ia earlier than $-17$ days. It is important to note that the first detection, or first light, is not equivalent to the explosion epoch for an SN observed with a given instrument. The first detection merely indicates the moment the SN becomes just bright enough to surpass the detection limit. Our program demonstrated this in several cases: even with rapid triggering, the SN was not always at an early phase in forced-photometry light curves, as this method typically provides deeper limits. The discrepancy between the explosion time and the first light imposes significant constraints on acquiring spectra at the earliest SN epochs. This, in turn, limits our ability to study the first few hours after the explosion and to place constraints on the properties of progenitor systems.

\begin{figure*}
\centering
    \includegraphics[width=\textwidth]{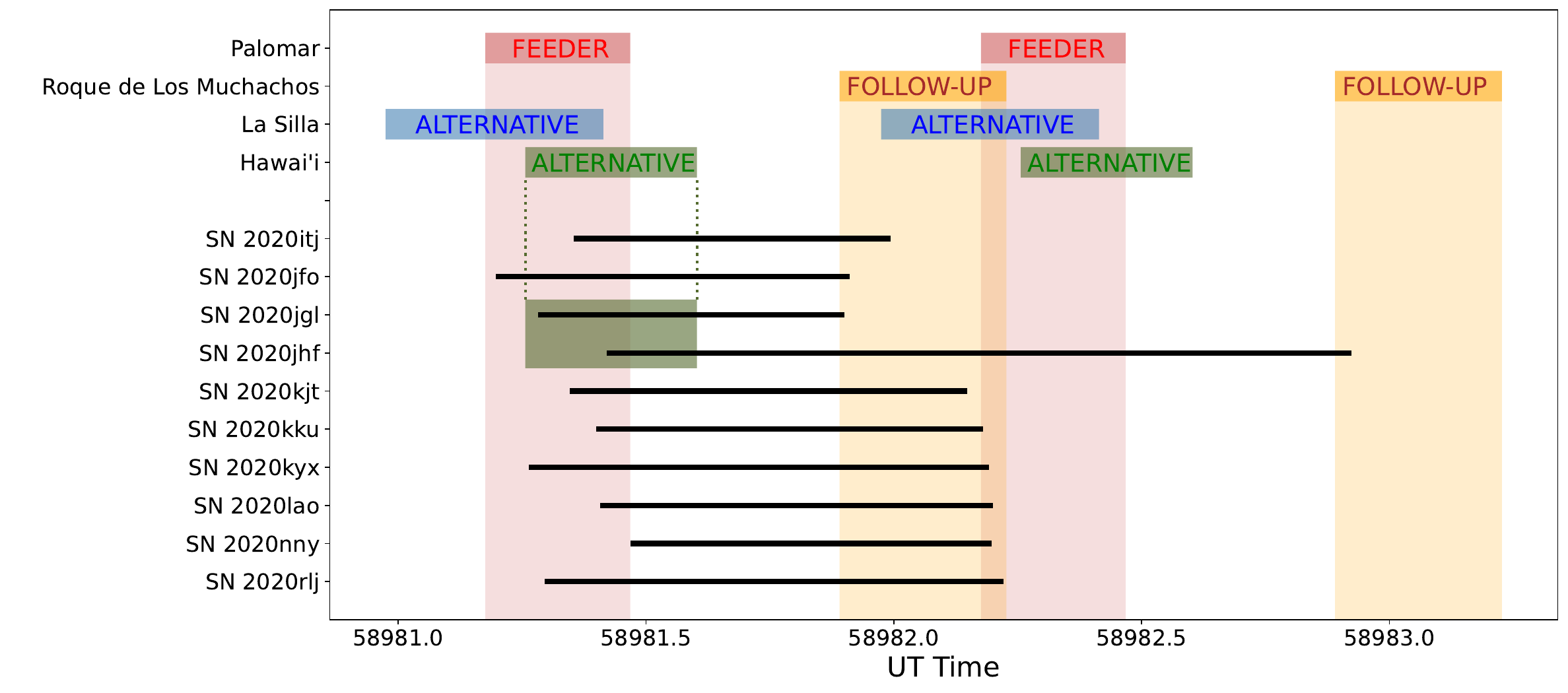}
    \caption{ 
    Representation of the nighttime hours at Palomar Observatory, where the ZTF survey discovers new SN candidates, and at La Palma, the location of the GTC telescope used for our early spectroscopy program. SN~2020jgl and SN~2020jhf were discovered by the ATLAS unit on Mauna Loa, Hawai'i, and we have also shaded the nighttime hours for this location for these two objects. Horizontal lines for each of the ten SNe in our sample connect the time of discovery with the time the spectra were obtained. Additionally, the nighttime hours at La Silla in Chile and Hawai'i are shown in blue and green, respectively, illustrating how faster follow-up observations could have been achieved from those locations. This representation corresponds to May 2020, when our GTC program began.
    }
\label{fig:schedule}
\end{figure*}

On the other hand, aside from the ability to detect the faint initial epochs of an SN, the physical location of the telescope performing the spectroscopic observation relative to the discovery telescope plays a significant role. In our case, most targets were identified by ZTF, a survey conducted with the P48 telescope at the Palomar Observatory in California, USA, with the exception of a few reported by the ATLAS unit at Mauna Loa, Hawai'i, USA. The spectroscopic classification was performed in all cases using the GTC at the Roque de Los Muchachos Observatory in La Palma, Spain. Owing to the longitude difference between these observatories, and depending on the time of year, there is a delay of 7 to 12\,hr between the end of the night at Palomar and the start of the night at La Palma. During the period our program was active, between May and August 2020, this delay was at least 10\,hr. For Mauna Loa, the difference is about 3\,hr less. In any case, even for an SN candidate that exploded in a very nearby galaxy and was discovered faint and early, it would not have been possible to obtain a spectrum earlier than 7--10\,hr after discovery. Furthermore, the sky visible at the end of the night at Palomar or Hawai'i is not visible at the beginning of the night at La Palma, adding additional time to the minimum response window. In Figure \ref{fig:schedule}, we illustrate the nighttime periods at Palomar, Hawai'i and El Roque using vertical shaded regions. We also plot lines for our 10 SNe, connecting the time of the science image in which the SN was discovered by the feeder survey (ZTF or ATLAS) to the time when the spectrum was obtained with the GTC. The only way to achieve shorter response times would be to use a telescope located at a similar longitude or as close to the west as possible relative to the discovery telescope. In the figure, we include examples of the nighttime ranges for the La Silla Observatory in Chile and the Maunakea Observatory, also in Hawai'i, USA. In this scenario, and also because La Silla is in the Southern hemisphere, any of the observatories in Hawai'i would have been the most suitable location for obtaining very early spectra of the objects in our sample.

This work served as a proof of concept for the Precision Observations of Infant Supernova Explosions (POISE; \citep{2021ATel14441....1B}) programme. POISE is using the Swope Telescope at the Las Campanas Observatory in Chile to confirm SN candidates from ZTF and ATLAS and obtains spectroscopy with several other larger-aperture telescopes, including those at Las Campanas, Maunakea, and La Palma. Results from POISE will be presented elsewhere. Our program also demonstrated the need for a rapid-response spectroscopic program at an optimal longitude, likely coupled with more stringent criteria for triggering (e.g., magnitude limits relative to redshift) than those applied here. Such an approach will be essential for future discovery surveys, such as the La Silla Southern Supernova Survey (LS4)\footnote{\href{https://sites.northwestern.edu/ls4/}{https://sites.northwestern.edu/ls4/}} and the Legacy Survey of Space and Time (LSST), both based in Chile, to build a sample of SN spectra obtained within a day of explosion.

\acknowledgments

The SNICE research group acknowledges financial support from the Spanish Ministerio de Ciencia, Innovaci\'on y Universidades (MCIU) and the Agencia Estatal de Investigaci\'on (AEI) 10.13039/501100011033 under the PID2020-115253GA-I00 HOSTFLOWS and PID2023-151307NB-I00 SNNEXT projects, from Centro Superior de Investigaciones Cient\'ificas (CSIC) under the projects PIE 20215AT016, ILINK23001, COOPB2304, and the program Unidad de Excelencia Mar\'ia de Maeztu CEX2020-001058-M, and from the Departament de Recerca i Universitats de la Generalitat de Catalunya through the 2021-SGR-01270 grant.
C.P.G. acknowledges financial support from the Secretariat of Universities and Research of the Department of Research and Universities of the Generalitat of Catalonia and by the Horizon 2020 Research and Innovation Programme of the European Union under the Marie Sk\l{}odowska-Curie and the Beatriu de Pin\'os 2021 BP 00168 programme.
L.P. acknowledges financial support from the CSIC project JAEICU-21-ICE-09.
D.N.C. and K.P. acknowledge support from the predoctoral program AGAUR Joan Or\'o of the Secretariat of Universities and Research of the Department of Research and Universities of the Generalitat of Catalonia and the European Social Plus Fund under fellowships 2022 FI-B-00526 and 2023 FI-1-00683, respectively.
C.J.-P. acknowledges financial support from grant PRE2021-096988 funded by AEI 10.13039/501100011033 and ESF Investing in your future.
M.K. acknowledges financial support from MCIU through the programme Juan de la Cierva-Incorporaci\'on JC2022-049447-I.
T.E.M.B. is funded by Horizon Europe ERC grant no. 101125877.
M.D. Stritzinger is funded by the Independent Research Fund Denmark (IRFD, grant number  10.46540/2032-00022B) and by an Aarhus University Research Foundation Nova project (AUFF-E-2023-9-28).
The Las Cumbres Observatory team is supported by NSF grants AST-1911225 and AST-1911151.
Generous financial support was provided to A.V.F.'s supernova group at U.C. Berkeley by the Christopher R. Redlich Fund, William Draper, Timothy and Melissa Draper, Briggs and Kathleen Wood, and Sanford Robertson (T.G.B. is a Draper-Wood-Robertson Specialist in Astronomy). 

Based on observations made with the Gran Telescopio Canarias (GTC), installed at the Spanish Observatorio del Roque de los Muchachos of the Instituto de Astrofísica de Canarias, on the island of La Palma.
The SALT data presented here were obtained through Rutgers University program 2020-1-MLT-007 (PI: S. W. Jha).
Based on observations obtained with the Samuel Oschin 48-inch Telescope at the Palomar Observatory as part of the Zwicky Transient Facility project. ZTF is supported by the NSF under grant AST-1440341 and a
collaboration including Caltech, IPAC, the Weizmann Institute for Science, the Oskar Klein Center at Stockholm University, the University of Maryland, the University of Washington, Deutsches Elektronen-Synchrotron and Humboldt University, Los Alamos National Laboratories, the TANGO Consortium of Taiwan, the University of Wisconsin at Milwaukee, and Lawrence Berkeley National Laboratories. Operations are conducted by COO, IPAC, and UW. 
This work has made use of data from the Asteroid Terrestrial-impact Last Alert System (ATLAS) project. The Asteroid Terrestrial-impact Last Alert System (ATLAS) project is primarily funded to search for near-earth asteroids through NASA grants NN12AR55G, 80NSSC18K0284, and 80NSSC18K1575; byproducts of the NEO search include images and catalogs from the survey area. This work was partially funded by Kepler/K2 grant J1944/80NSSC19K0112 and HST GO-15889, and STFC grants ST/T000198/1 and ST/S006109/1. The ATLAS science products have been made possible through the contributions of the University of Hawaii Institute for Astronomy, the Queen’s University Belfast, the Space Telescope Science Institute, the South African Astronomical Observatory, and The Millennium Institute of Astrophysics (MAS), Chile.
Based in part on observations made with the Nordic Optical Telescope, owned in collaboration by the University of Turku and Aarhus University, and operated jointly by Aarhus University, the University of Turku and the University of Oslo, representing Denmark, Finland and Norway, the University of Iceland and Stockholm University at the Observatorio del Roque de los Muchachos, La Palma, Spain, of the Instituto de Astrofisica de Canarias. The NOT data were obtained under program ID P62-505 
This work is based in part on observations obtained with the Las Cumbres Observatory network.
This includes a time allocation (2020A/031 PI Stritzinger) funded from the European Union's Horizon 2020 research and innovation programme under grant agreement \#730890 through OPTICON RadioNet Pilot.
A major upgrade of the Kast spectrograph on the Shane 3\,m telescope at Lick Observatory, led by Brad Holden, was made possible through generous gifts from the Heising-Simons Foundation, William and Marina Kast, and the University of California Observatories.
Research at Lick Observatory is partially supported by a gift from Google. 

\bibliographystyle{aasjournal}
\bibliography{gtc} 

\begin{thebibliography}{}
\expandafter\ifx\csname natexlab\endcsname\relax\def\natexlab#1{#1}\fi
\providecommand{\url}[1]{\href{#1}{#1}}
\providecommand{\dodoi}[1]{doi:~\href{http://doi.org/#1}{\nolinkurl{#1}}}
\providecommand{\doeprint}[1]{\href{http://ascl.net/#1}{\nolinkurl{http://ascl.net/#1}}}
\providecommand{\doarXiv}[1]{\href{https://arxiv.org/abs/#1}{\nolinkurl{https://arxiv.org/abs/#1}}}

\bibitem[{{Ailawadhi} {et~al.}(2023){Ailawadhi}, {Dastidar}, {Misra}, {Roy},
  {Hiramatsu}, {Howell}, {Brink}, {Zheng}, {Galbany}, {Shahbandeh}, {Arcavi},
  {Ashall}, {Bostroem}, {Burke}, {Chapman}, {Dimple}, {Filippenko},
  {Gangopadhyay}, {Ghosh}, {Hoffman}, {Hosseinzadeh}, {Jennings}, {Jha},
  {Kumar}, {Karamehmetoglu}, {McCully}, {McGinness}, {M{\"u}ller-Bravo},
  {Murakami}, {Pandey}, {Pellegrino}, {Piscarreta}, {Rho}, {Stritzinger},
  {Sunseri}, {Van Dyk}, \& {Yadav}}]{Ailawadhi_2023}
{Ailawadhi}, B., {Dastidar}, R., {Misra}, K., {et~al.} 2023, \mnras, 519, 248,
  \dodoi{10.1093/mnras/stac3234}

\bibitem[{{Aldering} {et~al.}(2002){Aldering}, {Adam}, {Antilogus}, {Astier},
  {Bacon}, {Bongard}, {Bonnaud}, {Copin}, {Hardin}, {Henault}, {Howell},
  {Lemonnier}, {Levy}, {Loken}, {Nugent}, {Pain}, {Pecontal}, {Pecontal},
  {Perlmutter}, {Quimby}, {Schahmaneche}, {Smadja}, \&
  {Wood-Vasey}}]{2002SPIE.4836...61A}
{Aldering}, G., {Adam}, G., {Antilogus}, P., {et~al.} 2002, in Society of
  Photo-Optical Instrumentation Engineers (SPIE) Conference Series, Vol. 4836,
  Survey and Other Telescope Technologies and Discoveries, ed. J.~A. {Tyson} \&
  S.~{Wolff}, 61--72, \dodoi{10.1117/12.458107}

\bibitem[{{Ashall} {et~al.}(2022){Ashall}, {Lu}, {Shappee}, {Burns}, {Hsiao},
  {Kumar}, {Morrell}, {Phillips}, {Shahbandeh}, {Baron}, {Boutsia}, {Brown},
  {DerKacy}, {Galbany}, {Hoeflich}, {Krisciunas}, {Mazzali}, {Piro},
  {Stritzinger}, \& {Suntzeff}}]{2022ApJ...932L...2A}
{Ashall}, C., {Lu}, J., {Shappee}, B.~J., {et~al.} 2022, \apjl, 932, L2,
  \dodoi{10.3847/2041-8213/ac7235}

\bibitem[{{Barbary} {et~al.}(2016){Barbary}, {Barclay}, {Biswas}, {Craig},
  {Feindt}, {Friesen}, {Goldstein}, {Jha}, {Rodney}, {Sofiatti}, {Thomas}, \&
  {Wood-Vasey}}]{Barbary_2016_sncosmo}
{Barbary}, K., {Barclay}, T., {Biswas}, R., {et~al.} 2016, {SNCosmo: Python
  library for supernova cosmology}, Astrophysics Source Code Library, record
  ascl:1611.017.
\newblock \doeprint{1611.017}

\bibitem[{{Barbon} {et~al.}(1995){Barbon}, {Benetti}, {Cappellaro}, {Patat},
  {Turatto}, \& {Iijima}}]{Barbon95P}
{Barbon}, R., {Benetti}, S., {Cappellaro}, E., {et~al.} 1995, \aaps, 110, 513

\bibitem[{{Bellm} {et~al.}(2019){Bellm}, {Kulkarni}, {Graham}, {Dekany},
  {Smith}, {Riddle}, {Masci}, {Helou}, {Prince}, {Adams}, {Barbarino},
  {Barlow}, {Bauer}, {Beck}, {Belicki}, {Biswas}, {Blagorodnova}, {Bodewits},
  {Bolin}, {Brinnel}, {Brooke}, {Bue}, {Bulla}, {Burruss}, {Cenko}, {Chang},
  {Connolly}, {Coughlin}, {Cromer}, {Cunningham}, {De}, {Delacroix}, {Desai},
  {Duev}, {Eadie}, {Farnham}, {Feeney}, {Feindt}, {Flynn}, {Franckowiak},
  {Frederick}, {Fremling}, {Gal-Yam}, {Gezari}, {Giomi}, {Goldstein},
  {Golkhou}, {Goobar}, {Groom}, {Hacopians}, {Hale}, {Henning}, {Ho}, {Hover},
  {Howell}, {Hung}, {Huppenkothen}, {Imel}, {Ip}, {Ivezi{\'c}}, {Jackson},
  {Jones}, {Juric}, {Kasliwal}, {Kaspi}, {Kaye}, {Kelley}, {Kowalski},
  {Kramer}, {Kupfer}, {Landry}, {Laher}, {Lee}, {Lin}, {Lin}, {Lunnan},
  {Giomi}, {Mahabal}, {Mao}, {Miller}, {Monkewitz}, {Murphy}, {Ngeow},
  {Nordin}, {Nugent}, {Ofek}, {Patterson}, {Penprase}, {Porter}, {Rauch},
  {Rebbapragada}, {Reiley}, {Rigault}, {Rodriguez}, {van Roestel}, {Rusholme},
  {van Santen}, {Schulze}, {Shupe}, {Singer}, {Soumagnac}, {Stein}, {Surace},
  {Sollerman}, {Szkody}, {Taddia}, {Terek}, {Van Sistine}, {van Velzen},
  {Vestrand}, {Walters}, {Ward}, {Ye}, {Yu}, {Yan}, \&
  {Zolkower}}]{2019PASP..131a8002B}
{Bellm}, E.~C., {Kulkarni}, S.~R., {Graham}, M.~J., {et~al.} 2019, \pasp, 131,
  018002, \dodoi{10.1088/1538-3873/aaecbe}

\bibitem[{{Bianco} {et~al.}(2014){Bianco}, {Modjaz}, {Hicken}, {Friedman},
  {Kirshner}, {Bloom}, {Challis}, {Marion}, {Wood-Vasey}, \& {Rest}}]{Bianco14}
{Bianco}, F.~B., {Modjaz}, M., {Hicken}, M., {et~al.} 2014, \apjs, 213, 19,
  \dodoi{10.1088/0067-0049/213/2/19}

\bibitem[{{Blondin} \& {Tonry}(2011)}]{SNID}
{Blondin}, S., \& {Tonry}, J.~L. 2011, {SNID: Supernova Identification},
  Astrophysics Source Code Library, record ascl:1107.001.
\newblock \doeprint{1107.001}

\bibitem[{{Blondin} {et~al.}(2006){Blondin}, {Dessart}, {Leibundgut}, {Branch},
  {H{\"o}flich}, {Tonry}, {Matheson}, {Foley}, {Chornock}, {Filippenko},
  {Sollerman}, {Spyromilio}, {Kirshner}, {Wood-Vasey}, {Clocchiatti},
  {Aguilera}, {Barris}, {Becker}, {Challis}, {Covarrubias}, {Davis},
  {Garnavich}, {Hicken}, {Jha}, {Krisciunas}, {Li}, {Miceli}, {Miknaitis},
  {Pignata}, {Prieto}, {Rest}, {Riess}, {Salvo}, {Schmidt}, {Smith}, {Stubbs},
  \& {Suntzeff}}]{2006AJ....131.1648B}
{Blondin}, S., {Dessart}, L., {Leibundgut}, B., {et~al.} 2006, \aj, 131, 1648,
  \dodoi{10.1086/498724}

\bibitem[{{Blondin} {et~al.}(2012){Blondin}, {Matheson}, {Kirshner}, {Mandel},
  {Berlind}, {Calkins}, {Challis}, {Garnavich}, {Jha}, {Modjaz}, {Riess}, \&
  {Schmidt}}]{2012AJ....143..126B}
{Blondin}, S., {Matheson}, T., {Kirshner}, R.~P., {et~al.} 2012, \aj, 143, 126,
  \dodoi{10.1088/0004-6256/143/5/126}

\bibitem[{{Bose} \& {Kumar}(2014)}]{Bose_Kumar_2014}
{Bose}, S., \& {Kumar}, B. 2014, \apj, 782, 98,
  \dodoi{10.1088/0004-637X/782/2/98}

\bibitem[{{Branch} {et~al.}(2006){Branch}, {Dang}, {Hall}, {Ketchum},
  {Melakayil}, {Parrent}, {Troxel}, {Casebeer}, {Jeffery}, \&
  {Baron}}]{2006PASP..118..560B}
{Branch}, D., {Dang}, L.~C., {Hall}, N., {et~al.} 2006, \pasp, 118, 560,
  \dodoi{10.1086/502778}

\bibitem[{{Bruch} {et~al.}(2023){Bruch}, {Gal-Yam}, {Yaron}, {Chen},
  {Strotjohann}, {Irani}, {Zimmerman}, {Schulze}, {Yang}, {Kim}, {Bulla},
  {Sollerman}, {Rigault}, {Ofek}, {Soumagnac}, {Masci}, {Fremling}, {Perley},
  {Nordin}, {Cenko}, {Ho}, {Adams}, {Adreoni}, {Bellm}, {Blagorodnova},
  {Burdge}, {De}, {Dekany}, {Dhawan}, {Drake}, {Duev}, {Graham}, {Graham},
  {Jencson}, {Karamehmetoglu}, {Kasliwal}, {Kulkarni}, {Miller}, {Neill},
  {Prince}, {Riddle}, {Rusholme}, {Sharma}, {Smith}, {Sravan}, {Taggart},
  {Walters}, \& {Yan}}]{2023ApJ...952..119B}
{Bruch}, R.~J., {Gal-Yam}, A., {Yaron}, O., {et~al.} 2023, \apj, 952, 119,
  \dodoi{10.3847/1538-4357/acd8be}

\bibitem[{{Burke} {et~al.}(2020){Burke}, {Hiramatsu}, {Howell}, {McCully},
  {Gonzalez}, \& {Pellegrino}}]{tns_2020lao}
{Burke}, J., {Hiramatsu}, D., {Howell}, D.~A., {et~al.} 2020, Transient Name
  Server Classification Report, 2020-1666, 1

\bibitem[{{Burke} {et~al.}(2021){Burke}, {Howell}, {Sarbadhicary}, {Sand},
  {Amaro}, {Hiramatsu}, {McCully}, {Pellegrino}, {Andrews}, {Brown}, {Itagaki},
  {Shahbandeh}, {Bostroem}, {Chomiuk}, {Hsiao}, {Smith}, \&
  {Valenti}}]{2021ApJ...919..142B}
{Burke}, J., {Howell}, D.~A., {Sarbadhicary}, S.~K., {et~al.} 2021, \apj, 919,
  142, \dodoi{10.3847/1538-4357/ac126b}

\bibitem[{{Burke} {et~al.}(2022){Burke}, {Howell}, {Sand}, {Amaro}, {Brown},
  {Andrews}, {Bostroem}, {Dong}, {Haislip}, {Hiramatsu}, {Hosseinzadeh},
  {Kouprianov}, {Lundquist}, {McCully}, {Pellegrino}, {Reichart}, {Tartaglia},
  {Valenti}, \& {Yang}}]{2022arXiv220707681B}
{Burke}, J., {Howell}, D.~A., {Sand}, D.~J., {et~al.} 2022, arXiv e-prints,
  arXiv:2207.07681, \dodoi{10.48550/arXiv.2207.07681}

\bibitem[{{Burns} {et~al.}(2021){Burns}, {Hsiao}, {Suntzeff}, {Baron},
  {Shappee}, {Aldoroty}, {Anderson}, {Ashall}, {Bersten}, {Brown}, {Burrow},
  {Clochiatti}, {Davis}, {DerKacy}, {Do}, {Folatelli}, {Forster Buron},
  {Galbany}, {Hoeflich}, {Holmbo}, {Karamehmetoglu}, {Krisciunas}, {Kumar},
  {Lu}, {Mazzali}, {Morrell}, {Pessi}, {Phillips}, {Pignata}, {Piro}, {Polin},
  {Shahbandeh}, {Stangl}, {Stritzinger}, {Teffs}, {Tonry}, {Tucker}, {Uddin},
  \& {Yang}}]{2021ATel14441....1B}
{Burns}, C., {Hsiao}, E., {Suntzeff}, N., {et~al.} 2021, The Astronomer's
  Telegram, 14441, 1

\bibitem[{{Burns} {et~al.}(2011){Burns}, {Stritzinger}, {Phillips}, {Kattner},
  {Persson}, {Madore}, {Freedman}, {Boldt}, {Campillay}, {Contreras},
  {Folatelli}, {Gonzalez}, {Krzeminski}, {Morrell}, {Salgado}, \&
  {Suntzeff}}]{Burns_2011_SNooPy}
{Burns}, C.~R., {Stritzinger}, M., {Phillips}, M.~M., {et~al.} 2011, \aj, 141,
  19, \dodoi{10.1088/0004-6256/141/1/19}

\bibitem[{{Burns} {et~al.}(2014){Burns}, {Stritzinger}, {Phillips}, {Hsiao},
  {Contreras}, {Persson}, {Folatelli}, {Boldt}, {Campillay}, {Castell{\'o}n},
  {Freedman}, {Madore}, {Morrell}, {Salgado}, \& {Suntzeff}}]{Burns_2014}
---. 2014, \apj, 789, 32, \dodoi{10.1088/0004-637X/789/1/32}

\bibitem[{{Burrow} {et~al.}(2020){Burrow}, {Baron}, {Ashall}, {Burns},
  {Morrell}, {Stritzinger}, {Brown}, {Folatelli}, {Freedman}, {Galbany},
  {Hoeflich}, {Hsiao}, {Krisciunas}, {Phillips}, {Piro}, {Suntzeff}, \&
  {Uddin}}]{2020ApJ...901..154B}
{Burrow}, A., {Baron}, E., {Ashall}, C., {et~al.} 2020, \apj, 901, 154,
  \dodoi{10.3847/1538-4357/abafa2}

\bibitem[{{Burrows}(2000)}]{2000Natur.403..727B}
{Burrows}, A. 2000, \nat, 403, 727, \dodoi{10.1038/35001501}

\bibitem[{{Cartier} {et~al.}(2017){Cartier}, {Sullivan}, {Firth}, {Pignata},
  {Mazzali}, {Maguire}, {Childress}, {Arcavi}, {Ashall}, {Bassett}, {Crawford},
  {Frohmaier}, {Galbany}, {Gal-Yam}, {Hosseinzadeh}, {Howell}, {Inserra},
  {Johansson}, {Kasai}, {McCully}, {Prajs}, {Prentice}, {Schulze}, {Smartt},
  {Smith}, {Smith}, {Valenti}, \& {Young}}]{2017MNRAS.464.4476C}
{Cartier}, R., {Sullivan}, M., {Firth}, R.~E., {et~al.} 2017, \mnras, 464,
  4476, \dodoi{10.1093/mnras/stw2678}

\bibitem[{{Chambers} {et~al.}(2016){Chambers}, {Magnier}, {Metcalfe},
  {Flewelling}, {Huber}, {Waters}, {Denneau}, {Draper}, {Farrow}, {Finkbeiner},
  {Holmberg}, {Koppenhoefer}, {Price}, {Rest}, {Saglia}, {Schlafly}, {Smartt},
  {Sweeney}, {Wainscoat}, {Burgett}, {Chastel}, {Grav}, {Heasley}, {Hodapp},
  {Jedicke}, {Kaiser}, {Kudritzki}, {Luppino}, {Lupton}, {Monet}, {Morgan},
  {Onaka}, {Shiao}, {Stubbs}, {Tonry}, {White}, {Ba{\~n}ados}, {Bell},
  {Bender}, {Bernard}, {Boegner}, {Boffi}, {Botticella}, {Calamida},
  {Casertano}, {Chen}, {Chen}, {Cole}, {Deacon}, {Frenk}, {Fitzsimmons},
  {Gezari}, {Gibbs}, {Goessl}, {Goggia}, {Gourgue}, {Goldman}, {Grant},
  {Grebel}, {Hambly}, {Hasinger}, {Heavens}, {Heckman}, {Henderson}, {Henning},
  {Holman}, {Hopp}, {Ip}, {Isani}, {Jackson}, {Keyes}, {Koekemoer}, {Kotak},
  {Le}, {Liska}, {Long}, {Lucey}, {Liu}, {Martin}, {Masci}, {McLean}, {Mindel},
  {Misra}, {Morganson}, {Murphy}, {Obaika}, {Narayan}, {Nieto-Santisteban},
  {Norberg}, {Peacock}, {Pier}, {Postman}, {Primak}, {Rae}, {Rai}, {Riess},
  {Riffeser}, {Rix}, {R{\"o}ser}, {Russel}, {Rutz}, {Schilbach}, {Schultz},
  {Scolnic}, {Strolger}, {Szalay}, {Seitz}, {Small}, {Smith}, {Soderblom},
  {Taylor}, {Thomson}, {Taylor}, {Thakar}, {Thiel}, {Thilker}, {Unger},
  {Urata}, {Valenti}, {Wagner}, {Walder}, {Walter}, {Watters}, {Werner},
  {Wood-Vasey}, \& {Wyse}}]{2016arXiv161205560C}
{Chambers}, K.~C., {Magnier}, E.~A., {Metcalfe}, N., {et~al.} 2016, arXiv
  e-prints, arXiv:1612.05560, \dodoi{10.48550/arXiv.1612.05560}

\bibitem[{{Cs{\"o}rnyei} {et~al.}(2023){Cs{\"o}rnyei}, {Vogl}, {Taubenberger},
  {Fl{\"o}rs}, {Blondin}, {Cudmani}, {Holas}, {Kressierer}, {Leibundgut}, \&
  {Hillebrandt}}]{Geza_2023}
{Cs{\"o}rnyei}, G., {Vogl}, C., {Taubenberger}, S., {et~al.} 2023, \aap, 672,
  A129, \dodoi{10.1051/0004-6361/202245379}

\bibitem[{{Dark Energy Survey Collaboration} {et~al.}(2016){Dark Energy Survey
  Collaboration}, {Abbott}, {Abdalla}, {Aleksi{\'c}}, {Allam}, {Amara},
  {Bacon}, {Balbinot}, {Banerji}, {Bechtol}, {Benoit-L{\'e}vy}, {Bernstein},
  {Bertin}, {Blazek}, {Bonnett}, {Bridle}, {Brooks}, {Brunner}, {Buckley-Geer},
  {Burke}, {Caminha}, {Capozzi}, {Carlsen}, {Carnero-Rosell}, {Carollo},
  {Carrasco-Kind}, {Carretero}, {Castander}, {Clerkin}, {Collett}, {Conselice},
  {Crocce}, {Cunha}, {D'Andrea}, {da Costa}, {Davis}, {Desai}, {Diehl},
  {Dietrich}, {Dodelson}, {Doel}, {Drlica-Wagner}, {Estrada}, {Etherington},
  {Evrard}, {Fabbri}, {Finley}, {Flaugher}, {Foley}, {Fosalba}, {Frieman},
  {Garc{\'\i}a-Bellido}, {Gaztanaga}, {Gerdes}, {Giannantonio}, {Goldstein},
  {Gruen}, {Gruendl}, {Guarnieri}, {Gutierrez}, {Hartley}, {Honscheid}, {Jain},
  {James}, {Jeltema}, {Jouvel}, {Kessler}, {King}, {Kirk}, {Kron}, {Kuehn},
  {Kuropatkin}, {Lahav}, {Li}, {Lima}, {Lin}, {Maia}, {Makler}, {Manera},
  {Maraston}, {Marshall}, {Martini}, {McMahon}, {Melchior}, {Merson}, {Miller},
  {Miquel}, {Mohr}, {Morice-Atkinson}, {Naidoo}, {Neilsen}, {Nichol}, {Nord},
  {Ogando}, {Ostrovski}, {Palmese}, {Papadopoulos}, {Peiris}, {Peoples},
  {Percival}, {Plazas}, {Reed}, {Refregier}, {Romer}, {Roodman}, {Ross},
  {Rozo}, {Rykoff}, {Sadeh}, {Sako}, {S{\'a}nchez}, {Sanchez}, {Santiago},
  {Scarpine}, {Schubnell}, {Sevilla-Noarbe}, {Sheldon}, {Smith}, {Smith},
  {Soares-Santos}, {Sobreira}, {Soumagnac}, {Suchyta}, {Sullivan}, {Swanson},
  {Tarle}, {Thaler}, {Thomas}, {Thomas}, {Tucker}, {Vieira}, {Vikram},
  {Walker}, {Wechsler}, {Weller}, {Wester}, {Whiteway}, {Wilcox}, {Yanny},
  {Zhang}, \& {Zuntz}}]{2016MNRAS.460.1270D}
{Dark Energy Survey Collaboration}, {Abbott}, T., {Abdalla}, F.~B., {et~al.}
  2016, \mnras, 460, 1270, \dodoi{10.1093/mnras/stw641}

\bibitem[{{de Jaeger} {et~al.}(2020){de Jaeger}, {Galbany},
  {Gonz{\'a}lez-Gait{\'a}n}, {Kessler}, {Filippenko}, {F{\"o}rster}, {Hamuy},
  {Brown}, {Davis}, {Guti{\'e}rrez}, {Inserra}, {Lewis}, {M{\"o}ller},
  {Scolnic}, {Smith}, {Brout}, {Carollo}, {Foley}, {Glazebrook}, {Hinton},
  {Macaulay}, {Nichol}, {Sako}, {Sommer}, {Tucker}, {Abbott}, {Aguena},
  {Allam}, {Annis}, {Avila}, {Bertin}, {Bhargava}, {Brooks}, {Burke}, {Carnero
  Rosell}, {Carrasco Kind}, {Carretero}, {Costanzi}, {Crocce}, {da Costa}, {De
  Vicente}, {Desai}, {Diehl}, {Doel}, {Drlica-Wagner}, {Eifler}, {Estrada},
  {Everett}, {Flaugher}, {Fosalba}, {Frieman}, {Garc{\'\i}a-Bellido},
  {Gaztanaga}, {Gruen}, {Gruendl}, {Gschwend}, {Gutierrez}, {Hartley},
  {Hollowood}, {Honscheid}, {James}, {Kuehn}, {Kuropatkin}, {Li}, {Lima},
  {Maia}, {Menanteau}, {Miquel}, {Palmese}, {Paz-Chinch{\'o}n}, {Plazas},
  {Romer}, {Roodman}, {Sanchez}, {Scarpine}, {Schubnell}, {Serrano},
  {Sevilla-Noarbe}, {Soares-Santos}, {Suchyta}, {Swanson}, {Tarle}, {Thomas},
  {Tucker}, {Varga}, {Walker}, {Weller}, {Wilkinson}, \& {DES
  Collaboration}}]{2020MNRAS.495.4860D}
{de Jaeger}, T., {Galbany}, L., {Gonz{\'a}lez-Gait{\'a}n}, S., {et~al.} 2020,
  \mnras, 495, 4860, \dodoi{10.1093/mnras/staa1402}

\bibitem[{{Dessart} \& {Jacobson-Gal{\'a}n}(2023)}]{2023A&A...677A.105D}
{Dessart}, L., \& {Jacobson-Gal{\'a}n}, W.~V. 2023, \aap, 677, A105,
  \dodoi{10.1051/0004-6361/202346754}

\bibitem[{{Fang} {et~al.}(2022){Fang}, {Maeda}, {Kuncarayakti}, {Tanaka},
  {Kawabata}, {Hattori}, {Aoki}, {Moriya}, \& {Yamanaka}}]{2022ApJ...928..151F}
{Fang}, Q., {Maeda}, K., {Kuncarayakti}, H., {et~al.} 2022, \apj, 928, 151,
  \dodoi{10.3847/1538-4357/ac4f60}

\bibitem[{{Filippenko} {et~al.}(2001){Filippenko}, {Li}, {Treffers}, \&
  {Modjaz}}]{2001ASPC..246..121F}
{Filippenko}, A.~V., {Li}, W.~D., {Treffers}, R.~R., \& {Modjaz}, M. 2001, in
  Astronomical Society of the Pacific Conference Series, Vol. 246, IAU Colloq.
  183: Small Telescope Astronomy on Global Scales, ed. B.~{Paczynski}, W.-P.
  {Chen}, \& C.~{Lemme}, 121

\bibitem[{{Filippenko} {et~al.}(1992){Filippenko}, {Richmond}, {Branch},
  {Gaskell}, {Herbst}, {Ford}, {Treffers}, {Matheson}, {Ho}, {Dey}, {Sargent},
  {Small}, \& {van Breugel}}]{1992AJ....104.1543F}
{Filippenko}, A.~V., {Richmond}, M.~W., {Branch}, D., {et~al.} 1992, \aj, 104,
  1543, \dodoi{10.1086/116339}

\bibitem[{{Finneran} {et~al.}(2024){Finneran}, {Cotter}, \&
  {Martin-Carrillo}}]{2024arXiv241111503F}
{Finneran}, G., {Cotter}, L., \& {Martin-Carrillo}, A. 2024, arXiv e-prints,
  arXiv:2411.11503, \dodoi{10.48550/arXiv.2411.11503}

\bibitem[{{Firth} {et~al.}(2015){Firth}, {Sullivan}, {Gal-Yam}, {Howell},
  {Maguire}, {Nugent}, {Piro}, {Baltay}, {Feindt}, {Hadjiyksta}, {McKinnon},
  {Ofek}, {Rabinowitz}, \& {Walker}}]{Firth_2015}
{Firth}, R.~E., {Sullivan}, M., {Gal-Yam}, A., {et~al.} 2015, \mnras, 446,
  3895, \dodoi{10.1093/mnras/stu2314}

\bibitem[{{Folatelli} {et~al.}(2013){Folatelli}, {Morrell}, {Phillips},
  {Hsiao}, {Campillay}, {Contreras}, {Castell{\'o}n}, {Hamuy}, {Krzeminski},
  {Roth}, {Stritzinger}, {Burns}, {Freedman}, {Madore}, {Murphy}, {Persson},
  {Prieto}, {Suntzeff}, {Krisciunas}, {Anderson}, {F{\"o}rster}, {Maza},
  {Pignata}, {Rojas}, {Boldt}, {Salgado}, {Wyatt}, {Olivares E.}, {Gal-Yam}, \&
  {Sako}}]{2013ApJ...773...53F}
{Folatelli}, G., {Morrell}, N., {Phillips}, M.~M., {et~al.} 2013, \apj, 773,
  53, \dodoi{10.1088/0004-637X/773/1/53}

\bibitem[{{Folatelli} {et~al.}(2014){Folatelli}, {Bersten}, {Kuncarayakti},
  {Olivares Estay}, {Anderson}, {Holmbo}, {Maeda}, {Morrell}, {Nomoto},
  {Pignata}, {Stritzinger}, {Contreras}, {F{\"o}rster}, {Hamuy}, {Phillips},
  {Prieto}, {Valenti}, {Afonso}, {Altenm{\"u}ller}, {Elliott}, {Greiner},
  {Updike}, {Haislip}, {LaCluyze}, {Moore}, \& {Reichart}}]{Folatelli14}
{Folatelli}, G., {Bersten}, M.~C., {Kuncarayakti}, H., {et~al.} 2014, \apj,
  792, 7, \dodoi{10.1088/0004-637X/792/1/7}

\bibitem[{{Foley} {et~al.}(2012){Foley}, {Challis}, {Filippenko},
  {Ganeshalingam}, {Landsman}, {Li}, {Marion}, {Silverman}, {Beaton},
  {Bennert}, {Cenko}, {Childress}, {Guhathakurta}, {Jiang}, {Kalirai},
  {Kirshner}, {Stockton}, {Tollerud}, {Vink{\'o}}, {Wheeler}, \&
  {Woo}}]{2012ApJ...744...38F}
{Foley}, R.~J., {Challis}, P.~J., {Filippenko}, A.~V., {et~al.} 2012, \apj,
  744, 38, \dodoi{10.1088/0004-637X/744/1/38}

\bibitem[{{F{\"o}rster} {et~al.}(2018){F{\"o}rster}, {Moriya}, {Maureira},
  {Anderson}, {Blinnikov}, {Bufano}, {Cabrera-Vives}, {Clocchiatti}, {de
  Jaeger}, {Est{\'e}vez}, {Galbany}, {Gonz{\'a}lez-Gait{\'a}n}, {Gr{\"a}fener},
  {Hamuy}, {Hsiao}, {Huentelemu}, {Huijse}, {Kuncarayakti}, {Mart{\'\i}nez},
  {Medina}, {Olivares E.}, {Pignata}, {Razza}, {Reyes}, {San Mart{\'\i}n},
  {Smith}, {Vera}, {Vivas}, {de Ugarte Postigo}, {Yoon}, {Ashall}, {Fraser},
  {Gal-Yam}, {Kankare}, {Le Guillou}, {Mazzali}, {Walton}, \&
  {Young}}]{2018NatAs...2..808F}
{F{\"o}rster}, F., {Moriya}, T.~J., {Maureira}, J.~C., {et~al.} 2018, Nature
  Astronomy, 2, 808, \dodoi{10.1038/s41550-018-0563-4}

\bibitem[{{Forster} {et~al.}(2020){Forster}, {Bauer}, {Galbany}, {Pignata},
  {Camacho}, {Silva-Farfan}, {Arredondo}, {Cabrera-Vives}, {Carrasco-Davis},
  {Estevez}, {Huijse}, {Reyes}, {Reyes}, {Sanchez-Saez}, {Valenzuela},
  {Castillo}, {Ruz-Mieres}, {Rodriguez-Mancini}, {Bauer}, {Catelan},
  {Eyheramendy}, \& {Graham}}]{2020TNSTR1395....1F}
{Forster}, F., {Bauer}, F.~E., {Galbany}, L., {et~al.} 2020, Transient Name
  Server Discovery Report, 2020-1395, 1

\bibitem[{{F{\"o}rster} {et~al.}(2021){F{\"o}rster}, {Cabrera-Vives},
  {Castillo-Navarrete}, {Est{\'e}vez}, {S{\'a}nchez-S{\'a}ez}, {Arredondo},
  {Bauer}, {Carrasco-Davis}, {Catelan}, {Elorrieta}, {Eyheramendy}, {Huijse},
  {Pignata}, {Reyes}, {Reyes}, {Rodr{\'\i}guez-Mancini}, {Ruz-Mieres},
  {Valenzuela}, {{\'A}lvarez-Maldonado}, {Astorga}, {Borissova}, {Clocchiatti},
  {De Cicco}, {Donoso-Oliva}, {Hern{\'a}ndez-Garc{\'\i}a}, {Graham},
  {Jord{\'a}n}, {Kurtev}, {Mahabal}, {Maureira}, {Mu{\~n}oz-Arancibia},
  {Molina-Ferreiro}, {Moya}, {Palma}, {P{\'e}rez-Carrasco}, {Protopapas},
  {Romero}, {Sabatini-Gacitua}, {S{\'a}nchez}, {San Mart{\'\i}n},
  {Sep{\'u}lveda-Cobo}, {Vera}, \& {Vergara}}]{2021AJ....161..242F}
{F{\"o}rster}, F., {Cabrera-Vives}, G., {Castillo-Navarrete}, E., {et~al.}
  2021, \aj, 161, 242, \dodoi{10.3847/1538-3881/abe9bc}

\bibitem[{{Fraser} {et~al.}(2021){Fraser}, {Stritzinger}, {Brennan},
  {Pastorello}, {Cai}, {Piro}, {Ashall}, {Brown}, {Burns}, {Elias-Rosa},
  {Kotak}, {Filippenko}, {Galbany}, {Hsiao}, {Jha}, {Reguitti}, {Zhang},
  {Moran}, {Morrell}, {Shappee}, {Tomasella}, {Anderson}, {Barna}, {Ochner},
  {Phillips}, {Tucker}, {Wang}, {Baron}, {Benetti}, {Bersten}, {Brink},
  {Camacho-Neves}, {Davis}, {Dettman}, {Folatelli}, {Gutierrez}, {Hoflich},
  {Holoien}, {Kankare}, {Kumar}, {Lu}, {Mazzali}, {Taubenberger}, {Tinyanont},
  {Kuncarayakti}, {Kwok}, {Shahbandeh}, {Suntzeff}, {Yan}, {Yang}, \&
  {Zheng}}]{2021arXiv210807278F}
{Fraser}, M., {Stritzinger}, M.~D., {Brennan}, S.~J., {et~al.} 2021, arXiv
  e-prints, arXiv:2108.07278, \dodoi{10.48550/arXiv.2108.07278}

\bibitem[{{Frieman} {et~al.}(2008){Frieman}, {Bassett}, {Becker}, {Choi},
  {Cinabro}, {DeJongh}, {Depoy}, {Dilday}, {Doi}, {Garnavich}, {Hogan},
  {Holtzman}, {Im}, {Jha}, {Kessler}, {Konishi}, {Lampeitl}, {Marriner},
  {Marshall}, {McGinnis}, {Miknaitis}, {Nichol}, {Prieto}, {Riess}, {Richmond},
  {Romani}, {Sako}, {Schneider}, {Smith}, {Takanashi}, {Tokita}, {van der
  Heyden}, {Yasuda}, {Zheng}, {Adelman-McCarthy}, {Annis}, {Assef},
  {Barentine}, {Bender}, {Blandford}, {Boroski}, {Bremer}, {Brewington},
  {Collins}, {Crotts}, {Dembicky}, {Eastman}, {Edge}, {Edmondson}, {Elson},
  {Eyler}, {Filippenko}, {Foley}, {Frank}, {Goobar}, {Gueth}, {Gunn},
  {Harvanek}, {Hopp}, {Ihara}, {Ivezi{\'c}}, {Kahn}, {Kaplan}, {Kent},
  {Ketzeback}, {Kleinman}, {Kollatschny}, {Kron}, {Krzesi{\'n}ski}, {Lamenti},
  {Leloudas}, {Lin}, {Long}, {Lucey}, {Lupton}, {Malanushenko}, {Malanushenko},
  {McMillan}, {Mendez}, {Morgan}, {Morokuma}, {Nitta}, {Ostman}, {Pan},
  {Rockosi}, {Romer}, {Ruiz-Lapuente}, {Saurage}, {Schlesinger}, {Snedden},
  {Sollerman}, {Stoughton}, {Stritzinger}, {Subba Rao}, {Tucker}, {Vaisanen},
  {Watson}, {Watters}, {Wheeler}, {Yanny}, \& {York}}]{2008AJ....135..338F}
{Frieman}, J.~A., {Bassett}, B., {Becker}, A., {et~al.} 2008, \aj, 135, 338,
  \dodoi{10.1088/0004-6256/135/1/338}

\bibitem[{{Galbany}(2020)}]{sirah}
{Galbany}, L. 2020, in XIV.0 Scientific Meeting (virtual) of the Spanish
  Astronomical Society, 37

\bibitem[{{Galbany} {et~al.}(2020{\natexlab{a}}){Galbany}, {Lavers}, {Ashall},
  {Karamehmetoglu}, {Stritzinger}, {Morales-Garoffolo}, \&
  {Rosa}}]{tns_2020kyx}
{Galbany}, L., {Lavers}, A.~L.~C., {Ashall}, C., {et~al.} 2020{\natexlab{a}},
  Transient Name Server Classification Report, 2020-1560, 1

\bibitem[{{Galbany} {et~al.}(2020{\natexlab{b}}){Galbany}, {Lavers}, {Jha},
  {Ashall}, {Stritzinger}, {Morales-Garoffolo}, \& {Rosa}}]{tns_2020rlj}
{Galbany}, L., {Lavers}, A.~L.~C., {Jha}, S., {et~al.} 2020{\natexlab{b}},
  Transient Name Server Classification Report, 2020-2523, 1

\bibitem[{{Galbany} {et~al.}(2020{\natexlab{c}}){Galbany}, {Lavers},
  {Karamehmetoglu}, {Stritzinger}, {Ashall}, {Morales-Garoffolo}, \&
  {Rosa}}]{tns_2020nny}
{Galbany}, L., {Lavers}, A.~L.~C., {Karamehmetoglu}, E., {et~al.}
  2020{\natexlab{c}}, Transient Name Server Classification Report, 2020-1971, 1

\bibitem[{{Galbany} {et~al.}(2020{\natexlab{d}}){Galbany}, {Lavers},
  {Stritzinger}, {Ashall}, {Morales-Garoffolo}, \& {Rosa}}]{tns_2020itj}
{Galbany}, L., {Lavers}, A.~L.~C., {Stritzinger}, M., {et~al.}
  2020{\natexlab{d}}, Transient Name Server Classification Report, 2020-1269, 1

\bibitem[{{Galbany} {et~al.}(2020{\natexlab{e}}){Galbany}, {Lavers},
  {Stritzinger}, {Ashall}, {Morales-Garoffolo}, \& {Rosa}}]{tns_2020jhf}
---. 2020{\natexlab{e}}, Transient Name Server Classification Report,
  2020-1280, 1

\bibitem[{{Galbany} {et~al.}(2020{\natexlab{f}}){Galbany}, {Lavers},
  {Stritzinger}, {Ashall}, {Morales-Garoffolo}, \&
  {Rosa}}]{tns_2020kjt_2020kku}
---. 2020{\natexlab{f}}, Transient Name Server Classification Report,
  2020-1420, 1

\bibitem[{{Galbany} {et~al.}(2020{\natexlab{g}}){Galbany}, {Lavers}, {Foley},
  {Jha}, {Ashall}, {Stritzinger}, {Morales-Garoffolo}, {Rosa}, \&
  {Project}}]{tns_2020jgl}
{Galbany}, L., {Lavers}, A.~L.~C., {Foley}, R., {et~al.} 2020{\natexlab{g}},
  Transient Name Server Classification Report, 2020-1270, 1

\bibitem[{{Gonz{\'a}lez-Gait{\'a}n}
  {et~al.}(2015{\natexlab{a}}){Gonz{\'a}lez-Gait{\'a}n}, {Tominaga}, {Molina},
  {Galbany}, {Bufano}, {Anderson}, {Gutierrez}, {F{\"o}rster}, {Pignata},
  {Bersten}, {Howell}, {Sullivan}, {Carlberg}, {de Jaeger}, {Hamuy},
  {Baklanov}, \& {Blinnikov}}]{GonzalezGaitan_2015}
{Gonz{\'a}lez-Gait{\'a}n}, S., {Tominaga}, N., {Molina}, J., {et~al.}
  2015{\natexlab{a}}, \mnras, 451, 2212, \dodoi{10.1093/mnras/stv1097}

\bibitem[{{Gonz{\'a}lez-Gait{\'a}n}
  {et~al.}(2015{\natexlab{b}}){Gonz{\'a}lez-Gait{\'a}n}, {Tominaga}, {Molina},
  {Galbany}, {Bufano}, {Anderson}, {Gutierrez}, {F{\"o}rster}, {Pignata},
  {Bersten}, {Howell}, {Sullivan}, {Carlberg}, {de Jaeger}, {Hamuy},
  {Baklanov}, \& {Blinnikov}}]{2015MNRAS.451.2212G}
---. 2015{\natexlab{b}}, \mnras, 451, 2212, \dodoi{10.1093/mnras/stv1097}

\bibitem[{{Graham} {et~al.}(2019){Graham}, {Kulkarni}, {Bellm}, {Adams},
  {Barbarino}, {Blagorodnova}, {Bodewits}, {Bolin}, {Brady}, {Cenko}, {Chang},
  {Coughlin}, {De}, {Eadie}, {Farnham}, {Feindt}, {Franckowiak}, {Fremling},
  {Gezari}, {Ghosh}, {Goldstein}, {Golkhou}, {Goobar}, {Ho}, {Huppenkothen},
  {Ivezi{\'c}}, {Jones}, {Juric}, {Kaplan}, {Kasliwal}, {Kelley}, {Kupfer},
  {Lee}, {Lin}, {Lunnan}, {Mahabal}, {Miller}, {Ngeow}, {Nugent}, {Ofek},
  {Prince}, {Rauch}, {van Roestel}, {Schulze}, {Singer}, {Sollerman}, {Taddia},
  {Yan}, {Ye}, {Yu}, {Barlow}, {Bauer}, {Beck}, {Belicki}, {Biswas}, {Brinnel},
  {Brooke}, {Bue}, {Bulla}, {Burruss}, {Connolly}, {Cromer}, {Cunningham},
  {Dekany}, {Delacroix}, {Desai}, {Duev}, {Feeney}, {Flynn}, {Frederick},
  {Gal-Yam}, {Giomi}, {Groom}, {Hacopians}, {Hale}, {Helou}, {Henning},
  {Hover}, {Hillenbrand}, {Howell}, {Hung}, {Imel}, {Ip}, {Jackson}, {Kaspi},
  {Kaye}, {Kowalski}, {Kramer}, {Kuhn}, {Landry}, {Laher}, {Mao}, {Masci},
  {Monkewitz}, {Murphy}, {Nordin}, {Patterson}, {Penprase}, {Porter},
  {Rebbapragada}, {Reiley}, {Riddle}, {Rigault}, {Rodriguez}, {Rusholme}, {van
  Santen}, {Shupe}, {Smith}, {Soumagnac}, {Stein}, {Surace}, {Szkody}, {Terek},
  {Van Sistine}, {van Velzen}, {Vestrand}, {Walters}, {Ward}, {Zhang}, \&
  {Zolkower}}]{Graham2019}
{Graham}, M.~J., {Kulkarni}, S.~R., {Bellm}, E.~C., {et~al.} 2019, \pasp, 131,
  078001, \dodoi{10.1088/1538-3873/ab006c}

\bibitem[{{Guti{\'e}rrez} {et~al.}(2017){Guti{\'e}rrez}, {Anderson}, {Hamuy},
  {Morrell}, {Gonz{\'a}lez-Gaitan}, {Stritzinger}, {Phillips}, {Galbany},
  {Folatelli}, {Dessart}, {Contreras}, {Della Valle}, {Freedman}, {Hsiao},
  {Krisciunas}, {Madore}, {Maza}, {Suntzeff}, {Prieto}, {Gonz{\'a}lez},
  {Cappellaro}, {Navarrete}, {Pizzella}, {Ruiz}, {Smith}, \&
  {Turatto}}]{Gutierrez_2017}
{Guti{\'e}rrez}, C.~P., {Anderson}, J.~P., {Hamuy}, M., {et~al.} 2017, \apj,
  850, 89, \dodoi{10.3847/1538-4357/aa8f52}

\bibitem[{{Guy} {et~al.}(2005){Guy}, {Astier}, {Nobili}, {Regnault}, \&
  {Pain}}]{Guy_2005}
{Guy}, J., {Astier}, P., {Nobili}, S., {Regnault}, N., \& {Pain}, R. 2005,
  \aap, 443, 781, \dodoi{10.1051/0004-6361:20053025}

\bibitem[{Guy {et~al.}(2007)Guy, Astier, Baumont, Hardin, Pain, Regnault, Basa,
  Carlberg, Conley, Fabbro, Fouchez, Hook, Howell, Perrett, Pritchet, Rich,
  Sullivan, Antilogus, Aubourg, Bazin, Bronder, Filiol, Palanque-Delabrouille,
  Ripoche, \& Ruhlmann-Kleider}]{Guy_2007}
Guy, J., Astier, P., Baumont, S., {et~al.} 2007, A\& A, 466, 11,
  \dodoi{10.1051/0004-6361:20066930}

\bibitem[{{Hicken} {et~al.}(2009){Hicken}, {Challis}, {Jha}, {Kirshner},
  {Matheson}, {Modjaz}, {Rest}, {Wood-Vasey}, {Bakos}, {Barton}, {Berlind},
  {Bragg}, {Brice{\~n}o}, {Brown}, {Caldwell}, {Calkins}, {Cho}, {Ciupik},
  {Contreras}, {Dendy}, {Dosaj}, {Durham}, {Eriksen}, {Esquerdo}, {Everett},
  {Falco}, {Fernandez}, {Gaba}, {Garnavich}, {Graves}, {Green}, {Groner},
  {Hergenrother}, {Holman}, {Hradecky}, {Huchra}, {Hutchison}, {Jerius},
  {Jordan}, {Kilgard}, {Krauss}, {Luhman}, {Macri}, {Marrone}, {McDowell},
  {McIntosh}, {McNamara}, {Megeath}, {Mochejska}, {Munoz}, {Muzerolle},
  {Naranjo}, {Narayan}, {Pahre}, {Peters}, {Peterson}, {Rines}, {Ripman},
  {Roussanova}, {Schild}, {Sicilia-Aguilar}, {Sokoloski}, {Smalley}, {Smith},
  {Spahr}, {Stanek}, {Barmby}, {Blondin}, {Stubbs}, {Szentgyorgyi}, {Torres},
  {Vaz}, {Vikhlinin}, {Wang}, {Westover}, {Woods}, \&
  {Zhao}}]{2009ApJ...700..331H}
{Hicken}, M., {Challis}, P., {Jha}, S., {et~al.} 2009, \apj, 700, 331,
  \dodoi{10.1088/0004-637X/700/1/331}

\bibitem[{{Hicken} {et~al.}(2012){Hicken}, {Challis}, {Kirshner}, {Rest},
  {Cramer}, {Wood-Vasey}, {Bakos}, {Berlind}, {Brown}, {Caldwell}, {Calkins},
  {Currie}, {de Kleer}, {Esquerdo}, {Everett}, {Falco}, {Fernandez},
  {Friedman}, {Groner}, {Hartman}, {Holman}, {Hutchins}, {Keys}, {Kipping},
  {Latham}, {Marion}, {Narayan}, {Pahre}, {Pal}, {Peters}, {Perumpilly},
  {Ripman}, {Sipocz}, {Szentgyorgyi}, {Tang}, {Torres}, {Vaz}, {Wolk}, \&
  {Zezas}}]{2012ApJS..200...12H}
{Hicken}, M., {Challis}, P., {Kirshner}, R.~P., {et~al.} 2012, \apjs, 200, 12,
  \dodoi{10.1088/0067-0049/200/2/12}

\bibitem[{{Hoogendam} {et~al.}(2025){Hoogendam}, {Jones}, {Ashall}, {Shappee},
  {Foley}, {Tucker}, {Huber}, {Auchettl}, {Desai}, {Do}, {Hinkle}, {Romagnoli},
  {Shi}, {Syncatto}, {Angus}, {Chambers}, {Coulter}, {Davis}, {de Boer},
  {Gagliano}, {Kong}, {Lin}, {Lowe}, {Magnier}, {Minguez}, {Pan}, {Patra},
  {Severson}, {Taggart}, {Wasserman}, \& {Yadavalli}}]{Hoogendam25}
{Hoogendam}, W.~B., {Jones}, D.~O., {Ashall}, C., {et~al.} 2025, arXiv
  e-prints, arXiv:2502.17556, \dodoi{10.48550/arXiv.2502.17556}

\bibitem[{{Hosseinzadeh} {et~al.}(2017){Hosseinzadeh}, {Sand}, {Valenti},
  {Brown}, {Howell}, {McCully}, {Kasen}, {Arcavi}, {Bostroem}, {Tartaglia},
  {Hsiao}, {Davis}, {Shahbandeh}, \& {Stritzinger}}]{2017ApJ...845L..11H}
{Hosseinzadeh}, G., {Sand}, D.~J., {Valenti}, S., {et~al.} 2017, \apjl, 845,
  L11, \dodoi{10.3847/2041-8213/aa8402}

\bibitem[{{Hosseinzadeh} {et~al.}(2022){Hosseinzadeh}, {Sand}, {Lundqvist},
  {Andrews}, {Bostroem}, {Dong}, {Janzen}, {Jencson}, {Lundquist}, {Meza
  Retamal}, {Pearson}, {Valenti}, {Wyatt}, {Burke}, {Howell}, {McCully},
  {Newsome}, {Gonzalez}, {Pellegrino}, {Terreran}, {Kwok}, {Jha}, {Strader},
  {Kundu}, {Ryder}, {Haislip}, {Kouprianov}, \&
  {Reichart}}]{2022ApJ...933L..45H}
{Hosseinzadeh}, G., {Sand}, D.~J., {Lundqvist}, P., {et~al.} 2022, \apjl, 933,
  L45, \dodoi{10.3847/2041-8213/ac7cef}

\bibitem[{{Howell} {et~al.}(2016){Howell}, {Arcavi}, {Hosseinzadeh}, {McCully},
  {Valenti}, \& {LCOGT Supernova Key Project}}]{LCOGT_2016}
{Howell}, D.~A., {Arcavi}, I., {Hosseinzadeh}, G., {et~al.} 2016, in American
  Astronomical Society Meeting Abstracts, Vol. 228, American Astronomical
  Society Meeting Abstracts \#228, 219.12

\bibitem[{{Hoyle} \& {Fowler}(1960)}]{1960ApJ...132..565H}
{Hoyle}, F., \& {Fowler}, W.~A. 1960, \apj, 132, 565, \dodoi{10.1086/146963}

\bibitem[{Hsiao {et~al.}(2007)Hsiao, Conley, Howell, Sullivan, Pritchet,
  Carlberg, Nugent, \& Phillips}]{Hsiao_2007}
Hsiao, E.~Y., Conley, A., Howell, D.~A., {et~al.} 2007, The Astrophysical
  Journal, 663, 1187–1200, \dodoi{10.1086/518232}

\bibitem[{{Hsiao} {et~al.}(2007){Hsiao}, {Conley}, {Howell}, {Sullivan},
  {Pritchet}, {Carlberg}, {Nugent}, \& {Phillips}}]{2007ApJ...663.1187H}
{Hsiao}, E.~Y., {Conley}, A., {Howell}, D.~A., {et~al.} 2007, \apj, 663, 1187,
  \dodoi{10.1086/518232}

\bibitem[{{Ivezi{\'c}} {et~al.}(2019){Ivezi{\'c}}, {Kahn}, {Tyson}, {Abel},
  {Acosta}, {Allsman}, {Alonso}, {AlSayyad}, {Anderson}, {Andrew}, {Angel},
  {Angeli}, {Ansari}, {Antilogus}, {Araujo}, {Armstrong}, {Arndt}, {Astier},
  {Aubourg}, {Auza}, {Axelrod}, {Bard}, {Barr}, {Barrau}, {Bartlett}, {Bauer},
  {Bauman}, {Baumont}, {Bechtol}, {Bechtol}, {Becker}, {Becla}, {Beldica},
  {Bellavia}, {Bianco}, {Biswas}, {Blanc}, {Blazek}, {Blandford}, {Bloom},
  {Bogart}, {Bond}, {Booth}, {Borgland}, {Borne}, {Bosch}, {Boutigny},
  {Brackett}, {Bradshaw}, {Brandt}, {Brown}, {Bullock}, {Burchat}, {Burke},
  {Cagnoli}, {Calabrese}, {Callahan}, {Callen}, {Carlin}, {Carlson},
  {Chandrasekharan}, {Charles-Emerson}, {Chesley}, {Cheu}, {Chiang}, {Chiang},
  {Chirino}, {Chow}, {Ciardi}, {Claver}, {Cohen-Tanugi}, {Cockrum}, {Coles},
  {Connolly}, {Cook}, {Cooray}, {Covey}, {Cribbs}, {Cui}, {Cutri}, {Daly},
  {Daniel}, {Daruich}, {Daubard}, {Daues}, {Dawson}, {Delgado}, {Dellapenna},
  {de Peyster}, {de Val-Borro}, {Digel}, {Doherty}, {Dubois},
  {Dubois-Felsmann}, {Durech}, {Economou}, {Eifler}, {Eracleous}, {Emmons},
  {Fausti Neto}, {Ferguson}, {Figueroa}, {Fisher-Levine}, {Focke}, {Foss},
  {Frank}, {Freemon}, {Gangler}, {Gawiser}, {Geary}, {Gee}, {Geha}, {Gessner},
  {Gibson}, {Gilmore}, {Glanzman}, {Glick}, {Goldina}, {Goldstein}, {Goodenow},
  {Graham}, {Gressler}, {Gris}, {Guy}, {Guyonnet}, {Haller}, {Harris},
  {Hascall}, {Haupt}, {Hernandez}, {Herrmann}, {Hileman}, {Hoblitt}, {Hodgson},
  {Hogan}, {Howard}, {Huang}, {Huffer}, {Ingraham}, {Innes}, {Jacoby}, {Jain},
  {Jammes}, {Jee}, {Jenness}, {Jernigan}, {Jevremovi{\'c}}, {Johns}, {Johnson},
  {Johnson}, {Jones}, {Juramy-Gilles}, {Juri{\'c}}, {Kalirai}, {Kallivayalil},
  {Kalmbach}, {Kantor}, {Karst}, {Kasliwal}, {Kelly}, {Kessler}, {Kinnison},
  {Kirkby}, {Knox}, {Kotov}, {Krabbendam}, {Krughoff}, {Kub{\'a}nek},
  {Kuczewski}, {Kulkarni}, {Ku}, {Kurita}, {Lage}, {Lambert}, {Lange},
  {Langton}, {Le Guillou}, {Levine}, {Liang}, {Lim}, {Lintott}, {Long},
  {Lopez}, {Lotz}, {Lupton}, {Lust}, {MacArthur}, {Mahabal}, {Mandelbaum},
  {Markiewicz}, {Marsh}, {Marshall}, {Marshall}, {May}, {McKercher}, {McQueen},
  {Meyers}, {Migliore}, {Miller}, \& {Mills}}]{2019ApJ...873..111I}
{Ivezi{\'c}}, {\v{Z}}., {Kahn}, S.~M., {Tyson}, J.~A., {et~al.} 2019, \apj,
  873, 111, \dodoi{10.3847/1538-4357/ab042c}

\bibitem[{{Jacobson-Gal{\'a}n} {et~al.}(2022){Jacobson-Gal{\'a}n}, {Dessart},
  {Jones}, {Margutti}, {Coppejans}, {Dimitriadis}, {Foley}, {Kilpatrick},
  {Matthews}, {Rest}, {Terreran}, {Aleo}, {Auchettl}, {Blanchard}, {Coulter},
  {Davis}, {de Boer}, {DeMarchi}, {Drout}, {Earl}, {Gagliano}, {Gall},
  {Hjorth}, {Huber}, {Ibik}, {Milisavljevic}, {Pan}, {Rest}, {Ridden-Harper},
  {Rojas-Bravo}, {Siebert}, {Smith}, {Taggart}, {Tinyanont}, {Wang}, \&
  {Zenati}}]{2022ApJ...924...15J}
{Jacobson-Gal{\'a}n}, W.~V., {Dessart}, L., {Jones}, D.~O., {et~al.} 2022,
  \apj, 924, 15, \dodoi{10.3847/1538-4357/ac3f3a}

\bibitem[{{Jacobson-Gal{\'a}n} {et~al.}(2024){Jacobson-Gal{\'a}n}, {Dessart},
  {Davis}, {Kilpatrick}, {Margutti}, {Foley}, {Chornock}, {Terreran},
  {Hiramatsu}, {Newsome}, {Padilla Gonzalez}, {Pellegrino}, {Howell},
  {Filippenko}, {Anderson}, {Angus}, {Auchettl}, {Bostroem}, {Brink},
  {Cartier}, {Coulter}, {de Boer}, {Drout}, {Earl}, {Ertini}, {Farah},
  {Farias}, {Gall}, {Gao}, {Gerlach}, {Guo}, {Haynie}, {Hosseinzadeh}, {Ibik},
  {Jha}, {Jones}, {Langeroodi}, {LeBaron}, {Magnier}, {Piro}, {Raimundo},
  {Rest}, {Rest}, {Rich}, {Rojas-Bravo}, {Sears}, {Taggart}, {Villar},
  {Wainscoat}, {Wang}, {Wasserman}, {Yan}, {Yang}, {Zhang}, \&
  {Zheng}}]{2024ApJ...970..189J}
{Jacobson-Gal{\'a}n}, W.~V., {Dessart}, L., {Davis}, K.~W., {et~al.} 2024,
  \apj, 970, 189, \dodoi{10.3847/1538-4357/ad4a2a}

\bibitem[{{Jha} {et~al.}(2020){Jha}, {Avelino}, {Burns}, {Camacho-Neves},
  {Dai}, {Dettman}, {Dhawan}, {Filippenko}, {Foley}, {Friedman}, {Galbany},
  {Garnavich}, {Hlozek}, {Holoien}, {Hounsell}, {Hsiao}, {Jones}, {Kelly},
  {Kessler}, {Kirshner}, {Mandel}, {Matheson}, {Narayan}, {Phillips}, {Ponder},
  {Rest}, {Riess}, {Pierel}, {Rodney}, {Sand}, {Scolnic}, {Stritzinger},
  {Strolger}, \& {Valenti}}]{2020hst..prop16234J}
{Jha}, S.~W., {Avelino}, A., {Burns}, C., {et~al.} 2020, {Supernovae in the
  Infrared avec Hubble}, HST Proposal. Cycle 28, ID. \#16234

\bibitem[{{Jones} {et~al.}(2022){Jones}, {Mandel}, {Kirshner}, {Thorp},
  {Challis}, {Avelino}, {Brout}, {Burns}, {Foley}, {Pan}, {Scolnic}, {Siebert},
  {Chornock}, {Freedman}, {Friedman}, {Frieman}, {Galbany}, {Hsiao}, {Kelsey},
  {Marion}, {Nichol}, {Nugent}, {Phillips}, {Rest}, {Riess}, {Sako}, {Smith},
  {Wiseman}, \& {Wood-Vasey}}]{2022ApJ...933..172J}
{Jones}, D.~O., {Mandel}, K.~S., {Kirshner}, R.~P., {et~al.} 2022, \apj, 933,
  172, \dodoi{10.3847/1538-4357/ac755b}

\bibitem[{{Khazov} {et~al.}(2016){Khazov}, {Yaron}, {Gal-Yam}, {Manulis},
  {Rubin}, {Kulkarni}, {Arcavi}, {Kasliwal}, {Ofek}, {Cao}, {Perley},
  {Sollerman}, {Horesh}, {Sullivan}, {Filippenko}, {Nugent}, {Howell}, {Cenko},
  {Silverman}, {Ebeling}, {Taddia}, {Johansson}, {Laher}, {Surace},
  {Rebbapragada}, {Wozniak}, \& {Matheson}}]{2016ApJ...818....3K}
{Khazov}, D., {Yaron}, O., {Gal-Yam}, A., {et~al.} 2016, \apj, 818, 3,
  \dodoi{10.3847/0004-637X/818/1/3}

\bibitem[{{Kilpatrick} {et~al.}(2023){Kilpatrick}, {Izzo}, {Bentley},
  {Chambers}, {Coulter}, {Drout}, {de Boer}, {Foley}, {Gall}, {Halford},
  {Jones}, {Langeroodi}, {Lin}, {Magnier}, {McGill}, {O'Grady}, {Pan},
  {Ramirez-Ruiz}, {Rest}, {Swift}, {Tinyanont}, {Villar}, {Wainscoat},
  {Wasserman}, {Yadavalli}, \& {Yang}}]{Kilpatrick_2023}
{Kilpatrick}, C.~D., {Izzo}, L., {Bentley}, R.~O., {et~al.} 2023, \mnras, 524,
  2161, \dodoi{10.1093/mnras/stad1954}

\bibitem[{{Kim} {et~al.}(2016){Kim}, {Lee}, {Park}, {Kim}, {Cha}, {Lee}, {Han},
  {Chun}, \& {Yuk}}]{2016JKAS...49...37K}
{Kim}, S.-L., {Lee}, C.-U., {Park}, B.-G., {et~al.} 2016, Journal of Korean
  Astronomical Society, 49, 37, \dodoi{10.5303/JKAS.2016.49.1.37}

\bibitem[{{Krisciunas} {et~al.}(2017){Krisciunas}, {Contreras}, {Burns},
  {Phillips}, {Stritzinger}, {Morrell}, {Hamuy}, {Anais}, {Boldt}, {Busta},
  {Campillay}, {Castell{\'o}n}, {Folatelli}, {Freedman}, {Gonz{\'a}lez},
  {Hsiao}, {Krzeminski}, {Persson}, {Roth}, {Salgado}, {Ser{\'o}n}, {Suntzeff},
  {Torres}, {Filippenko}, {Li}, {Madore}, {DePoy}, {Marshall}, {Rheault}, \&
  {Villanueva}}]{2017AJ....154..211K}
{Krisciunas}, K., {Contreras}, C., {Burns}, C.~R., {et~al.} 2017, \aj, 154,
  211, \dodoi{10.3847/1538-3881/aa8df0}

\bibitem[{{Kumar} {et~al.}(2013)}]{Kumar13}
{Kumar}, B., {et~al.} 2013, \mnras, 431, 308, \dodoi{10.1093/mnras/stt162}

\bibitem[{{Liu} {et~al.}(2016){Liu}, {Modjaz}, {Bianco}, \& {Graur}}]{Liu16}
{Liu}, Y.-Q., {Modjaz}, M., {Bianco}, F.~B., \& {Graur}, O. 2016, \apj, 827,
  90, \dodoi{10.3847/0004-637X/827/2/90}

\bibitem[{{Maguire} {et~al.}(2012){Maguire}, {Jerkstrand}, {Smartt},
  {Fransson}, {Pastorello}, {Benetti}, {Valenti}, {Bufano}, \&
  {Leloudas}}]{2012MNRAS.420.3451M}
{Maguire}, K., {Jerkstrand}, A., {Smartt}, S.~J., {et~al.} 2012, \mnras, 420,
  3451, \dodoi{10.1111/j.1365-2966.2011.20276.x}

\bibitem[{Masci {et~al.}(2018)Masci, Laher, Rusholme, Shupe, Groom, Surace,
  Jackson, Monkewitz, Beck, Flynn, Terek, Landry, Hacopians, Desai, Howell,
  Brooke, Imel, Wachter, Ye, Lin, Cenko, Cunningham, Rebbapragada, Bue, Miller,
  Mahabal, Bellm, Patterson, Jurić, Golkhou, Ofek, Walters, Graham, Kasliwal,
  Dekany, Kupfer, Burdge, Cannella, Barlow, Sistine, Giomi, Fremling,
  Blagorodnova, Levitan, Riddle, Smith, Helou, Prince, \& Kulkarni}]{ZTF}
Masci, F.~J., Laher, R.~R., Rusholme, B., {et~al.} 2018, Publications of the
  Astronomical Society of the Pacific, 131, 018003,
  \dodoi{10.1088/1538-3873/aae8ac}

\bibitem[{{Miller} {et~al.}(2020{\natexlab{a}}){Miller}, {Yao}, {Bulla},
  {Pankow}, {Bellm}, {Cenko}, {Dekany}, {Fremling}, {Graham}, {Kupfer},
  {Laher}, {Mahabal}, {Masci}, {Nugent}, {Riddle}, {Rusholme}, {Smith},
  {Shupe}, {van Roestel}, \& {Kulkarni}}]{2020ApJ...902...47M}
{Miller}, A.~A., {Yao}, Y., {Bulla}, M., {et~al.} 2020{\natexlab{a}}, \apj,
  902, 47, \dodoi{10.3847/1538-4357/abb13b}

\bibitem[{{Miller} {et~al.}(2020{\natexlab{b}}){Miller}, {Magee}, {Polin},
  {Maguire}, {Zimmerman}, {Yao}, {Sollerman}, {Schulze}, {Perley}, {Kromer},
  {Dhawan}, {Bulla}, {Andreoni}, {Bellm}, {De}, {Dekany}, {Delacroix},
  {Fremling}, {Gal-Yam}, {Goldstein}, {Golkhou}, {Goobar}, {Graham}, {Irani},
  {Kasliwal}, {Kaye}, {Kim}, {Laher}, {Mahabal}, {Masci}, {Nugent}, {Ofek},
  {Phinney}, {Prentice}, {Riddle}, {Rigault}, {Rusholme}, {Schweyer}, {Shupe},
  {Soumagnac}, {Terreran}, {Walters}, {Yan}, {Zolkower}, \&
  {Kulkarni}}]{2020ApJ...898...56M}
{Miller}, A.~A., {Magee}, M.~R., {Polin}, A., {et~al.} 2020{\natexlab{b}},
  \apj, 898, 56, \dodoi{10.3847/1538-4357/ab9e05}

\bibitem[{{Modjaz} {et~al.}(2014){Modjaz}, {Blondin}, {Kirshner}, {Matheson},
  {Berlind}, {Bianco}, {Calkins}, {Challis}, {Garnavich}, {Hicken}, {Jha},
  {Liu}, \& {Marion}}]{Modjaz14}
{Modjaz}, M., {Blondin}, S., {Kirshner}, R.~P., {et~al.} 2014, \aj, 147, 99,
  \dodoi{10.1088/0004-6256/147/5/99}

\bibitem[{{Morales-Garoffolo} {et~al.}(2015){Morales-Garoffolo}, {Elias-Rosa},
  {Bersten}, {Jerkstrand}, {Taubenberger}, {Benetti}, {Cappellaro}, {Kotak},
  {Pastorello}, \& {Bufano}}]{Morales-Garoffolo15}
{Morales-Garoffolo}, A., {Elias-Rosa}, N., {Bersten}, M., {et~al.} 2015,
  \mnras, 454, 95, \dodoi{10.1093/mnras/stv1972}

\bibitem[{{Morrell} {et~al.}(2024){Morrell}, {Phillips}, {Folatelli},
  {Stritzinger}, {Hamuy}, {Suntzeff}, {Hsiao}, {Taddia}, {Burns}, {Hoeflich},
  {Ashall}, {Contreras}, {Galbany}, {Lu}, {Piro}, {Anais}, {Baron}, {Burrow},
  {Busta}, {Campillay}, {Castell{\'o}n}, {Corco}, {Diamond}, {Freedman},
  {Gonzalez}, {Krisciunas}, {Kumar}, {Persson}, {Ser{\'o}n}, {Shahbandeh},
  {Torres}, {Uddin}, {Anderson}, {Baltay}, {Gall}, {Goobar}, {Hadjiyska},
  {Holmbo}, {Kasliwal}, {Lidman}, {Marion}, {Mazzali}, {Nugent}, {Perlmutter},
  {Pignata}, {Rabinowitz}, {Roth}, {Ryder}, {Shappee}, {Vink{\'o}}, {Wheeler},
  {de Jaeger}, {Lira}, {Ruiz}, {Rich}, {Prieto}, {Di Mille}, {Osip}, {Blanc},
  \& {Palunas}}]{2024ApJ...967...20M}
{Morrell}, N., {Phillips}, M.~M., {Folatelli}, G., {et~al.} 2024, \apj, 967,
  20, \dodoi{10.3847/1538-4357/ad38af}

\bibitem[{{Müller-Bravo} {et~al.}(2022){Müller-Bravo}, {Sullivan}, {Smith},
  {Frohmaier}, {Guti{\'e}rrez}, {Wiseman}, \& {Zontou}}]{2022MNRAS.512.3266M}
{Müller-Bravo}, T.~E., {Sullivan}, M., {Smith}, M., {et~al.} 2022, \mnras,
  512, 3266, \dodoi{10.1093/mnras/stab3065}

\bibitem[{{Ni} {et~al.}(2022){Ni}, {Moon}, {Drout}, {Polin}, {Sand},
  {Gonz{\'a}lez-Gait{\'a}n}, {Kim}, {Lee}, {Park}, {Howell}, {Nugent}, {Piro},
  {Brown}, {Galbany}, {Burke}, {Hiramatsu}, {Hosseinzadeh}, {Valenti},
  {Afsariardchi}, {Andrews}, {Antoniadis}, {Arcavi}, {Beaton}, {Bostroem},
  {Carlberg}, {Cenko}, {Cha}, {Dong}, {Gal-Yam}, {Haislip}, {Holoien},
  {Johnson}, {Kouprianov}, {Lee}, {Matzner}, {Morrell}, {McCully}, {Pignata},
  {Reichart}, {Rich}, {Ryder}, {Smith}, {Wyatt}, \&
  {Yang}}]{2022NatAs...6..568N}
{Ni}, Y.~Q., {Moon}, D.-S., {Drout}, M.~R., {et~al.} 2022, Nature Astronomy, 6,
  568, \dodoi{10.1038/s41550-022-01603-4}

\bibitem[{{Oke} \& {Gunn}(1983)}]{1983ApJ...266..713O}
{Oke}, J.~B., \& {Gunn}, J.~E. 1983, \apj, 266, 713, \dodoi{10.1086/160817}

\bibitem[{{Papadogiannakis}(2019)}]{2019PhDT.......134P}
{Papadogiannakis}, S. 2019, PhD thesis, Stockholm University

\bibitem[{{Pastorello} {et~al.}(2008)}]{Pastorello08}
{Pastorello}, A., {et~al.} 2008, \mnras, 389, 955,
  \dodoi{10.1111/j.1365-2966.2008.13618.x}

\bibitem[{{Pellegrino} {et~al.}(2020){Pellegrino}, {Howell}, {Sarbadhicary},
  {Burke}, {Hiramatsu}, {McCully}, {Milne}, {Andrews}, {Brown}, {Chomiuk},
  {Hsiao}, {Sand}, {Shahbandeh}, {Smith}, {Valenti}, {Vink{\'o}}, {Wheeler},
  {Wyatt}, \& {Yang}}]{2020ApJ...897..159P}
{Pellegrino}, C., {Howell}, D.~A., {Sarbadhicary}, S.~K., {et~al.} 2020, \apj,
  897, 159, \dodoi{10.3847/1538-4357/ab8e3f}

\bibitem[{{Perley} {et~al.}(2020){Perley}, {Barbarino}, {Sollerman},
  {Schweyer}, {Schulze}, \& {Yang}}]{tns_2020jfo}
{Perley}, D., {Barbarino}, C., {Sollerman}, J., {et~al.} 2020, Transient Name
  Server Classification Report, 2020-1259, 1

\bibitem[{{Perlmutter} {et~al.}(1999){Perlmutter}, {Aldering}, {Goldhaber},
  {Knop}, {Nugent}, {Castro}, {Deustua}, {Fabbro}, {Goobar}, {Groom}, {Hook},
  {Kim}, {Kim}, {Lee}, {Nunes}, {Pain}, {Pennypacker}, {Quimby}, {Lidman},
  {Ellis}, {Irwin}, {McMahon}, {Ruiz-Lapuente}, {Walton}, {Schaefer}, {Boyle},
  {Filippenko}, {Matheson}, {Fruchter}, {Panagia}, {Newberg}, {Couch}, \&
  {Project}}]{1999ApJ...517..565P}
{Perlmutter}, S., {Aldering}, G., {Goldhaber}, G., {et~al.} 1999, \apj, 517,
  565, \dodoi{10.1086/307221}

\bibitem[{{Phillips} {et~al.}(2019){Phillips}, {Contreras}, {Hsiao}, {Morrell},
  {Burns}, {Stritzinger}, {Ashall}, {Freedman}, {Hoeflich}, {Persson}, {Piro},
  {Suntzeff}, {Uddin}, {Anais}, {Baron}, {Busta}, {Campillay}, {Castell{\'o}n},
  {Corco}, {Diamond}, {Gall}, {Gonzalez}, {Holmbo}, {Krisciunas}, {Roth},
  {Ser{\'o}n}, {Taddia}, {Torres}, {Anderson}, {Baltay}, {Folatelli},
  {Galbany}, {Goobar}, {Hadjiyska}, {Hamuy}, {Kasliwal}, {Lidman}, {Nugent},
  {Perlmutter}, {Rabinowitz}, {Ryder}, {Schmidt}, {Shappee}, \&
  {Walker}}]{2019PASP..131a4001P}
{Phillips}, M.~M., {Contreras}, C., {Hsiao}, E.~Y., {et~al.} 2019, \pasp, 131,
  014001, \dodoi{10.1088/1538-3873/aae8bd}

\bibitem[{{Pignata} {et~al.}(2009){Pignata}, {Maza}, {Antezana}, {Cartier},
  {Folatelli}, {Forster}, {Gonzalez}, {Gonzalez}, {Hamuy}, {Iturra}, {Lopez},
  {Silva}, {Conuel}, {Crain}, {Foster}, {Ivarsen}, {Lacluyze}, {Nysewander}, \&
  {Reichart}}]{2009AIPC.1111..551P}
{Pignata}, G., {Maza}, J., {Antezana}, R., {et~al.} 2009, in American Institute
  of Physics Conference Series, Vol. 1111, Probing Stellar Populations Out to
  the Distant Universe: Cefalu 2008, Proceedings of the International
  Conference, ed. G.~{Giobbi}, A.~{Tornambe}, G.~{Raimondo}, M.~{Limongi},
  L.~A. {Antonelli}, N.~{Menci}, \& E.~{Brocato}, 551--554,
  \dodoi{10.1063/1.3141608}

\bibitem[{{Prochaska} {et~al.}(2020{\natexlab{a}}){Prochaska}, {Hennawi},
  {Westfall}, {Cooke}, {Wang}, {Hsyu}, {Davies}, \& {Farina}}]{pypeit_1}
{Prochaska}, J.~X., {Hennawi}, J.~F., {Westfall}, K.~B., {et~al.}
  2020{\natexlab{a}}, arXiv e-prints, arXiv:2005.06505.
\newblock \doarXiv{2005.06505}

\bibitem[{{Prochaska} {et~al.}(2020{\natexlab{b}}){Prochaska}, {Hennawi},
  {Cooke}, {Westfall}, {Wang}, {EmAstro}, {Tiffanyhsyu}, {Wasserman},
  {Villaume}, {Marijana777}, {Schindler}, {Young}, {Simha}, {Wilde}, {Tejos},
  {Isbell}, {Fl{\"o}rs}, {Sandford}, {Vasovi{\'c}}, {Betts}, \&
  {Holden}}]{pypeit_2}
{Prochaska}, J.~X., {Hennawi}, J., {Cooke}, R., {et~al.} 2020{\natexlab{b}},
  {pypeit/PypeIt: Release 1.0.0}, v1.0.0,  Zenodo,
  \dodoi{10.5281/zenodo.3743493}

\bibitem[{{Rasmussen} \& {Williams}(2006)}]{Rasmussen2006}
{Rasmussen}, C.~E., \& {Williams}, C. K.~I. 2006, {Gaussian Processes for
  Machine Learning} (The MIT Press)

\bibitem[{{Rau} {et~al.}(2009){Rau}, {Kulkarni}, {Law}, {Bloom}, {Ciardi},
  {Djorgovski}, {Fox}, {Gal-Yam}, {Grillmair}, {Kasliwal}, {Nugent}, {Ofek},
  {Quimby}, {Reach}, {Shara}, {Bildsten}, {Cenko}, {Drake}, {Filippenko},
  {Helfand}, {Helou}, {Howell}, {Poznanski}, \&
  {Sullivan}}]{2009PASP..121.1334R}
{Rau}, A., {Kulkarni}, S.~R., {Law}, N.~M., {et~al.} 2009, \pasp, 121, 1334,
  \dodoi{10.1086/605911}

\bibitem[{{Richmond} {et~al.}(1994){Richmond}, {Treffers}, {Filippenko},
  {Paik}, {Leibundgut}, {Schulman}, \& {Cox}}]{Richmond94}
{Richmond}, M.~W., {Treffers}, R.~R., {Filippenko}, A.~V., {et~al.} 1994, \aj,
  107, 1022, \dodoi{10.1086/116915}

\bibitem[{{Riess} {et~al.}(1998){Riess}, {Filippenko}, {Challis},
  {Clocchiatti}, {Diercks}, {Garnavich}, {Gilliland}, {Hogan}, {Jha},
  {Kirshner}, {Leibundgut}, {Phillips}, {Reiss}, {Schmidt}, {Schommer},
  {Smith}, {Spyromilio}, {Stubbs}, {Suntzeff}, \&
  {Tonry}}]{1998AJ....116.1009R}
{Riess}, A.~G., {Filippenko}, A.~V., {Challis}, P., {et~al.} 1998, \aj, 116,
  1009, \dodoi{10.1086/300499}

\bibitem[{{Rose} {et~al.}(2021){Rose}, {Baltay}, {Hounsell}, {Macias}, {Rubin},
  {Scolnic}, {Aldering}, {Bohlin}, {Dai}, {Deustua}, {Foley}, {Fruchter},
  {Galbany}, {Jha}, {Jones}, {Joshi}, {Kelly}, {Kessler}, {Kirshner}, {Mandel},
  {Perlmutter}, {Pierel}, {Qu}, {Rabinowitz}, {Rest}, {Riess}, {Rodney},
  {Sako}, {Siebert}, {Strolger}, {Suzuki}, {Thorp}, {Van Dyk}, {Wang}, {Ward},
  \& {Wood-Vasey}}]{2021arXiv211103081R}
{Rose}, B.~M., {Baltay}, C., {Hounsell}, R., {et~al.} 2021, arXiv e-prints,
  arXiv:2111.03081, \dodoi{10.48550/arXiv.2111.03081}

\bibitem[{{Sahu} {et~al.}(2013){Sahu}, {Anupama}, \& {Chakradhari}}]{Sahu13}
{Sahu}, D.~K., {Anupama}, G.~C., \& {Chakradhari}, N.~K. 2013, \mnras, 433, 2,
  \dodoi{10.1093/mnras/stt647}

\bibitem[{{Schlafly} \& {Finkbeiner}(2011)}]{Schlafly_Finkbeiner_2011}
{Schlafly}, E.~F., \& {Finkbeiner}, D.~P. 2011, \apj, 737, 103,
  \dodoi{10.1088/0004-637X/737/2/103}

\bibitem[{{Shappee} {et~al.}(2014){Shappee}, {Prieto}, {Grupe}, {Kochanek},
  {Stanek}, {De Rosa}, {Mathur}, {Zu}, {Peterson}, {Pogge}, {Komossa}, {Im},
  {Jencson}, {Holoien}, {Basu}, {Beacom}, {Szczygie{\l}}, {Brimacombe},
  {Adams}, {Campillay}, {Choi}, {Contreras}, {Dietrich}, {Dubberley},
  {Elphick}, {Foale}, {Giustini}, {Gonzalez}, {Hawkins}, {Howell}, {Hsiao},
  {Koss}, {Leighly}, {Morrell}, {Mudd}, {Mullins}, {Nugent}, {Parrent},
  {Phillips}, {Pojmanski}, {Rosing}, {Ross}, {Sand}, {Terndrup}, {Valenti},
  {Walker}, \& {Yoon}}]{2014ApJ...788...48S}
{Shappee}, B.~J., {Prieto}, J.~L., {Grupe}, D., {et~al.} 2014, \apj, 788, 48,
  \dodoi{10.1088/0004-637X/788/1/48}

\bibitem[{{Shivvers} {et~al.}(2019){Shivvers}, {Filippenko}, {Silverman},
  {Zheng}, {Foley}, {Chornock}, {Barth}, {Cenko}, {Clubb}, {Fox},
  {Ganeshalingam}, {Graham}, {Kelly}, {Kleiser}, {Leonard}, {Li}, {Matheson},
  {Mauerhan}, {Modjaz}, {Serduke}, {Shields}, {Steele}, {Swift}, {Wong}, \&
  {Yuk}}]{Shivvers19}
{Shivvers}, I., {Filippenko}, A.~V., {Silverman}, J.~M., {et~al.} 2019, \mnras,
  482, 1545, \dodoi{10.1093/mnras/sty2719}

\bibitem[{{Silverman} {et~al.}(2012){Silverman}, {Ganeshalingam}, {Cenko},
  {Filippenko}, {Li}, {Barth}, {Carson}, {Childress}, {Clubb}, {Cucchiara},
  {Graham}, {Marion}, {Nguyen}, {Pei}, {Tucker}, {Vinko}, {Wheeler}, \&
  {Worseck}}]{2012ApJ...756L...7S}
{Silverman}, J.~M., {Ganeshalingam}, M., {Cenko}, S.~B., {et~al.} 2012, \apjl,
  756, L7, \dodoi{10.1088/2041-8205/756/1/L7}

\bibitem[{{Smartt}(2009)}]{2009ARA&A..47...63S}
{Smartt}, S.~J. 2009, \araa, 47, 63,
  \dodoi{10.1146/annurev-astro-082708-101737}

\bibitem[{{Smartt} {et~al.}(2015){Smartt}, {Valenti}, {Fraser}, {Inserra},
  {Young}, {Sullivan}, {Pastorello}, {Benetti}, {Gal-Yam}, {Knapic},
  {Molinaro}, {Smareglia}, {Smith}, {Taubenberger}, {Yaron}, {Anderson},
  {Ashall}, {Balland}, {Baltay}, {Barbarino}, {Bauer}, {Baumont}, {Bersier},
  {Blagorodnova}, {Bongard}, {Botticella}, {Bufano}, {Bulla}, {Cappellaro},
  {Campbell}, {Cellier-Holzem}, {Chen}, {Childress}, {Clocchiatti},
  {Contreras}, {Dall'Ora}, {Danziger}, {de Jaeger}, {De Cia}, {Della Valle},
  {Dennefeld}, {Elias-Rosa}, {Elman}, {Feindt}, {Fleury}, {Gall},
  {Gonzalez-Gaitan}, {Galbany}, {Morales Garoffolo}, {Greggio}, {Guillou},
  {Hachinger}, {Hadjiyska}, {Hage}, {Hillebrandt}, {Hodgkin}, {Hsiao}, {James},
  {Jerkstrand}, {Kangas}, {Kankare}, {Kotak}, {Kromer}, {Kuncarayakti},
  {Leloudas}, {Lundqvist}, {Lyman}, {Hook}, {Maguire}, {Manulis}, {Margheim},
  {Mattila}, {Maund}, {Mazzali}, {McCrum}, {McKinnon}, {Moreno-Raya},
  {Nicholl}, {Nugent}, {Pain}, {Pignata}, {Phillips}, {Polshaw}, {Pumo},
  {Rabinowitz}, {Reilly}, {Romero-Ca{\~n}izales}, {Scalzo}, {Schmidt},
  {Schulze}, {Sim}, {Sollerman}, {Taddia}, {Tartaglia}, {Terreran},
  {Tomasella}, {Turatto}, {Walker}, {Walton}, {Wyrzykowski}, {Yuan}, \&
  {Zampieri}}]{2015A&A...579A..40S}
{Smartt}, S.~J., {Valenti}, S., {Fraser}, M., {et~al.} 2015, \aap, 579, A40,
  \dodoi{10.1051/0004-6361/201425237}

\bibitem[{{Smith} {et~al.}(2020{\natexlab{a}}){Smith}, {Smartt}, {Young},
  {Tonry}, {Denneau}, {Flewelling}, {Heinze}, {Weiland}, {Stalder}, {Rest},
  {Stubbs}, {Anderson}, {Chen}, {Clark}, {Do}, {F{\"o}rster}, {Fulton},
  {Gillanders}, {McBrien}, {O'Neill}, {Srivastav}, \& {Wright}}]{Smith2020}
{Smith}, K.~W., {Smartt}, S.~J., {Young}, D.~R., {et~al.} 2020{\natexlab{a}},
  \pasp, 132, 085002, \dodoi{10.1088/1538-3873/ab936e}

\bibitem[{{Smith} {et~al.}(2020{\natexlab{b}}){Smith}, {Smartt}, {Young},
  {Tonry}, {Denneau}, {Flewelling}, {Heinze}, {Weiland}, {Stalder}, {Rest},
  {Stubbs}, {Anderson}, {Chen}, {Clark}, {Do}, {F{\"o}rster}, {Fulton},
  {Gillanders}, {McBrien}, {O'Neill}, {Srivastav}, \& {Wright}}]{ATLAS_2}
---. 2020{\natexlab{b}}, \pasp, 132, 085002, \dodoi{10.1088/1538-3873/ab936e}

\bibitem[{{Sollerman} {et~al.}(2021){Sollerman}, {Yang}, {Schulze},
  {Strotjohann}, {Jerkstrand}, {Van Dyk}, {Kool}, {Barbarino}, {Brink},
  {Bruch}, {De}, {Filippenko}, {Fremling}, {Patra}, {Perley}, {Yan}, {Yang},
  {Andreoni}, {Campbell}, {Coughlin}, {Kasliwal}, {Kim}, {Rigault}, {Shin},
  {Tzanidakis}, {Ashley}, {Moore}, \& {Travouillon}}]{Sollerman_2021}
{Sollerman}, J., {Yang}, S., {Schulze}, S., {et~al.} 2021, \aap, 655, A105,
  \dodoi{10.1051/0004-6361/202141374}

\bibitem[{{Stritzinger} {et~al.}(2018){Stritzinger}, {Shappee}, {Piro},
  {Ashall}, {Baron}, {Hoeflich}, {Holmbo}, {Holoien}, {Phillips}, {Burns},
  {Contreras}, {Morrell}, \& {Tucker}}]{2018ApJ...864L..35S}
{Stritzinger}, M.~D., {Shappee}, B.~J., {Piro}, A.~L., {et~al.} 2018, \apjl,
  864, L35, \dodoi{10.3847/2041-8213/aadd46}

\bibitem[{{Stritzinger} {et~al.}(2020){Stritzinger}, {Taddia}, {Holmbo},
  {Baron}, {Contreras}, {Karamehmetoglu}, {Phillips}, {Sollerman}, {Suntzeff},
  {Vinko}, {Ashall}, {Avila}, {Burns}, {Campillay}, {Castellon}, {Folatelli},
  {Galbany}, {Hoeflich}, {Hsiao}, {Marion}, {Morrell}, \&
  {Wheeler}}]{2020A&A...634A..21S}
{Stritzinger}, M.~D., {Taddia}, F., {Holmbo}, S., {et~al.} 2020, \aap, 634,
  A21, \dodoi{10.1051/0004-6361/201936619}

\bibitem[{{Taddia} {et~al.}(2019){Taddia}, {Sollerman}, {Fremling},
  {Barbarino}, {Karamehmetoglu}, {Arcavi}, {Cenko}, {Filippenko}, {Gal-Yam},
  {Hiramatsu}, {Hosseinzadeh}, {Howell}, {Kulkarni}, {Laher}, {Lunnan},
  {Masci}, {Nugent}, {Nyholm}, {Perley}, {Quimby}, \&
  {Silverman}}]{2019A&A...621A..71T}
{Taddia}, F., {Sollerman}, J., {Fremling}, C., {et~al.} 2019, \aap, 621, A71,
  \dodoi{10.1051/0004-6361/201834429}

\bibitem[{{Taubenberger} {et~al.}(2011){Taubenberger}, {Navasardyan}, {Maurer},
  {Zampieri}, {Chugai}, {Benetti}, {Agnoletto}, {Bufano}, {Elias-Rosa},
  {Turatto}, {Patat}, {Cappellaro}, {Mazzali}, {Iijima}, {Valenti},
  {Harutyunyan}, {Claudi}, \& {Dolci}}]{Taubenberger11}
{Taubenberger}, S., {Navasardyan}, H., {Maurer}, J.~I., {et~al.} 2011, \mnras,
  413, 2140, \dodoi{10.1111/j.1365-2966.2011.18287.x}

\bibitem[{{Teja} {et~al.}(2022){Teja}, {Singh}, {Sahu}, {Anupama}, {Kumar}, \&
  {A.~J.}}]{Teja_2022}
{Teja}, R.~S., {Singh}, A., {Sahu}, D.~K., {et~al.} 2022, \apj, 930, 34,
  \dodoi{10.3847/1538-4357/ac610b}

\bibitem[{{Tody}(1986)}]{IRAF}
{Tody}, D. 1986, in Society of Photo-Optical Instrumentation Engineers (SPIE)
  Conference Series, Vol. 627, Instrumentation in astronomy VI, ed. D.~L.
  {Crawford}, 733, \dodoi{10.1117/12.968154}

\bibitem[{{Tonry} {et~al.}(2018){Tonry}, {Denneau}, {Heinze}, {Stalder},
  {Smith}, {Smartt}, {Stubbs}, {Weiland}, \& {Rest}}]{2018PASP..130f4505T}
{Tonry}, J.~L., {Denneau}, L., {Heinze}, A.~N., {et~al.} 2018, \pasp, 130,
  064505, \dodoi{10.1088/1538-3873/aabadf}

\bibitem[{{Van Dyk}(2017)}]{2017RSPTA.37560277V}
{Van Dyk}, S.~D. 2017, Philosophical Transactions of the Royal Society of
  London Series A, 375, 20160277, \dodoi{10.1098/rsta.2016.0277}

\bibitem[{{Wang} {et~al.}(2024){Wang}, {Rest}, {Dimitriadis}, {Ridden-Harper},
  {Siebert}, {Magee}, {Angus}, {Auchettl}, {Davis}, {Foley}, {Fox}, {Gomez},
  {Jencson}, {Jones}, {Kilpatrick}, {Pierel}, {Piro}, {Polin}, {Politsch},
  {Rojas-Bravo}, {Shahbandeh}, {Villar}, {Zenati}, {Ashall}, {Chambers},
  {Coulter}, {de Boer}, {DiLullo}, {Gall}, {Gao}, {Hsiao}, {Huber}, {Izzo},
  {Khetan}, {LeBaron}, {Magnier}, {Mandel}, {McGill}, {Miao}, {Pan}, {Stevens},
  {Swift}, {Taggart}, \& {Yang}}]{2024ApJ...962...17W}
{Wang}, Q., {Rest}, A., {Dimitriadis}, G., {et~al.} 2024, \apj, 962, 17,
  \dodoi{10.3847/1538-4357/ad0edb}

\bibitem[{{Williams} {et~al.}(2024){Williams}, {Kotak}, {Lundqvist}, {Mattila},
  {Mazzali}, {Pastorello}, {Reguitti}, {Stritzinger}, {Fiore}, {Hook}, {Moran},
  \& {Salmaso}}]{2024A&A...685A.135W}
{Williams}, S.~C., {Kotak}, R., {Lundqvist}, P., {et~al.} 2024, \aap, 685,
  A135, \dodoi{10.1051/0004-6361/202348130}

\bibitem[{{Yesmin} {et~al.}(2025){Yesmin}, {Pellegrino}, {Modjaz}, {Baer-Way},
  {Howell}, {Arcavi}, {Farah}, {Hiramatsu}, {Hosseinzadeh}, {McCully},
  {Newsome}, {Padilla Gonzalez}, {Terreran}, \& {Jha}}]{2025A&A...693A.307Y}
{Yesmin}, N., {Pellegrino}, C., {Modjaz}, M., {et~al.} 2025, \aap, 693, A307,
  \dodoi{10.1051/0004-6361/202452214}

\bibitem[{{Zhang} {et~al.}(2016){Zhang}, {Wang}, {Zhang}, {Zhang},
  {Ganeshalingam}, {Li}, {Filippenko}, {Zhao}, {Zheng}, {Bai}, {Chen}, {Chen},
  {Huang}, {Mo}, {Rui}, {Song}, {Sai}, {Li}, {Wang}, \&
  {Wu}}]{2016ApJ...820...67Z}
{Zhang}, K., {Wang}, X., {Zhang}, J., {et~al.} 2016, \apj, 820, 67,
  \dodoi{10.3847/0004-637X/820/1/67}

\bibitem[{{Zhang} {et~al.}(2022){Zhang}, {Zhang}, {Danzengluobu}, {Li}, {Zhao},
  {Zhang}, {Du}, {Zhu}, \& {Wu}}]{2022PASP..134g4201Z}
{Zhang}, Y., {Zhang}, T., {Danzengluobu}, {et~al.} 2022, \pasp, 134, 074201,
  \dodoi{10.1088/1538-3873/ac7583}

\bibitem[{{Zheng} {et~al.}(2013){Zheng}, {Silverman}, {Filippenko}, {Kasen},
  {Nugent}, {Graham}, {Wang}, {Valenti}, {Ciabattari}, {Kelly}, {Fox},
  {Shivvers}, {Clubb}, {Cenko}, {Balam}, {Howell}, {Hsiao}, {Li}, {Marion},
  {Sand}, {Vinko}, {Wheeler}, \& {Zhang}}]{2013ApJ...778L..15Z}
{Zheng}, W., {Silverman}, J.~M., {Filippenko}, A.~V., {et~al.} 2013, \apjl,
  778, L15, \dodoi{10.1088/2041-8205/778/1/L15}

\bibitem[{{Zheng} {et~al.}(2022){Zheng}, {Stahl}, {de Jaeger}, {Filippenko},
  {Wang}, {Gan}, {Brink}, {Altunin}, {Baer-Way}, {Bigley}, {Blanchard},
  {Blanchard}, {Bradley}, {Cargill}, {Casper}, {Chapman}, {Chander}, {Channa},
  {Choi}, {Choksi}, {Chu}, {Clubb}, {Cohen}, {Dalba}, {deGraw}, {de
  Kouchkovsky}, {Ellison}, {Falcon}, {Fox}, {Fuller}, {Ganeshalingam},
  {Girish}, {Gould}, {Halevi}, {Halle}, {Hayakawa}, {Hardy}, {Hestenes},
  {Hoffman}, {Hyland}, {Jeffers}, {Jennings}, {Kandrashoff}, {Khodanian},
  {Kim}, {Kim}, {Kislak}, {Krishnan}, {Kumar}, {Kumar}, {Leja}, {Leonard},
  {Li}, {Li}, {Lian}, {Liu}, {Lowe}, {Lu}, {Ma}, {Mason}, {May}, {McAllister},
  {McGinness}, {Modak}, {Molloy}, {Murakami}, {Nayak}, {Perera}, {Pina},
  {Punjabi}, {Rikhter}, {Ross}, {Sipple}, {Soler}, {Stegman}, {Stephens},
  {Sunseri}, {Tang}, {Taylor}, {Thrasher}, {Van Dyk}, {Wang}, {Wayland},
  {Wilkins}, {Yagubyan}, {Yuk}, {Yunus}, \& {Zhang}}]{Zheng22}
{Zheng}, W., {Stahl}, B.~E., {de Jaeger}, T., {et~al.} 2022, \mnras, 512, 3195,
  \dodoi{10.1093/mnras/stac723}

\end{thebibliography}

\newpage

\appendix

\section{Additional Figures and Tables} \label{app}

\begin{table*}
\centering
\small
\caption{Spectroscopic Observation Information. All spectra taken with the OSIRIS instrument and the R1000B and R1000R grisms with a 1.0" slit.}
\begin{center}
\begin{tabular}{lcccc}
\hline \hline
\begin{tabular}[c]{@{}l@{}}
Name\\ \\ (1)\end{tabular} & 
\multicolumn{1}{c}{\begin{tabular}[c]{@{}c@{}}Exp. time\\ \\ (2)\end{tabular}} & 
\multicolumn{1}{c}{\begin{tabular}[c]{@{}c@{}}UTC Date\\ \\ (3)\end{tabular}} & 
\multicolumn{1}{c}{\begin{tabular}[c]{@{}c@{}}MJD\\ \\ (4)\end{tabular}} & 
\multicolumn{1}{c}{\begin{tabular}[c]{@{}c@{}}Airmass\\ \\ (5)\end{tabular}} \\ \hline
2020itj & 900 & 2020-05-01T23:43:23.268 & 58970.99 & 1.14 \\
2020jfo & 500 & 2020-05-06T21:43:06.823 & 58975.91 & 1.12 \\
2020jgl & 900 & 2020-05-07T21:29:06.883 & 58976.90 & 1.60 \\
2020jhf & 600 & 2020-05-09T22:02:11.533 & 58978.92 & 1.04 \\
2020kjt & 900 & 2020-05-19T03:25:04.636 & 58988.14 & 1.37 \\
2020kku & 900 & 2020-05-19T04:10:33.937 & 58988.17 & 1.08 \\
2020kyx & 800 & 2020-05-25T04:29:14.015 & 58994.19 & 1.39 \\
2020lao & 800 & 2020-05-26T04:40:38.404 & 58995.20 & 1.21 \\
2020nny & 800 & 2020-06-27T04:36:01.170 & 59027.19 & 1.07 \\
2020rlj & 850 & 2020-08-17T05:09:44.299 & 59078.22 & 1.24 \\
\hline
\end{tabular}
\end{center}
Columns: (1) IAU name; (2) total exposure time in seconds in each grism; (3) UTC date of the observation; (4) Modified Julian date of the observation; (5) average airmass during the observation.
\label{tab:obs_info}
\end{table*}

\begin{table*}
\centering
\small
\caption{Spectroscopic observations of SN~2020itj}
\label{tab:spec20itj}
\begin{center}
\begin{tabular}[t]{cccccccccc}
\hline
\hline
UTC date  &	   MJD    &       Phase    	&  Range       &  Telescope   & Grism/Grating  \\
         &            & (days)$^{\ast}$ &  (\AA)       & +Instrument  &                \\
\hline
\hline
20200501 &  58970.99  &     9.48        & $3630-10400$ &  GTC+OSIRIS  & R1000B/R1000R  \\
20200506 &  58976.05  &     14.38       & $3400-9680$  &  NOT+ALFOSC  & Grism\#4       \\
20200510 &  58980.10  &     18.30       & $3400-9680$  &  NOT+ALFOSC  & Grism\#4       \\
20200515 &  58984.35  &     22.41       & $3500-10000$ & Las Cumbres+FLOYDS &  red/blue      \\
20200520 &  58989.40  &     27.30       & $3500-10000$ & Las Cumbres+FLOYDS &  red/blue      \\
20200525 &  58994.38  &     32.12       & $3500-10000$ & Las Cumbres+FLOYDS &  red/blue      \\
20200622 &  59022.02  &     58.88       & $3600-9690$  &  NOT+ALFOSC  &  Grism\#4      \\
\hline 
\end{tabular}
\end{center}
$^{\ast}$ Rest-frame phase in days from the explosion, JD$=2,458,961.20\pm0.7$.
\textbf{Telescope code:} \textbf{GTC:} Gran Telescopio Canarias; \textbf{NOT:} Nordic Optical Telescope; \textbf{Las Cumbres:} Las Cumbres Observatory Faulkes North telescope. 
\end{table*}

\begin{table*}
\centering
\caption{Follow-up photometry of SN~2020jgl.}
\label{tab:phot20jgl}
\resizebox{\textwidth}{!}{ 
\begin{tabular}[t]{ccc|ccc|ccc|ccc|ccc|ccc}
\hline
\hline
\multicolumn{3}{c|}{U}     & \multicolumn{3}{c|}{B}     & \multicolumn{3}{c|}{V}     & \multicolumn{3}{c|}{g}     & \multicolumn{3}{|c}{r}     & \multicolumn{3}{c}{i}      \\
MJD       & m   &$\Delta m$&MJD        & m   &$\Delta m$& MJD      & m    &$\Delta m$& MJD      & m    &$\Delta m$& MJD      & m    &$\Delta m$\\
\hline
58976.72 & 18.569 & 0.110 & 58976.72 & 17.902 & 0.058 & 58976.72 & 17.272 & 0.033 & 58976.73 & 17.554 & 0.025 & 58976.73 & 17.223 & 0.027 & 58977.46 & 17.299 & 0.038 \\
58977.45 & 18.004 & 0.108 & 58977.45 & 17.427 & 0.164 & 58977.46 & 16.721 & 0.049 & 58977.46 & 16.998 & 0.026 & 58977.46 & 16.760 & 0.025 & 58977.73 & 17.114 & 0.037 \\
58978.80 & 16.945 & 0.085 & 58977.73 & 17.094 & 0.030 & 58977.73 & 16.649 & 0.024 & 58978.81 & 16.174 & 0.021 & 58978.81 & 16.043 & 0.029 & 58978.76 & 16.530 & 0.037 \\
58979.83 & 16.336 & 0.093 & 58978.75 & 16.403 & 0.031 & 58978.75 & 16.091 & 0.022 & 58979.84 & 15.716 & 0.016 & 58979.84 & 15.587 & 0.019 & 58978.81 & 16.503 & 0.041 \\
58980.80 & 15.926 & 0.097 & 58978.80 & 16.367 & 0.030 & 58978.81 & 16.068 & 0.031 & 58980.81 & 15.340 & 0.017 & 58980.81 & 15.211 & 0.021 & 58979.42 & 16.374 & 0.085 \\
58982.75 & 15.228 & 0.090 & 58979.42 & 16.042 & 0.038 & 58979.42 & 15.785 & 0.036 & 58982.47 & 14.926 & 0.040 & 58982.47 & 14.691 & 0.046 & 58979.84 & 16.012 & 0.033 \\
58983.76 & 14.891 & 0.079 & 58979.83 & 15.830 & 0.028 & 58979.84 & 15.639 & 0.020 & 58982.76 & 14.739 & 0.020 & 58982.76 & 14.641 & 0.028 & 58980.72 & 15.638 & 0.026 \\
58985.42 & 14.700 & 0.087 & 58980.71 & 15.441 & 0.038 & 58980.72 & 15.337 & 0.016 & 58983.77 & 14.511 & 0.017 & 58983.78 & 14.410 & 0.020 & 58980.81 & 15.607 & 0.038 \\
58986.78 & 14.496 & 0.096 & 58980.81 & 15.407 & 0.029 & 58980.81 & 15.290 & 0.024 & 58983.78 & 14.507 & 0.015 & 58985.36 & 14.134 & 0.043 & 58981.71 & 15.288 & 0.031 \\
58988.34 & 14.373 & 0.235 & 58981.70 & 15.080 & 0.032 & 58981.71 & 15.011 & 0.021 & 58985.35 & 14.300 & 0.052 & 58985.43 & 14.116 & 0.021 & 58982.48 & 14.977 & 0.049 \\
58988.42 & 14.303 & 0.088 & 58982.47 & 14.798 & 0.059 & 58982.47 & 14.792 & 0.040 & 58985.43 & 14.219 & 0.021 & 58986.36 & 14.011 & 0.041 & 58982.70 & 15.003 & 0.034 \\
58989.77 & 14.466 & 0.089 & 58982.70 & 14.797 & 0.033 & 58982.70 & 14.766 & 0.023 & 58986.35 & 14.144 & 0.052 & 58986.79 & 13.952 & 0.029 & 58982.76 & 14.955 & 0.040 \\
58991.76 & 14.228 & 0.085 & 58982.75 & 14.765 & 0.031 & 58982.75 & 14.719 & 0.026 & 58986.79 & 13.983 & 0.024 & 58988.42 & 13.860 & 0.019 & 58983.70 & 14.750 & 0.033 \\
58993.44 & 14.251 & 0.098 & 58983.70 & 14.554 & 0.031 & 58982.76 & 14.713 & 0.026 & 58988.42 & 13.899 & 0.019 & 58988.77 & 13.782 & 0.075 & 58983.78 & 14.725 & 0.034 \\
58996.37 & 14.522 & 0.098 & 58983.77 & 14.514 & 0.030 & 58983.70 & 14.531 & 0.024 & 58988.77 & 13.889 & 0.092 & 58989.78 & 13.784 & 0.021 & 58984.73 & 14.527 & 0.042 \\
58998.36 & 14.714 & 0.072 & 58984.72 & 14.298 & 0.033 & 58983.77 & 14.486 & 0.035 & 58989.78 & 13.798 & 0.018 & 58989.81 & 13.717 & 0.083 & 58985.36 & 14.358 & 0.047 \\
59001.73 & 15.051 & 0.077 & 58985.35 & 14.323 & 0.060 & 58984.73 & 14.291 & 0.028 & 58989.80 & 13.803 & 0.104 & 58990.41 & 13.758 & 0.033 & 58985.43 & 14.438 & 0.033 \\
59004.70 & 15.290 & 0.094 & 58985.43 & 14.261 & 0.028 & 58985.35 & 14.231 & 0.050 & 58990.41 & 13.822 & 0.064 & 58991.76 & 13.697 & 0.021 & 58985.70 & 14.410 & 0.042 \\
59007.75 & 15.763 & 0.076 & 58985.70 & 14.169 & 0.031 & 58985.43 & 14.198 & 0.017 & 58991.76 & 13.736 & 0.016 & 58993.45 & 13.670 & 0.020 & 58986.36 & 14.272 & 0.043 \\
59017.40 & 16.876 & 0.137 & 58986.35 & 14.173 & 0.065 & 58985.70 & 14.145 & 0.035 & 58993.45 & 13.754 & 0.019 & 58996.38 & 13.600 & 0.060 & 58986.76 & 14.276 & 0.048 \\
59019.69 & 16.908 & 0.093 & 58986.76 & 14.018 & 0.033 & 58986.35 & 14.097 & 0.045 & 58996.38 & 13.737 & 0.040 & 58998.37 & 13.758 & 0.022 & 58986.79 & 14.272 & 0.044 \\
59023.71 & 17.483 & 0.092 & 58986.78 & 14.030 & 0.033 & 58986.76 & 13.993 & 0.036 & 58998.37 & 13.898 & 0.018 & 59001.73 & 13.962 & 0.019 & 58988.43 & 14.222 & 0.028 \\
59029.37 & 17.613 & 0.176 & 58988.34 & 13.956 & 0.041 & 58986.79 & 13.972 & 0.035 & 59001.73 & 14.030 & 0.017 & 59004.71 & 14.223 & 0.032 & 58988.77 & 14.176 & 0.064 \\
 $-$     & $-$    & $-$   & 58988.42 & 13.922 & 0.028 & 58988.42 & 13.871 & 0.017 & 59004.70 & 14.294 & 0.027 & 59007.75 & 14.322 & 0.021 & 58989.78 & 14.228 & 0.034 \\
 $-$     & $-$    & $-$   & 58988.77 & 13.820 & 0.140 & 58988.77 & 13.791 & 0.120 & 59007.75 & 14.563 & 0.018 & 59013.75 & 14.417 & 0.057 & 58989.81 & 14.169 & 0.085 \\
 $-$     & $-$    & $-$   & 58989.78 & 13.816 & 0.028 & 58989.78 & 13.771 & 0.023 & 59013.75 & 15.092 & 0.061 & 59017.40 & 14.464 & 0.018 & 58990.41 & 14.227 & 0.036 \\
 $-$     & $-$    & $-$   & 58989.80 & 13.760 & 0.110 & 58989.80 & 13.692 & 0.118 & 59017.40 & 15.404 & 0.021 & 59019.70 & 14.557 & 0.020 & 58991.76 & 14.244 & 0.030 \\
 $-$     & $-$    & $-$   & 58990.40 & 13.839 & 0.068 & 58990.41 & 13.771 & 0.046 & 59019.70 & 15.634 & 0.019 & 59023.71 & 14.803 & 0.022 & 58993.45 & 14.287 & 0.027 \\
 $-$     & $-$    & $-$   & 58991.76 & 13.772 & 0.028 & 58991.76 & 13.688 & 0.023 & 59023.71 & 15.923 & 0.023 & 59029.38 & 15.168 & 0.023 & 58996.38 & 14.276 & 0.087 \\
 $-$     & $-$    & $-$   & 58993.45 & 13.822 & 0.025 & 58993.45 & 13.671 & 0.020 & 59029.38 & 16.151 & 0.023 & 59032.70 & 15.293 & 0.024 & 58998.37 & 14.450 & 0.028 \\
 $-$     & $-$    & $-$   & 58996.37 & 13.903 & 0.030 & 58996.37 & 13.644 & 0.068 & 59037.36 & 16.398 & 0.015 & 59037.36 & 15.497 & 0.018 & 58998.42 & 14.421 & 0.031 \\
 $-$     & $-$    & $-$   & 58998.36 & 14.055 & 0.028 & 58998.36 & 13.724 & 0.020 & 59047.70 & 16.455 & 0.029 & 59047.70 & 15.750 & 0.028 & 58999.38 & 14.470 & 0.037 \\
 $-$     & $-$    & $-$   & 58998.41 & 14.053 & 0.029 & 58998.41 & 13.723 & 0.021 &  $-$     & $-$    & $-$   &  $-$     & $-$    & $-$   & 59000.36 & 14.550 & 0.042 \\
 $-$     & $-$    & $-$   & 58999.37 & 14.095 & 0.027 & 58999.38 & 13.724 & 0.030 &  $-$     & $-$    & $-$   &  $-$     & $-$    & $-$   & 59001.73 & 14.719 & 0.029 \\
 $-$     & $-$    & $-$   & 59000.35 & 14.195 & 0.031 & 59000.36 & 13.778 & 0.028 &  $-$     & $-$    & $-$   &  $-$     & $-$    & $-$   & 59003.41 & 14.808 & 0.026 \\
 $-$     & $-$    & $-$   & 59001.73 & 14.269 & 0.033 & 59001.73 & 13.838 & 0.021 &  $-$     & $-$    & $-$   &  $-$     & $-$    & $-$   & 59004.35 & 14.886 & 0.035 \\
 $-$     & $-$    & $-$   & 59003.40 & 14.491 & 0.024 & 59003.40 & 13.950 & 0.019 &  $-$     & $-$    & $-$   &  $-$     & $-$    & $-$   & 59004.71 & 14.899 & 0.039 \\
 $-$     & $-$    & $-$   & 59004.34 & 14.622 & 0.029 & 59004.35 & 14.037 & 0.022 &  $-$     & $-$    & $-$   &  $-$     & $-$    & $-$   & 59005.37 & 14.860 & 0.034 \\
 $-$     & $-$    & $-$   & 59004.70 & 14.622 & 0.034 & 59004.70 & 14.056 & 0.027 &  $-$     & $-$    & $-$   &  $-$     & $-$    & $-$   & 59006.44 & 14.959 & 0.029 \\
 $-$     & $-$    & $-$   & 59005.37 & 14.704 & 0.029 & 59005.37 & 14.028 & 0.022 &  $-$     & $-$    & $-$   &  $-$     & $-$    & $-$   & 59007.44 & 14.904 & 0.028 \\
 $-$     & $-$    & $-$   & 59006.43 & 14.881 & 0.028 & 59006.44 & 14.189 & 0.021 &  $-$     & $-$    & $-$   &  $-$     & $-$    & $-$   & 59007.75 & 14.909 & 0.030 \\
 $-$     & $-$    & $-$   & 59007.43 & 14.996 & 0.030 & 59007.44 & 14.241 & 0.019 &  $-$     & $-$    & $-$   &  $-$     & $-$    & $-$   & 59012.36 & 14.779 & 0.030 \\
 $-$     & $-$    & $-$   & 59007.75 & 14.988 & 0.035 & 59007.75 & 14.226 & 0.021 &  $-$     & $-$    & $-$   &  $-$     & $-$    & $-$   & 59013.76 & 14.710 & 0.129 \\
 $-$     & $-$    & $-$   & 59012.35 & 15.508 & 0.025 & 59012.36 & 14.433 & 0.022 &  $-$     & $-$    & $-$   &  $-$     & $-$    & $-$   & 59014.70 & 14.748 & 0.034 \\
 $-$     & $-$    & $-$   & 59013.75 & 15.650 & 0.086 & 59013.75 & 14.514 & 0.040 &  $-$     & $-$    & $-$   &  $-$     & $-$    & $-$   & 59015.71 & 14.720 & 0.036 \\
 $-$     & $-$    & $-$   & 59014.69 & 15.751 & 0.030 & 59014.69 & 14.583 & 0.036 &  $-$     & $-$    & $-$   &  $-$     & $-$    & $-$   & 59016.70 & 14.690 & 0.030 \\
 $-$     & $-$    & $-$   & 59015.71 & 15.819 & 0.038 & 59015.71 & 14.625 & 0.022 &  $-$     & $-$    & $-$   &  $-$     & $-$    & $-$   & 59017.41 & 14.669 & 0.025 \\
 $-$     & $-$    & $-$   & 59016.69 & 15.912 & 0.037 & 59016.70 & 14.669 & 0.019 &  $-$     & $-$    & $-$   &  $-$     & $-$    & $-$   & 59017.70 & 14.700 & 0.035 \\
 $-$     & $-$    & $-$   & 59017.40 & 15.937 & 0.024 & 59017.70 & 14.720 & 0.026 &  $-$     & $-$    & $-$   &  $-$     & $-$    & $-$   & 59018.71 & 14.676 & 0.034 \\
 $-$     & $-$    & $-$   & 59017.69 & 15.996 & 0.037 & 59018.71 & 14.799 & 0.023 &  $-$     & $-$    & $-$   &  $-$     & $-$    & $-$   & 59019.39 & 14.661 & 0.033 \\
 $-$     & $-$    & $-$   & 59018.71 & 16.093 & 0.032 & 59019.38 & 14.828 & 0.020 &  $-$     & $-$    & $-$   &  $-$     & $-$    & $-$   & 59019.70 & 14.674 & 0.033 \\
 $-$     & $-$    & $-$   & 59019.38 & 16.128 & 0.028 & 59019.70 & 14.875 & 0.024 &  $-$     & $-$    & $-$   &  $-$     & $-$    & $-$   & 59020.71 & 14.700 & 0.026 \\
 $-$     & $-$    & $-$   & 59019.69 & 16.179 & 0.034 & 59020.71 & 14.913 & 0.041 &  $-$     & $-$    & $-$   &  $-$     & $-$    & $-$   & 59022.70 & 14.804 & 0.038 \\
 $-$     & $-$    & $-$   & 59020.71 & 16.222 & 0.036 & 59022.70 & 15.079 & 0.021 &  $-$     & $-$    & $-$   &  $-$     & $-$    & $-$   & 59023.70 & 14.884 & 0.032 \\
 $-$     & $-$    & $-$   & 59022.69 & 16.380 & 0.030 & 59023.70 & 15.123 & 0.022 &  $-$     & $-$    & $-$   &  $-$     & $-$    & $-$   & 59023.72 & 14.865 & 0.030 \\
 $-$     & $-$    & $-$   & 59023.69 & 16.403 & 0.037 & 59023.71 & 15.112 & 0.022 &  $-$     & $-$    & $-$   &  $-$     & $-$    & $-$   & 59024.70 & 14.942 & 0.036 \\
 $-$     & $-$    & $-$   & 59023.71 & 16.392 & 0.039 & 59024.70 & 15.203 & 0.024 &  $-$     & $-$    & $-$   &  $-$     & $-$    & $-$   & 59025.71 & 15.033 & 0.035 \\
 $-$     & $-$    & $-$   & 59024.69 & 16.468 & 0.027 & 59025.71 & 15.272 & 0.021 &  $-$     & $-$    & $-$   &  $-$     & $-$    & $-$   & 59027.34 & 15.124 & 0.031 \\
 $-$     & $-$    & $-$   & 59025.71 & 16.511 & 0.030 & 59027.34 & 15.374 & 0.022 &  $-$     & $-$    & $-$   &  $-$     & $-$    & $-$   & 59028.34 & 15.213 & 0.032 \\
 $-$     & $-$    & $-$   & 59027.34 & 16.622 & 0.030 & 59028.34 & 15.395 & 0.025 &  $-$     & $-$    & $-$   &  $-$     & $-$    & $-$   & 59029.38 & 15.224 & 0.030 \\
 $-$     & $-$    & $-$   & 59028.34 & 16.637 & 0.028 & 59029.38 & 15.440 & 0.019 &  $-$     & $-$    & $-$   &  $-$     & $-$    & $-$   & 59032.70 & 15.449 & 0.032 \\
 $-$     & $-$    & $-$   & 59029.38 & 16.622 & 0.030 & 59032.70 & 15.579 & 0.024 &  $-$     & $-$    & $-$   &  $-$     & $-$    & $-$   & 59037.36 & 15.656 & 0.038 \\
 $-$     & $-$    & $-$   & 59037.35 & 16.832 & 0.026 & 59037.36 & 15.707 & 0.022 &  $-$     & $-$    & $-$   &  $-$     & $-$    & $-$   & 59047.70 & 16.087 & 0.047 \\
 $-$     & $-$    & $-$   & 59047.70 & 16.822 & 0.044 & 59047.70 & 15.879 & 0.023 &  $-$     & $-$    & $-$   &  $-$     & $-$    & $-$   &  $-$     & $-$    & $-$   \\
\hline 
\end{tabular}}
\end{table*}

\begin{table*}
\centering
\small
\caption{Spectroscopic observations of SN~2020jgl}
\label{tab:spec20jgl}
\begin{center}
\begin{tabular}[t]{ccc}
\hline
\hline
UTC date      &       Phase   &  Telescope           \\
             &(days)$^{\ast}$&  +Instrument         \\
\hline
\hline
2020-05-08T19:46:30 & $-$15.25 & SALT+RSS             \\
2020-05-08T21:02:44 & $-$15.20 & GTC+OSIRIS           \\
2020-05-09T21:28:22 & $-$14.19 & GTC+OSIRIS           \\
2020-05-11T09:14:42 & $-$12.71 & Las Cumbres+FLOYDS   \\
2020-05-11T11:15:36 & $-$12.63 & Las Cumbres+FLOYDS   \\
2020-05-13T09:22:39 & $-$10.72 & Las Cumbres+FLOYDS   \\
2020-05-14T11:27:24 &  $-$9.64 & Las Cumbres+FLOYDS   \\
2020-05-15T09:15:10 &  $-$8.74 & Las Cumbres+FLOYDS   \\
2020-05-20T06:13:21 &  $-$3.89 & Las Cumbres+FLOYDS   \\
2020-05-20T20:53:17 &  $-$3.29 & LT+SPRAT             \\
2020-05-23T20:56:25 &  $-$0.31 & LT+SPRAT             \\
2020-05-24T09:19:26 &   0.21 & Las Cumbres+FLOYDS   \\
2020-05-24T18:32:08 &   0.59 & SALT+RSS             \\
2020-05-25T04:19:27 &   0.99 & Shane+Kast           \\
2020-05-28T09:47:30 &   4.20 & Las Cumbres+FLOYDS   \\
2020-05-29T04:23:31 &   4.97 & Shane+Kast           \\
2020-06-02T18:01:27 &   9.51 & SALT+RSS             \\
2020-06-04T09:12:39 &  11.13 & Las Cumbres+FLOYDS   \\
2020-06-06T17:42:26 &  13.47 & SALT+RSS             \\
2020-06-12T09:15:15 &  19.08 & Las Cumbres+FLOYDS   \\
2020-06-30T08:28:13 &  36.92 & Las Cumbres+FLOYDS   \\
\hline
\end{tabular}
\end{center}
$^{\ast}$ Rest-frame phase in days from maximum light, JD$=2,458,993.18\pm2$.
\textbf{Telescope code:} \textbf{SALT:} Southern African Large Telescope; \textbf{GTC:} Gran Telescopio Canarias; \textbf{LT:} Liverpool Telescope; \textbf{Las Cumbres:} Las Cumbres Observatory; \textbf{Shane:} Lick Observatory 3\,m telescope
\end{table*}

\begin{table*}
\centering
\small
\caption{Velocities and pseudoequivalent widths of the main calcium and silicon spectral features measured in all spectra obtained of SN~2020jgl.}
\label{tab:velpew20jgl}
\resizebox{\textwidth}{!}{
\begin{tabular}[t]{ccccccccc}
\hline
\hline
Epoch & $v_{\rm Ca H\& K}$ & pEW$_{\rm Ca H\& K}$ & $v_{\rm Si \lambda 5972}$ & pEW$_{\rm Si \lambda 5972}$ & $v_{\rm Si \lambda 6355}$ & pEW$_{\rm Si \lambda 6355}$ & $v_{\rm Ca NIR}$ & pEW$_{\rm Ca NIR}$ \\
~[days] & [km~s$^{-1}$] & [\AA] & [km~s$^{-1}$] & [\AA] & [km~s$^{-1}$] & [\AA] & [km~s$^{-1}$] & [\AA] \\ 
\hline
-16.17 &       $-$      &   $-$         &       $-$      &  $-$         & 21,247$\pm$120 & 117.4$\pm$3.7 & 35,444$\pm$1177 & 746.4$\pm$2.5 \\
-15.25 &  1,056$\pm$245 &  13.5$\pm$4.8 & 18,041$\pm$634 &  9.3$\pm$3.1 & 19,936$\pm$420 & 106.0$\pm$3.4 & 33,921$\pm$654  & 666.0$\pm$6.4 \\
-15.20 &       $-$      &   $-$         &       $-$      &  $-$         & 19,235$\pm$300 & 112.2$\pm$2.9 & 34,397$\pm$467  & 737.4$\pm$2.0 \\
-14.19 &       $-$      &   $-$         & 15,863$\pm$509 & 20.4$\pm$2.8 & 18,401$\pm$298 & 127.3$\pm$3.0 & 32,450$\pm$1068 & 714.5$\pm$2.0 \\
-12.71 &  0,445$\pm$135 &   5.3$\pm$5.0 & 15,125$\pm$434 & 25.7$\pm$2.3 & 17,775$\pm$452 & 144.3$\pm$2.6 & 29,571$\pm$861  & 483.1$\pm$2.2 \\
-12.63 & 33,035$\pm$465 & 229.2$\pm$4.2 & 14,850$\pm$847 & 28.1$\pm$1.9 & 17,397$\pm$889 & 157.4$\pm$2.1 & 28,690$\pm$2513 & 528.3$\pm$2.2 \\
-10.72 &       $-$      &        $-$    & 14,145$\pm$287 & 25.9$\pm$2.2 & 16,390$\pm$458 & 149.9$\pm$2.3 & 25,134$\pm$1094 & 460.5$\pm$2.0 \\
 -9.64 & 27,557$\pm$570 & 181.9$\pm$5.4 & 12,648$\pm$636 & 18.1$\pm$2.2 & 14,883$\pm$464 & 146.6$\pm$2.3 & 24,656$\pm$1243 & 427.0$\pm$3.0 \\
 -8.74 &       $-$      &        $-$    & 13,307$\pm$214 & 20.0$\pm$1.9 & 14,376$\pm$464 & 145.1$\pm$1.9 & 24,750$\pm$845  & 354.5$\pm$1.5 \\
 -3.89 & 20,002$\pm$249 & 171.6$\pm$5.0 & 11,466$\pm$120 &  7.7$\pm$2.8 & 13,480$\pm$271 & 147.3$\pm$2.9 & 15,855$\pm$469  & 275.8$\pm$2.3 \\
 -3.29 &       $-$      &        $-$    & 11,765$\pm$196 & 10.2$\pm$1.4 & 13,217$\pm$273 & 145.9$\pm$1.3 &       $-$       &      $-$      \\
 -0.31 &       $-$      &        $-$    & 11,967$\pm$413 & 12.8$\pm$1.7 & 13,218$\pm$277 & 150.9$\pm$1.6 &       $-$       &      $-$      \\
  0.21 & 19,287$\pm$253 & 156.4$\pm$4.8 & 11,493$\pm$249 & 10.6$\pm$2.9 & 12,801$\pm$189 & 149.3$\pm$2.9 & 14,634$\pm$482  & 288.1$\pm$1.9 \\
  0.59 & 18,440$\pm$162 & 152.1$\pm$4.6 & 11,760$\pm$109 & 13.6$\pm$2.7 & 13,015$\pm$126 & 144.5$\pm$2.6 & 14,510$\pm$50   & 250.7$\pm$1.8 \\
  0.99 & 18,615$\pm$159 & 130.0$\pm$5.2 & 11,625$\pm$177 & 18.3$\pm$3.2 & 13,127$\pm$117 & 146.1$\pm$3.0 & 14,009$\pm$323  & 243.1$\pm$1.0 \\
  4.20 &       $-$      &   $-$         & 11,370$\pm$210 & 11.6$\pm$2.5 & 12,551$\pm$197 & 144.6$\pm$2.5 & 14,460$\pm$1099 & 323.8$\pm$1.5 \\
  4.97 & 17,355$\pm$342 & 124.3$\pm$4.9 & 11,696$\pm$190 & 16.8$\pm$3.6 & 12,498$\pm$209 & 145.1$\pm$3.4 & 14,678$\pm$494  & 279.8$\pm$1.1 \\
  9.51 & 16,241$\pm$167 &  99.2$\pm$3.3 & 11,995$\pm$352 & 43.1$\pm$3.0 & 12,227$\pm$96  & 137.0$\pm$2.8 & 14,554$\pm$34   & 326.9$\pm$2.2 \\
 11.13 &       $-$      &   $-$         & 13,761$\pm$458 & 56.3$\pm$3.5 & 11,931$\pm$371 & 146.4$\pm$3.5 & 12,219$\pm$1015 & 384.2$\pm$3.6 \\
 13.47 & 16,161$\pm$279 &  87.6$\pm$2.8 & 13,647$\pm$159 & 78.3$\pm$3.5 & 11,912$\pm$164 & 156.7$\pm$3.8 & 14,548$\pm$46   & 336.1$\pm$2.5 \\
 19.08 &       $-$      &   $-$         & 13,026$\pm$494 & 75.1$\pm$4.5 & 10,939$\pm$311 & 187.1$\pm$5.0 & 11,805$\pm$621  & 506.3$\pm$5.5 \\
 36.92 & 12,266$\pm$905 &  93.0$\pm$1.7 & 12,476$\pm$669 & 52.4$\pm$5.3 & 10,567$\pm$370 & 246.6$\pm$5.1 & 14,629$\pm$1214 & 495.2$\pm$8.2 \\
\hline
\hline 
\end{tabular}}
\end{table*}

\begin{figure*}
\centering
    \includegraphics[width=0.4\textwidth]{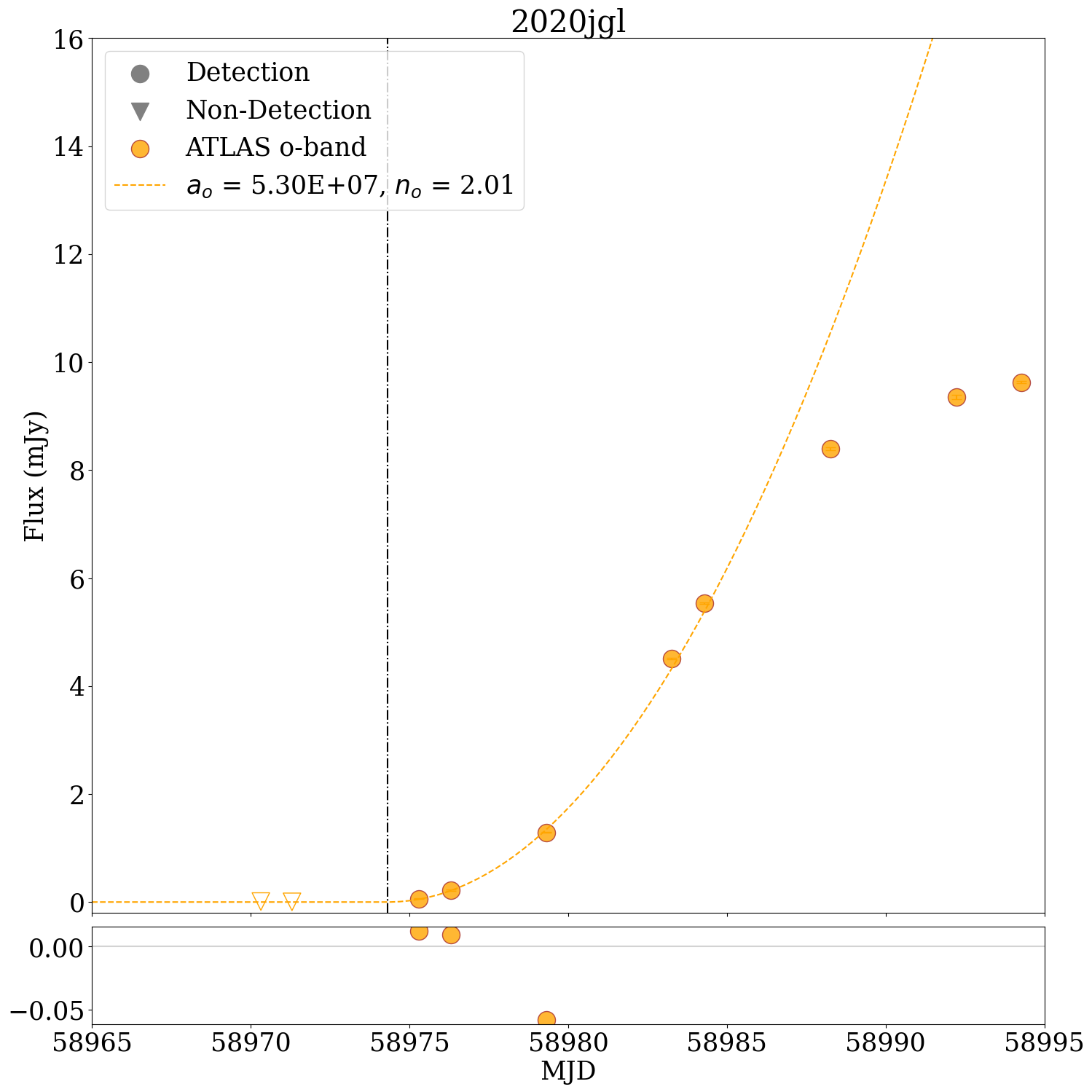}\qquad
    \includegraphics[width=0.4\textwidth]{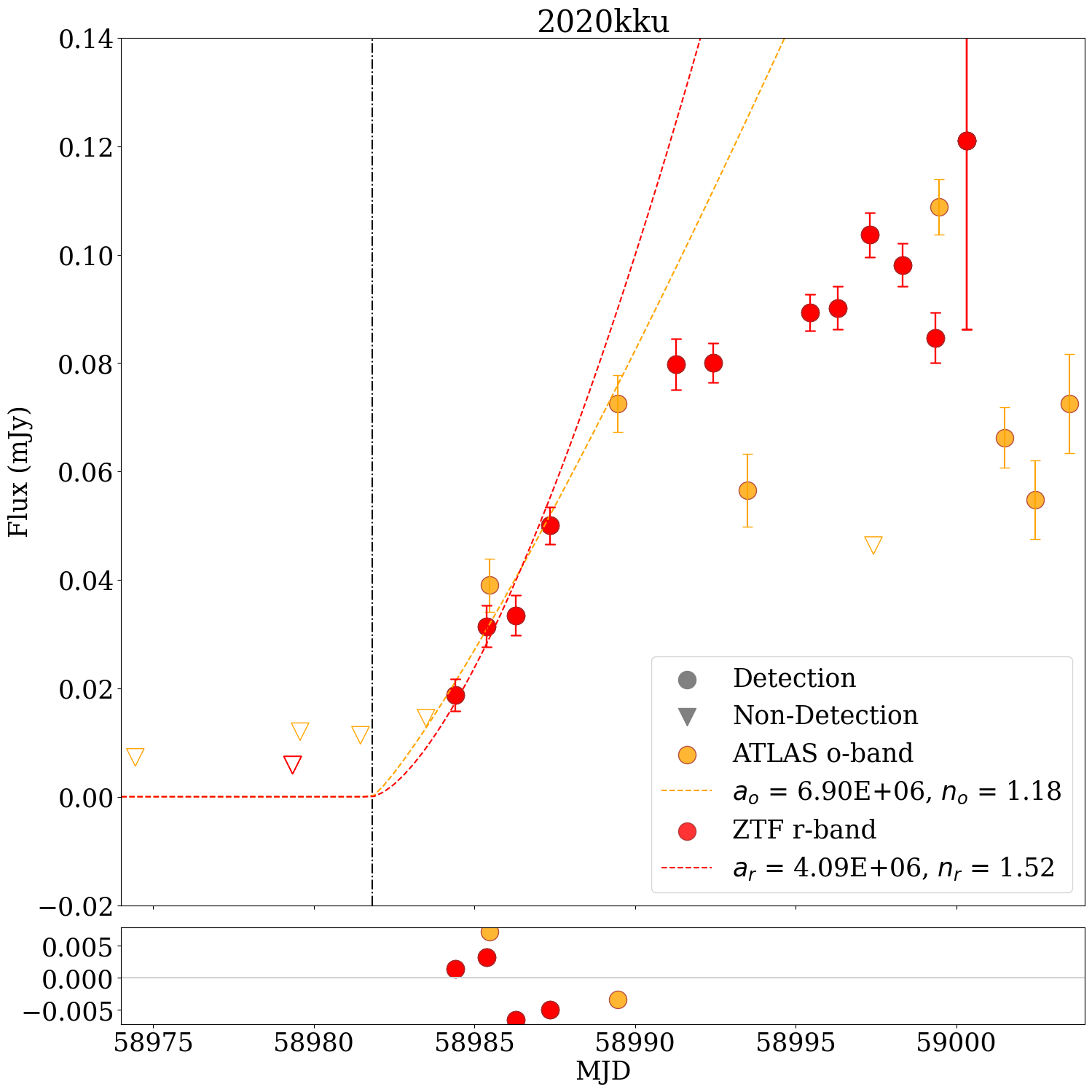}\vspace{0.5cm}
    \includegraphics[width=0.4\textwidth]{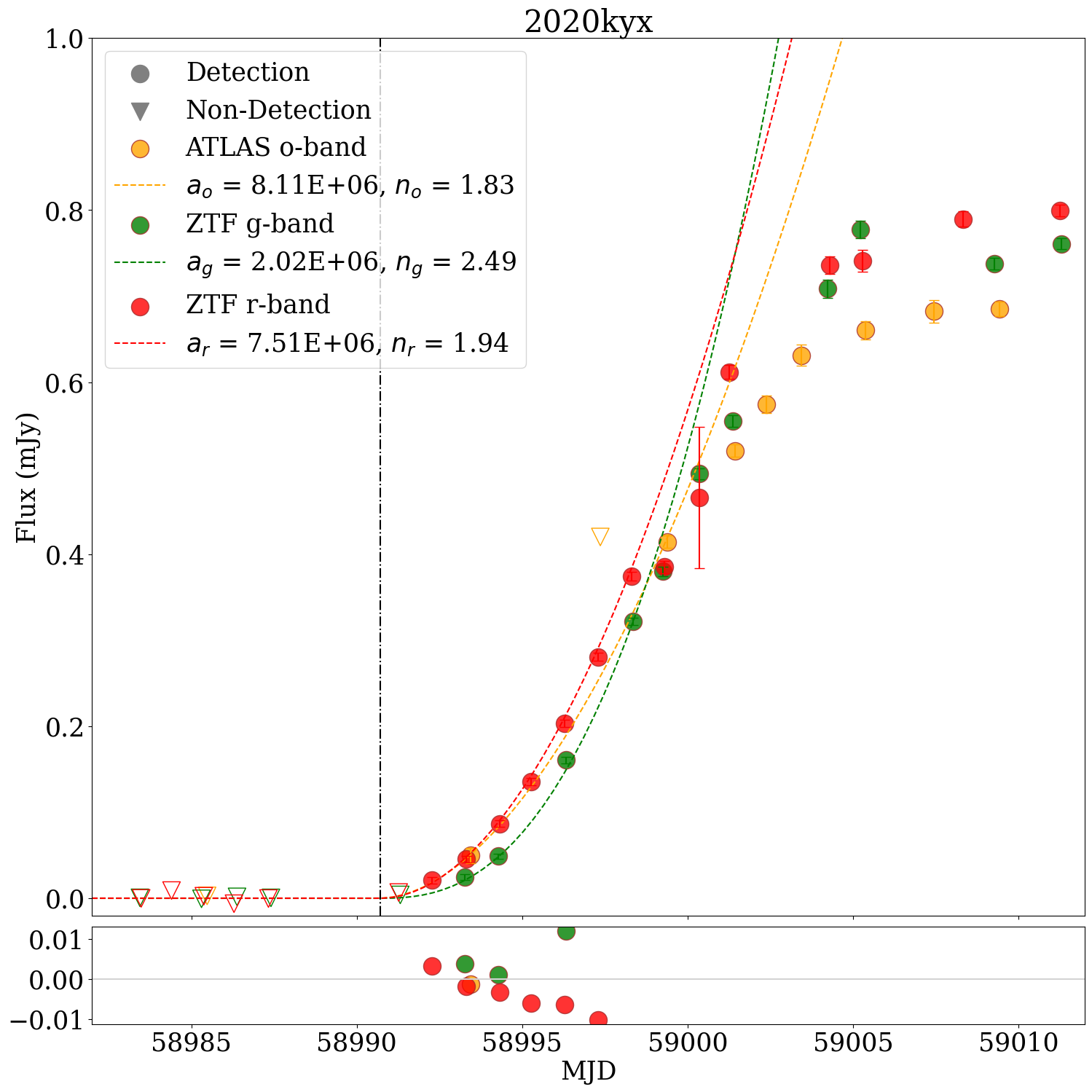}
    \includegraphics[width=0.4\textwidth]{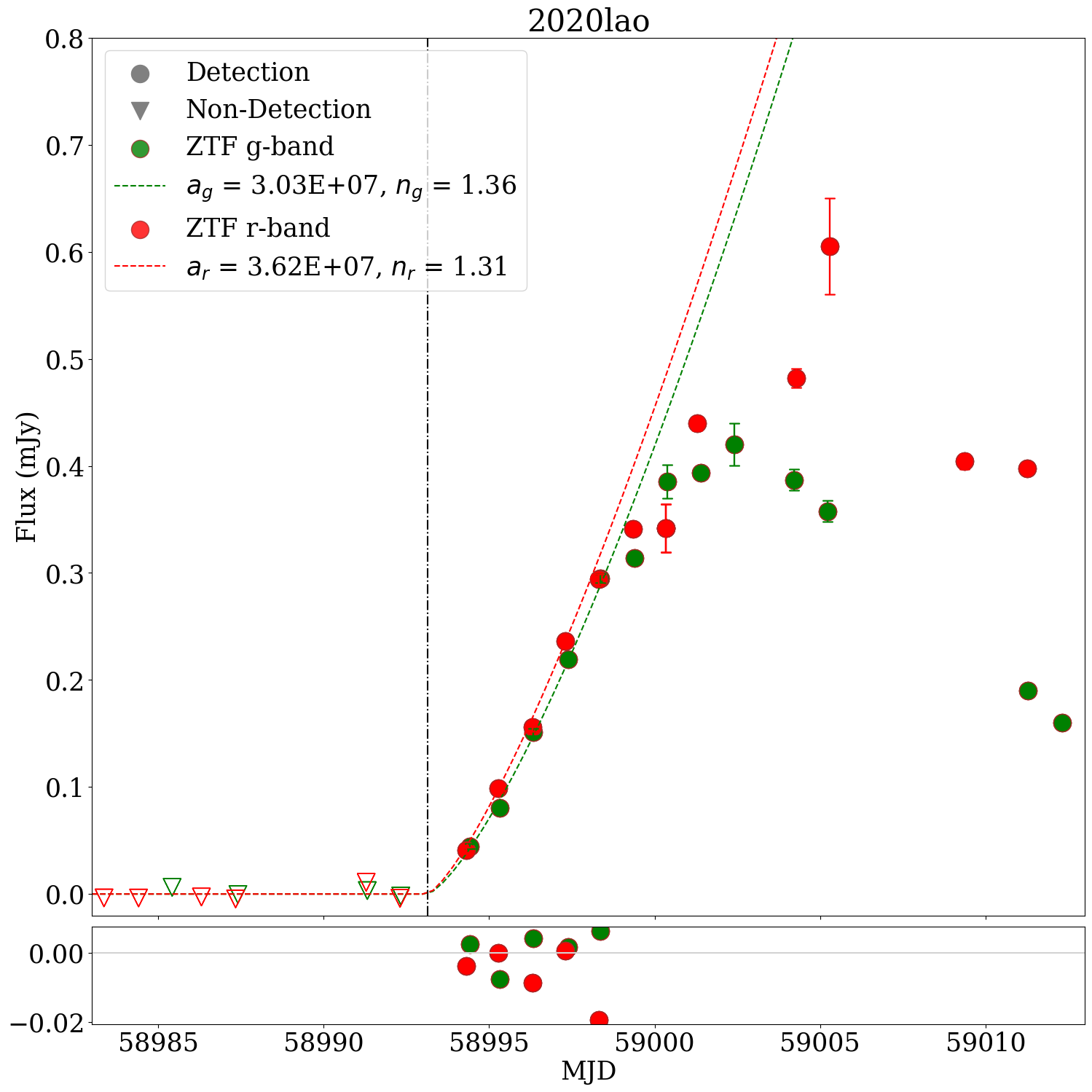}\vspace{0.5cm}
    \includegraphics[width=0.4\textwidth]{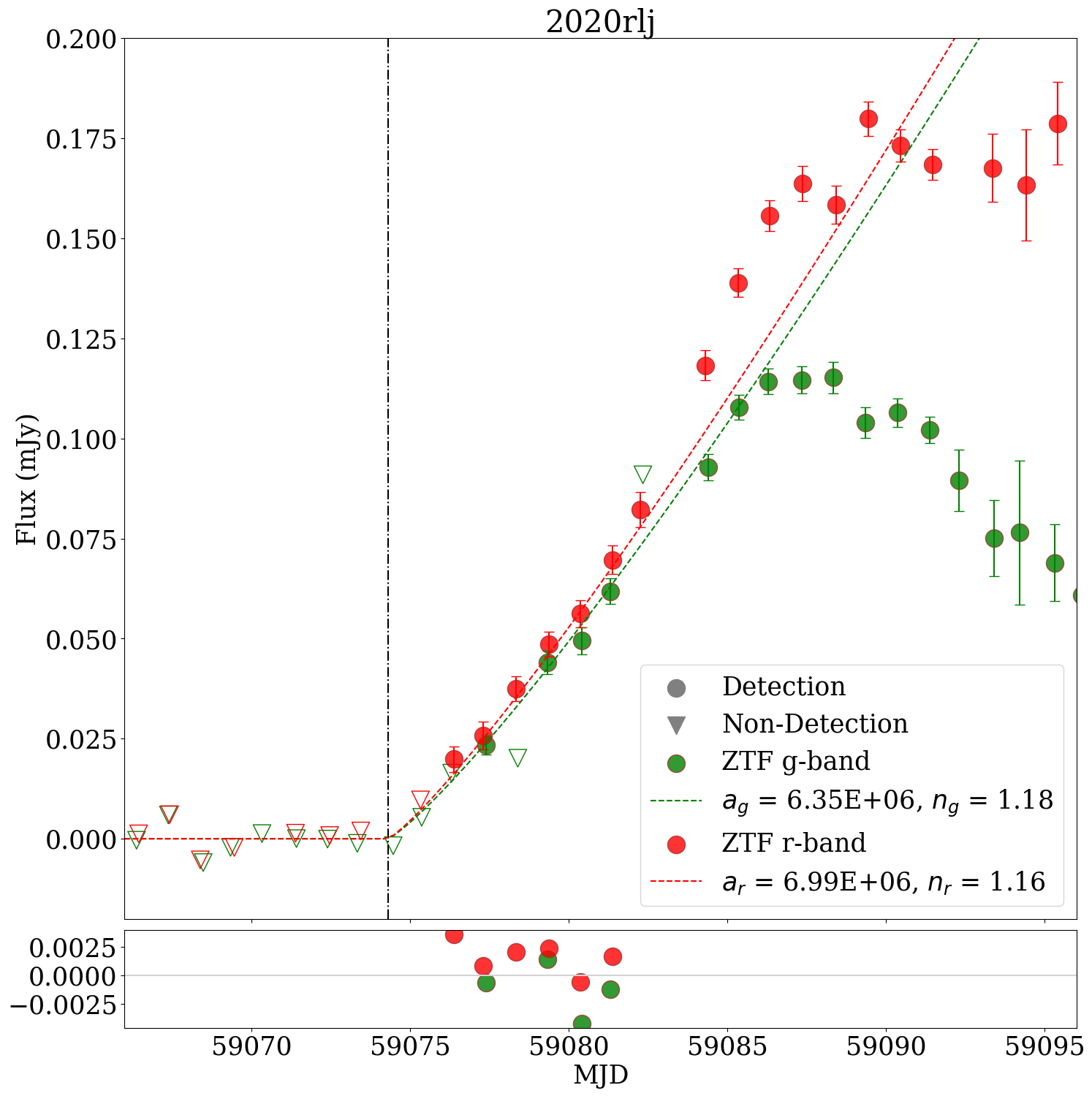}
\caption{Fireball fits to the light-curve rise for the remaining four SNe Ia and the SN Ic-BL.}
\label{fig:risefits}
\end{figure*}

\begin{figure*}
\centering
    \includegraphics[width=\textwidth]{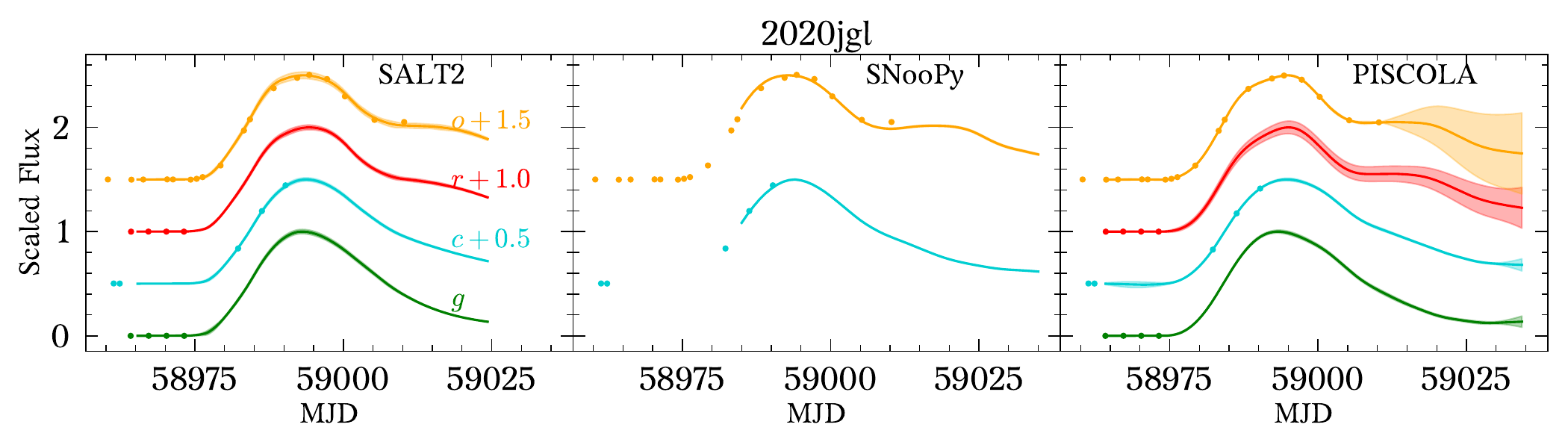}
    \includegraphics[width=\textwidth]{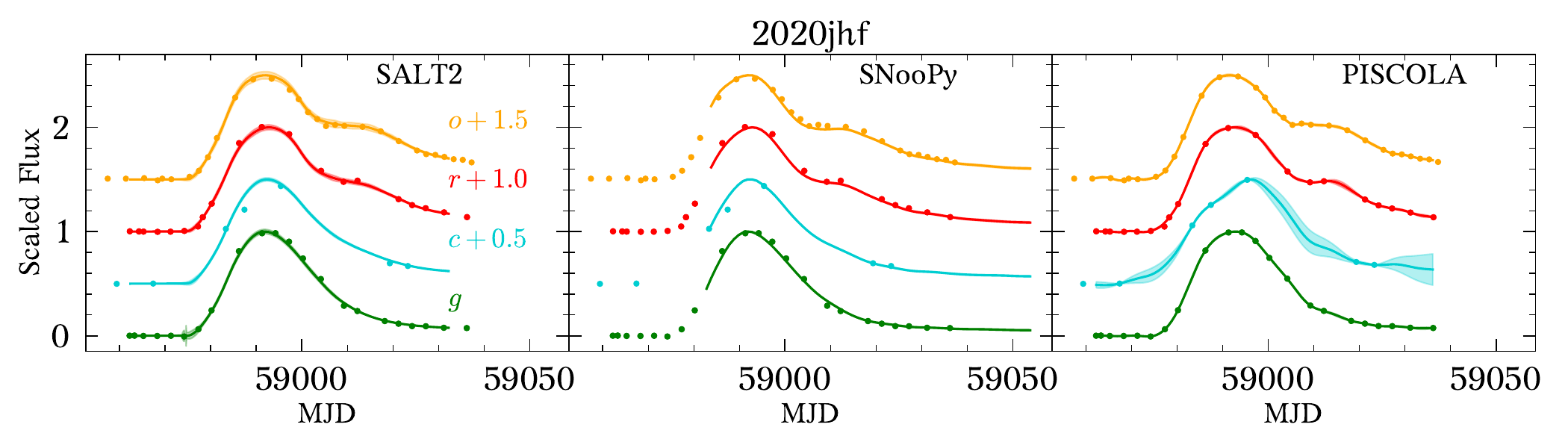}
    \includegraphics[width=\textwidth]{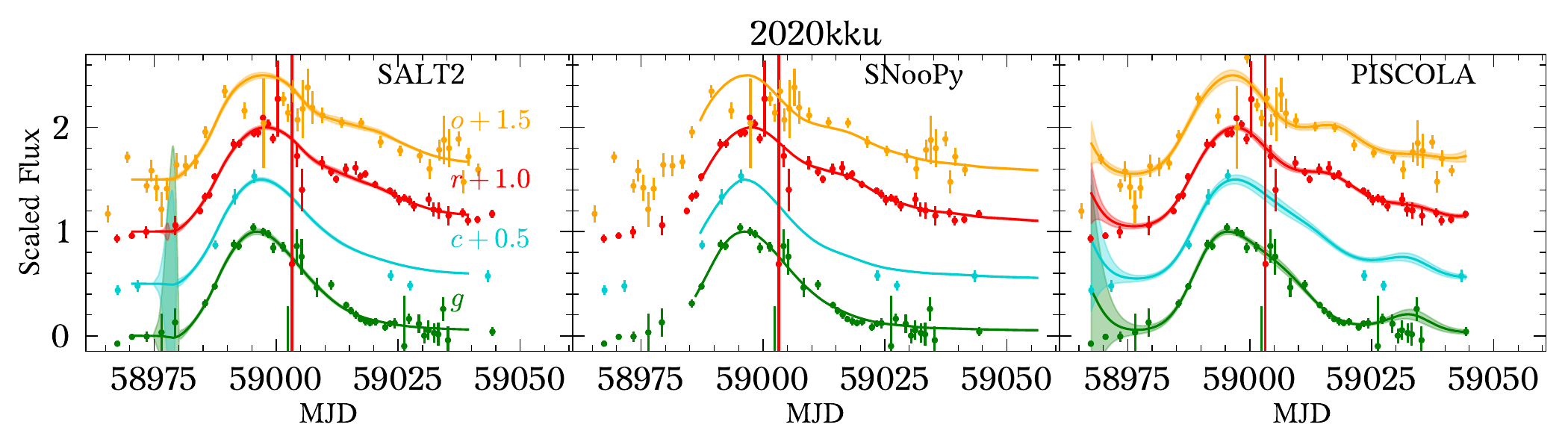}
    \includegraphics[width=\textwidth]{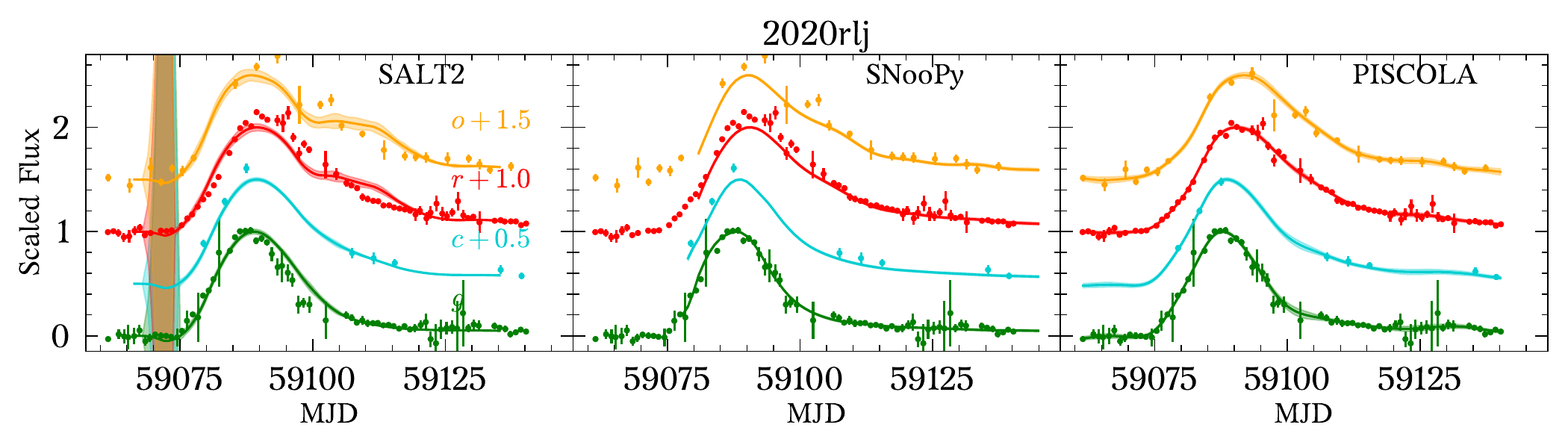}
    \caption{Similar to Figure \ref{fig:snia_fits}. Summary of the light-curve fits for the remaining four SNe Ia. In the left column we show SALT2 fits, in the central column SNooPy fits, and in the right column fits from PISCOLA. More details in section \ref{sec:lc}.}
\label{fig:snialc} 
\end{figure*}

\begin{figure*}
\centering
    \includegraphics[width=0.925\textwidth]{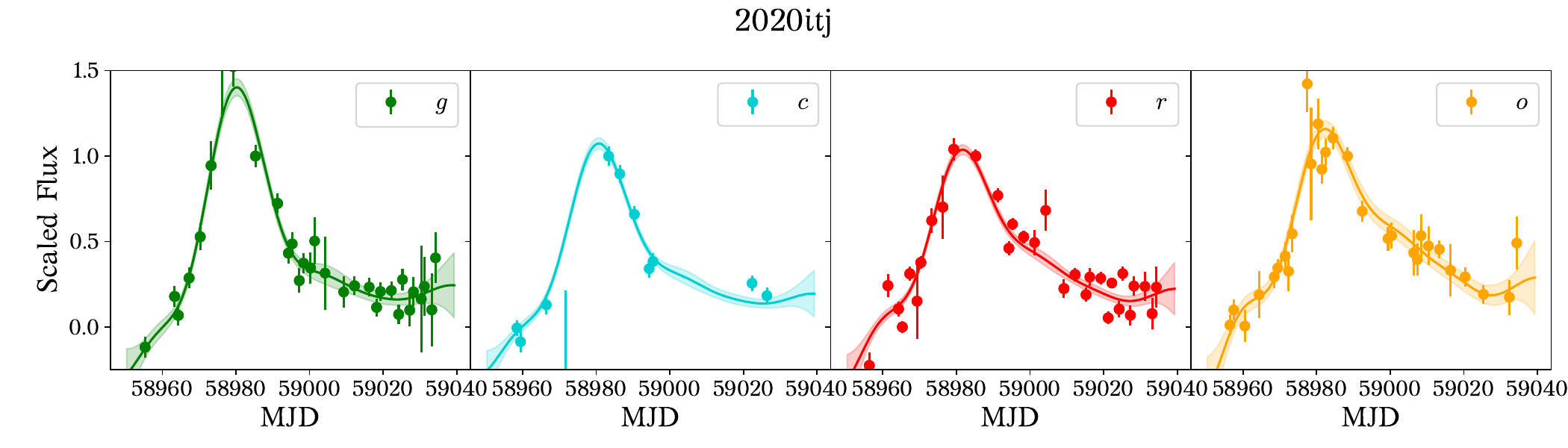}
    \includegraphics[width=0.925\textwidth]{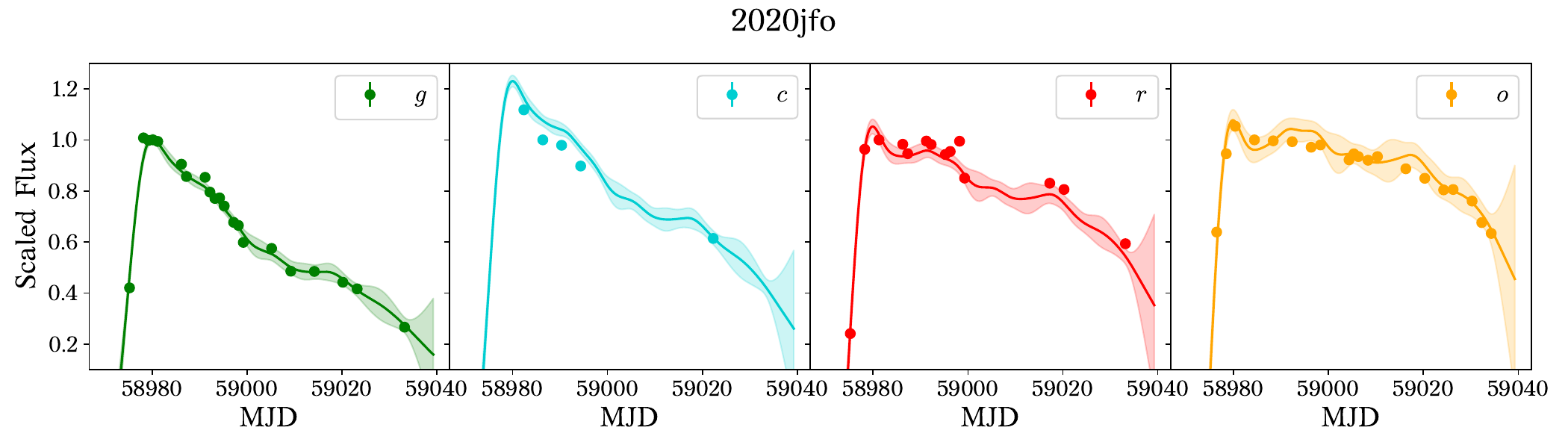}
    \includegraphics[width=0.925\textwidth]{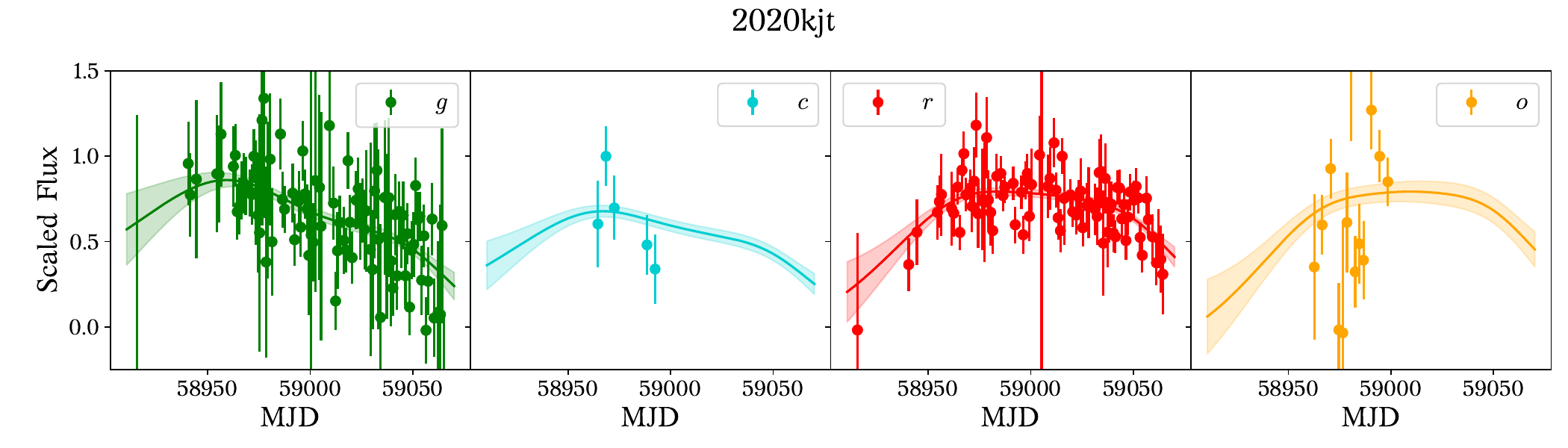}
    \includegraphics[width=0.925\textwidth]{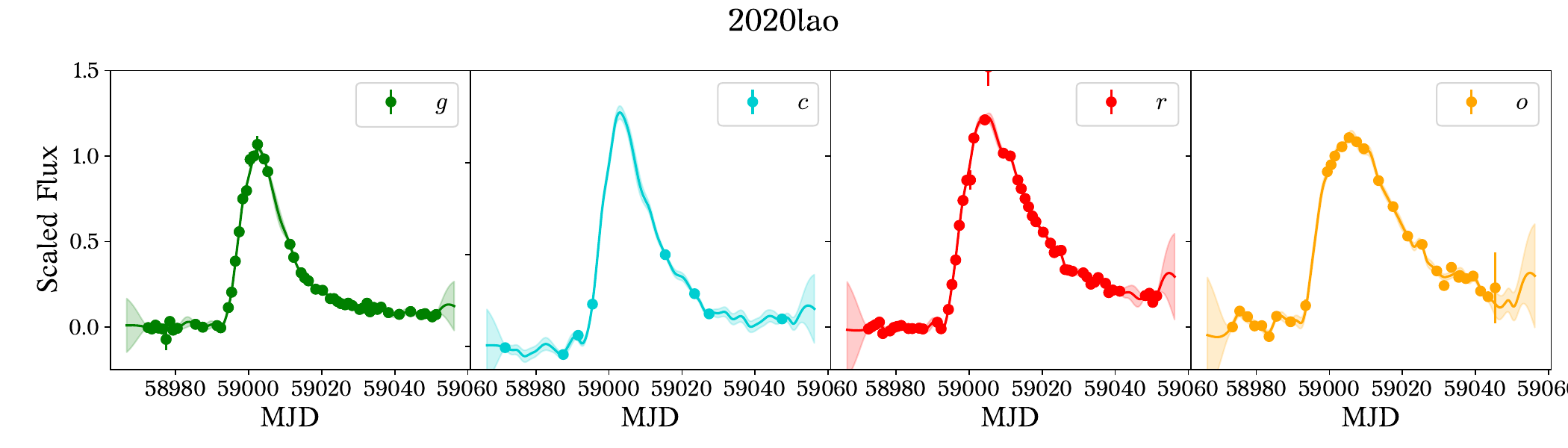}
    \includegraphics[width=0.925\textwidth]{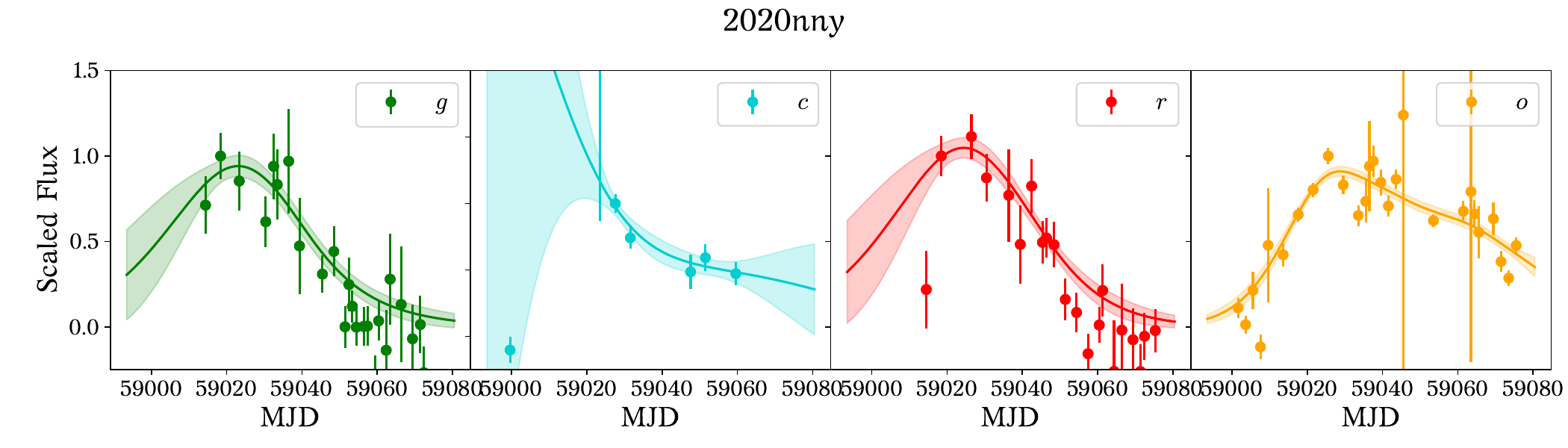}
\caption{PISCOLA fits to all five CC SNe in our sample.}
\label{fig:ccsnpiscola}
\end{figure*}

\section{Interesting objects} \label{app:sne}

Although not fulfilling our initial goals, we consider SN~2020itj and SN2020jgl the most interesting SNe found through this programme. For these objects, we complemented the ATLAS and ZTF photometric data with the \textit{UBVg'r'i'} bands from Las Cumbres Observatory \citep{LCOGT_2016}. Below, we present these multiband light curves along with extended spectral sequences compiled from additional sources.
 
\subsection{SN 2020itj} \label{sec:itj}

\begin{figure}
\centering
\includegraphics[width=0.8\textwidth]{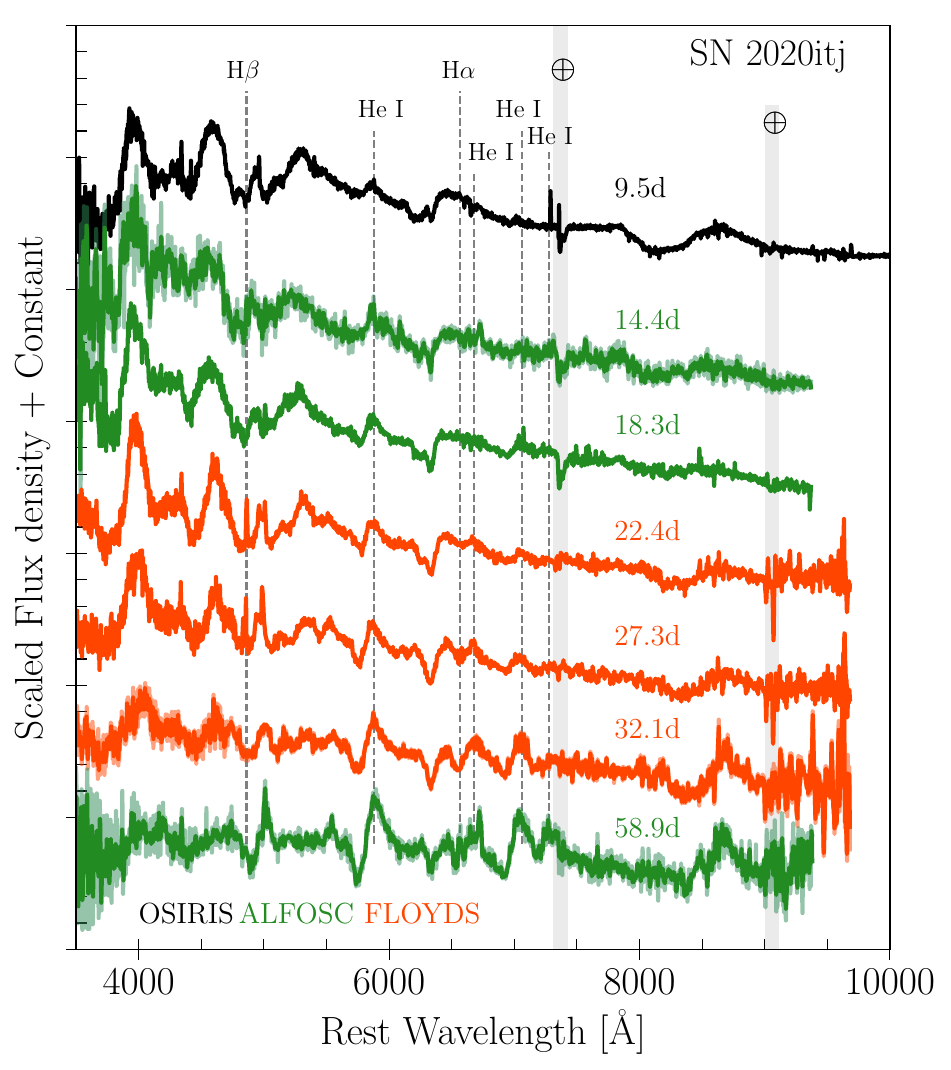}
\caption{Spectral sequence of the Type IIb SN 2020itj from 9.5 to 58.9\,d from explosion in the rest frame. The phases are labelled on the right. Each spectrum has been corrected for Milky Way reddening, put in the rest frame, and shifted vertically by an arbitrary amount for clarity. The color of the spectra represents the different instruments used to obtain the data. The vertical dashed lines indicate the rest position of the strongest lines. Telluric absorption is highlighted with grey lines ($\oplus$ symbol).}
\label{fig:spec20itj}
\end{figure}

\begin{figure}[t]
\centering
\includegraphics[width=0.8\textwidth]{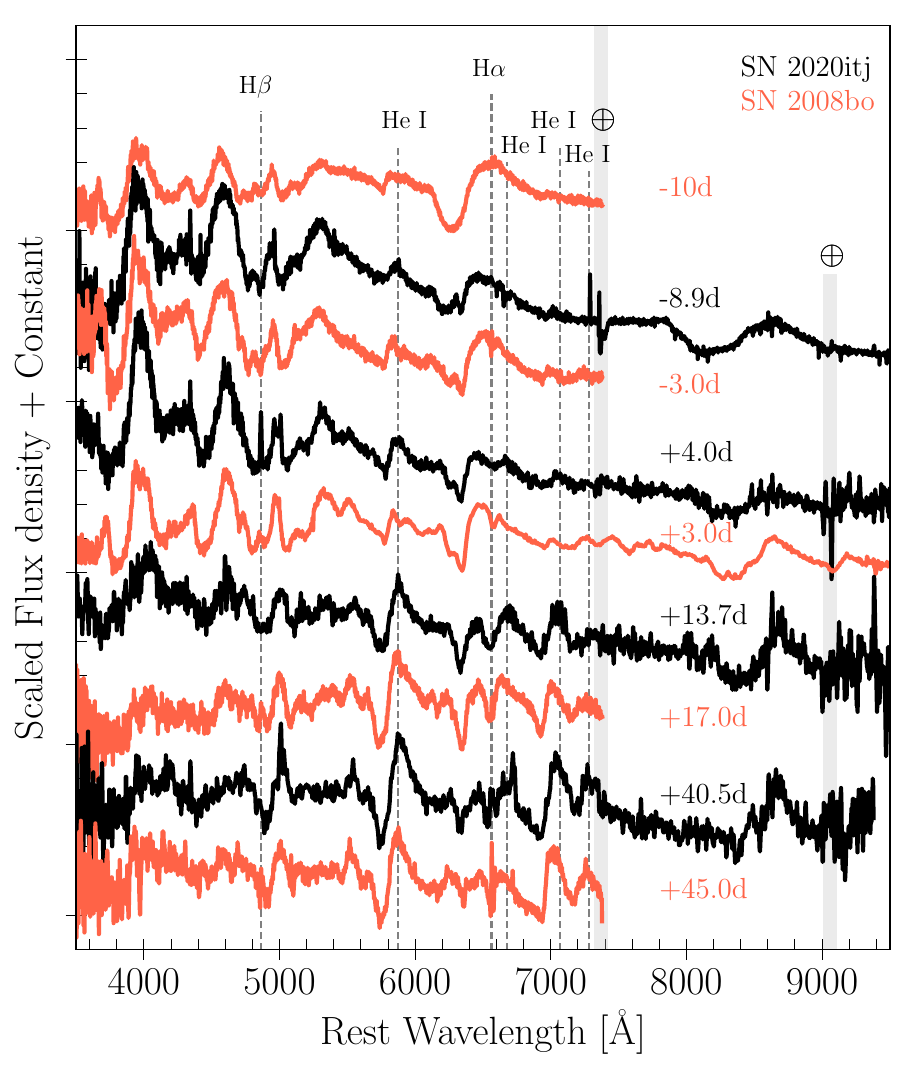}
\caption{Spectral comparison of SN~2020itj with the Type IIb SN~2008bo \citep{Modjaz14, Shivvers19}. Each spectrum has been corrected for Milky Way reddening and shifted vertically by an arbitrary amount for clarity. The vertical dashed lines indicate the rest position of the strongest lines. Telluric absorption is highlighted with grey lines ($\oplus$ symbol).}
\label{fig:comp20itj}
\end{figure}

\begin{figure}[t]
\centering
\includegraphics[width=0.7\textwidth]{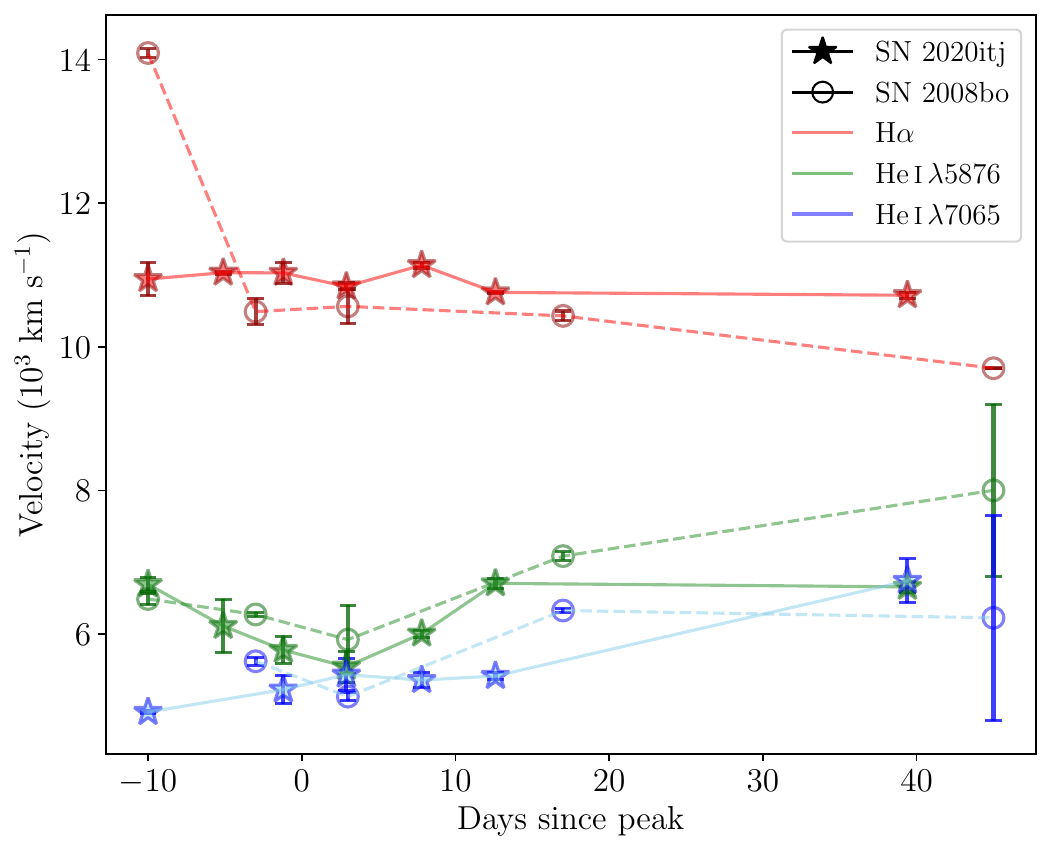}
\caption{Evolution of expansion velocities of SN~2020itj and SN~2008bo derived from the absorptions of the strongest lines.}
\label{fig:vel_evolution_itj}
\end{figure}

\begin{figure}[t]
\centering
\includegraphics[width=0.7\textwidth]{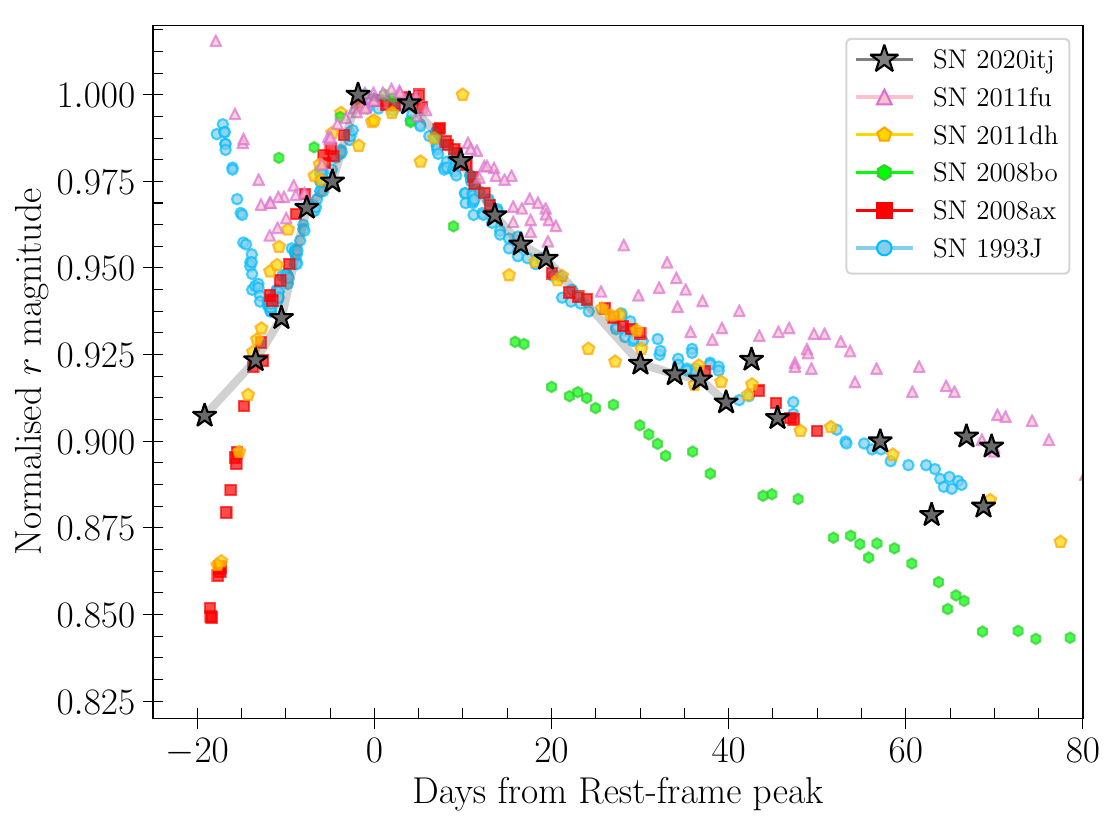}
\caption{Photometric comparison of SN~2020itj with well-studied SNe~IIb. The light curves have been normalized to the peak. The photometry for SNe~1993J, 2011dh and 2011fu is in the $R$ band.  }
\label{fig:LCcomp}
\end{figure}

SN~2020itj was initially classified as a Type Ib SN based on spectral matching \citep{tns_2020itj}, but a more detailed analysis of the spectra and light curves presented below suggests that it is, in fact, a Type IIb SN. The middle panel of Figure~\ref{fig:finders_lcs} shows that the SN~2020itj GTC spectrum was obtained within 24\,hr from the reported nondetection, which turned out to be about 9.5 days after the time of first light, as estimated from forced photometry. This implies some critical information was missed in the early phases. However, the light curve in the $r$ band shows a hint of an initial peak (a double-peaked structure), while as we show below, the spectral evolution goes from a spectrum with H lines (at 9.5 days) to a spectrum dominated by \ion{He}{I} (at $\sim59$ days). These two properties are commonly seen in SNe~IIb. 

Figure~\ref{fig:spec20itj} shows our spectrum and six further follow-up spectra obtained from 14.4  to 59.8 days after the explosion. Details of the spectroscopic observations of SN~2020itj can be found in Table~\ref{tab:spec20itj}. Our first spectrum at 9.5 days is dominated by strong lines of H$\alpha$, \ion{He}{I}  $\lambda5876$, \ion{Ca}{II}  (H\&K and NIR triplet), and \ion{Fe}{II}  (the ``W" feature produced by the blending of \ion{Fe}{II}; \citep{Liu16}). H$\alpha$ is visible at all  times, from 9.5 to 58.9 days. At 24 days, \ion{He}{I}  $\lambda6678$ and 7065 start to be visible. As time passes, H$\alpha$ becomes weak, and the \ion{He}{I} lines become stronger. Finally, after 30 days, the spectra start to be dominated by \ion{He}{I} lines. In Figure~\ref{fig:comp20itj}, we compare SN~2020itj and the best spectral match object obtained with a typical well-studied Type IIb SN~2008bo \citep{Modjaz14, Shivvers19}. The first spectrum of SN~2020itj (at $-9$ days around peak) is identical to SN~2008bo at $-3$ days. Both objects have the ``W" \ion{Fe}{II} feature, \ion{He}{I} $\lambda5876$ and a weak detection of H$\alpha$. As can be seen, at all epochs, the evolution of these objects is similar. Although SN~2020itj does not have spectra before $-10$ days, we expect that at early times, it behaves similarly to SN~2008bo,  a spectrum dominated by H lines (first spectrum in the figure). 

The expansion velocities of H$\alpha$, He I $\lambda6678$, and He I $\lambda7065$, derived from the minima of the P-Cygni absorptions for SN~2020itj and SN~2008bo, are presented in Figure~\ref{fig:vel_evolution_itj}. As expected, the highest velocity for both objects corresponds to H$\alpha$, with an average velocity of $\sim11000$\,km\,s$^{-1}$, while the lowest velocities correspond to He~{\sc I} lines with values between 5000 and 6000\,km\,s$^{-1}$. The velocities for these SNe are much lower than those measured for well-observed SNe~IIb, such as SN~1993J and SN~2008ax \citep[see][]{Taubenberger11}, but similar to those found by \citet{Folatelli14} for flat-velocity SNe~IIb. Comparing these flat-velocity SNe with SN~2020itj and SN~2008bo, we see that their velocity evolution is identical: flat velocities before 10 days from the peak and an increased velocity after that. These flat velocities have been attributed to a dense shell in the ejecta \citep{Folatelli14}. 

In Figure~\ref{fig:LCcomp}, we compare the $r$-band light curve of SN~2020itj with the well-observed Type IIb SNe~1993J \citep{Richmond94, Barbon95P}, 2008ax \citep{Pastorello08}, 2008bo \citep{Bianco14}, 2011dh \citep{Sahu13, Zheng22}, 2011fu \citep{Kumar13, Morales-Garoffolo15}. To analyze the photometric behavior of SN~2020itj and the comparison sample, we normalize their magnitudes. Overall, the light curve of SN~2020itj has a similar shape to SNe~2008ax and 2011dh; however, the first point of SN 2020itj deviates from the expected rise, suggesting that this first point corresponds to an initial peak (shock cooling), which may resemble those observed in SN~1993J and SN~2011fu. In fact, the light curves of SN~1993J and SN~2020itj match at all epochs. The only difference is the pronounced initial peak observed in the former. Despite SN 2008bo's spectral evolution being similar to SN 2020itj's, their photometric behavior is slightly different. After the peak, SN~2008bo has a faster decline rate. However, the spectral and photometric evolution of SN~2020itj is consistent with that of SNe~IIb.  

\subsection{SN 2020jgl}\label{ref:jgl}

\begin{figure}[t]
\centering
\includegraphics[width=0.75\textwidth]{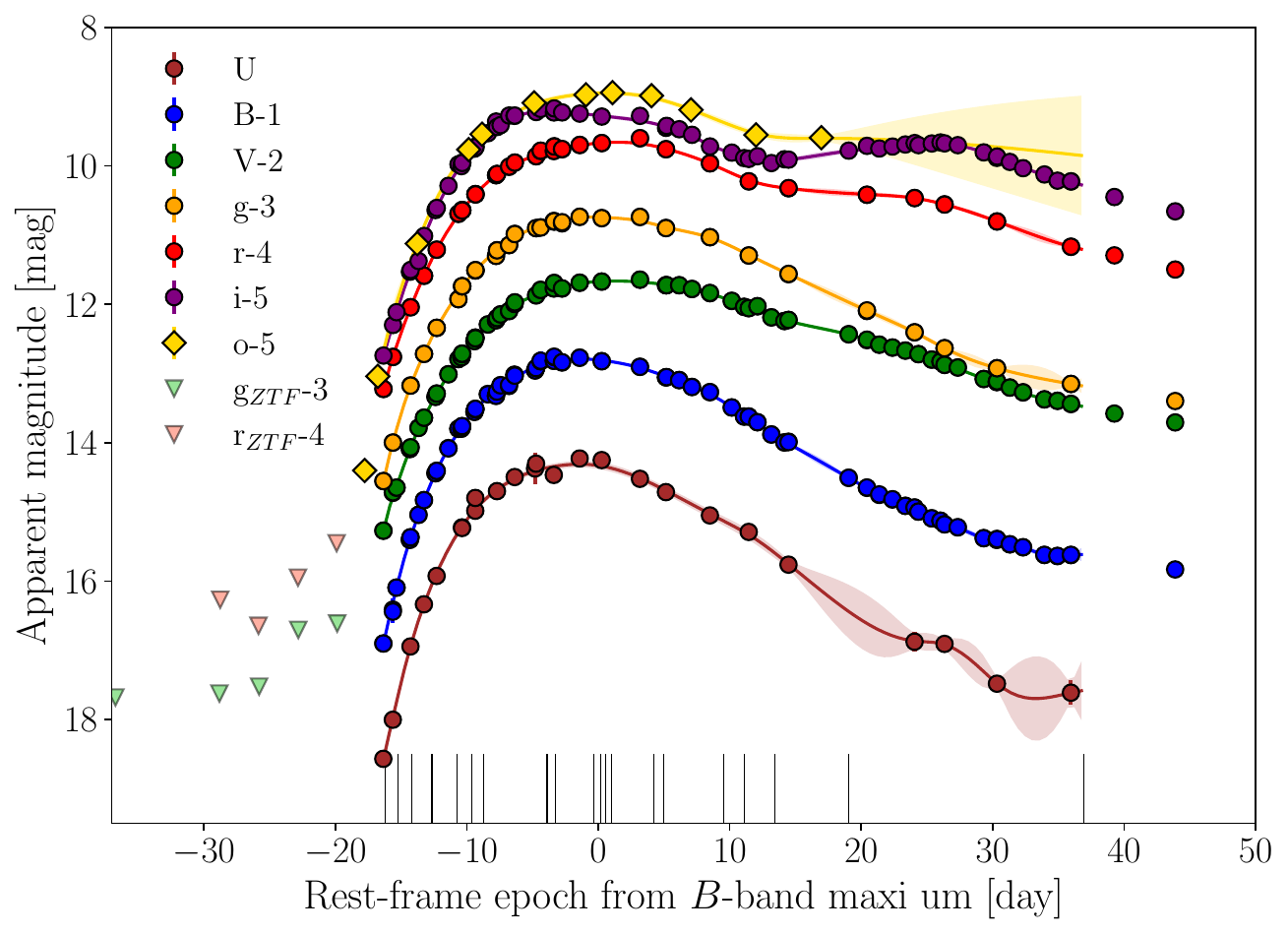}
\includegraphics[width=0.75\textwidth]{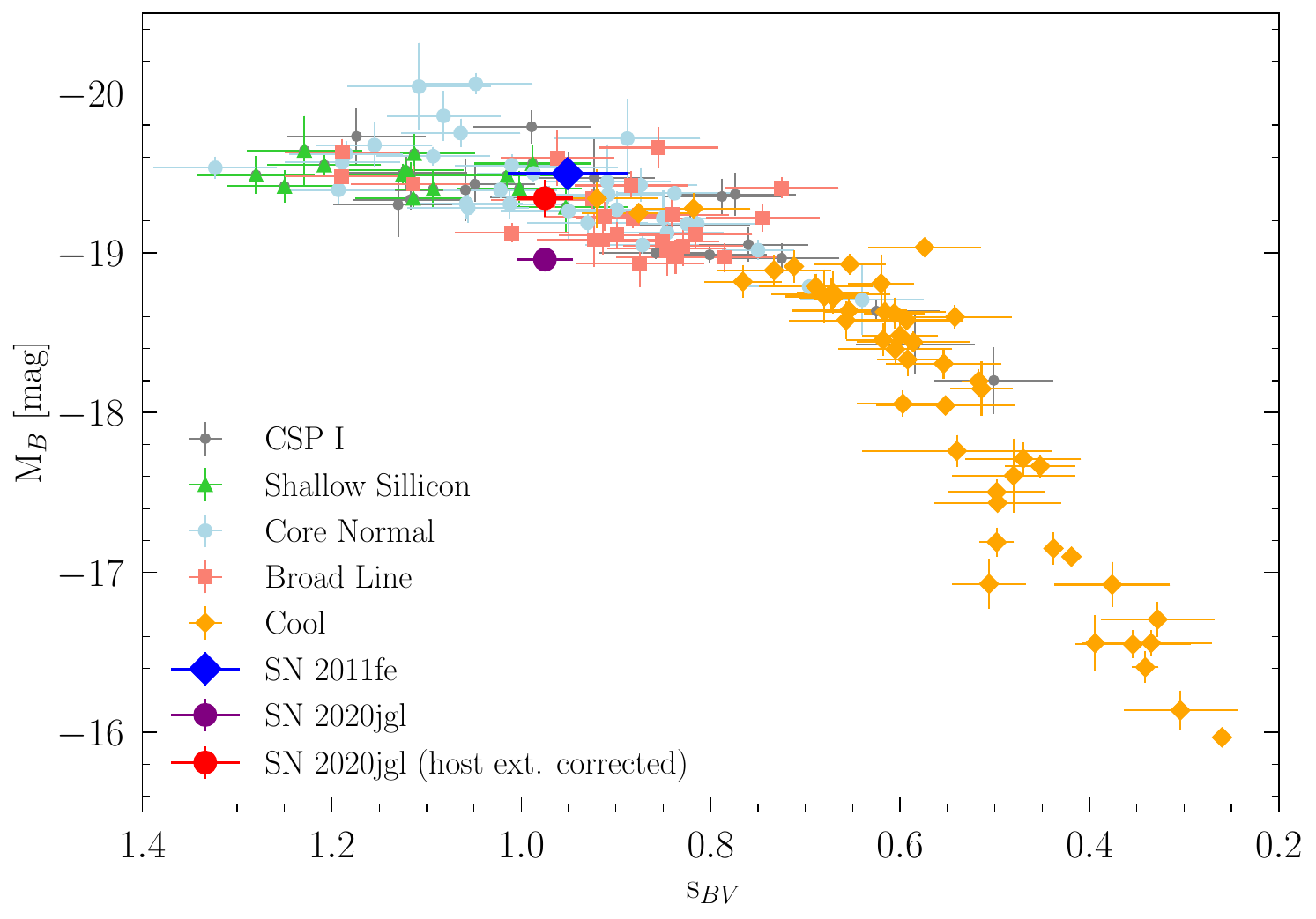}
\caption{{\it Top:} SN~2020jgl multiband light curve from ZTF, ATLAS, and Las Cumbres Observatory network of telescopes together with a GP interpolation for clarity. Vertical lines at the bottom represent the epochs at which we obtained optical spectra.
{\it Bottom:} Absolute magnitude vs. color-stretch parameter diagram populated with CSP-I SNe Ia. Colored by their position in the Branch diagram (see Figure \ref{fig:specBranch}). The normal SN2011fe is indicated with a blue dot, while SN~2020jgl is represented with a purple dot or a red dot, depending on whether the host-galaxy extinction correction is applied.}
\label{fig:LCjgl}
\end{figure}

\begin{figure}[t]
\centering
\includegraphics[width=0.49\textwidth]{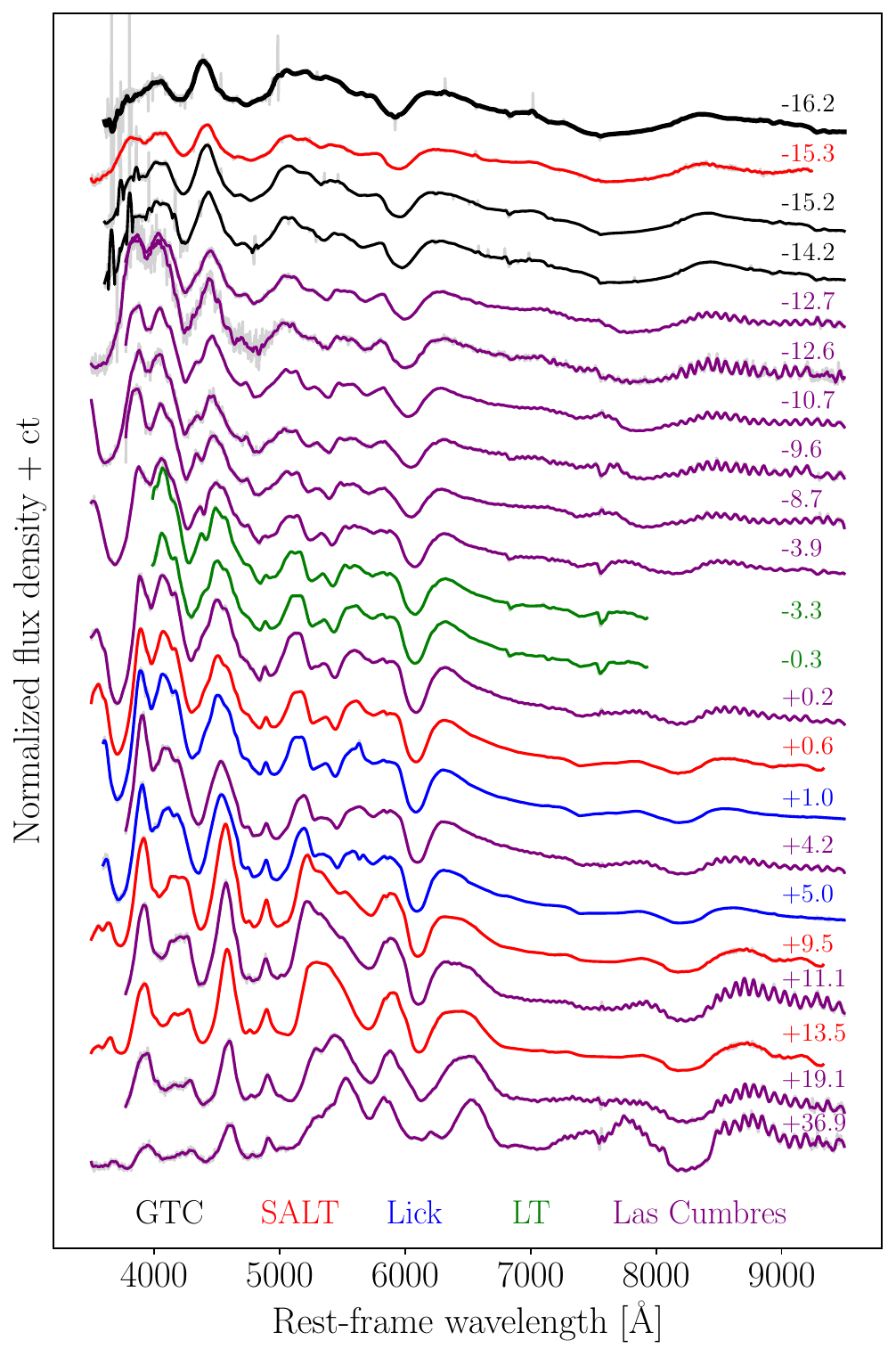}
\includegraphics[width=0.49\textwidth]{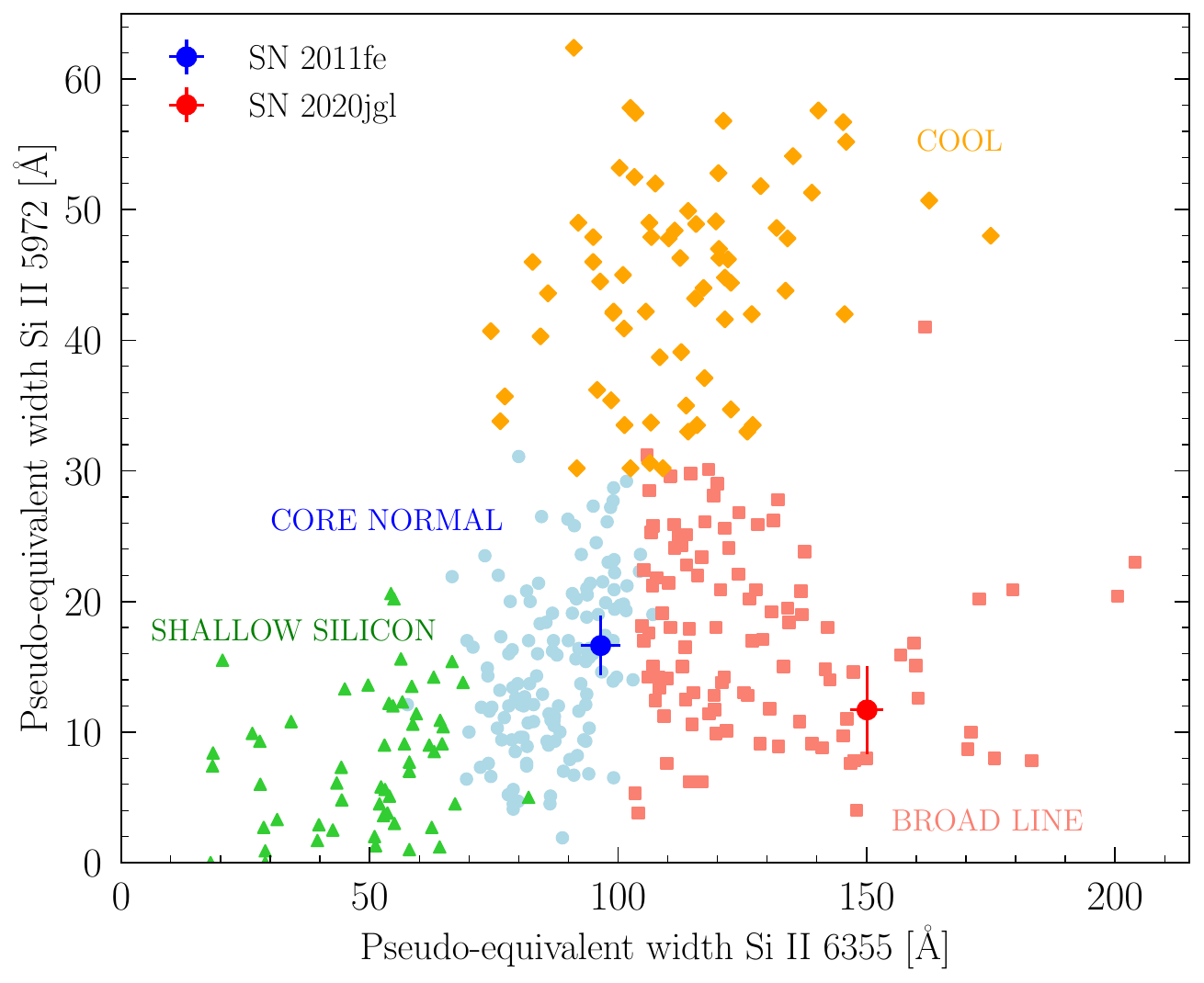}
\caption{{\it Left:} Spectral sequence of SN~2020jgl from various ground-based telescopes, sorted by epoch (earliest to latest).
{\it Right:} Branch diagram of \ion{Si}{II} $\lambda$5972 and $\lambda$6355 pseudoequivalent widths (pEWs), including CSP and CfA SN Ia samples. SN2020jgl is in the broad-line region; SN~2011fe included for reference.}
\label{fig:specBranch}
\end{figure}

SN~2020jgl is an SN~Ia with average light-curve parameters, as reported in Table~\ref{tab:SNeIa_parameters}. To refine and provide more precise measurements, we supplemented ZTF and ATLAS photometry with $UBVgri$ photometry from the Las Cumbres Observatory network of telescopes and repeated the light-curve fits with SNooPy using the \verb|EBV_model2|. All the photometry available and the new SNooPy fit are shown in Figure \ref{fig:LCjgl} and reported in Table \ref{tab:phot20jgl}. This new fit gives a $T_{\rm max} = 58992.924  \pm  0.34$, which is 0.1 day earlier than the value previously reported for SNooPy in Table \ref{tab:SNeIa_parameters}, and light-curve stretch parameters $\Delta m_{15} (B)$ = 1.026  $\pm$  0.061 mag and $s_{BV}$ = 0.975  $\pm$ 0.030, also consistent with previous results. The rise time to optical maximum with this new time of maximum is $t_{\rm rise} = 18.49$\,days, consistent with the average values measured for SNe~Ia \citep[$18.98 \pm 0.54$\,days;][]{Firth_2015}. This time, with a light curve for the $B$ band available, the peak apparent magnitude is $m_B = 13.522 \pm 0.034$ once corrected for Galactic extinction and H-corrected using the updated \cite{2007ApJ...663.1187H} SN~Ia template. At the redshift of the SN and considering the distance modulus obtained from SNooPy, 32.480 $\pm$ 0.093 mag, it corresponds to an absolute peak magnitude of $M_B =  -18.958 \pm 0.034$. Once a correction for the host-galaxy extinction of $E(B-V)_{\rm host} = 0.192 \pm 0.060$\,mag and an $R_V=2$ is included, the peak absolute magnitude is $M_B =  -19.342 \pm 0.116$\,mag. Figure~\ref{fig:LCjgl} also shows a peak magnitude vs. stretch-color parameter diagram populated with all CSP-I SNe~Ia \citep[CSP;][]{2017AJ....154..211K}, the typical normal SN~Ia 2011fe, and our two measurements for SN~2020jgl with and without accounting for host-galaxy extinction.

As mentioned above, we started a spectroscopic follow-up campaign for SN~2020jgl using a number of telescopes: the 10.4\,m GTC and the 2\,m Liverpool Telescope (LT) at the Roque de Los Muchachos observatory, the 3\,m Shane telescope at Lick Observatory, the 9.2\,m Southern African Large Telescope (SALT), and the 2\,m Las Cumbres Observatory telescope at the Siding Spring Observatory through the Global Supernova Project. Figure \ref{fig:specBranch} shows all 22 optical spectra obtained for the SN, ranging from $-$16.2 to +36.9 days from the epoch of maximum light. Details of the spectroscopic observations of SN~2020itj can be found in Table~\ref{tab:spec20jgl}. 

The main spectroscopic features defined in \cite{2013ApJ...773...53F} were analyzed using {\sc Spextractor} \citep{2019PhDT.......134P,2020ApJ...901..154B} providing an estimate of the pEW, velocity ($v$), and depth ($d$). Results are reported in Table \ref{tab:velpew20jgl}. Figure \ref{fig:specBranch} also presents the \cite{2006PASP..118..560B} diagram, using the pEW of the two main silicon features in the optical. The diagram includes the results for SN~2011fe and SN~2020jgl, and is populated with CSP-I and II measurements from \cite{2024ApJ...967...20M} and Center for Astrophysics (CfA) measurements from \cite{2012AJ....143..126B}. While SN~2011fe falls in the ``core-normal'' region, SN~2020jgl is classified as a broad-line SN Ia according to the diagram. The absolute magnitude vs. light-curve width diagram previously presented in Figure \ref{fig:LCjgl} also shows the CSP-I SNe Ia colored by the spectral subgroups defined in this Branch diagram. We can clearly see how the groups fall in different regions of the relation, with some overlap. SN~2011fe falls in the middle of the core normal SNe, and SN~2020jgl is surrounded by both core-normal and broad-line dots. 

SN~2020jgl was initially triggered for being a good candidate for the {\it Hubble Space Telescope (HST)} Supernovae in the near-InfraRed Avec Hubble (SIRAH; \citep{2020hst..prop16234J}) project, under which two WFC3/IR spectra were obtained. We additionally obtained one NIR spectrum with the  Espectr\'ografo Multiobjeto Infra-Rojo (EMIR) mounted on the GTC, and two more spectra with the SpeX spectrograph mounted on the 3.2\,m NASA Infrared Telescope Facility (IRTF) on Maunakea Observatory, to complement {\it HST} data. Since this is not the goal of this paper, these NIR spectra will not be presented here but elsewhere (Pierel et al., in prep.).

\end{document}